\documentclass[12pt]{article}

\usepackage{amssymb}
\usepackage{amsmath}
\usepackage{amscd}
\usepackage{latexsym}
\usepackage{graphicx}

\usepackage{cite}
\numberwithin{equation}{section}

\newcommand{\be}{\begin{equation}}
\newcommand{\ee}{\end{equation}}
\newcommand{\Dlt}{\Delta}
\newcommand{\dlt}{\delta}
\newcommand{\prt}{\partial}
\newcommand{\br}{{\bf r}}
\newcommand{\bk}{{\bf k}}
\newcommand{\bfe}{{\bf e}}
\newcommand{\ba}{{\bf a}}
\newcommand{\bS}{{\bf S}}
\newcommand{\bp}{{\bf p}}
\newcommand{\bu}{{\bf u}}
\newcommand{\bt}{\beta}
\newcommand{\vp}{\varphi}
\newcommand{\ep}{\varepsilon}
\newcommand{\al}{\alpha}
\newcommand{\ra}{\rightarrow}
\newcommand{\sgm}{\sigma}

\newcommand{\gm}{\gamma}
\newcommand{\om}{\omega}
\newcommand{\Om}{\Omega}
\newcommand{\Gm}{\Gamma}
\newcommand{\dgr}{\dagger}
\newcommand{\lbd}{\lambda}
\newcommand{\Lbd}{\Lambda}
\newcommand{\rgl}{\rangle}
\newcommand{\lgl}{\langle}
\newcommand{\cH}{{\cal H}}
\newcommand{\cA}{{\cal A}}
\newcommand{\cD}{{\cal D}}

\begin{document}

\begin{center}

{\Large{\bf Models of Mixed Matter} \\ [5mm]

V.I. Yukalov$^{1,2}$ and E.P. Yukalova$^{3}$ }  \\ [3mm]

{\it
$^1$Bogolubov Laboratory of Theoretical Physics, \\
Joint Institute for Nuclear Research, Dubna 141980, Russia \\ [2mm]

$^2$Instituto de Fisica de S\~ao Carlos, Universidade de S\~ao Paulo, \\
CP 369, S\~ao Carlos 13560-970, S\~ao Paulo, Brazil \\ [2mm]

$^3$Laboratory of Information Technologies, \\
Joint Institute for Nuclear Research, Dubna 141980, Russia } \\ [3mm]

{\bf E-mails}: {\it yukalov@theor.jinr.ru}, ~~ {\it yukalova@theor.jinr.ru}

\end{center}

\vskip 1cm

\begin{abstract}

The review considers statistical systems composed of several phases that are 
intermixed in space at mesoscopic scale and systems representing a mixture of 
several components of microscopic objects. These types of mixtures should be 
distinguished from the Gibbs phase mixture, where the system is filled by 
macroscopic pieces of phases. The description of the macroscopic Gibbs mixture 
is rather simple, consisting in the consideration of pure phases separated by 
a surface, whose contribution becomes negligible in thermodynamic limit. The 
properties of mixtures, where phases are intermixed at mesoscopic scale, are 
principally different. The emphasis in the review is on the matter with phases 
mixed at mesoscopic scale. Heterogeneous materials composed of mesoscopic 
mixtures are ubiquitous in nature. A general theory of such mesoscopic mixtures 
is presented and illustrated by several condensed matter models. A mixture of 
several components of microscopic objects is illustrated by clustering 
quark-hadron matter.   

\end{abstract}

\vskip 3mm


\vskip 1.5cm

 
\newpage

\begin{center}
{\bf CONTENTS}
\end{center}

\vskip 1cm

{\bf 1}. Types of Phase Mixture

\vskip 3mm
{\bf 2}. Mesoscopic Heterophase Mixture

\vskip 3mm
{\bf 3}. Examples of Heterophase Materials

\vskip 1mm
\hspace{0.5cm}   3.1. Mixture of Ferromagnetic and Antiferromagnetic Phases

\vskip 1mm
\hspace{0.5cm}   3.2. Mixture of Magnetic and Paramagnetic Phases

\vskip 1mm
\hspace{0.5cm}   3.3. Mixture of Magnetic and Spin-Glass Phases

\vskip 1mm
\hspace{0.5cm}   3.4. Mixture of Phases with Different Magnetic Orientations

\vskip 1mm
\hspace{0.5cm}   3.5. Mixture of Ferroelectric and Paraelectric Phases

\vskip 1mm
\hspace{0.5cm}   3.6. Mixture of Different Crystalline Structures

\vskip 1mm
\hspace{0.5cm}   3.7. Mixture of Gaseous and Liquid Phases

\vskip 1mm
\hspace{0.5cm}   3.8. Mixture of Liquid and Solid Phases

\vskip 1mm
\hspace{0.5cm}   3.9. Mixture of Metallic and Nonmetallic Phases

\vskip 1mm
\hspace{0.5cm}   3.10. Mixture of Superconducting and Normal Phases

\vskip 1mm
\hspace{0.5cm}   3.11. Mixture of Metastable Amorphous Phases

\vskip 1mm
\hspace{0.5cm}   3.12. Mixture of Nonequilibrium Phases

\vskip 3mm
{\bf 4}. Theory of Heterophase Systems

\vskip 1mm
\hspace{0.5cm}   4.1. Spontaneous Breaking of Equilibrium

\vskip 1mm
\hspace{0.5cm}   4.2. Statistical Ensembles and States

\vskip 1mm
\hspace{0.5cm}   4.3. Methods of Symmetry Breaking

\vskip 1mm
\hspace{0.5cm}   4.4. Weighted Hilbert Space

\vskip 1mm
\hspace{0.5cm}   4.5. Spatial Phase Separation

\vskip 1mm
\hspace{0.5cm}   4.6. Statistical Operator of Mixture

\vskip 1mm
\hspace{0.5cm}   4.7. Averaging over Phase Configurations

\vskip 1mm
\hspace{0.5cm}   4.8. Thermodynamic Potential of Mixture

\vskip 1mm
\hspace{0.5cm}   4.9. Observable Quantities of Mixture

\vskip 1mm
\hspace{0.5cm}   4.10. Statistics of Heterophase Systems

\vskip 1mm
\hspace{0.5cm}   4.11. Interphase Surface States
 
\vskip 1mm   
\hspace{0.5cm}   4.12. Geometric Phase Probabilities

\vskip 3mm
{\bf 5}. Models of Heterophase Systems

\vskip 1mm
\hspace{0.5cm}   5.1. Heterophase Heisenberg Model

\vskip 1mm
\hspace{0.5cm}   5.2. Role of Spin Waves 

\vskip 1mm
\hspace{0.5cm}   5.3. Heterophase Ising Model

\vskip 1mm
\hspace{0.5cm}   5.4. Heterophase Nagle Model

\vskip 1mm
\hspace{0.5cm}   5.5. Model of Heterophase Antiferromagnet

\vskip 1mm
\hspace{0.5cm}   5.6. Heterophase Hubbard Model

\vskip 1mm
\hspace{0.5cm}   5.7. Heterophase Vonsovsky-Zener Model

\vskip 1mm
\hspace{0.5cm}   5.8. Heterophase Spin Glass

\vskip 1mm
\hspace{0.5cm}   5.9. Systems with Magnetic Reorientations

\vskip 1mm
\hspace{0.5cm}   5.10. Model of Heterophase Superconductor

\vskip 1mm
\hspace{0.5cm}   5.11. Stability of Heterophase States

\vskip 1mm
\hspace{0.5cm}   5.12. Uniform Heterophase Superconductor

\vskip 1mm
\hspace{0.5cm}   5.13. Anisotropic Heterophase Superconductor

\vskip 1mm
\hspace{0.5cm}   5.14. Model of Heterophase Ferroelectric

\vskip 1mm
\hspace{0.5cm}   5.15. Heterophase Crystalline Structure

\vskip 1mm
\hspace{0.5cm}   5.16. Structural Phase Transition

\vskip 1mm
\hspace{0.5cm}   5.17. Stability of Heterophase Solids

\vskip 1mm
\hspace{0.5cm}   5.18. Solids with Nanoscale Defects

\vskip 1mm
\hspace{0.5cm}   5.19. Theory of Melting and Crystallization

\vskip 1mm
\hspace{0.5cm}   5.20. Model of Superfluid Solid

\vskip 1mm
\hspace{0.5cm}   5.21. Relations between Chemical Potentials

\vskip 1mm
\hspace{0.5cm}   5.22. Hamiltonian of Superfluid Solid

\vskip 1mm
\hspace{0.5cm}   5.23. Possibility of Superfluid Crystals 

\vskip 3mm
{\bf 6}. Mixture of Microscopic Components

\vskip 1mm
\hspace{0.5cm}   6.1. Mixed Quark-Hadron Matter

\vskip 1mm
\hspace{0.5cm}   6.2. Stability of Multicomponent Mixture

\vskip 1mm
\hspace{0.5cm}   6.3. Theory of Clustering Matter

\vskip 1mm
\hspace{0.5cm}   6.4. Clustering Quark-Hadron Mixture

\vskip 1mm
\hspace{0.5cm}   6.5. Thermodynamics of Quark-Hadron Matter 

\vskip 3mm  
{\bf 7}. Conclusion

\newpage

\section{Types of Phase Mixture}

There are three types of mixtures consisting of several phases or components,
macroscopic, mesoscopic, and microscopic. Macroscopic, or Gibbs, mixture consists 
of pieces of different pure phases having macroscopic sizes, for instance as is 
shown in Fig. 1. To form a thermodynamic phase, the substance has to have sizes
$l$ much larger than the mean interparticle distance $a$. The size of a phase $l$ 
is called macroscopic, when it is of the order of the system size $L$, so that 
$a \ll l \sim L$. The phases are separated by a surface whose influence becomes 
negligibly small in the thermodynamic limit. The description of this kind of mixture
reduces to the consideration of separate pure phases \cite{Kubo_1} complimented
by the conditions of phase equilibrium. This simple case is not considered in 
the review. 

\begin{figure}[ht]
\centerline{
\includegraphics[width=10cm]{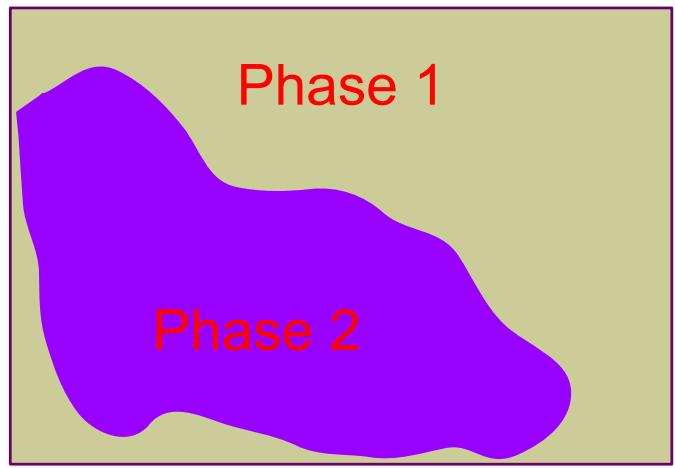} }
\caption{Macroscopic Gibbs mixture.
}
\label{fig:Fig.1}
\end{figure}

Much more interesting is the case of mesoscopic mixtures, where the phases are 
intermixed in space so that at least one of the phases is randomly distributed 
inside another phase in the form of regions of mesoscopic size that is between 
the mean interparticle distance and the size of the system, $a\ll l\ll L$. 
Usually, mesoscopic size in condensed matter corresponds to nanoscale. This 
situation is schematically shown in Fig. 2. Many materials in nature are formed 
by such mesoscopic mixtures, as will be discussed below. This kind of materials 
is of the main interest in the present review. 

\begin{figure}[ht]
\centerline{
\includegraphics[width=10cm]{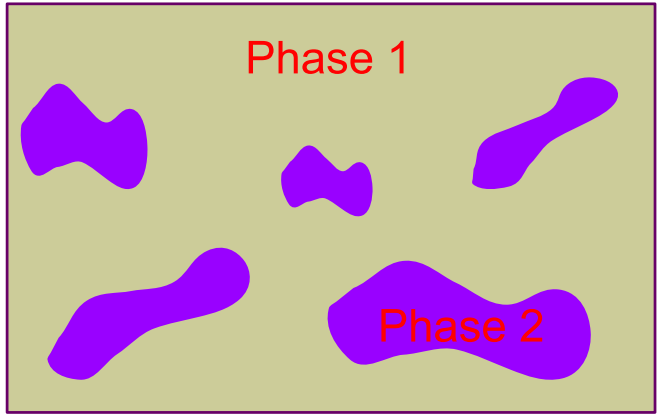} }
\caption{Mesoscopic mixture of two phases.
}
\label{fig:Fig.2}
\end{figure}

The third type of mixed systems is presented by microscopic mixtures, where the
particles that could form separate phases are intermixed on microscopic scales,
as is shown in Fig. 3, so that the sample becomes a multicomponent composition, 
but not a multiphase system. However, such a multicomponent matter can exhibit 
mesoscopic fluctuations and even separate into phases composed of different 
particles. The microscopic mixture will be touched upon in the last section of 
the review and illustrated by the example of mixed quark-hadron matter.   

\begin{figure}[ht]
\centerline{
\includegraphics[width=10cm]{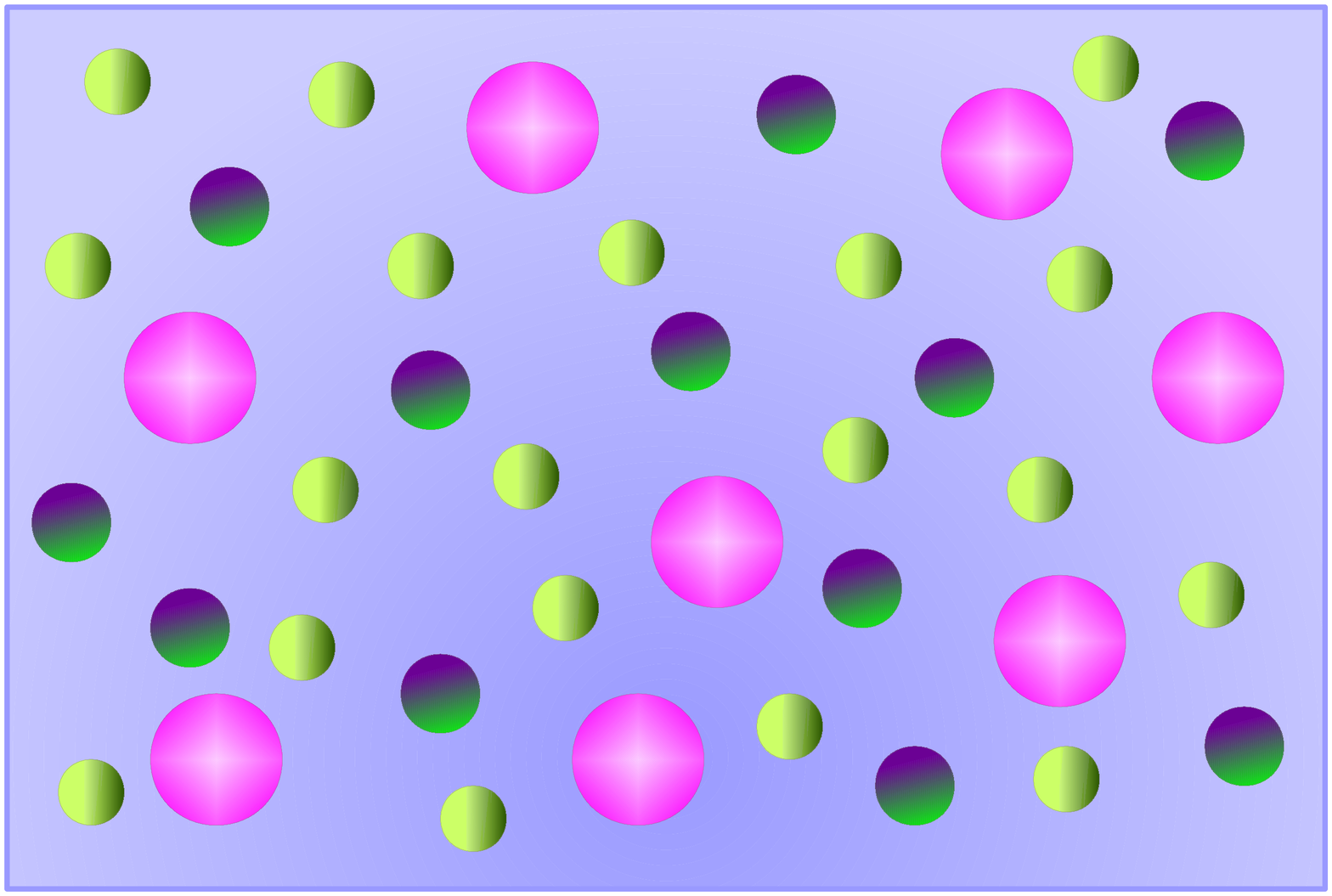} }
\caption{Microscopic mixture composed of several kinds of particles.
}
\label{fig:Fig.3}
\end{figure}
   
Throughout the paper, we shall, as a rule, use the system of units, where the 
Planck constant and the Boltzmann constant are set to one. This will be done 
everywhere, except those places where concrete numerical values are evaluated.

\section{Mesoscopic Heterophase Mixture}

In order to characterize a mesoscopic mixture, it is necessary to recollect the 
main spatial and temporal scales typical of condensed matter. 

The range of particle interactions is described by the interaction radius 
$r_{int}$. The mean interparticle distance $a$ is connected with the mean particle 
density $\rho$ as $\rho a^3 = 1$. The mean free path can be evaluated as
\be
\label{2.1}
\lbd \sim \frac{1}{\rho r^2_{int}} \sim \frac{a^3}{r_{int}^2} \;   .
\ee
The largest spacial scale is prescribed by the linear system size or the length 
of the region subject to the experimental observation, $l_{exp}$. 

These spatial scales define the related temporal scales: The interaction time reads 
as
\be
\label{2.2}
 t_{int} \sim \frac{r_{int}}{v} \sim \frac{m r^2_{int} }{\hbar} 
\qquad
\left( v \sim \frac{\hbar}{mr_{int}} \right) \; ,
\ee
where $v$ is the characteristic particle velocity. The local equilibration time 
is
\be
\label{2.3}
 t_{loc} \sim \frac{\lbd}{v} \sim \frac{m a^3 }{\hbar r_{int}} \; .
\ee

In condensed matter, $r_{int}\sim a\sim 10^{-8}$ cm, while $\lbd$ is of the order 
of $a$ or only slightly larger. Then the interaction time is $t_{int}\sim10^{-14}$ s
and the local equilibration time is $t_{loc}\sim 10^{-13}$ s. 

In dilute gas, where $r_{int}$ can be much shorter than $a$, instead of $r_{int}$, 
one considers the scattering length $a_s$. Typical times could be  
$t_{int}\sim 10^{-8}$ s and $t_{loc}\sim 10^{-3}$ s.  

In nuclear matter arising, e.g., in fireballs formed under heavy-ion collisions, 
the strong-interaction time is $t_{int}\sim 10^{-24}$ s and the local-equilibration 
time $t_{loc}\sim 10^{-23}$ s, while the fireball lifetime is $t_{exp}\sim 10^{-22}$ s.
 
Heterophase inclusions, arising inside the host phase, are mesoscopic, since their 
size is between the mean interparticle distance and the size of the experimentally 
studied region,
\be
\label{2.4}
 a \ll l_{het} \ll l_{exp} \; .
\ee
Usually, these inclusions are not frozen but can appear and disappear, because 
of which they are named heterophase fluctuations. Their lifetime has also to be 
mesoscopic, being between the local equilibration time and the time of experimental 
observation,
\be
\label{2.5}
 t_{loc} \ll t_{het} \ll t_{exp} \;  .
\ee
To form an embryo of a phase, the size of a mesoscopic fluctuation $l_{het}$ has 
to be at least an order larger than the mean interparticle distance which, for 
condensed matter, gives $l_{het} \sim 10 a \sim 10^{-7}$ cm. Respectively, the 
lifetime of a phase fluctuation has to be essentially longer than the local 
equilibration time which results in $t_{het}\sim 10 t_{loc}\sim 10^{-12}$ s. 
Thus typical mesoscopic fluctuations in condensed matter are of nanosize. However 
the size $l_{het} \sim 10^{-7}$ cm and lifetime $t_{het} \sim 10^{-12}$ s of 
heterophase fluctuations should be considered as low-boundary estimates. 

To summarize, the main features of a mesoscopic heterophase mixture are as 
follows.

\vskip 2mm
(i) The phase arising inside another phase appears in the form of mesoscopic 
embryos, with the sizes being between the interparticle distance and the size of 
the experimentally studied spatial region. This implies that not necessarily all, 
but at least one of the sizes has to be mesoscopic. For example, quantum vortices 
can be treated as embryos of the normal phase inside the superfluid phase. Although 
their length can be close to the system size, but the vortex radius is mesoscopic. 
Another example are dislocations in crystals, whose length can be comparable to the 
whole crystal sample, but whose radius is mesoscopic. 

\vskip 2mm
(ii) Heterophase fluctuations are not frozen, in the sense that their fractions 
are not prescribed, as for the case of fixed admixtures, but are defined by the 
material parameters and external conditions. Often, the fluctuations are of final 
lifetime, appearing and disappearing. But this is not compulsory. The main is that 
their concentrations are self-consistently defined by the system parameters and 
external conditions.    

\vskip 2mm
(iii) Usually, heterophase fluctuations are randomly distributed in space. In some 
cases they can form spatial structures. The most important is that the weight, or 
probability, of these heterophase fluctuations be defined self-consistently in the 
sense explained above.    

\vskip 2mm
(iv) A system with heterophase fluctuations, strictly speaking, is 
quasi-equilibrium. This is because the notion of a phase requires the existence 
of at least local equilibrium. At the same time, dynamically, the appearance of 
such fluctuations assumes the occurrence of local instability. Overall, on the 
average, the system can be treated as equilibrium, since its space-averaged 
characteristics are constant.  

\vskip 2mm
Heterophase materials with mesoscopic fluctuations is not an exotic object but 
rather the typical situation in condensed matter as can be inferred from the 
books \cite{Frenkel_2,Frenkel_3,Bakai_101} and reviews 
\cite{Khait_4,Yukalov_5,Yukalov_6,Yukalov_7,Bakai_178,Bakai_179,Kagan_261}. 
Below some concrete examples are listed.

\section{Examples of Heterophase Materials}

There are numerous examples of materials that are formed not by a single phase 
but by a mixture of several phases, or where inside the host phase there occur 
nanosize regions of a competing phase. We do not plan to give an exhaustive 
enumeration of all available references on experimental data where the mesoscopic 
heterophase coexistence has been observed. There are thousands of such works. 
Here we mention only some typical situations, while many more references can be 
found in the reviews 
\cite{Khait_4,Yukalov_5,Yukalov_6,Yukalov_7,Bakai_178,Bakai_179,Kagan_261}.

\subsection{Mixture of Ferromagnetic and Antiferromagnetic Phases}

In the pioneering article, Wollan and Koehler \cite{Wollan_8} reported 
their neutron diffraction study of the magnetic properties of the series 
of Perovskite-type compounds Ca$_x$La$_{1-x}$MnO$_3$. They found that the 
samples are mainly ferromagnetic, but also containing a small admixture of 
antiferromagnetic inclusions dispersed in the matrix as metastable bubbles. 
A number of other materials exhibit the coexistence of ferromagnetic and 
antiferromagnetic clusters, for example in micromagnetic MnBi alloys 
\cite{Grazhdankina_9}, in many magnetic semiconductors, such as 
La$_{1-x}$Ca$_x$MnO$_3$, La$_{1-x}$Sr$_x$MnO$_3$, \cite{Nagaev_10}, 
in disordered Au$_4$Mn and Cu$_3$Mn \cite{Suzuki_11}, in alloys MnZn, 
Mn$_x$Cr$_{1-x}$Sb, FePd$_x$Pt$_{1-x}$, Sc$_{1-x}$Ti$_x$Fe$_2$, 
Zr$_x$Nb$_{1-x}$Fe$_2$, and (Mn$_{1-x}$Ni$_x$)$_3$B$_4$ \cite{Moriya_12}, 
in manganites \cite{Marder_13}, such as La$_{1-x}$Ca$_x$MnO$_3$ 
\cite{Allodi_14,Allodi_15}, La$_{1-x}$Sr$_x$MnO$_3$, La$_{1-x}$Ba$_x$MnO$_3$ 
\cite{Yi_16,Yunoki_17,Yunoki_18,Dagotto_19,Balagurov_20,Yi_21,Kapusta_22}, 
La$_{5/8-x}$Pr$_x$Ca$_{3/8}$MnO$_3$ \cite{Podzorov_23}, 
Pr$_{0.5}$Ca$_{0.5-x}$Sr$_x$MnO$_3$ \cite{Krupicka_24}, Pr$_{1-x}$Ca$_x$MnO$_3$ 
\cite{Sha_25}, Pr$_{1-x}$Sr$_{x}$MnO$_3$ \cite{Pramanik_26}, and in many other 
colossal magnetoresistance materials \cite{Nagaev_27,Sudheendra_28,Shenoy_29}.

\subsection{Mixture of Magnetic and Paramagnetic Phases}

Many materials exhibit the coexistence of magnetic (ferromagnetic or 
antiferromagnetic) and paramagnetic phases. Thus, using the M\"{o}ssbaur 
effect, the coexistence of antiferromagnetic and paramagnetic phases is 
observed in FeF$_3$ \cite{Bertelsen_30}, in CaFe$_2$O$_4$ \cite{Yamamoto_31}, 
and in a number of orthoferrites, such as LaFeO$_3$, PrFeO$_3$, NdFeO$_3$, 
SmFeO$_3$, EuFeO$_3$, GdFeO$_3$, TbFeO$_3$, DyFeO$_3$, YFeO$_3$, HoFeO$_3$, 
ErFeO$_3$, TmFeO$_3$, YbFeO$_3$ \cite{Eibschutz_32,Levinson_33}. Ferromagnetic 
cluster fluctuations, called ferrons or fluctuons, can arise inside a 
paramagnetic matrix of some semiconductors 
\cite{Krivoglaz_34,Nagaev_35,Belov_36,Belov_37,Reissner_36}. In magnetic 
materials, magnetic cluster excitations can occur in the paramagnetic region 
above $T_c$ or above $T_N$ 
\cite{Goldman_39,Reimann_40,Bhargava_41,Uen_42,Srivastava_43,Halg_44}, causing 
the appearance of spin waves in the paramagnetic phase, for instance in Ni, Fe, 
EuO, EuS, Pd$_3$Fe, and Gd \cite{Lynn_45,Liu_46,Lynn_47,Cable_48,Lynn_49,Cable_50}. 
The coexistence of ferromagnetic and nonmagnetic phases were also observed in 
Y$_2$Co$_7$, YCo$_3$, Co(S$_x$Se$_{1-x}$)$_2$, Co(Ti$_x$Al$_{1-x}$)$_2$, and 
Lu(Co$_{1-x}$Al$_x$)$_2$ \cite{Goto_51,Shinogi_52}. In colossal magnetoresistance 
materials, such as La$_{1-x}$Ca$_x$MnO$_3$ and La$_{1-x}$Sr$_x$CoO$_3$, one 
observes the coexistence of a paramagnetic insulating, or semiconducting, phase 
and a ferromagnetic metallic phase \cite{Jaime_53,Merithew_54,Baio_55}, while in 
La$_{0.67-x}$Bi$_x$Ca$_{0.33}$MnO$_3$, paramagnetic and antiferromagnetic phases 
coexist \cite{Sun_56}. Nanoscale phase separation into ferromagnetic and 
paramagnetic regions has been observed in the colossal magnetoresistance compound 
EuB$_{5.99}$C$_{0.01}$ \cite{Batko_68}. The regions of competing phases are of 
mesoscopic size between $10$ \AA \; and $100$ \AA.

\subsection{Mixture of Magnetic and Spin-Glass Phases}

Ferromagnetic and spin-glass phases coexist in many alloys, for instance in Au-Fe
\cite{Gabay_57}, Pd-Ni \cite{Cheung_58}, Ni-Mn \cite{Abdul_59} alloys and in solid 
solutions, such as (CuCr$_2$Se$_4$)$_x$(Cu$_{0.5}$In$_{0.5}$Cr$_2$Se$_4$)$_{1-x}$ 
and (CuCr$_2$Se$_4$)$_x$(Cu$_{0.5}$GaCr$_2$Se$_4$)$_{1-x}$ \cite{Koroleva_60}. 
Spin-glass phase can also coexist with an antiferromagnetic phase or with a 
paramagnetic phase \cite{Matsuda_61,Li_62}.

\subsection{Mixture of Phases with Different Magnetic Orientations}

Magnetic phases with different orientations of magnetic moments coexist in 
rare-earth magnets \cite{Belov_63}, in Fe$_{1-x}$Co$_x$Cl$_2$ $\cdot$ 2H$_2$O 
\cite{Miwa_64}, in yttrium iron garnets with substitution of Ru$^4$ 
\cite{Balestrino_65}, and in Mn-Cu alloys \cite{Long_66}. Such mixtures occur 
around spin-reorientation transitions \cite{Belyaeva_67}. The regions of 
competing magnetization directions remind fluctuating domains or droplets, 
because of which they are called precursor fluctuations or local configuration 
fluctuations.

\subsection{Mixture of Ferroelectric and Paraelectric Phases}

In many materials around ferroelectric-paraelectric phase transitions there 
exist pretransitional effects caused by the arising clusters of competing phases 
\cite{Cook_69,Rigamonti_70,Brookeman_71,Bruce_72,Gordon_73,Gordon_74,Muller_75}. 
This happens, e.g., in HCl, HCl-DCl, RbCaF$_3$, BaTiO$_3$, and SbSI. It is 
believed that these pretransitional fluctuations are responsible for the 
characteristic saggings of the M\"{o}ssbauer-effect factor at the point of 
ferroelectric phase transitions 
\cite{Bhide_76,Bokov_77,Gleason_78,Jain_79,Canner_80,Bhide_81,Suzdalev_82}. 

Similar anomalous saggings of the M\"{o}ssbauer factor happen at the Morin 
magnetic reorientation phase transition \cite{Nikolov_83}, structural transitions 
\cite{Dezsi_84,Tsurin_85}, structural transitions accompanying superconducting 
transitions \cite{Cherepanov_86,Cherepanov_87,Cherepanov_88,Egorushkin_89}, and 
at structural transitions in macromolecular systems \cite{Nowik_90}. The typical 
depth of the M\"{o}ssbauer-effect factor sagging is about $30\%$ as compared to its 
value at the temperature above the phase transition.

\subsection{Mixture of Different Crystalline Structures}

In the vicinity of structural phase transitions, there appear the embryos of 
competing structures. Thus in He$^3$-He$^4$ solid solutions, in a wide range 
around the structural transition between the body-centered cubic (bcc) structural 
phase and hexagonal close packed (hcp) structure there exists a mixture of both 
these phases randomly intermixed in space \cite{Eselson_91,Eselson_92}. This type 
of coexistence of different phases around first-order crystallographic transitions 
is typical of martensitic transformations 
\cite{Kosilov_93,Pushin_94,Gibaud_95,Van_96}. The clusters of competing phases 
have the linear sizes of order $10-100$\AA. Similar pretransitional structural 
fluctuations exist around other structural transitions 
\cite{Emery_97,Morris_98,Zhu_99,Gibaud_100,Bakai_101} and in liquid crystals 
\cite{Kizel_102}.

\subsection{Mixture of Gaseous and Liquid Phases}

A typical illustration of a two-phase mixture is the mixture of a gas and 
a liquid close to the evaporation-condensation point 
\cite{Volmer_103,Akulichev_104,Kertez_105,Bakai_180,Bakai_106,Bakai_181,Bakai_182}.  
Before evaporation, there appear heterophase fluctuations in a liquid in the form 
of gas bubbles, and before condensation, there develop liquid droplets in a gas. 
A similar liquid-gas coexistence happens in exciton (electron-hole) systems in 
semiconductors, where one phase is formed by a more dense exciton liquid and the 
other phase is a less dense exciton gas \cite{Bakai_106,Rice_107}. Another close 
example is the existence of vacancy rich regions in emulsion bilayers 
\cite{Exerowa_108}. A general description of the nucleation dynamics can be 
found in Refs. \cite{Kelton_109,Dubrovskii_110}.

\subsection{Mixture of Liquid and Solid Phases}

R\"{o}ntgen \cite{Rontgen_111} was, probably, the first who has proposed that 
liquid water is not a single-phase fluid but a mixture of two components, a bulky 
icelike component and a less bulky normal liquid. This idea was developed by Brody 
\cite{Brody_112} and Bernal and Fowler \cite{Bernal_113}. According to this picture, 
in the solid state below the melting point there occurs a fluctuational appearance 
of local regions of liquid phase and above the melting temperature there develop 
fluctuational crystalline clusters. Frenkel \cite{Frenkel_2,Frenkel_3} emphasized 
that such heterophase fluctuations are common for condensed matter and happen around 
almost all phase transitions. The role of heterophase fluctuations in the vicinity 
of melting points was studied by Bartenev 
\cite{Bartenev_114,Bartenev_115,Bartenev_116,Bartenev_117} who stressed that the 
existence of these heterophase fluctuations explains thermodynamic anomalies 
occurring around the points of melting phase transitions. 

A great number of experimental data confirm that liquids above the crystallization 
point contain heterophase fluctuations in the form of quasi-crystalline clusters
\cite{Fisher_118,Samoilov_119,Dass_120,Gilra_121,Gilra_122,Gilra_123}. The sizes 
of the clusters range from about $10$ to $100$ molecules \cite{Fletcher_124}. The 
quasi-crystalline structure changes very quickly. The clusters themselves are not 
permanent entities but continuously form and dissociate under the influence of 
thermal fluctuations. The fluctuating cluster lifetime is much longer in comparison 
with molecular vibration periods of order of $10^{-13}$ s, but at the same time it 
must be shorter than the typical experimental measurement time of about $10^{-11}$ s, 
so that
$$
10^{-13} s \sim t_{loc} \ll t_{het} \ll t_{exp} \sim 10^{-11} s \; ,
$$
hence the reasonable estimate for the heterophase fluctuation lifetime is 
$10^{-12}$ s. The linear cluster size is typically of order of $10^{-7}$ cm 
\cite{Fletcher_124,Skripov_125,Hayes_126,Gabuda_127}. Icelike heterophase 
fluctuations are especially noticeable in supercooled liquids
\cite{Rasmussen_128,Speedy_129,Trinh_130,Arrigo_131,Arrigo_132}, although 
they do exist on both sides of the usual melting-crystallization transition. 
Below the melting point, heterophase clusters are represented by regions of 
disorder created by defects, such as vacancies, interstitials, dislocations, 
and disclinations \cite{Ubbelohde_133,Ziman_134,Wadati_135,Borsa_136}
and above the melting point, heterophase fluctuations are formed by 
quasi-crystalline clusters 
\cite{Fisher_118,Samoilov_119,Dass_120,Gilra_121,Gilra_122,Gilra_123,Steffen_137}. 
The effects caused by the appearance of heterophase fluctuations around 
the melting or crystallization point, are termed premelting and prefreezing, 
respectively \cite{Anisimov_138,Bilgram_139,Fontana_140,Rousset_141}. 

Melting and crystallization have been studied by means of computer simulations 
using the Monte Carlo method \cite{Barker_142} and molecular dynamic calculations 
\cite{Lagarkov_143,Choquard_144,Hiwatari_145}. These studies confirmed that around 
the point of the crystal-liquid phase transition there is a region of coexistence 
of solid-like and liquid-like phases. It is important that the melting transition 
is a first-order phase transition in either $3$ or $2$ dimensions 
\cite{Lagarkov_143,Choquard_144,Hiwatari_145,Stishov_146,Zollweg_147}. This is 
contrary to speculations on the possible second-order melting transition in two 
dimensions \cite{Kosterlitz_148,Barber_149}.   
 
Heterophase fluctuations appear even without any external influence, although 
they arise more easily if in the system there exist impurities. Then around these 
impurities in solids there happens local premelting 
\cite{Stern_150,Schechter_151,Stern_152}.

\subsection{Mixture of Metallic and Nonmetallic Phases}

The coexistence of solid and liquid phases can be accompanied by the 
coexistence of metallic and nonmetallic (dielectric or semiconductor) phases, 
where the metallic phase is liquid while the nonmetallic phase is solid. This 
type of coexistence occurs in liquid Te and Hg, in the liquid solutions 
In$_2$Te$_3$, Ga$_2$Te$_3$, Tl$_2$Te, Al$_2$Te$_3$, Mg$_x$Bi$_{1-x}$, 
Te$_x$Se$_{1-x}$, Hg$_{1-x}$Cd$_x$Te, in the metal-ammonia solutions Li-NH$_3$ 
and Na-NH$_3$ \cite{Cohen_153,Cohen_154,Jortner_155,Cohen_156,Tsuchiya_157,
Tsuchiya_158,Tsuchiya_159,Chandra_160,Takeda_161,Regel_162}, in sulfides of $3$d 
metals \cite{Loseva_163}, in bimetallic alloy clusters, such as Pd$_6$Ni$_7$ 
\cite{Lopez_164}, and in manganites, e.g., in La$_x$MnO$_{1-\delta}$ \cite{Gao_165}. 
The inhomogeneity in these materials is due to local density fluctuations, which 
causes a random spatial variation of conductivity. The metal-semiconductor 
transition is a continuous second-order phase transition.

\subsection{Mixture of Superconducting and Normal Phases}

A model representing low-temperature superconductors as a mixture of 
superconducting and ``normal" components was suggested by Gorter and Casimir 
\cite{Gorter_166,Schrieffer_167,Lynton_168}. However, for the low-temperature 
superconductors, the two-fluid model is just an effective representation of a 
single superconducting phase, where the so-called ``normal" component describes 
excitations above the ground state. The same concerns the two-fluid model of 
superfluid helium, where there exists a single superfluid phase and the ``normal" 
component corresponds to particle excitations above the coherent ground state 
\cite{Yukalov_169,Yukalov_170}.  

Real mixtures of superconducting and normal phases have been observed in 
high-temperature superconductors 
\cite{Hizhnyakov_171,Phillips_172,Hizhnyakov_173,Benedek_174,Sigmund_175,
Kivelson_176}. The occurrence of such mixtures sometimes is called mesoscopic 
phase separation \cite{Gorkov_177}. The normal phase is formed by insulating 
clusters or droplets. The phase separation is dynamic, the insulating and 
superconducting phases change their locations and shapes. The phase mixture 
often arises close to the structural instability of the lattice.

\subsection{Mixture of Metastable Amorphous Phases}

There exists a class of the so-called glass-forming liquids that solidify into 
a glassy state 
\cite{Bakai_101,Bakai_178,Bakai_179,Spielberg_183,Klinger_184,Bakai_185,
Bakai_186,Lazarev_187}. Numerous experiments have found that both these liquid 
and solid states are, actually, heterophase systems consisting of a mixture of 
liquidlike and solidlike clusters. The glassification is a transition that can 
be discontinuous (first order), although more often it is a continuous transition. 

Amorphous solids and glasses are metastable objects. Respectively, the mixtures 
of solidlike and liquidlike phases, forming these objects, are examples of 
metastable heterophase mixtures.

\subsection{Mixture of Nonequilibrium Phases}

Heterophase mixtures can arise as well in nonequilibrium systems. For example, 
electric current in a superconductor can display the so-called resistive states, 
where superconductivity coexists with normal state. This coexistence is not 
stationary. The gap, that is the order parameter for superconductivity, 
fluctuationally becomes zero for some period of time at random spatial regions. 
These regions of zero gap are the nuclei of normal phase \cite{Ivlev_188}.

Inside a laminar liquid flow, there can appear turbulent regions, arising 
stochastically in space and time, then disappearing, and then again spontaneously 
arising in random areas \cite{Rabinovich_189}.     

By subjecting a system of trapped bosonic atoms to an alternating external field, 
it is possible to create several nonequilibrium states housing the mixtures of 
Bose-condensed and uncondensed phases. Thus, a vortex turbulent state can be 
formed, where inside the Bose-condensed phase there exists a random bunch of 
quantum vortices playing the role of normal phase nuclei 
\cite{Yukalov_190,Shiozaki_191,Seman_192,Bagnato_193}. The other heterophase 
state occurs when inside the normal incoherent fluid there happens a cloud of 
randomly located Bose-condensed droplets
\cite{Bagnato_193,Yukalov_194,Yukalov_195,Yukalov_196,Yukalov_197,Yukalov_198}. 
This state represents {\it grain turbulence} or {\it droplet turbulence}.

\section{Theory of Heterophase Systems}

In this chapter, the basic ideas of the theory for describing mesoscopic 
heterophase mixtures are presented. Because of the importance of this topic and 
in order to avoid the following questions of how the concrete models are defined, 
the exposition of the basic theory is sufficiently comprehensive.

\subsection{Spontaneous Breaking of Equilibrium}

That was, probably, Boltzmann \cite{Boltzmann_199} who first advanced the idea 
that in a large system, that on average looks equilibrium, there can develop 
strongly nonequilibrium local fluctuations. This idea is called the Boltzmann 
fluctuational hypothesis. Actually, Boltzmann was talking about the Universe, 
but the same is applicable to any large system. 

Mesoscopic heterophase fluctuations can emerge spontaneously 
\cite{Frenkel_2,Frenkel_3,Yukalov_5,Boltzmann_199}, because of which this effect 
can be termed {\it spontaneous breaking of local equilibrium} \cite{Yukalov_5}. 
Local defects and external fields can facilitate the emergence of these 
fluctuations. 

There also exist systems, called stochasticity amplifiers, where even small 
external perturbations can be drastically strengthened \cite{Neimark_200}. Also, 
there are systems, where any weak initial noise can be transformed and result in 
strong fluctuations, characterized by the system properties, at any further times, 
being independent from the initial noise. These systems are termed stochasticity 
generators \cite{Neimark_200,Neimark_244}.   
  
Suppose, we are studying an observable quantity $f(t,\xi)$ as a function of time 
and depending on an external noise of strength $\xi$. If the limits $t\ra\infty$ 
and $\xi\ra 0$ are not commutative, so that
$$
\lim_{t\ra\infty} \; \lim_{\xi\ra 0} \; f(t,\xi) \neq 
\lim_{\xi\ra 0} \; \lim_{t\ra\infty} \; f(t,\xi) \;  ,
$$
this property is called {\it stochastic instability} \cite{Yukalov_5,Yukalov_201}. 
The property of stochastic instability is responsible for the irreversibility of 
time \cite{Yukalov_202,Yukalov_203,Yukalov_204}.   

In this way, there can exist two origins of heterophase fluctuations, even 
when the system as a whole looks equilibrium on average. These fluctuations can 
be produced by the system itself. Or they can be triggered by some weak external 
noise, that always exists, as far as there are no absolutely isolated systems, 
but only quasi-isolated \cite{Yukalov_205,Yukalov_206}. In both these cases, the 
properties of the fluctuations are completely characterized by the system parameters, 
under the given external conditions. Therefore in both the cases we can say that 
there occurs spontaneous breaking of local equilibrium. When the fluctuations 
correspond to a phase with a symmetry different from the surrounding matter, we 
can say that there happens {\it spontaneous local symmetry breaking or restoration} 
\cite{Yukalov_5,Yukalov_7,Yukalov_207}.

\subsection{Statistical Ensembles and States}

Before going to the specification of heterophase systems, we need to briefly 
recollect the main notions employed for describing statistical systems, keeping 
in mind the general case of quantum statistical systems. Here we give a brief 
account of notions that will be used in the following sections. More details can 
be found in Refs. \cite{Yukalov_5,Ruelle_208,Dixmier_209,Emch_210,Bratteli_211}.  

First, we have to define a Hilbert space of microstates
\be
\label{4.1}
\cH = {\rm span}_n \; \{ \vp_n \}
\ee
that is a closed linear envelope over an orthonormalized basis. The system state 
is described by a statistical operator $\hat\rho$ that is a semi-positive trace-one 
operator on $\mathcal{H}$. The pair $\{\mathcal{H}, \hat\rho\}$ is a {\it quantum 
statistical ensemble}.

Local observables are represented by operators on $\mathcal{H}$ forming the 
algebra of local observables $\mathcal{A}$. This is a von Neumann algebra that 
is a self-adjoint, closed in the weak operator topology subalgebra, containing 
the identity operator, of the algebra of all bounded operators on a Hilbert space. 
Observable quantities are the averages
\be
\label{4.2}
\lgl \; \hat A \; \rgl \equiv {\rm Tr}\; \hat\rho \; \hat A = 
\sum_n ( \vp_n,\hat\rho \; \hat A\; \vp_n )
\ee
of operators $\hat{A} \in \mathcal{A}$. The trace is over $\mathcal{H}$. The 
collection of the averages of all observable quantities is the {\it statistical 
state}
\be
\label{4.3}
\lgl \; \hat \cA \; \rgl = \{ \lgl \; \hat A \; \rgl \} \;  .
\ee

A special role is played by the order-parameter operator $\hat{\eta}\in\cA$ 
that yields the order parameter
\be
\label{4.4}
\eta = \lgl \; \hat\eta \; \rgl
\ee
helping to distinguish different thermodynamic phases. Order parameters 
characterize the long-range order. A more general classification of thermodynamic 
phases, including those characterized by mid-range order, can be done by means of 
{\it order indices} \cite{Yukalov_212,Yukalov_213}.  

The notion of pure thermodynamic phases requires to consider the thermodynamic 
limit, when the number of particles in the system $N$ and the system volume $V$ 
tend to infinity, with their ratio, defining the particle density, tending to a 
constant,
\be
\label{4.5}
 N \ra \infty \; , \qquad V \ra \infty \; , \qquad 
\frac{N}{V} \ra const \; .
\ee
Then the averages of the operators of local observables are proportional to 
the number of particles, because of which one has to consider the ratio 
$\lgl\hat{A}\rgl/N$. If a system exhibits a phase transition, then there exists 
a region of parameters, where the thermodynamic limit yields the decomposition 
of the system state into pure states:
\be
\label{4.6}
\lim_{N\ra\infty} \; \frac{1}{N} \; \lgl \; \hat A \; \rgl = 
\sum_f \lbd_f \; \lim_{N\ra\infty} \; 
\frac{1}{N} \; \lgl \; \hat A_f \; \rgl \; ,
\ee
with the normalized coefficients
$$
\sum_f \lbd_f = 1 \; \qquad 0 \leq \lbd_f \leq 1 \;  .
$$
Here $\hat{A}_f$ is the representation of $\hat{A}$ on a subspace $\cH_f\subset\cH$ 
of microstates associated with the $f$-phase. The index $f=1,2,\ldots$ enumerates 
the phases. The limit $N\ra\infty$ implies the thermodynamic limit (\ref{4.5}). 
Respectively, the Hilbert space of microstates in that case becomes a direct sum 
of subspaces associated with the pure phases,
\be
\label{4.7}
 \cH \mapsto \bigoplus_f \cH_f \qquad ( N \ra \infty) \; .
\ee 

The state decomposition (\ref{4.6}) corresponds to the macroscopic Gibbs mixture, 
but not to a heterophase system with mesoscopic phase fluctuations. When the 
system possesses a symmetry described by a symmetry group, the decomposition into 
pure states (\ref{4.6}) is the decomposition over the symmetry subgroups. In order 
to describe a particular pure state, it is necessary to select the microstates 
characterizing this particular symmetry subgroup.

\subsection{Methods of Symmetry Breaking}

The state decomposition arises when the system Hamiltonian is invariant with 
respect to a symmetry group, while a pure phase corresponds to a broken symmetry. 
The selection of a pure phase can be done in several ways. In order to separate 
a pure phase, it is possible to break the symmetry of the Hamiltonian using the 
Bogolubov method of quasi-averages \cite{Bogolubov_214,Bogolubov_215,Bogolubov_216}.

Let the system be described by a Hamiltonian $\hat{H}$ that is invariant with 
respect to a transformation forming a group. One introduces a Hamiltonian
\be
\label{4.8}
 \hat H_{f\ep} = \hat H +\ep \hat\Gm_f \;  ,
\ee
by adding to the initial Hamiltonian a term containing an operator $\hat\Gm_f$ 
breaking the symmetry to a subgroup corresponding to the required phase $f$. Here 
$\ep$ is a real-valued parameter. For the operator of an observable $\hat{A}$, the 
quasi-average, selecting the representation corresponding to the $f$-phase, is  
\be
\label{4.9}
\lim_{N\ra\infty}\; \frac{1}{N} \;\lgl \; \hat A_f \; \rgl \equiv
\lim_{\ep\ra 0}\; \lim_{N\ra\infty}\; 
\frac{1}{N} \;\lgl \; \hat A \; \rgl_{f\ep} \;  .
\ee
In the right-hand side, the average is defined for the case of Hamiltonian 
(\ref{4.8}). The limit $N\ra\infty$ implies the thermodynamic limit (\ref{4.5}). 
It is useful to stress that the limits here are not commutative. The thermodynamic 
limit has to necessarily be taken before the limit $\ep\ra 0$. 

Instead of taking two limits, it is possible to define {\it thermodynamic 
quasi-averages} \cite{Yukalov_217,Yukalov_218} calculated with the Hamiltonian
\be
\label{4.10}
\hat H_f = \hat H + \frac{1}{N^\gm} \; \hat\Gm_f \;   ,
\ee
in which $0 < \gamma < 1$. Then the thermodynamic quasi-average is
\be
\label{4.11}
\lim_{N\ra\infty}\; \frac{1}{N} \;\lgl \; \hat A_f \; \rgl \equiv
\lim_{N\ra\infty}\; \frac{1}{N} \;\lgl \; \hat A \; \rgl_f \; ,
\ee
where the right-hand side is calculated with Hamiltonian (\ref{4.10}). 

Among other methods of symmetry breaking, it is possible to mention the method 
of restricted trace or restricted Hilbert space, the method of boundary conditions, 
breaking of commutation relations, use of canonical transformations, analytical 
continuation, imposing symmetry conditions for correlation functions or Green
functions, and mean-field approximations. The details can be found in Refs.
\cite{Brout_219,Sewell_220,Sinai_221,Shumovsky_222,Bogolubov_223,Shumovsky_224,
Yukalov_225}.

\subsection{Weighted Hilbert Space}

All methods of symmetry breaking can be summarized by formulating the notion of 
the weighted Hilbert space \cite{Yukalov_5,Yukalov_6,Yukalov_7}. Suppose, we are 
considering a system characterized by the Hilbert space (\ref{4.1}). Let each 
member $\vp_n$ of the basis be associated with a weight $p_f(\vp_n)$ describing 
how typical this basis member is for the phase $f$. The weights are normalized, 
so that
$$
 \sum_f p_f(\vp_n) = 1 \; , \qquad 0 \leq p_f(\vp_n) \leq 1 \;  .
$$
The set of all weights is denoted as
\be
\label{4.12}
p_f(\vp) \equiv \{ \; p_f(\vp_n) : \; \forall n \; \} \;   .
\ee
The weighted Hilbert space is the Hilbert space with a weighted basis,
\be
\label{4.13}
 \cH_f \equiv \{ \; \cH, p_f(\vp) \; \} \; .
\ee
The quantum statistical ensemble characterizing a phase $f$ is the pair 
$\{\cH_f,\hat\rho\}$. The statistical state associated with a phase $f$ is the
collection of the averages for the operators of local observables defined as
\be
\label{4.14}
 \lgl \; \hat A_f \; \rgl  \equiv {\rm Tr}_{\cH_f} \; \hat\rho \;\hat A 
\equiv \sum_n p_f(\vp_n) \; (\vp_n,\hat\rho\; \hat A \;\vp_n) \; .
\ee
The basis weights are to be such that the order parameter
\be
\label{4.15}
\eta_f = \lgl \; \hat \eta_f \; \rgl 
\ee
would have the symmetry properties typical of the considered phase. Concrete 
models exemplifying this procedure will be given in the following sections.

\subsection{Spatial Phase Separation}

When a system consists of several thermodynamic phases, its spatial geometry can 
be described following the Gibbs idea of imagining that the phases are divided by 
a thin surface \cite{Gibbs_226,Ono_227,Rusanov_228,Rusanov_229}. The Gibbs 
separating surface is standardly defined by considering thermodynamic quantities. 
In our case, we need to introduce a separating surface allowing for the additive, 
with respect to phases, representations of the operators of extensive observables. 
As we show below, the additivity of operators does not preclude from the possibility 
of defining interfacial effects on the macroscopic level. 
 
The number of particles in the system is the sum
\be
\label{4.16}
N = \sum_f N_f 
\ee
of the particles in the phases composing the system. The system real space can 
be associated with the spatial orthogonal covering
\be
\label{4.17}
 \mathbb{V} = \bigcup_f \mathbb{V}_f \; ,  \qquad  
\mathbb{V}_f \bigcap \mathbb{V}_g = \dlt_{fg} \mathbb{V}_f \; .
\ee  
Respectively, the system volume is the sum
\be
\label{4.18}
 V = \sum_f V_f  \qquad 
( V \equiv {\rm mes} \mathbb{V} \;, V_f \equiv {\rm mes} \mathbb{V}_f ) \;.
\ee

It is convenient to represent the topology of a composite system by using the 
manifold indicator functions \cite{Bourbaki_230} defined by the condition
\begin{eqnarray}
\label{4.19}
\xi_f(\br) = \left\{ 
\begin{array}{ll}
1 , ~ & \br \in \mathbb{V}_f \\
0 , ~ & \br \not\in \mathbb{V}_f 
\end{array} \; ,
\right.
\end{eqnarray}
with the properties
$$ 
 \sum_f \xi_f(\br) = 1 \; , \qquad 
\xi_f(\br) \xi_g(\br) = \dlt_{fg} \xi_f(\br)  
$$
and
$$
\int_\mathbb{V} \xi_f(\br) \; d\br = V_f \;   .
$$
The collection of all manifold indicator functions (\ref{4.19}) will be denoted 
as
\be
\label{4.20}
\xi \equiv \{ \xi_f(\br) : \; \br \in\mathbb{V} , \; f = 1,2,\ldots \} \; .
\ee

In order to consider different shapes and locations of the phases, the spatial 
covering can be represented as being composed of the orthogonal subcoverings
\be
\label{4.21}
 \mathbb{V}_f = \bigcup_{i=1}^{n_f} \mathbb{V}_{fi} \; , \qquad
\mathbb{V}_{fi} \bigcap \mathbb{V}_{gj} = 
\dlt_{fg}\; \dlt_{ij} \mathbb{V}_{fi} \;  . 
\ee
Then the manifold indicator functions(\ref{4.19}) can be written as the sums
\be
\label{4.22}
\xi_f(\br) = \sum_{i=1}^{n_f} \xi_{fi}(\br-\ba_{fi} ) \qquad
(\ba_{fi} \in \mathbb{V}_{fi} )
\ee
of the submanifold indicator functions
\begin{eqnarray}
\label{4.23}
\xi_{fi}(\br) \equiv \left\{ 
\begin{array}{ll}
1 , ~ & \br \in \mathbb{V}_{fi} \\
0 , ~ & \br \not\in \mathbb{V}_{fi} 
\end{array} \; .
\right.
\end{eqnarray}
Here $\ba_{fi}$ is a fixed vector associated with a spatial cell $\mathbb{V}_{fi}$. 
The collection of the manifold indicator functions (\ref{4.22}) for a given phase 
$f$ is denoted by 
\be
\label{4.24}
 \xi_f \equiv \{ \xi_f(\br) : \; \br \in \mathbb{V} \} \;  .
\ee

The Hilbert space of microscopic states is the tensor product
\be
\label{4.25}
\widetilde \cH = \bigotimes_f \; \cH_f   
\ee
of the weighted Hilbert spaces described in the previous section. Space 
(\ref{4.25}) can be called the {\it fiber space}.

In agreement with the additivity of extensive quantities, the related operators 
of observable quantities are additive, although being dependent on the spatial 
configuration of the phases. Thus the number-of-particle operator is
\be
\label{4.26}
\hat N(\xi) = \bigoplus_f \; \hat N_f(\xi_f) \; ,
\ee
and the energy operator is
\be
\label{4.27}
\hat H(\xi) = \bigoplus_f \; \hat H_f(\xi_f) \; .
\ee
Here $\hat{N}_f(\xi_f)$ and $\hat{H}_f(\xi_f)$ are the representations of the 
corresponding operators on the Hilbert space $\mathcal{H}_f$, the notation 
$\xi$ denotes the set of the manifold indicator functions (\ref{4.20}) and 
$\xi_f$ is the set (\ref{4.24}) of the manifold indicator functions for a fixed 
$f$.

\subsection{Statistical Operator of Mixture}

The statistical operator of a mixture formed by different thermodynamic phases 
can be found from the minimization of an information functional 
\cite{Yukalov_5,Yukalov_6,Yukalov_7}. The statistical operator is assumed to 
satisfy several conditions. First of all, this is the normalization condition
\be
\label{4.28}
 {\rm Tr} \int \hat\rho(\xi) \; \cD\xi = 1 \; ,
\ee
where the trace is over the fiber space (\ref{4.25}) and the integral over $\xi$ 
means a functional integral over the set (\ref{4.20}) of the manifold indicator 
functions. The integration over the manifold indicator functions implies the 
averaging over a random phase distribution in the system space. 

The system energy is given by the average
\be
\label{4.29}
 {\rm Tr} \int \hat\rho(\xi)\; \hat H(\xi) \; \cD\xi = E \; .
\ee
And the total number of particles in the system is fixed by the condition
\be
\label{4.30}
 {\rm Tr} \int \hat\rho(\xi)\; \hat N(\xi) \; \cD\xi = N \;  .
\ee

Taking account of these conditions, the information functional in the 
Kullback-Leibler form \cite{Kullback_231,Kullback_232} reads as
$$
I[\; \hat\rho \; ] = {\rm Tr} \int 
\hat\rho(\xi) \; \ln \; \frac{\hat\rho(\xi)}{\hat\rho_0(\xi)} \; \cD\xi 
\; + \; 
\al \left[ \; {\rm Tr} \int \hat\rho(\xi) \; \cD\xi - 1 \; \right] 
\; + 
$$
\be
\label{4.31}
+ \; 
\bt \left[ \; {\rm Tr} \int \hat\rho(\xi)\; \hat H(\xi)\; \cD\xi - E \; 
\right] \; + \;
\gm \left[ \; {\rm Tr} \int \hat\rho(\xi)\; \hat N(\xi)\; \cD\xi - N \; 
\right] \; ,
\ee
where $\al$, $\bt$, and $\gm$ are Lagrange multipliers and $\hat\rho_0(\xi)$
is a trial statistical operator, with $1/\hat\rho_0$ meaning $(\hat\rho_0)^{-1}$.   

Minimizing the information functional, we set $\gamma = -\beta \mu$ and introduce 
the grand Hamiltonian
\be
\label{4.32}
H(\xi) \equiv \hat H(\xi) - \mu \hat N(\xi) \;  .
\ee
The parameter $\beta = 1/T$ implies the inverse temperature. The grand Hamiltonian 
acquires the form
\be
\label{4.33}
 H(\xi) = \bigoplus_f \; H_f(\xi_f) = 
\bigoplus_f \; \left[ \; \hat H_f(\xi_f) - \mu \hat N_f(\xi_f) \; \right] \; .
\ee

Thus we find the statistical operator of the mixture
\be
\label{4.34} 
 \hat\rho(\xi) = \frac{\hat\rho_0(\xi)\exp\{-\bt H(\xi)\} }
{{\rm Tr}\int \hat\rho_0(\xi)\exp\{-\bt H(\xi)\} } \; .
\ee
If there is no any a priori information on the distribution of the heterophase 
regions, we have to set 
\be
\label{4.35}
 \hat\rho_0(\xi) = \left( {\rm Tr}\int \cD\xi \right)^{-1} \; .
\ee
As a result, we come to the statistical operator
\be
\label{4.36}
  \hat\rho(\xi) = \frac{1}{Z} \; \exp\{ -\bt H(\xi) \} \; ,
\ee
with the partition function
\be
\label{4.37}
Z =  {\rm Tr}\int \exp\{ -\bt H(\xi) \} \; \cD\xi \;  .
\ee

Specifying the operators of observables, we employ the second quantization 
representation and use the identity
\be
\label{4.38}
\int_{\mathbb{V}_f} d\br = \int_\mathbb{V}  \xi_f(\br) \; d\br \; .
\ee
The operator of energy for an $f$-th phase reads as
$$
\hat H_f(\xi_f) = \int \xi_f(\br) \; \psi_f^\dgr(\br) \;
\left[ \; - \; \frac{\nabla^2}{2m} + U(\br) \; 
\right] \; \psi_f(\br) \; d\br \; +
$$
\be
\label{4.39}
 +\; \frac{1}{2} \int \xi_f(\br) \; \xi_f(\br') \; 
\psi_f^\dgr(\br)\; \psi_f^\dgr(\br')\;
\Phi(\br-\br') \; \psi_f(\br')\; \psi_f(\br) \; d\br \; d\br' \; ,
\ee
where $\psi_f({\bf r})$ is the representation of a field operator on the Hilbert 
space $\mathcal{H}_f$, $U({\bf r})$ is an external potential and $\Phi({\bf r})$ 
is an interaction potential. The number-of-particle operator for the $f$-th phase 
is
\be
\label{4.40}
 \hat N_f(\xi_f) = 
\int \xi_f(\br) \; \psi_f^\dgr(\br)\; \psi_f(\br)\; \; d\br .
\ee
Here and in what follows, the integration over space, where the volume is not shown, 
assumes the integration over the whole system volume $\mathbb{V}$. The dependence of
the field operators on the spatial variable is shown explicitly, while the internal 
degrees of freedom, such as spin, isospin, or like that, can be taken into account by 
representing the field operators as columns whose rows are labeled by these internal 
degrees of freedom.

\subsection{Averaging over Phase Configurations}
 
The thermodynamic phases are assumed to be distributed in space randomly. The 
averaging over their locations and shapes is denoted by the integration over the 
manifold indicator functions. For each given phase configuration, the system is 
nonuniform. The main idea of the approach to describing heterophase fluctuations, 
advanced in Refs. \cite{Yukalov_233,Yukalov_234,Yukalov_235,Yukalov_236,Yukalov_237,
Yukalov_238,Yukalov_239}, is to reduce the nonuniform multiphase problem to a set 
of single-phase problems, which could be done by averaging over phase configurations. 
In the earlier papers, this averaging was assumed to lead to effective models. 
The explicit mathematical integration over the manifold indicator functions is 
accomplished in the papers\cite{Yukalov_240,Yukalov_241,Yukalov_242,Yukalov_243} 
and summarized in the reviews \cite{Yukalov_5,Yukalov_6,Yukalov_7}.
   
Concretely, the idea is as follows. Suppose we are able to find the renormalized 
grand Hamiltonian
\be
\label{4.41}
\widetilde H = - T \; \ln \int \exp\{ - \bt H(\xi)\} \; \cD\xi \;  ,
\ee
which requires to accomplish the averaging over phase configurations
\be
\label{4.42}
\int \exp\{ - \bt H(\xi)\} \; \cD\xi = \exp(-\bt \widetilde H ) \; ,
\ee
then the partition function (\ref{4.37}) becomes
\be
\label{4.43}
Z = {\rm Tr} \exp(-\bt \widetilde H ) \; .
\ee
The renormalized Hamiltonian (\ref{4.41}) already does not depend on the spatial 
phase distribution.

Now we need to define the functional integration over the manifold indicator 
functions. Let us introduce the variable
\be
\label{4.44}
x_f \equiv \frac{1}{V} \int \xi_f(\br) \; d\br
\ee
and the set of these variables for all phases
\be
\label{4.45}
 x \equiv \{ x_f : \; f = 1,2,\ldots \} \;  .
\ee
Quantity (\ref{4.44}), having the meaning of a varying geometric weight of an 
$f$-th phase, has the properties
\be
\label{4.46}
 \sum_f x_f  = 1 \; , \qquad 0 \leq x_f \leq 1 \; .
\ee 

The differential measure $\mathcal{D}\xi$ can be separated into two parts,
\be
\label{4.47}
 \cD\xi = \prod_f \cD\xi_f \; dx \; ,
\ee
characterizing the variation over the locations and shapes of the phases and 
over their geometric weights.   

For the subcovering, defined in Sec. 8, the volume of a cell $\mathbb{V}_{fi}$ 
is
\be
\label{4.48}
 v_{fi} \equiv \int \xi_{fi}(\br-\ba_{fi} ) \; d\br = 
{\rm mes}\mathbb{V}_{fi} \; .
\ee
From the equality
\be
\label{4.49}
 \sum_{i=1}^{n_f} v_{fi}  = V_f 
\ee
it follows that the number of small cells increases, so that
\be
\label{4.50}
 n_f \ra \infty \; , \qquad v_f \ra 0 \; .
\ee

For the averaging over the locations and shapes of the regions containing an 
$f$-th phase, the differential measure can be written as
\be
\label{4.51}
\cD\xi_f = 
\lim_{n_f\ra\infty}\; \prod_{i=1}^{n_f} \; \frac{d\ba_{fi}}{V} \;  ,
\ee
where the limit (\ref{4.50}) is understood. The averaging over the geometric 
weights of the phases corresponds to the differential measure
\be
\label{4.52}
dx = \dlt\left( \sum_f x_f -1 \right) \prod_f d x_f  \; ,
\ee
where the normalization condition (\ref{4.46}) is taken into account. Each 
weight can vary between $0$ and $1$. 

Consider a functional
\be
\label{4.53}
C_f(\xi_f) = 
\sum_{m=0}^\infty \int \xi_f(\br_1)\; \xi_f(\br_2) \ldots \xi_f(\br_m)\;
C_f(\br_1,\br_2,\ldots,\br_m) \; d\br_1 d\br_2 \ldots d\br_m \;  .
\ee
Replacing here $\xi_f$ by $x_f$, in the sense of the substitution 
\be
\label{4.54}
 C_f(x_f) = \lim_{\xi_f\ra x_f} \; C_f(\xi_f) \; ,
\ee
we come to the functional
\be
\label{4.55}
 C_f(x_f) = \sum_{m=0}^\infty x_f^m \; 
\int  C_f(\br_1,\br_2,\ldots,\br_m) \; d\br_1 d\br_2 \ldots d\br_m \; .
\ee
The following theorem holds \cite{Yukalov_241,Yukalov_242,Yukalov_243}. 

\vskip 2mm

{\bf Theorem 1}. {\it The averaging of functional (\ref{4.53}) over the spatial 
locations and shapes of the $f$-th phase yields functional (\ref{4.55})},
\be
\label{4.56}
\int  C_f(\xi_f)\; \cD\xi_f =C_f(x_f) \;  .
\ee

\vskip 2mm

{\it Proof}. To prove the theorem, it is possible to resort to the Dirichlet 
representation of the manifold indicator functions 
\cite{Yukalov_241,Yukalov_242,Yukalov_243} or one can take into account that
\be
\label{4.57}
\int \xi_{fi}(\br-\ba_{fi}) \; d\ba_{fi} = v_{fi}   
\ee
and 
\be
\label{4.58}
\int \xi_{fi}(\br-\ba_{fi}) \; \xi_{fi}(\br'-\ba_{fi}) \; d\ba_{fi} \; 
\leq \; v^2_{fi} \; .
\ee
In the latter inequality, the quantities ${\bf r}$ and $\br'$ are independent 
variables. Under the limit (\ref{4.50}), we have
$$
\sum_{i=1}^{n_f} \frac{v_{fi}^2}{V} \leq x_f \; \max_i v_{fi} \; \simeq \; 0 
\qquad 
(n_f \ra \infty) \; ,
$$
$$
\sum_{i\neq j}^{n_f} \frac{v_{fi}v_{fj}}{V^2} \; \simeq \; x^2_f 
\qquad 
(n_f \ra \infty) \; .
$$
Therefore we come to the equalities
\be
\label{4.59}
\int \xi_f(\br) \; \cD\xi_f = x_f
\ee
and 
\be
\label{4.60}
\int \xi_f(\br) \; \xi_f(\br') \; \cD\xi_f = x_f^2  \;  .
\ee
Continuing this procedure results in the general equality
\be
\label{4.61}
\int \xi_f(\br_1) \; \xi_f(\br_2) \ldots \xi_f(\br_m) \; \cD\xi_f = 
x_f^m  \; ,
\ee
from where Eq. (\ref{4.55}) follows. $\square$

\subsection{Thermodynamic Potential of Mixture}

The grand thermodynamic potential of a mixed heterophase system 
\be
\label{4.62}
\Om = - T\ln Z
\ee
is defined through the partition function (\ref{4.43}), which is expressed through 
the renormalized Hamiltonian (\ref{4.41}). Keeping in mind the method of averaging 
over phase configurations \cite{Yukalov_5,Yukalov_6} described in the previous 
section, the partition function reads as
\be
\label{4.63}
 Z = \int_0^1 \left[\; \prod_f {\rm Tr}_{\cH_f} 
\int \exp\{ - \bt H_f(\xi_f) \} \; \cD\xi_f \; \right] \; dx \; .   
\ee
Expanding the exponential in powers of the Hamiltonian and using Theorem 1 of the 
previous section yields
\be
\label{4.64}
 \int \exp\{ - \bt H_f(\xi_f) \} \; \cD\xi_f = 
\exp \{ - \bt H_f(x_f) \} \; .
\ee
For the grand Hamiltonian (\ref{4.32}) specified in Eqs. (\ref{4.37}) and 
(\ref{4.40}), this gives
$$
H_f(x_f) = 
x_f \int \psi_f^\dgr(\br) \; \left[\; - \; \frac{\nabla^2}{2m} + U(\br) - 
\mu \; \right] \; \psi_f(\br) \; d\br \; +
$$
\be
\label{4.65}
+ \; \frac{1}{2} \; x_f^2 \int \psi_f^\dgr(\br) \; \psi_f^\dgr(\br') \; 
\Phi(\br-\br') \; \psi_f(\br') \; \psi_f(\br) \; d\br d\br' \; .
\ee
Then the partition function (\ref{4.63}) becomes
\be
\label{4.66}
Z = \int_0^1 \prod_f {\rm Tr}_{\cH_f} \; \exp\{ -\bt H_f(x_f) \} \; dx \; .
\ee

Introducing the notation
\be
\label{4.67}
\Om_f(x_f) \equiv - T \ln \; {\rm Tr}_{\cH_f} \; \exp\{ -\bt H_f(x_f) \}
\ee
results in the equality
\be
\label{4.68}
{\rm Tr}_{\cH_f} \; \exp\{ -\bt H_f(x_f) \} = 
\exp\{ -\bt\; \Om_f(x_f) \} \; .
\ee
Then the partition function (\ref{4.66}) takes the form
\be
\label{4.69}
Z = \int_0^1  \exp\{ -\bt \Om(x) \} \; dx \; ,
\ee
in which
\be
\label{4.70}
\Om(x) \equiv \sum_f \Om_f(x_f) \;  .
\ee
If the number of thermodynamic phases in the mixture is $\nu$, then the 
partition function (\ref{4.69}) reads as
\be
\label{4.71}
Z = \int_0^1 e^{-\bt\Om(x)} \; \dlt\left( \sum_f x_f -1 \right) \;
\prod_{f=1}^\nu dx_f \;  .
\ee

The thermodynamic potential (\ref{4.70}) is an extensive quantity that in the
thermodynamic limit is proportional to the number of particles in the system 
$N$. It is convenient to define the reduced quantity
\be
\label{4.72}
\om(x) \equiv \frac{1}{N} \; \Om(x) \;  ,
\ee
which in the thermodynamic limit is finite. Therefore the partition function 
(\ref{4.71}) can be presented as
\be
\label{4.73}
Z = \int_0^1 e^{-N\bt\om(x)} \; \prod_{f=1}^{\nu-1} dx_f \; .
\ee
Defining the absolute minimum
\be
\label{4.74}
\om(w) \equiv {\rm abs}\;\min_x \; \om(x) \;  ,
\ee
where the set of weights $w_f$ is denoted as
\be
\label{4.75}
w \equiv \{ w_f : \; f = 1,2, \ldots, \nu \} \; .
\ee
By the Laplace method for large $N \gg 1$, we obtain
$$
\int_0^1 e^{-N\bt\om(x)} \; \prod_{f=1}^{\nu-1} dx_f \simeq 
e^{-N\bt\om(w)} \; \prod_{f=1}^{\nu-1} \sqrt{ \frac{2\pi}{N\om_f''} } \; ,
$$
where
$$
\om_f'' \equiv \frac{\prt^2\om_f(w_f)}{\prt w_f^2} > 0 \; .
$$
Therefore the partition function (\ref{4.73}) becomes
\be
\label{4.76}
Z = e^{-\bt\Om(w)} \; 
\prod_{f=1}^{\nu-1} \sqrt{ \frac{2\pi}{N\om_f''} } \;   .
\ee
Since
$$
 \frac{\ln(N\om_f'')}{N} \simeq 0 \qquad ( N \ra \infty) \; ,
$$
we find the grand thermodynamic potential
\be
\label{4.77}
 \Om(w) \equiv N \om(w) = \sum_f \Om_f(w_f) \; ,
\ee
in which
\be
\label{4.78}
 \Om_f(w_f) = - T \ln \; {\rm Tr}_{\cH_f} \;\exp\{ -\bt H_f(w_f) \} \; .
\ee

The weights $w_f$, by their definition, are the geometric probabilities of the 
phases, and they enjoy the properties
\be
\label{4.79}
\sum_f w_f = 1 \; , \qquad 0 \leq w_f \leq 1 \; .
\ee
In agreement with Eq. (\ref{4.74}), these weights are the thermodynamic potential 
minimizers,
\be
\label{4.80}
 \Om(w) = {\rm abs} \; \min_x \; \Om(x) \; . 
\ee
The results of the present section can be summarized as a theorem.

\vskip 2mm 

{\bf Theorem 2}. {\it The grand thermodynamic potential of a heterophase system in 
the thermodynamic limit has the form}
\be
\label{4.81}
\Om(w) = - T \ln \; {\rm Tr}\; \exp( -\bt \widetilde H ) \; ,
\ee
{\it with the renormalized Hamiltonian}
\be
\label{4.82}
\widetilde H = \bigoplus_f \; H_f(w_f) \; .
\ee

In the case of Hamiltonian (\ref{4.65}), the terms of the direct sum (\ref{4.82}) 
are
$$
H_f(w_f) = w_f \int \psi_f^\dgr(\br) \; \left[ \; - \; 
\frac{\nabla^2}{2m} + U(\br) - \mu \; \right] \; \psi_f(\br) \; d\br \; +
$$
\be
\label{4.83}
+ \; \frac{w_f^2}{2} \int \psi_f^\dgr(\br) \; \psi_f^\dgr(\br') \;
\Phi(\br - \br') \; \psi_f(\br')\; \psi_f(\br) \; d\br d\br' \; .
\ee
The geometric probabilities of the phases composing the system are the minimizers 
of the thermodynamic potential. $\square$

\subsection{Observable Quantities of Mixture}

The operators of observables $\hat{A}(\xi)$ act on the Hilbert space (\ref{4.25}). 
The related observable quantities are the statistical averages of these operators,
\be
\label{4.84} 
\lgl \;\hat A \; \rgl = {\rm Tr} \int \hat\rho(\xi)\; \hat A(\xi) \;
\cD \xi \;  ,
\ee
where the trace is over $\widetilde{\mathcal{H}}$. Similarly to the number-of-particle 
operator (\ref{4.26}) and the energy operator (\ref{4.27}), the operators of 
observables are given by the direct sums
\be
\label{4.85}
 \hat A(\xi) = \bigoplus_f \; \hat A_f(\xi_f) \; ,
\ee
whose terms have the form
\be
\label{4.86}
\hat A_f(\xi_f) = \sum_{m=0}^\infty \int \xi_f(\br_1) \; \xi_f(\br_2) \ldots
\xi_f(\br_m) \; 
A_f(\br_1,\br_2,\ldots,\br_m) \; d\br_1 d\br_2\ldots d\br_m \;   .
\ee

The statistical average (\ref{4.84}) reads as
$$
 \lgl \; \hat A \; \rgl = \frac{1}{Z} 
\int_0^1 \sum_f \int {\rm Tr}_{\cH_f} \; \exp\{ -\bt H_f(\xi_f)\} \;
 \hat A_f(\xi_f) \; \times
$$
\be
\label{4.87}
\times \;
 \prod_{g(\neq f)} \; 
{\rm Tr}_{\cH_g} \; \exp\{ -\bt H_g(\xi_g)\} \; \prod_f \cD\xi_f \; dx \; .
\ee
The averaging over heterophase configurations \cite{Yukalov_5,Yukalov_6} is 
accomplished using Theorem 1. As a result, we get the expression 
$$
 \lgl \; \hat A \; \rgl = 
\int_0^1 {\rm Tr} \; \hat\rho(x) \; \hat A(x) \; dx =
$$
\be
\label{4.88}
= \int_0^1 {\rm Tr} \; \hat\rho(x) \; \hat A(x) \; 
\dlt \left( \sum_f x_f - 1\right) \; \prod_{f=1}^{\nu_f} \; dx_f =
\int_0^1 {\rm Tr} \; \hat\rho(x) \; \hat A(x) \; 
\prod_{f=1}^{\nu_f-1} \; dx_f \;  ,
\ee
with the operator of observable
\be
\label{4.89}
 \hat A(x) = \bigoplus_f \;\hat A_f(x_f)  
\ee
and the statistical operator 
\be
\label{4.90}
 \hat\rho(x) = \frac{1}{Z} \; \exp\{ - \bt H(x) \} \; ,
\ee
in which the effective grand Hamiltonian is
\be
\label{4.91}
 H(x) \equiv \bigoplus_f \; H_f(x_f) \; .
\ee

In view of the identity
\be
\label{4.92}
 {\rm Tr} \; \exp\{ - \bt H(x) \} = \exp\{ - N\bt \om(x) \} \;  ,
\ee
we can write
\be
\label{4.93}
 {\rm Tr} \; \hat\rho(x) \; \hat A(x) = 
\frac{\exp\{-N\bt\om(x)\}\overline A(x)}{\int_0^1\exp\{-N\bt\om(x)\} dx } \; ,
\ee
where we use the notation
\be
\label{4.94}
\overline A(x) \equiv 
\frac{{\rm Tr}\exp\{-\bt H(x)\}\hat A(x)}{{\rm Tr}\exp\{-\bt H(x)\} } \; .
\ee
Then the average (\ref{4.88}) can be represented as
\be
\label{4.95}
\int_0^1 {\rm Tr}\; \hat\rho(x) \; \hat A(x)\; dx = 
\frac{\int_0^1\exp\{-N\bt\om(x)\}\overline A(x)dx}
{\int_0^1\exp\{-N\bt\om(x)\}dx} \; .
\ee
  
In the thermodynamic limit, when $N \ra \infty$, we have
\be
\label{4.96}
\int_0^1 \exp\{-N\bt\om(x)\} \; \overline A(x) \; dx \simeq
\exp\{-N\bt\om(w)\} \; \overline A(w) \; \prod_{f=1}^{\nu_f-1}
\sqrt{\frac{2\pi}{N\om_f''} } \;  ,
\ee
where $\omega(w)$ is the absolute minimum of $\omega(x)$, in agreement with 
definition (\ref{4.74}). Therefore we come to the equality
\be
\label{4.97}
\lgl \; \hat A \;\rgl =  \overline A(w) = 
\frac{ {\rm Tr}\exp\{-\bt \widetilde H(w)\}\hat A(w) }{{\rm Tr}\exp\{-\bt \widetilde H(w)\} } \; ,
\ee
in which
\be
\label{4.98}
 \hat A(w) = \bigoplus_f \; \hat A_f(w_f) \;  ,
\ee        
with 
\be
\label{4.99}
\hat A_f(w_f) = \sum_{m=0}^\infty w_f^m \int A_f(\br_1,\br_2,\ldots,\br_m) \;
d\br_1 d\br_2 \ldots d\br_m \;  .
\ee
These results can be summarized as a theorem

\vskip 2mm

{\bf Theorem 3}. {\it Observable quantities of a heterophase system in the thermodynamic 
limit can be represented by the statistical averages}
\be
\label{4.100}
\lgl \; \hat A \;\rgl = 
{\rm Tr} \int \hat\rho(\xi) \; \hat A(\xi) \; \cD\xi = 
{\rm Tr}\; \hat\rho(w) \;\hat A(w) \; ,
\ee
{\it with the statistical operator}
\be
\label{4.101}
 \hat\rho(w) = \frac{1}{Z} \; \exp\{ - \bt\widetilde H(w)\} \;  ,
\ee
{\it partition function}
\be
\label{4.102}
 Z =  {\rm Tr}\; \exp\{ -\bt\widetilde H(w)\} \; ,
\ee
{\it and the renormalized grand Hamiltonian} $\tilde{H} = \tilde{H}(w)$. 
$\square$

\subsection{Statistics of Heterophase Systems}

For the convenience of the reader, let us summarize the main formulas describing 
a heterophase system \cite{Yukalov_5,Yukalov_6}. The latter is characterized by a 
renormalized Hamiltonian
\be
\label{4.103}
\widetilde H = \widetilde H(w) = \bigoplus_f \; H_f(w_f) \;  .
\ee
Similarly, the operators of observable quantities have the form
\be
\label{4.104}
\hat A(w) = \bigoplus_f \; \hat A_f(w_f) \;   .
\ee
For instance, the number-of-particle operator is
\be
\label{4.105}
 \hat N(w) = \bigoplus_f \; \hat N_f(w_f) \;   ,
\ee
where
\be
\label{4.106}
 \hat N_f(w_f) = w_f \int \psi_f^\dgr(\br) \; \psi_f(\br) \; d\br \; .
\ee

The observable quantities are given by the statistical averages
\be
\label{4.107}
\lgl \; \hat A \; \rgl = {\rm Tr}\; \hat\rho(w) \; \hat A(w) \;  ,
\ee  
with the statistical operator
\be
\label{4.108}
  \hat\rho(w) = \frac{1}{Z} \; \exp\{ -\bt \widetilde H(w) \} =
\bigotimes_f \; \hat\rho_f(w_f)  
\ee
and the partition function
\be
\label{4.109}
Z = {\rm Tr} \;  \exp\{ -\bt \widetilde H(w) \} = \prod_f \; Z_f \;  ,
\ee
where
\be
\label{4.110}
\hat\rho_f(w_f) = \frac{1}{Z_f} \; \exp\{ -\bt H_f(w_f) \} \; , \qquad
Z_f = {\rm Tr}_{\cH_f} \;  \exp\{ -\bt H_f(w_f) \} \; .
\ee
Thus the observable quantities become
\be
\label{4.111}
\lgl \; \hat A \; \rgl = \sum_f \; \lgl \; \hat A_f \; \rgl  \; ,
\ee
with
\be
\label{4.112}
\lgl \; \hat A_f \; \rgl \equiv  
{\rm Tr}_{\cH_f}\; \hat\rho_f(w_f) \hat A_f(w_f) \; .
\ee

The grand thermodynamic potential reads as
\be
\label{4.113}
 \Om = - T\ln \; {\rm Tr}\; \exp(-\bt\widetilde H) =
\sum_f \Om_f(w_f) \; ,
\ee
where
\be
\label{4.114}
 \Om_f(w_f) = - T\ln \; {\rm Tr}_{\cH_f}\; \exp\{-\bt H_f(w_f)\} \; .
\ee
The number of particles in an $f$-th phase can be found from the derivatives
\be
\label{4.115}
 N_f(w_f) = -\; \frac{\prt\Om_f(w_f)}{\prt\mu} = - \; 
\left\lgl \; \frac{\prt H_f(w_f)}{\prt\mu} \; \right\rgl \; .
\ee

The particle density of an $f$-th phase is
\be
\label{4.116}
 \rho_f \equiv \frac{N_f(w_f)}{V_f} = 
\frac{1}{V} \int \lgl \; \psi_f^\dgr(\br) \; \psi_f(\br) \; \rgl \; d\br \;  ,
\ee
where we take into account that the geometric probability of a phase has the form
\be
\label{4.117}
 w_f = \frac{V_f}{V} \;  .
\ee
The average density of all particles in the system is
\be
\label{4.118} 
\rho \equiv \frac{N}{V} = \frac{1}{V} \sum_f N_f(w_f) = 
\sum_f w_f \; \rho_f \;  .
\ee

The phase probabilities are the minimizers of the thermodynamic potential 
$\Om$. This implies that the system state corresponds to the minimal between 
the thermodynamic potential of the mixture $\Om=\Om(w)$, with the phase 
probabilities defined by the equations
\be
\label{4.119}
 \frac{\prt\Om(w)}{\prt w_f} = 0 \; , \qquad  
\frac{\prt^2\Om(w)}{\prt w_f^2} > 0  \qquad \left(\sum_f w_f = 1\right) \; ,
\ee
and any of the pure states, so that
\be
\label{4.120}
 \Om = \min\{ \Om(w), \; \Om_f(1) : \; f = 1,2,\ldots \} \; .
\ee
In the same way, the phase probabilities can be defined as the minimizers of the 
system free energy $F=\Om+\mu N$. Then the most stable state corresponds to the 
free energy 
\be
\label{4.121}
 F = \min\{ F(w), \; F_f(1) : \; f = 1,2,\ldots \}  
\ee
providing the minimum for the set of the mixture free energy $F = F(w)$ and of the 
free energies $F_f(1)$ of possible pure phases occupying the whole system.

\subsection{Interphase Surface States}

It is important to stress that thermodynamic potentials as well as observables 
of a heterophase system are not simple linear combinations of the corresponding 
quantities for pure phases. Therefore all effects, related to the existence of 
surfaces separating the phases of the mixture, are taken into account. Actually, 
the introduction of separating surfaces acquires the meaning only on the 
thermodynamic level \cite{Ono_227,Rusanov_228,Rusanov_229,Rusanov_244} and does 
not contradict the formal additivity of operators (\ref{4.104}) at the operator 
level. The definition of surface effects can be clearly illustrated by using the
local-density approximation \cite{Kjelstrup_245,Bedeaux_246}. 

Let us start with the definition of the surface free energy that is given as the 
difference 
\be
\label{4.122}
F_{sur} \equiv F - F_G
\ee
between the real free energy of a heterophase system $F$ and the free energy of 
the Gibbs mixture 
\be
\label{4.123}
F_G = \sum_f F_f^G \;  .
\ee
Recall that the Gibbs phase mixture is a mixture of uniform phases occupying 
each its part of the volume $V_f$ of the whole system. Thus the surface free 
energy is the excess free energy caused by the nonuniformity of the system due 
to the coexistence of different phases,
\be
\label{4.124}
 F_{sur} =  F - \sum_f F_f^G \;  .
\ee

The free energy $F_f^G$ of a phase occupying the volume $V_f$ and the free energy 
$F_f(1)$ of this pure phase occupying the total system volume $V$ are connected by 
the relation
$$
 \frac{F_f^G}{V_f} = \frac{F_f(1)}{V} \; ,
$$  
which yields
\be
\label{4.125}
 F_f^G = w_f F_f(1) \; .
\ee
Hence the surface free energy is
\be
\label{4.126}
 F_{sur} = F - \sum_f w_f F_f(1) \;  .
\ee
The free energy of a heterophase system reads as
\be
\label{4.127}
 F = \sum_f F_f(w_f) \; .
\ee
Thus we come to the surface free energy
\be
\label{4.128}
 F_{sur} = \sum_f \; [\; F_f(w_f) - w_f F_f(1) \; ] \; .
\ee
Since the system free energy $F$ is not a linear combination of the energies 
$F_f(1)$, the surface free energy (\ref{4.128}) is not zero. 

Moreover, if a heterophase system is in an absolutely stable thermodynamic state, 
then the surface free energy is non-positive. This follows from the inequalities 
\be
\label{4.129}
 F = {\rm abs}\; \min \; F(w) \leq \min_f \; F_f(1) \; , \qquad
 \min_f \; F_f(1) \leq \sum_f w_f F_f(1) \; ,
\ee         
from where
\be
\label{4.130}
F_{sur} \leq F - \min_f \; F_f(1) \leq 0 \; .
\ee
Analogously, it is straightforward to introduce the surface grand potential or 
other thermodynamic potentials.  

In the same way, it is possible to define the excessive term of an observable 
describing the difference between the corresponding operator average (\ref{4.111}) 
in the case of a heterophase system and the sum of the related observables 
\be
\label{4.131}
 A_f(1) \equiv \lim_{w_f\ra 1} \; \lgl \; \hat A_f\; \rgl  
\ee
for the Gibbs mixture. Similarly to definition (\ref{4.126}), the excessive part 
of an observable, due to the phase separation in the space, is
\be
\label{4.132}
 \lgl \; \hat A \; \rgl_{sur} \equiv \lgl \; \hat A \; \rgl \; - \;
\sum_f w_f A_f(1) \; .
\ee
This, in view of Eq. (\ref{4.111}), results in the surface observable
\be  
\label{4.133}
  \lgl \; \hat A \; \rgl_{sur} = \sum_f \; \left[\; 
\lgl \; \hat A_f \; \rgl \; - \; w_f A_f(1) \; \right] \; .
\ee
The set of all surface observables (\ref{4.133}) forms the {\it interphase 
surface state}.

\subsection{Geometric Phase Probabilities}

In the description of heterophase systems, as compared to homogeneous systems, 
there appears a novel quantity, the geometric probability of phases 
\be
\label{4.134}
w_f = \frac{V_f}{V} \qquad ( f = 1,2,\ldots )
\ee
showing the fraction of the system volume occupied by the related thermodynamic 
phases. This quantity plays the role of an additional order parameter 
characterizing the system \cite{Yukalov_5,Yukalov_236,Yukalov_247,Yukalov_248}. 
The qualitative change of this probability parameter signifies that the system 
experiences a kind of a phase transition. The point of the standard phase 
transition is defined by the qualitative variation of a phase order parameter 
\cite{Kubo_1,Ruelle_208,Brout_219,Sinai_221,Yukalov_249} or of order indices 
\cite{Yukalov_212,Yukalov_213,Coleman_250,Coleman_251}. In addition to the usual 
phase transitions, there can arise the point where a phase probability changes 
as follows. For instance, there is a phase probability that below some point, say 
below a temperature $T_n$, equals to one, hence there exists a pure $f$-th phase,
\be
\label{4.135}
w_f(T) = 1 \qquad ( T < T_n ) \;  .
\ee
However, above this point, the admixture of at least one other phase appears, 
so that
\be
\label{4.136}
w_f(T) < 1 \qquad ( T > T_n ) \;   .
\ee
The point where a pure phase becomes mixed, due to the arising nuclei of another 
phase, can be named the {\it nucleation point}. In the above example, it is the 
{\it nucleation temperature}. 

The nucleation at the point $T_n$ can be either continuous, when
\be
\label{4.137}
 | \; w_f(T_n + 0 ) - w_f(T_n-0) \; | = 0  
\ee
or discontinuous, such that
\be
\label{4.138}
 | \; w_f(T_n + 0 ) - w_f(T_n-0) \; | > 0 \;  .
\ee
Contrary to this, in the case of a phase transition between two pure phases, 
the process of nucleation is discontinuous, with the phase probability jumping 
between $1$ and $0$, and the nucleation point coincides with the phase transition 
point.    
  
Phase probabilities enter in the expressions of observable quantities as well 
as in thermodynamic characteristics. For example, the number of particles in 
an $f$-th phase is
\be
\label{4.139}
N_f = w_f \int \lgl \; \psi_f^\dgr(\br) \; \psi_f(\br) \; \rgl \; d\br \; .
\ee
This defines the phase fraction 
\be
\label{4.140}
 n_f \equiv \frac{N_f}{N}  
\ee
satisfying the normalization
\be
\label{4.141}
\sum_f n_f = 1 \; , \qquad 0 \leq n_f \leq 1 \; .
\ee

The density of an $f$-th phase is
\be
\label{4.142}
\rho_f \equiv \frac{N_f}{V_f} = \frac{n_f}{w_f}\; \rho \; ,
\ee
where $\rho$ is the total average density
\be
\label{4.143}
\rho \equiv \frac{N}{V} = \sum_f{w_f}\; \rho_f \; .
\ee
Then relation (\ref{4.139}) becomes
\be
\label{4.144}
 \int \; \lgl \; \psi_f^\dgr(\br) \; \psi_f(\br) \rgl \; d\br =
\frac{\rho_f}{\rho} \; N \; .
\ee

If the phases possess the same density, while being distinguished by other 
properties, say magnetic, electric, or structural, then the phase probabilities 
and phase fractions coincide,
\be
\label{4.145}
 w_f =n_f \qquad ( \rho_f = \rho ) \; ,
\ee
which follows from equation (\ref{4.142}). In that case, expression (\ref{4.144}) 
reduces to
\be
\label{4.146}
 \int \; \lgl \; \psi_f^\dgr(\br) \; \psi_f(\br) \rgl \; d\br = N 
\qquad
 ( \rho_f = \rho ) \;  .
\ee

\section{Models of Heterophase Systems}

In the present section, some typical models of heterophase systems are considered. 
We keep in mind mesoscopic phase mixtures whose theory is exposed in the previous 
Chapter 4. Numerous examples of such systems are listed in Chapter 3. We start with 
the model that, in particular, describes ferromagnets with paramagnetic fluctuations. 
However this model is generic for a large class of order-disorder systems, because 
of which it is described in detail, since many other models are treated analogously.

\subsection{Heterophase Heisenberg Model}

A typical model for ferromagnets is the Heisenberg model \cite{Heisenberg_252}. 
To take into account paramagnetic heterophase fluctuations, we follow the theory 
of Chapter 4 and, averaging out phase configurations, we obtain the effective 
Hamiltonian describing the mixture of ferromagnetic and paramagnetic phases
\be
\label{5.1}
 \widetilde H = H_1 \; \bigoplus \; H_2 \; ,   
\ee
with the phase-replica Hamiltonians
\be
\label{5.2}
H_f = \frac{N}{2} \; w_f^2 U - 
w_f^2 \sum_{i\neq j} J_{ij} \; \bS_{if} \cdot \bS_{jf} \;  .
\ee
Here $U$ is a parameter characterizing the strength of effective direct interactions
between the particles forming the system, while $J_{ij}$ is an exchange interaction 
potential. In the standard Heisenberg model, the constant parameter $U$ is usually 
omitted, since it does not influence the thermodynamics of the system. However for 
a heterophase system, the first term, with the parameter $U$, contains the phase 
probability $w_f$ depending on thermodynamic variables, hence it cannot be neglected. 
Summation is over the lattice cites enumerated by the index $j=1,2,\ldots,N$. The 
exchange interaction $J_{ij}$ is of ferromagnetic type, which implies that $J_{ij}>0$. 
The operator ${\bf S}_{jf}$ is a representation of the spin operator for the $f$-th 
phase, acting on $\mathcal{H}_f$, and located at the lattice cite $j$. The lattice is 
assumed to be ideal. In the present section, we keep in mind spin $S = 1/2$.  
  
The phases are distinguished by the observable values of the average spin. The phase 
labeled by $f=1$ is assumed to correspond to ferromagnetic state, while that labeled 
by $f=2$, to paramagnetic state, so that
\be
\label{5.3}
 \left\lgl \; \frac{1}{N} \sum_j \bS_{j1} \; \right\rgl \neq 0 \; , 
\qquad
 \left\lgl \; \frac{1}{N} \sum_j \bS_{j2} \; \right\rgl = 0 \;  .
\ee
It is convenient to introduce the relative average spins
\be
\label{5.4}
{\bf s}_f \equiv \frac{1}{NS} \sum_j \; \lgl \; \bS_{jf} \; \rgl =
\frac{2}{N} \sum_j \; \lgl \; \bS_{jf} \; \rgl
\ee
and to distinguish the phases by the order parameters
\be
\label{5.5}
 s_1 \neq 0 \; , \qquad s_2 = 0 \qquad 
( s_f \equiv |\; {\bf s}_f \; | ) \; .
\ee

In the mean-field approximation
$$
 \bS_{if} \cdot \bS_{jf}  = 
\lgl \; \bS_{if} \; \rgl \cdot \bS_{jf}  + 
\bS_{if} \cdot \lgl \; \bS_{jf} \; \rgl - 
\lgl \; \bS_{if} \; \rgl \lgl \; \bS_{jf} \; \rgl \; ,
$$
Hamiltonian (\ref{5.2}) becomes
\be
\label{5.6}
 H_f = \frac{N}{2} \; w_f^2 \; \left( U + \frac{J}{2}\; s_f^2 \right) -
J w_f^2 \; {\bf s}_f \cdot \sum_j \bS_{jf} \;  ,
\ee
in which
\be
\label{5.7}
 J \equiv \frac{1}{N} \sum_{i\neq j} J_{ij} > 0 \; .
\ee
The free energy of the mixture is
\be
\label{5.8}
 F = F_1 + F_2 \; ,
\ee
where
\be
\label{5.9}
F_f = - \; \frac{T}{N} \; \ln \; {\rm Tr}\; e^{-\bt H_f} = 
\frac{1}{2} \; w_f^2 \; \left( U + \frac{J}{2} \; s_f^2 \right) -
T \ln \; \left[ 2 \cosh\left( \frac{J w_f^2 s_f}{2T}\right)\right] \; .
\ee

The order parameters can be found either directly from definition (\ref{5.4}) or 
by minimizing the free energy (\ref{5.8}) with respect to $s_f$. Both ways give 
the same equation
\be
\label{5.10}
s_f = \tanh \left( \frac{J w_f^2 s_f}{2T}\right) \; .
\ee
This equation possesses a nonzero solution as well as the zero solution, in 
agreement with condition (\ref{5.5}). In what follows, we measure the free 
energies and temperature in units of $J$ and use the notation
\be
\label{5.11}
 u \equiv \frac{U}{J} \; .
\ee
For the ferromagnetic state of the mixture, we have
\be
\label{5.12}
F_1 = \frac{1}{2} \; w_1^2 \left( u + \frac{1}{2}\; s_1^2\right) -
T\ln \; \left[ 2\cosh\left( \frac{w_1^2 s_1}{2T}\right) \right] \; ,
\ee
while for the paramagnetic state,
\be
\label{5.13}
F_2 = \frac{1}{2}\; w_2^2 u - T \ln 2 \;  .
\ee
Minimizing the free energy (\ref{5.8}) with respect to $w_f$, under the 
normalization condition $w_1+w_2=1$, we find the probability of the ferromagnetic 
phase
\be
\label{5.14}
 w_1 = \frac{2u}{4u-s_1^2} \qquad ( 4u >s_1^2 ) \; .
\ee
For simplicity, below we use the notation
\be
\label{5.15}
 w_1 \equiv w \; , \qquad w_2 = 1- w \; .
\ee

A state is stable, provided it has the minimal free energy and satisfies 
stability conditions. For this purpose, we compare the free energy (\ref{5.8}) 
of the mixed state, which takes the form
\be
\label{5.16}
F = \left( w^2 - w + \frac{1}{2} \right) u + \frac{1}{4} \; w^2 s_1^2 -
T \ln \; \left[ 4\cosh\left( \frac{w^2 s_1}{2T}\right) \right] \; ,
\ee
under the order parameter
\be
\label{5.17}
 s_1 = \tanh \left( \frac{w^2 s_1}{2T}\right) \; ,   
\ee
with probability (\ref{5.14}), the free energy of the pure ferromagnetic 
phase
\be
\label{5.18}
 F_{fer} \equiv F_1(w_1=1) = \frac{1}{2}\; u + \frac{1}{4} \; s^2 -
T \ln\; \left[ 2\cosh\left( \frac{s}{2T}\right) \right] \; ,
\ee
under the order parameter
\be
\label{5.19}
 s = \tanh \left( \frac{s}{2T}\right) \; ,
\ee
and the free energy of the pure paramagnetic phase
\be
\label{5.20}
 F_{par} \equiv F_2(w_2=1) = \frac{1}{2} \; u - T \ln 2 \; .
\ee

Also, we compare the free energy $F = F(w)$, given by Eq. (\ref{5.16}), under 
probability (\ref{5.14}) and the order parameter (\ref{5.17}), with the free 
energy 
\be
\label{5.21}
 F_0 \equiv F \left( \frac{1}{2} \right) = 
\frac{1}{4} \; u - T \ln 4 \;
\ee
of the degenerate paramagnetic state, with $w=1/2$ and $s_1=0$.
 
The stability conditions for the mixed state, characterized by the free energy 
(\ref{5.16}), are as follows. The state is an extremum, provided that the first 
derivatives are zero,
$$
\frac{\prt F}{\prt w} = w \left( 2u \; - \;
\frac{1}{2}\; s_1^2\right) - u = 0 \; ,
\qquad
\frac{\prt F}{\prt s_1} = 
\frac{w^2}{2} \; \left[\; s_1 - 
\tanh\left( \frac{w^2 s_1}{2T}\right) \; \right] = 0 \; .
$$
The necessary and sufficient condition for the potential $F$ to be minimal is the 
positivity of the Hessian matrix for all the variables $s_1$ and $w$. The elements 
of the Hessian matrix are
$$
\frac{\prt^2 F}{\prt w^2} = 2u \; - \; \frac{1}{2} \; s_1^2 \; - \;
\frac{w^2 s_1^2}{T} \; \left( 1 - s_1^2\right) \; ,
$$
$$
\frac{\prt^2 F}{\prt w\prt s_1} = -\; 
\frac{w^3 s_1}{2T} \; \left( 1 - s_1^2\right) \; ,
\qquad 
\frac{\prt^2 F}{\prt s_1^2} =  
\frac{w^2}{2} \; - \; \frac{w^4}{4T} \;\left( 1 - s_1^2\right) \; .
$$
The matrix is positive when all its principal minors are positive, which yields 
the stability conditions
$$
\frac{\prt^2 F}{\prt w^2} \; > \; 0 \; , 
\qquad
\frac{\prt^2 F}{\prt w^2} \; \cdot \; \frac{\prt^2 F}{\prt s_1^2} \; - \;
\left( \frac{\prt^2 F}{\prt w\prt s_1}\right)^2 \; > \; 0 \; .
$$

The behavior of heterophase ferromagnets has been studied in Refs. 
\cite{Yukalov_5,Yukalov_6,Shumovsky_253,Shumovsky_254,Shumovsky_255,Shumovsky_256,
Shumovsky_257,Shumovsky_258}. The stable state is described by the minimal 
thermodynamic potential among $F$, $F_{fer}$, $F_{par}$, and $F_0$. Generally, the 
potential $F$ can have two branches and, respectively, two types of solutions for 
$w$ and $s_1$ as is shown in Fig. 4. We have to choose the branch that is minimal. 
Overall, there exist the following qualitatively different types of behavior 
depending on the parameter $u$.

\begin{figure}[ht]
\centerline{
\hbox{ \includegraphics[width=7.5cm]{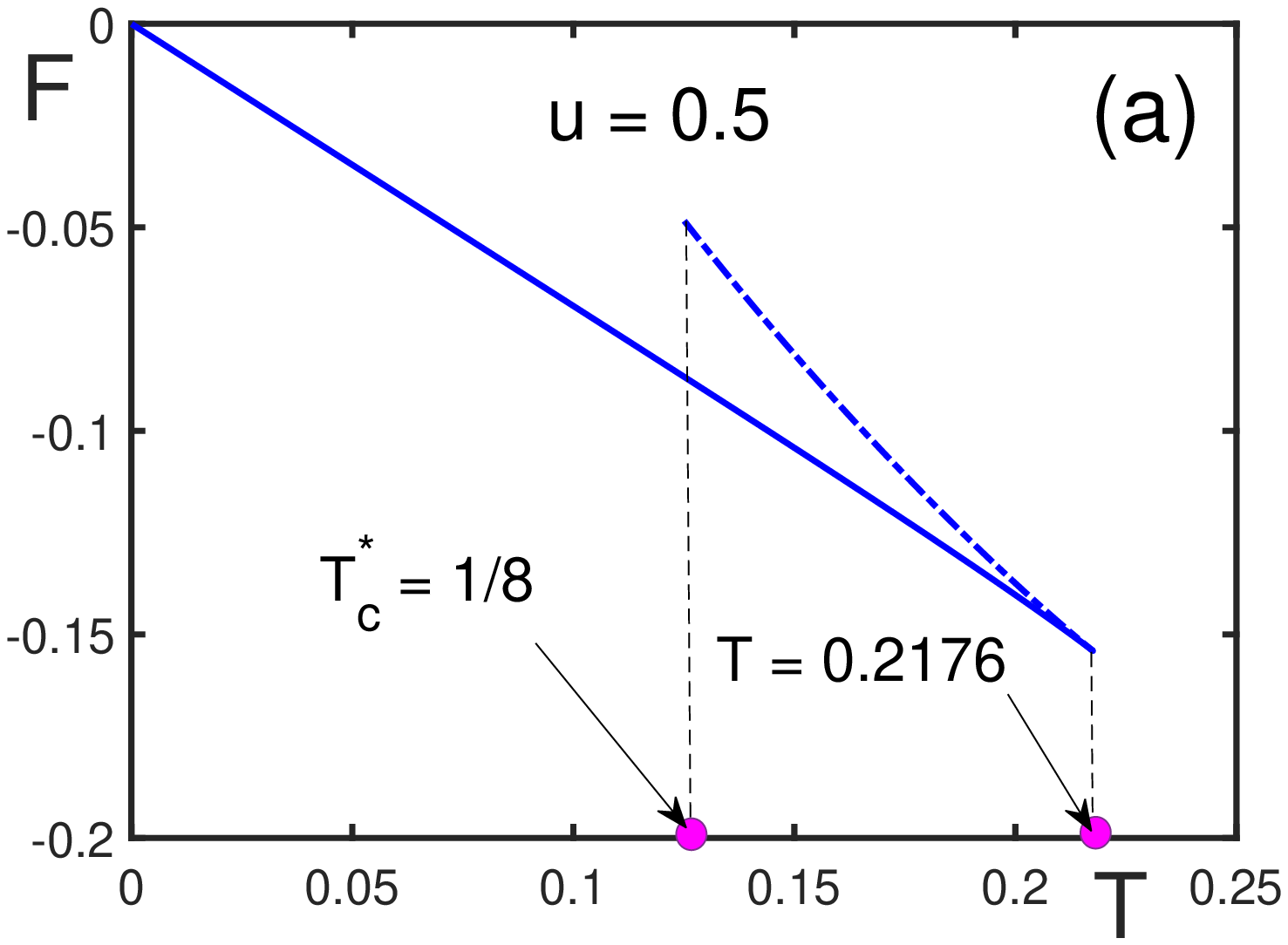} \hspace{1cm}
\includegraphics[width=7.5cm]{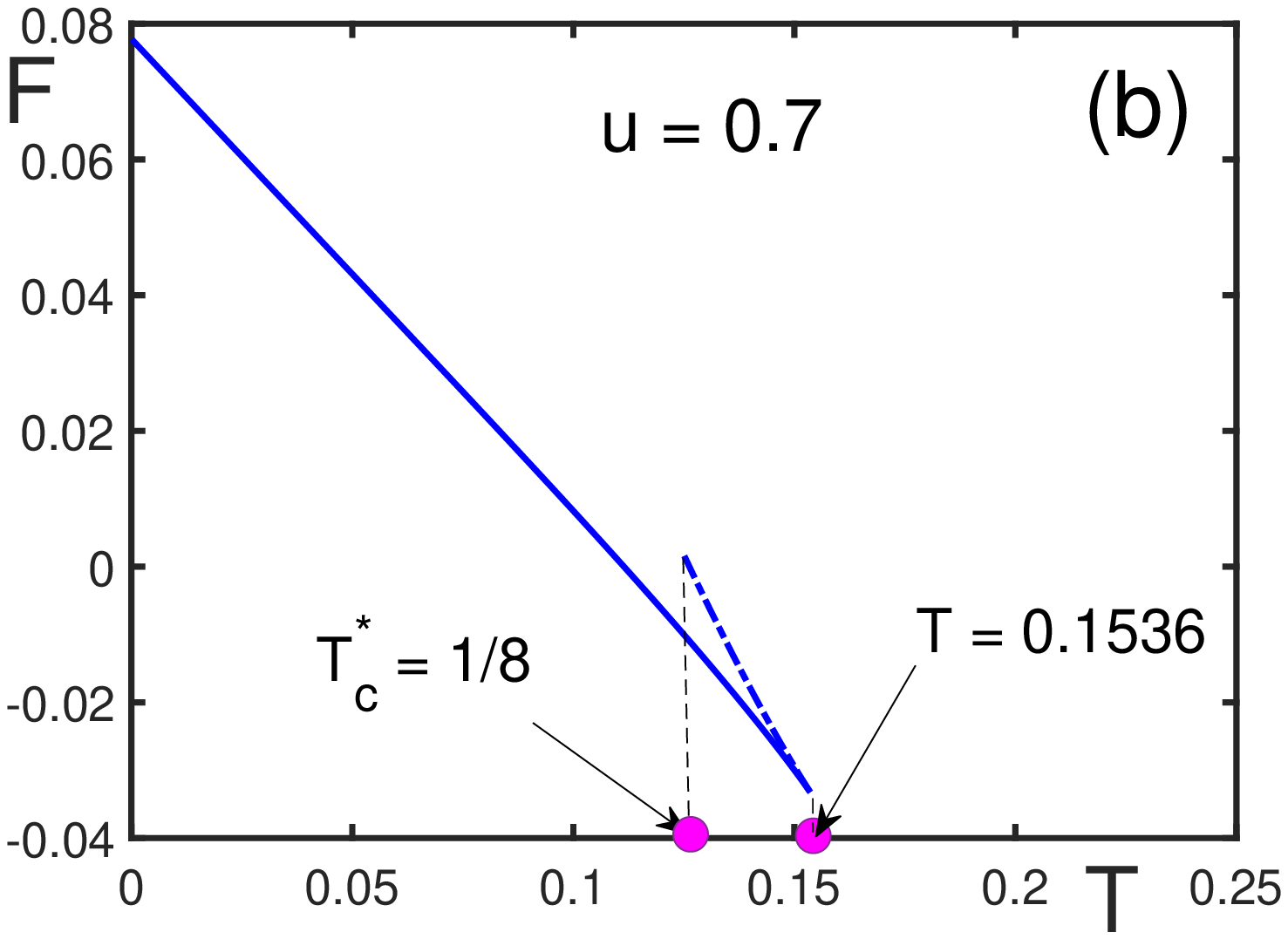}  } }
\vspace{12pt}
\centerline{
\hbox{ \includegraphics[width=7.5cm]{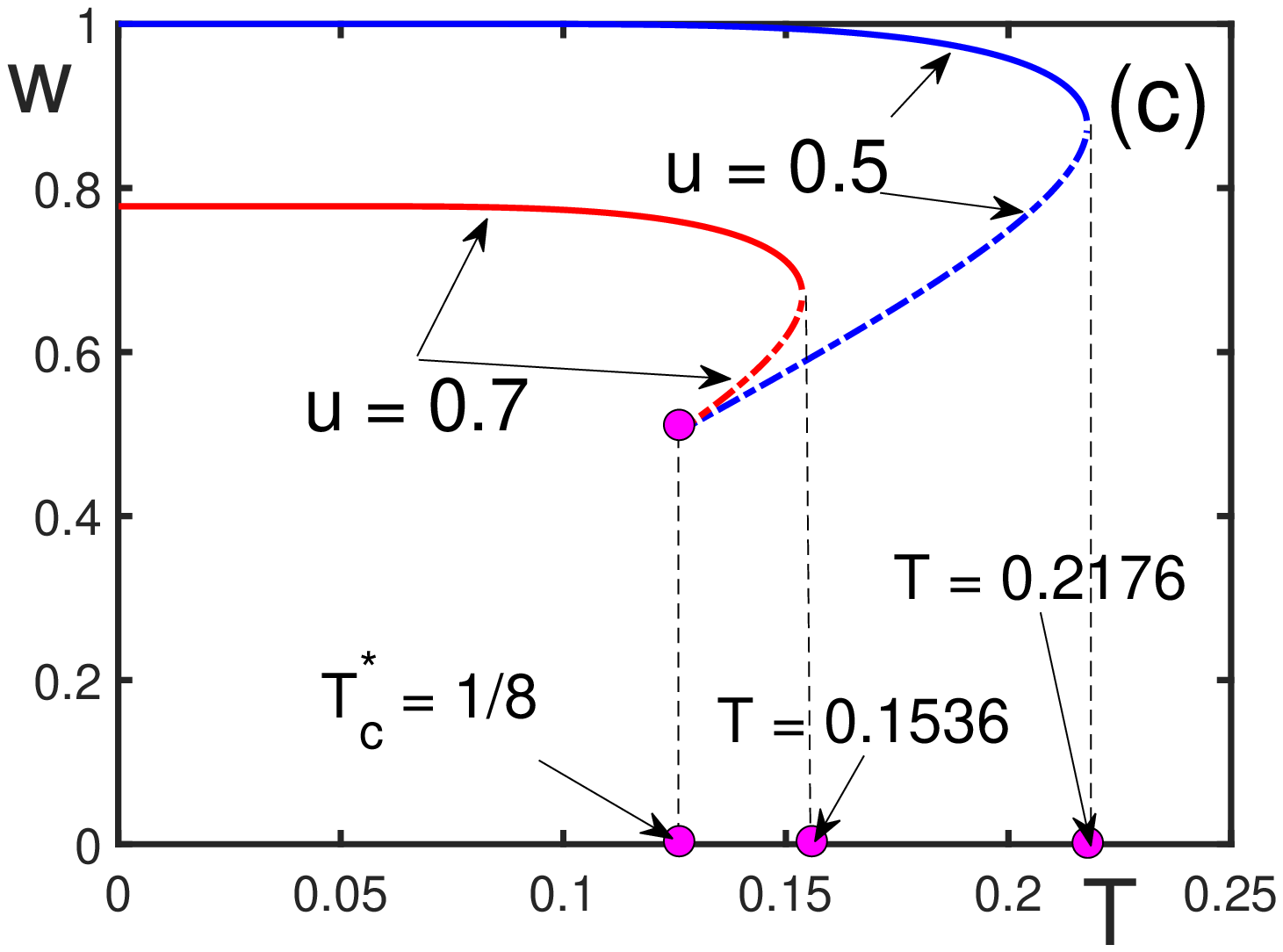} \hspace{1cm}
\includegraphics[width=7.5cm]{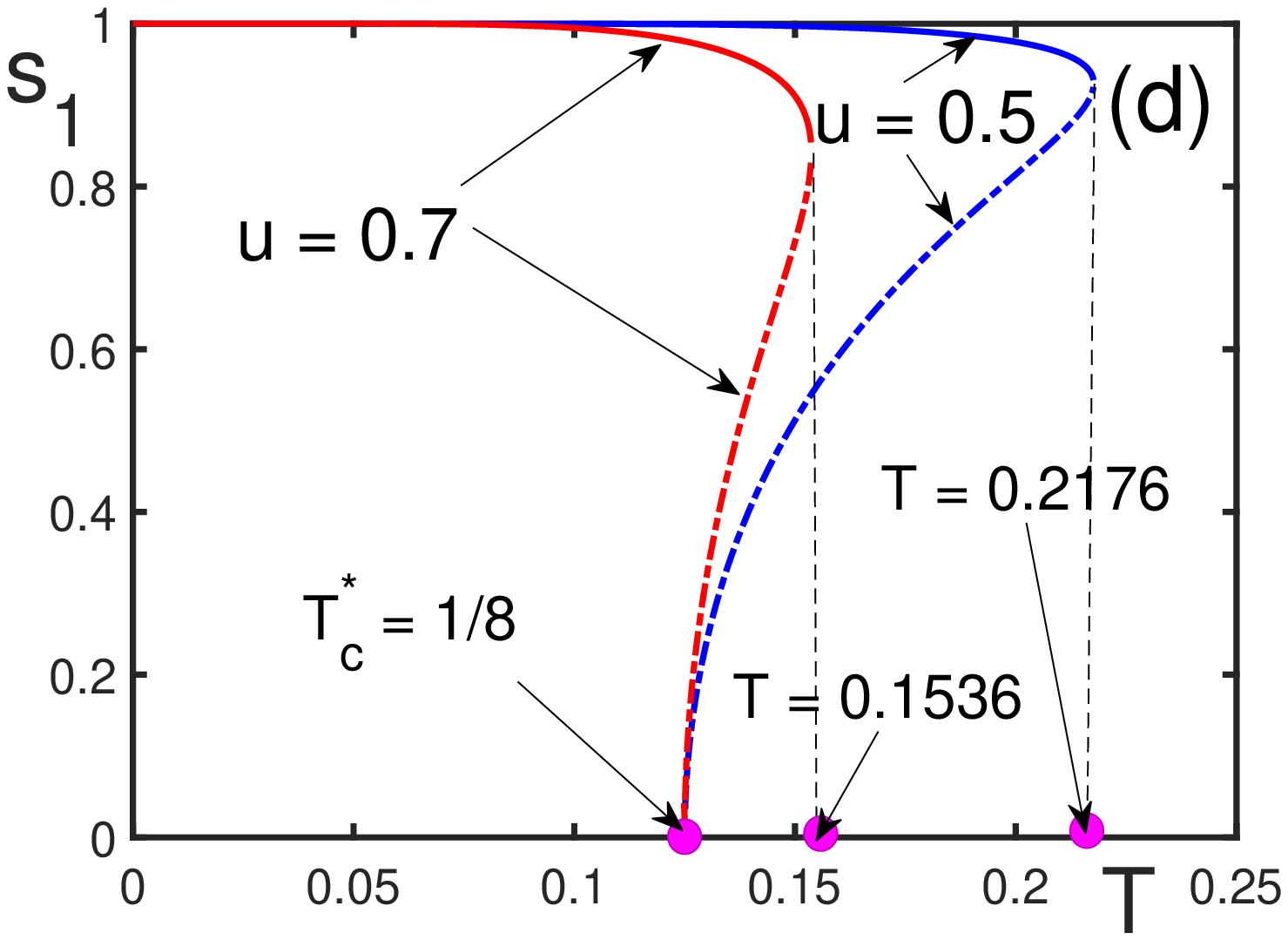} } }
\caption{Two branches of solutions for the thermodynamic potential $F$ as a
function of dimensionless temperature $T$ for the parameters:
(a) $u=0.5$ and (b) $u=0.7$. (c) The probability $w$ of the thermodynamic phase. 
(d) the order parameter $s_1$ as a function of temperature. 
The stable branch is shown by solid line and the unstable one is shown by 
dashed-dotted line. 
}
\label{fig:Fig.4}
\end{figure}

\vskip 2mm
(i) $u\leq 0$. No stable heterophase states exist. Heterophase ferromagnet can 
only be metastable. Below the critical temperature $T_c=1/2$, the pure ferromagnetic 
phase, with $w_1=1$, is absolutely stable. At the critical temperature $T_c=1/2$ 
the system becomes paramagnetic through the phase transition of second order. The 
behavior of the thermodynamic potentials $F_{fer}$ and $F_{par}$, the probability 
of the ferromagnetic fraction $w$, and of the order parameter $s$ as functions 
of temperature, are shown in Fig. 5.  

\begin{figure}[ht]
\centerline{
\hbox{ \includegraphics[width=7.5cm]{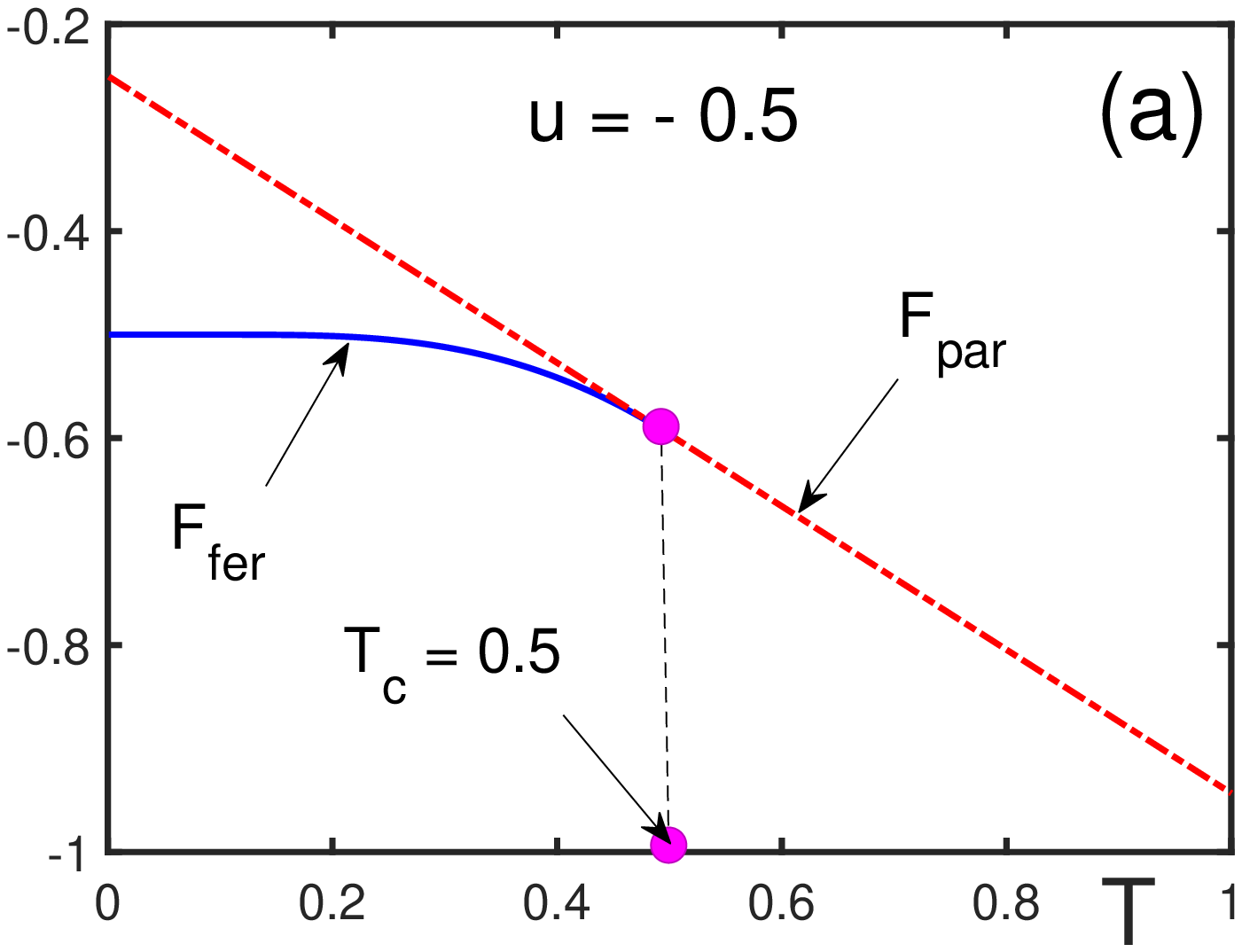}  } }
\vspace{12pt}
\centerline{
\hbox{ \includegraphics[width=7.5cm]{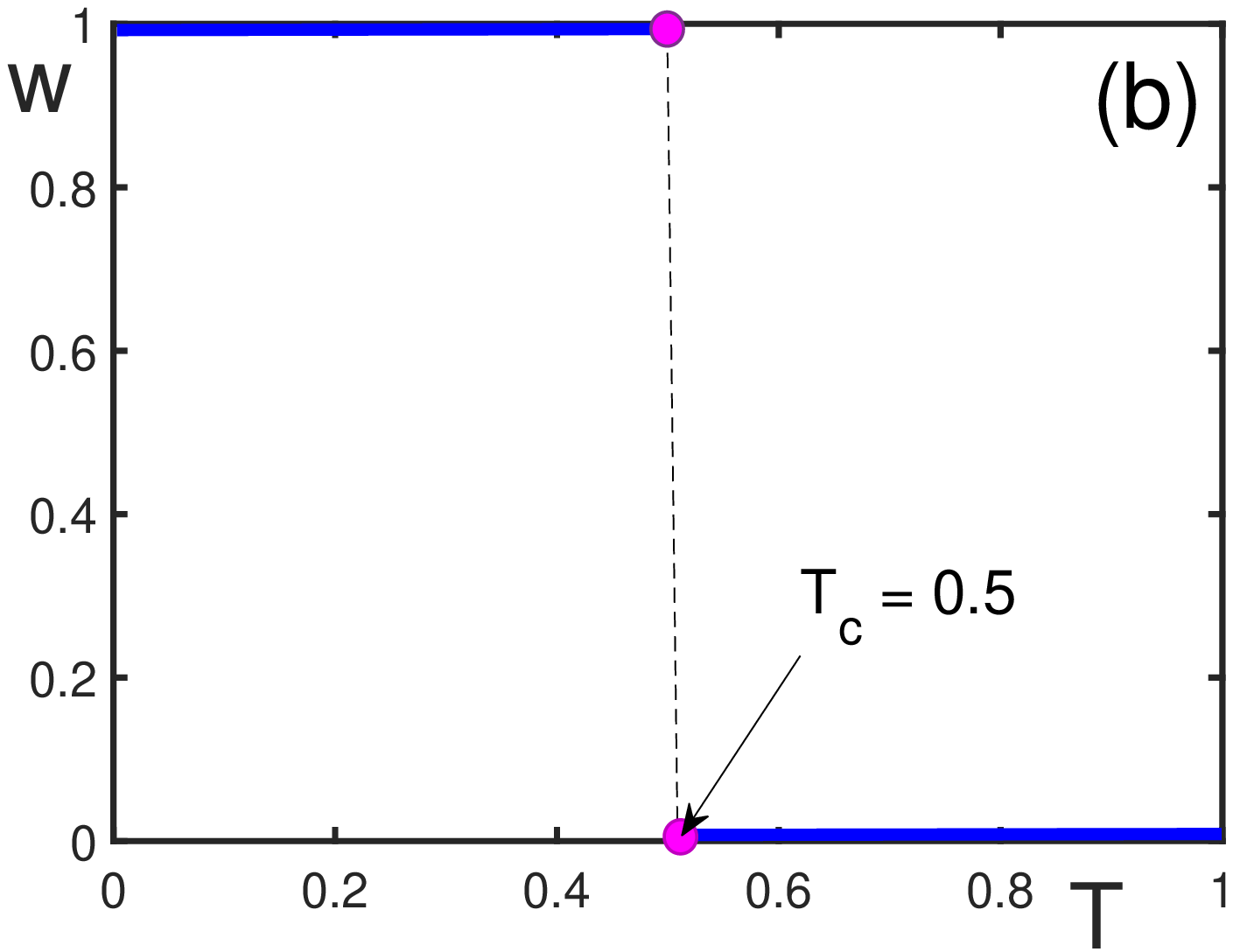} \hspace{1cm}
\includegraphics[width=7.5cm]{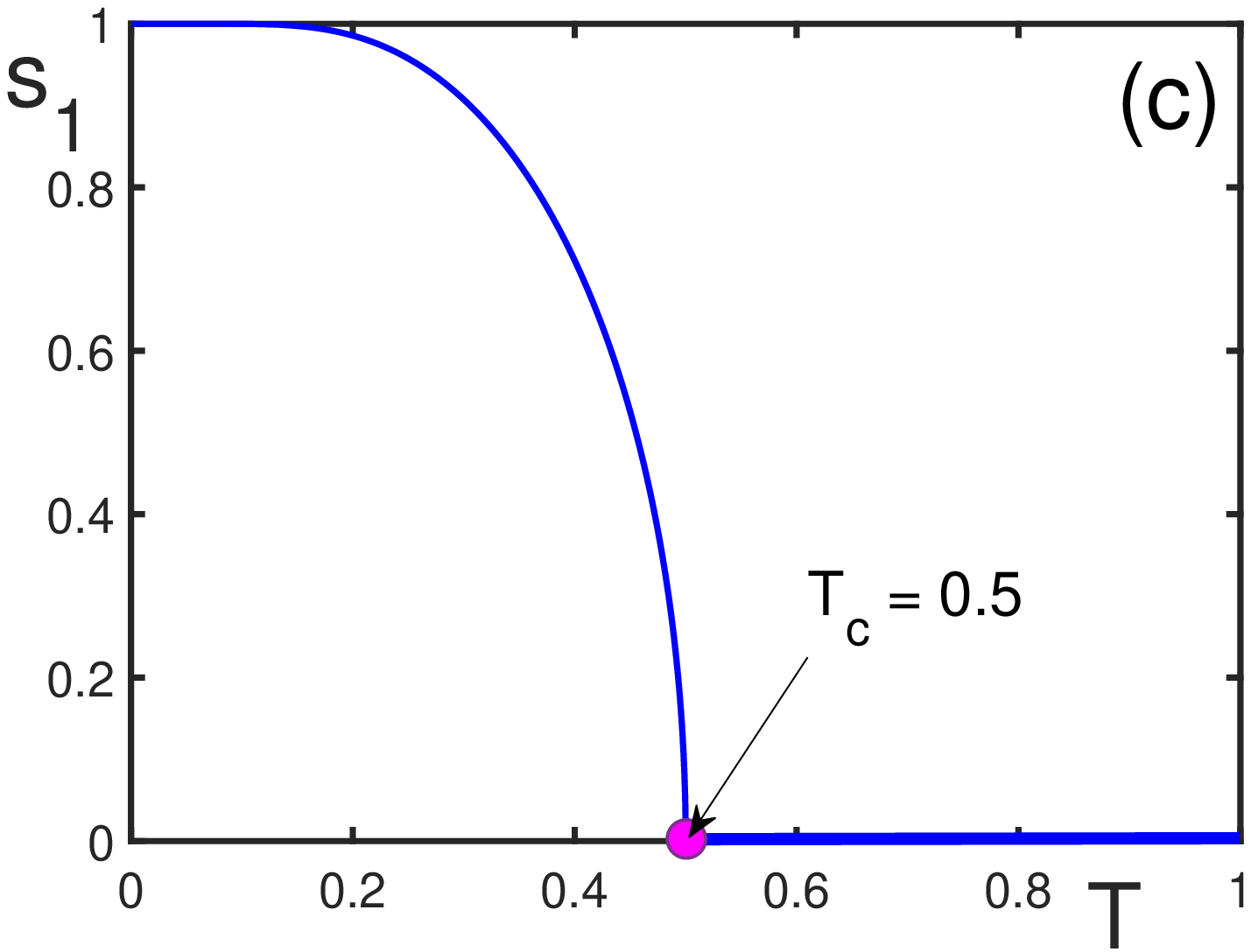} } }
\caption{Thermodynamic potentials $F_{fer}$ and $F_{par}$, the probability 
$w$ of the ferromagnetic phase, and the order parameter $s_1$ as functions 
of temperature $T$ for $u=-0.5$. Second order phase transition takes place 
at $T_c=0.5$.
}
\label{fig:Fig.5}
\end{figure}
 
\vskip 2mm
(ii) $0<u<0.5$. At low temperature, the free energy $F_{fer}$ is lower than $F_0$ 
up to the temperature $T_0$, where the first-order phase transition occurs from the pure 
ferromagnetic phase to the degenerate nonmagnetic phase with the free energy $F_0$. 
The transition temperature is in the interval $0.090 < T_0 < 0.181$. The overall 
behavior is presented in Fig. 6.   

\begin{figure}[ht]
\centerline{
\hbox{ \includegraphics[width=7.5cm]{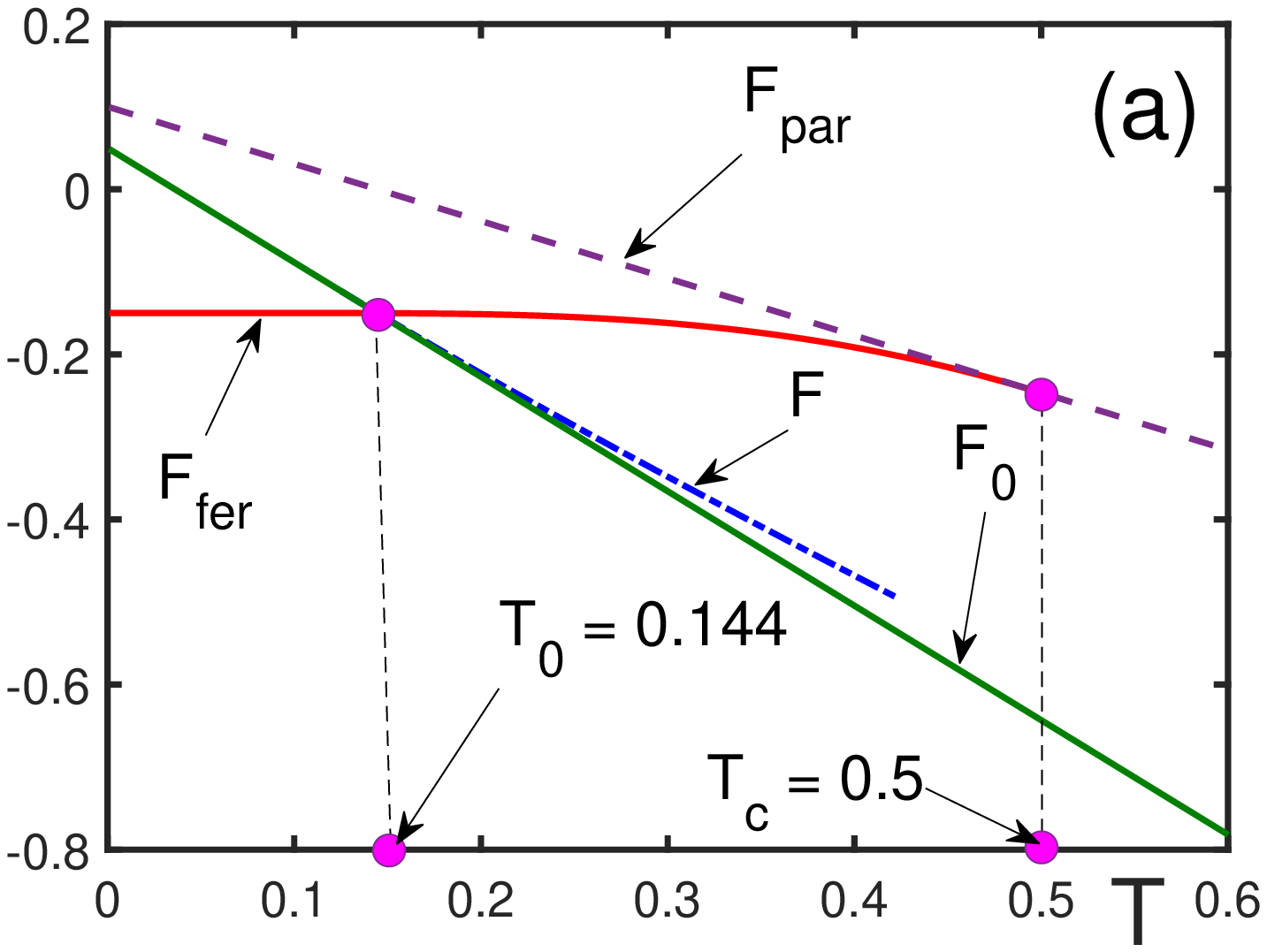}  } }
\vspace{12pt}
\centerline{
\hbox{ \includegraphics[width=7.5cm]{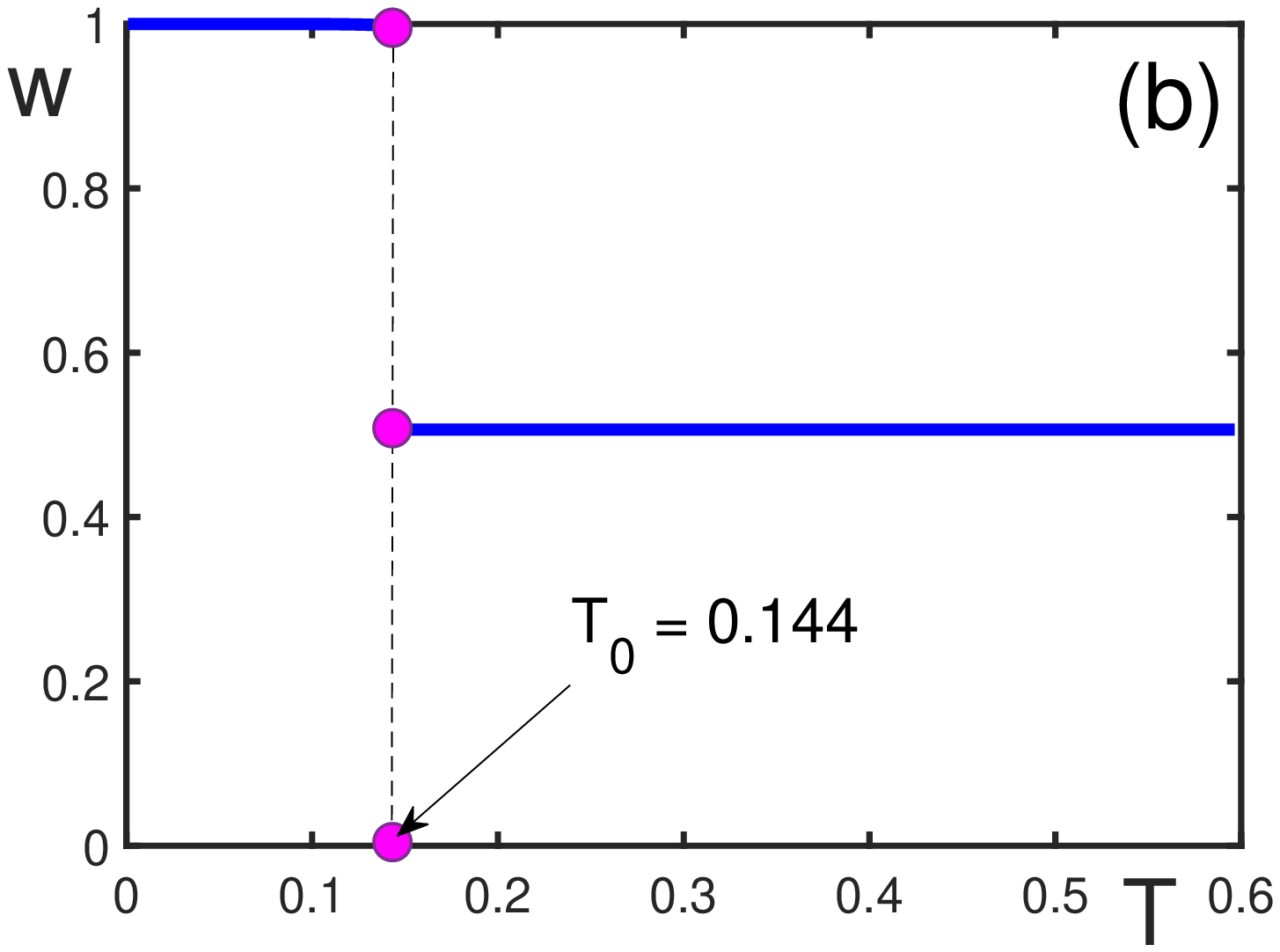} \hspace{1cm}
\includegraphics[width=7.5cm]{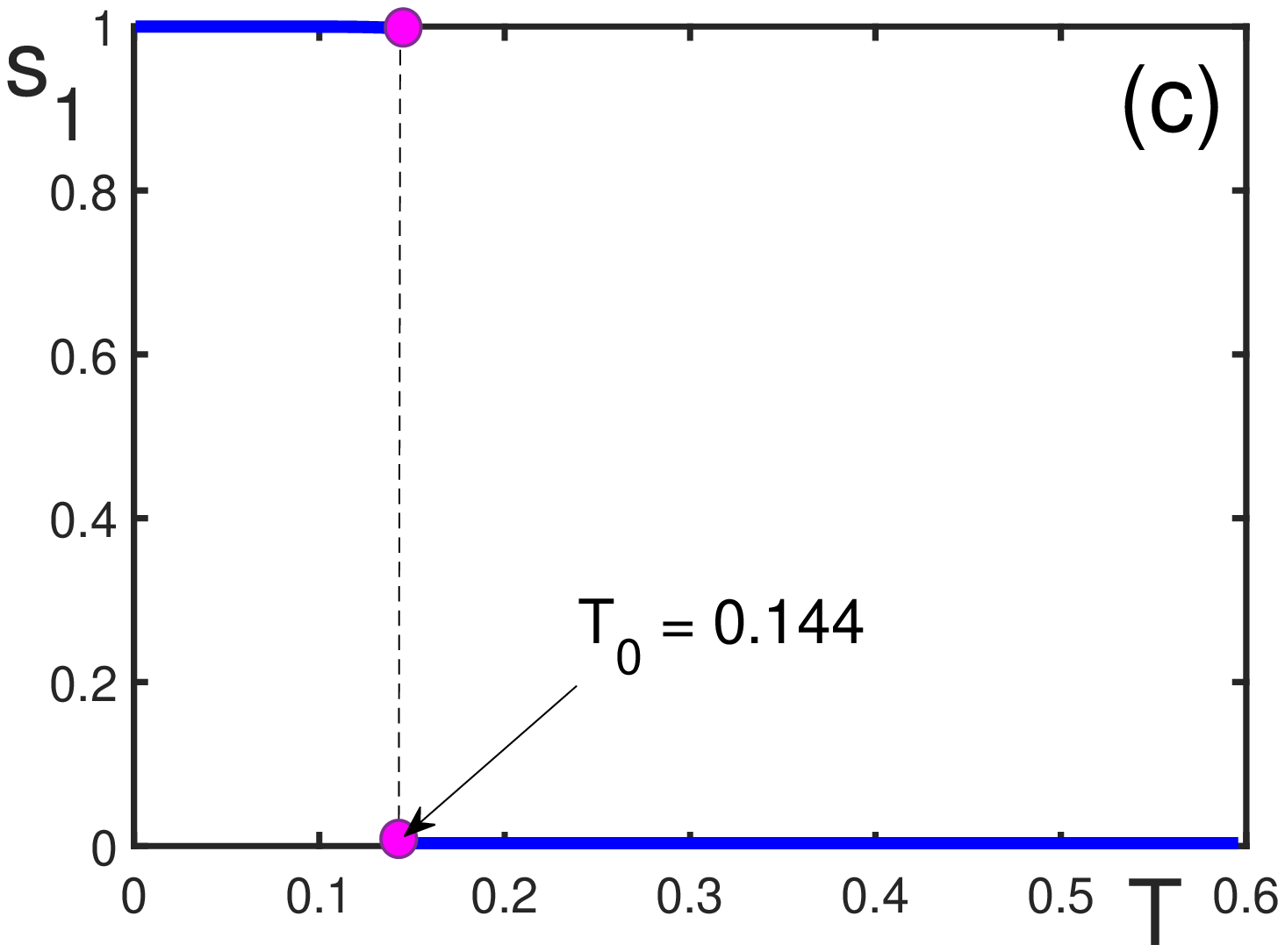} } }
\caption{Free energies, the probability of the ferromagnetic phase $w$, 
and the order parameter $s_1$ as functions of temperature $T$ for $u=0.2$.
First-order phase transition occurs at temperature $T_0=0.144$ between the
pure ferromagnetic phase with the free energy $F_{fer}$ and the nonmagnetic
phase with the free energy $F_0$.
}
\label{fig:Fig.6}
\end{figure}

\vskip 2mm
(iii) $0.5\leq u< 3/2$. In the region of temperatures $0<T<T_0$, the system is a 
mixture of ferromagnetic and paramagnetic phases. A first-order phase transition 
from the mixed state to the degenerate nonmagnetic state occurs at the temperature 
$T_0$ that lays in the interval $1/8<T_0<0.182$. The first-order transition occurs 
when the free energies $F$ and $F_0$ intersect. This is illustrated in Fig. 7. 

\begin{figure}[ht]
\centerline{
\hbox{ \includegraphics[width=7.5cm]{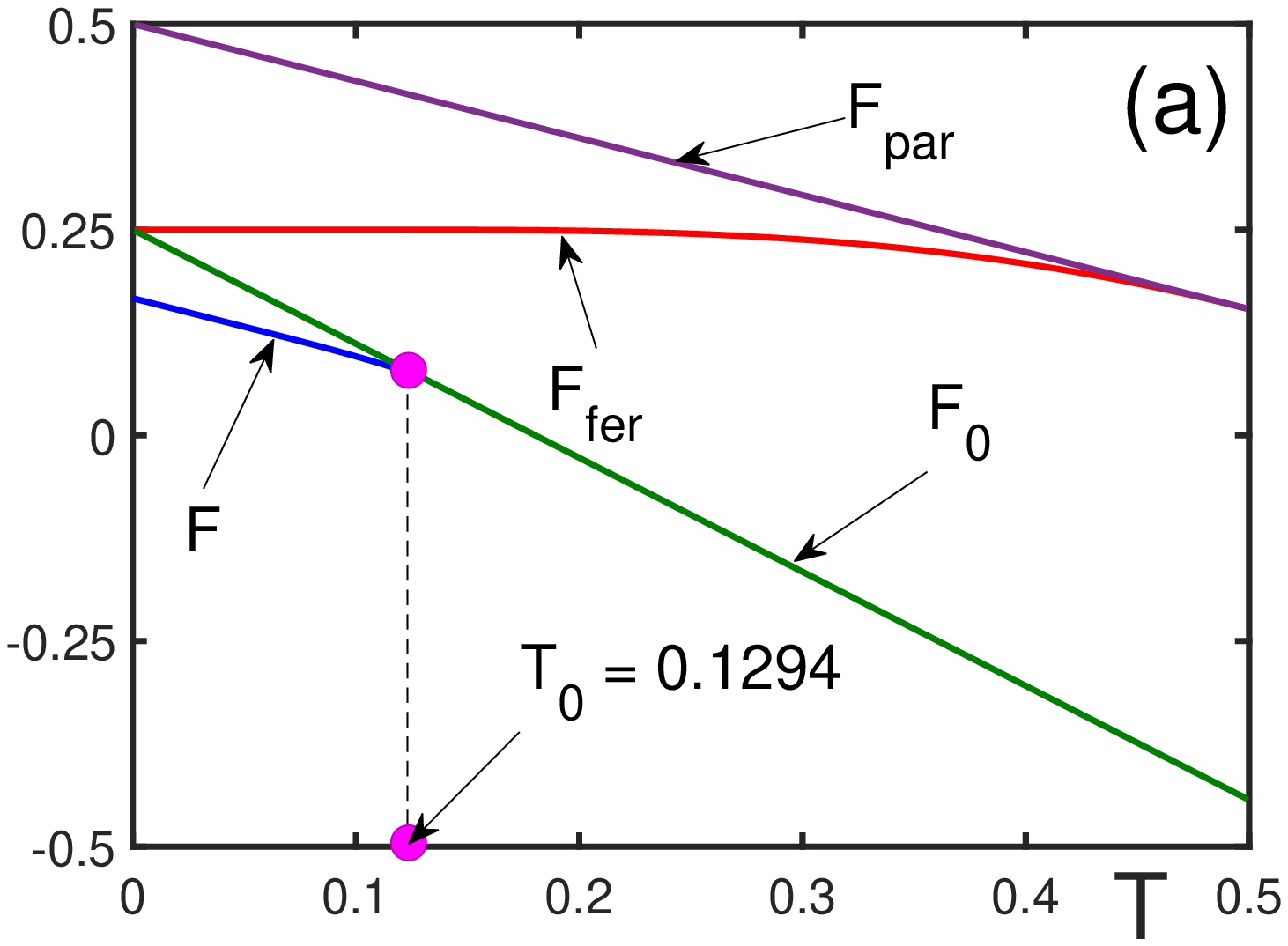}  } }
\vspace{12pt}
\centerline{
\hbox{ \includegraphics[width=7.5cm]{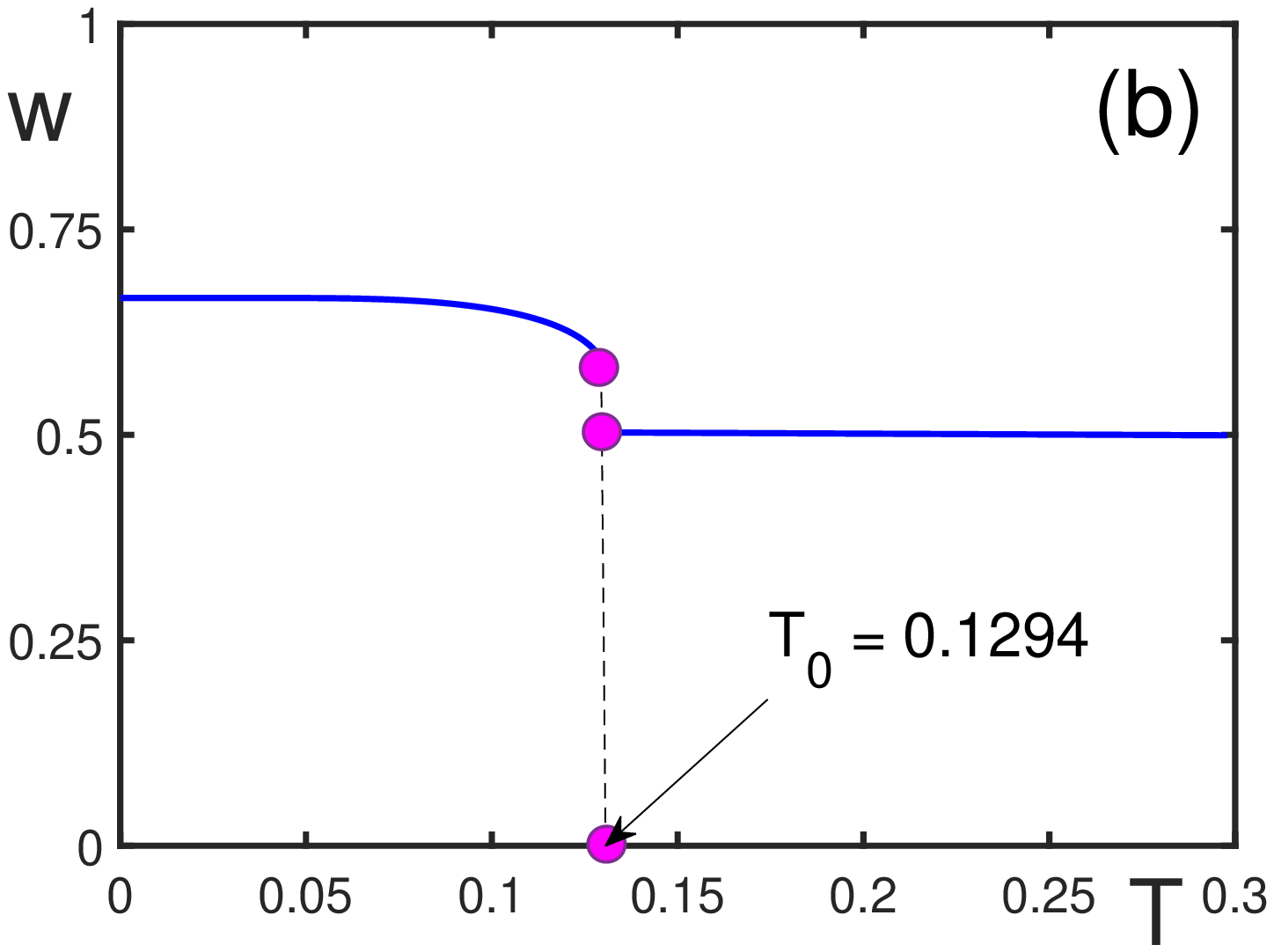} \hspace{1cm}
\includegraphics[width=7.5cm]{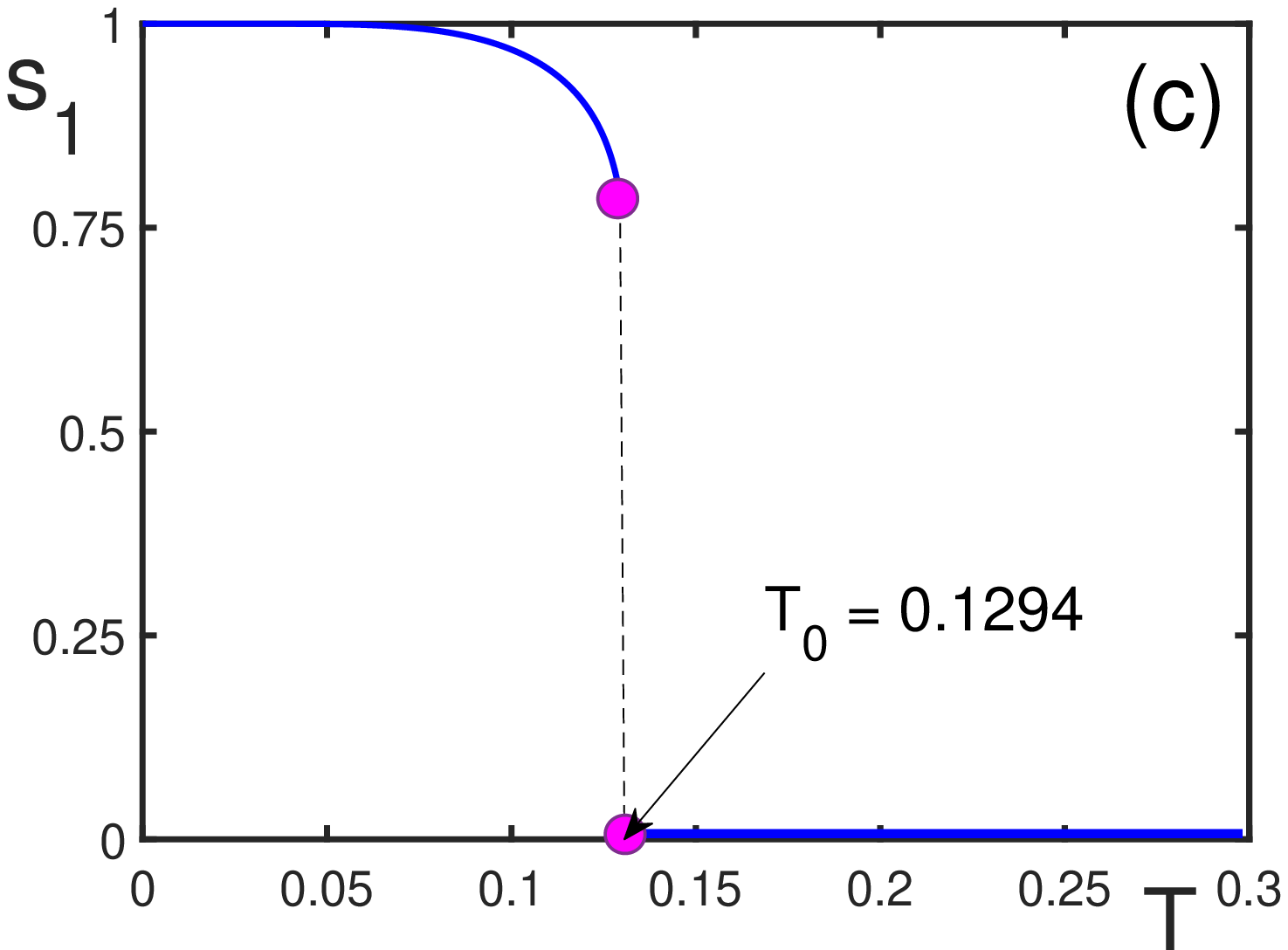} } }
\caption{Free energies, the probability of the ferromagnetic fraction $w$,
and the order parameter $s_1$ as functions of temperature $T$ for $u=1$.
First-order phase transition occurs at temperature $T_0=0.129$ between
the mixed phase with the free energy $F$ and the nonmagnetic phase with
the free energy $F_0$.
}
\label{fig:Fig.7}
\end{figure}

\vskip 2mm
(iv) $u=3/2$. At low temperature, the system is heterophase, with the free 
energy $F$, up to the transition temperature $T_c^*=1/8$, where it becomes 
nonmagnetic, with the free energy $F_0$. The temperature $T_c^*$ is a tricritical 
point separating the lines of first- and second-order phase transitions (see 
\cite{Lawrie_259,Yukalov_260}).

\vskip 2mm
(v) $u>3/2$. Heterophase ferromagnet, with the free energy $F$, is stable at 
low temperatures. The second-order phase transition to the nonmagnetic phase, 
with the free energy $F_0$, happens at the critical temperature $T_c^*=1/8$. 
The corresponding behavior is shown in Fig. 8. 

\begin{figure}[ht]
\centerline{
\hbox{ \includegraphics[width=7.5cm]{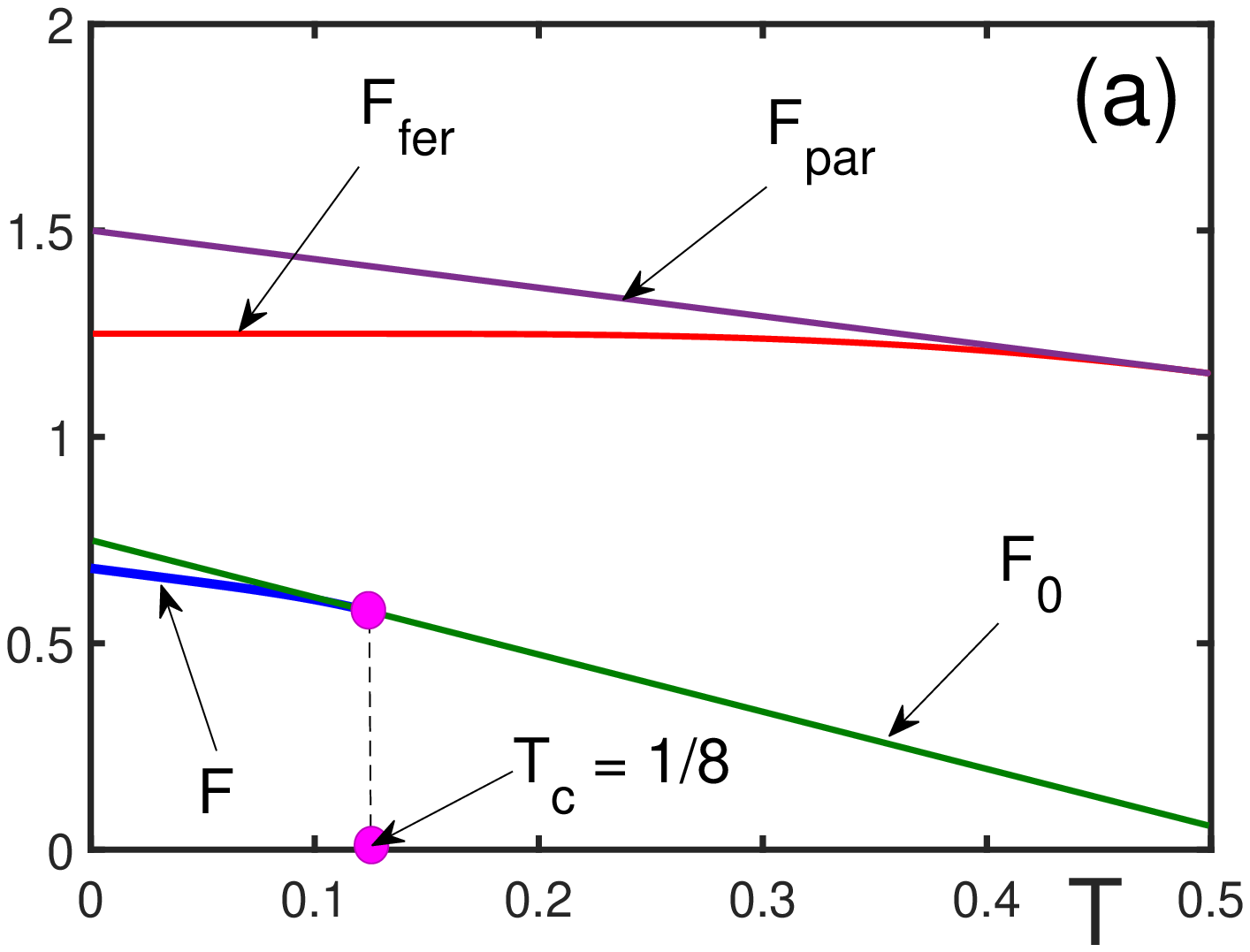}  } }
\vspace{12pt}
\centerline{
\hbox{ \includegraphics[width=7.5cm]{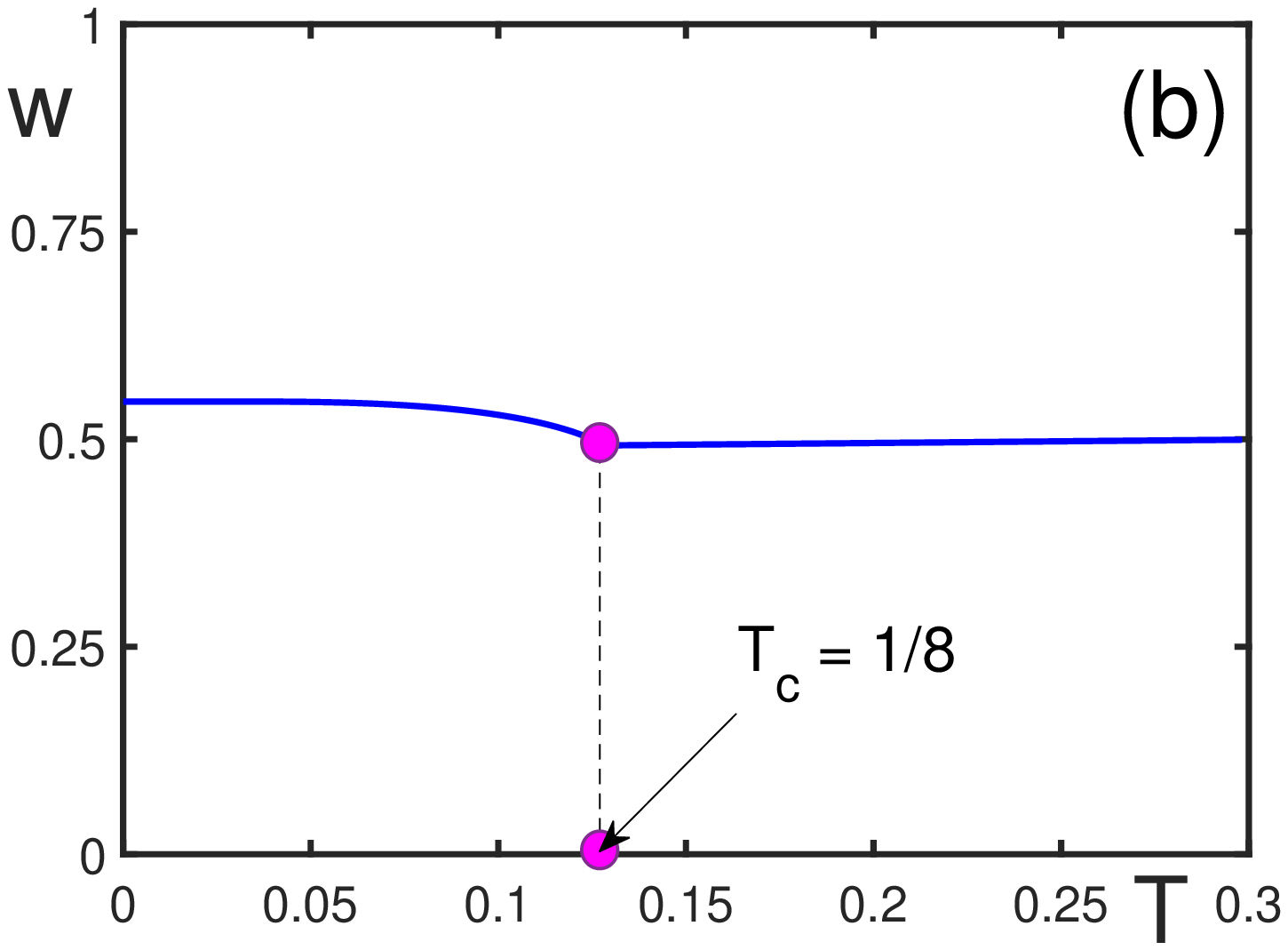} \hspace{1cm}
\includegraphics[width=7.5cm]{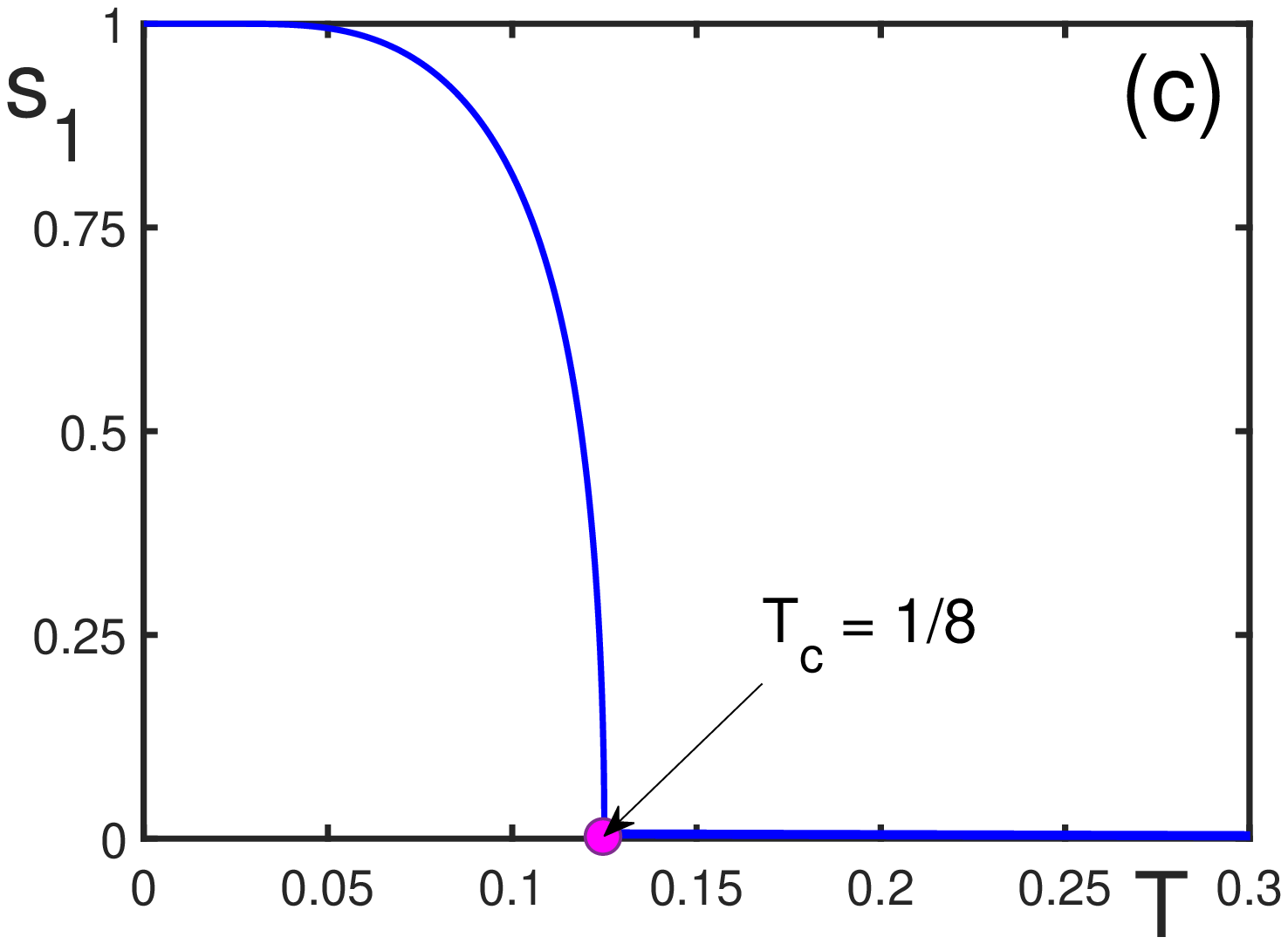} } }
\caption{Free energies, the probability of the ferromagnetic fraction $w$,
and the order parameter $s_1$ as functions of temperature $T$ for $u=3$.
Second-order phase transition between the heterophase state, with the free 
energy $F$, and the nonmagnetic phase, with the free energy $F_0$, occurs 
at the critical point $T_c=1/8$.
}
\label{fig:Fig.8}
\end{figure}

\vskip 2mm
Having the expression for the free energy, it is straightforward to find other 
thermodynamic characteristics. Thus for the heterophase system, the relative 
internal energy is
\be
\label{5.22}
E = \frac{\lgl \widetilde H\rgl}{N} = 
\left( w^2 - w + \frac{1}{2}\right) u \; - \; \frac{1}{4}\; w^2 s_1^2 \;  .
\ee
The relative entropy reads as
\be
\label{5.23}
S = \frac{E-F}{T} = -\; \frac{w^2 s_1^2}{2T} + 
\ln\;\left[ 4\cosh\left( \frac{w^2s_1}{2T}\right) \right] \;  .
\ee
At the critical temperature $T_c^*$ the critical exponents of the specific heat 
\be
\label{5.24}
C_H = \frac{\prt E}{\prt T} = 
T \; \frac{\prt S}{\prt T} \; \propto \; ( -\tau)^\al
\ee
and of the order parameter $s_1\propto (-\tau)^\bt$, where
$$
 \tau \equiv \frac{T-T_c^*}{T_c^*} \ra -0 \; ,
$$
experience a jump, when $T_c^*$ becomes a tricritical point,
\begin{eqnarray}
\nonumber
\al = \left\{ \begin{array}{ll}
0 , ~ & ~ u \neq 3/2 \\
1/2, ~ & ~ u = 3/2 \;  \end{array} \right. \; ,
\qquad
\bt = \left\{ \begin{array}{ll}
1/2 , ~ & ~ u \neq 3/2 \\
1/4 , ~ & ~ u = 3/2 \;  \end{array} \right. \; .
\end{eqnarray}
The property $\alpha + 2\beta = 1$ remains valid. This behavior is typical of 
tricritical points \cite{Lawrie_259}. 
 
The influence of an external magnetic field is considered in Ref. 
\cite{Yukalov_262}. At zero temperature, the system is in a pure ferromagnetic 
state. However, at finite temperatures, for some interaction parameters, the 
system can exhibit a zeroth-order nucleation transition between the pure 
ferromagnetic phase and the mixed state with coexisting ferromagnetic and 
paramagnetic phases.

\subsection{Role of Spin Waves}

In the previous section, a ferromagnetic system with paramagnetic fluctuations 
is treated in the mean-field approximation. Heterophase fluctuations are nonlinear 
and mesoscopic, which principally distinguishes them from homogenous microscopic 
fluctuations \cite{Yukalov_263}. Homogeneous fluctuations in ferromagnets are 
represented by spin waves \cite{Tyablikov_264,Akhiezer_265}. The characteristic 
time of spin fluctuations is defined by spin interactions, which gives 
$t_{int}\sim 10^{-14}-10^{-13}$ s. The lifetime of heterophase paramagnetic 
fluctuations is about $t_{het} \sim 10^{-12}$ s. In the present section, we 
consider the interplay between heterophase paramagnetic fluctuations and spin 
waves \cite{Yukalov_266} for the model with Hamiltonian (\ref{5.1}). 

Aiming at using the random-phase approximation, we define the magnon operators
\be
\label{5.25}
b_{jf} = S_{jf}^x + i S_{jf}^y \; , \qquad 
b_{jf}^\dgr = S_{jf}^x - i S_{jf}^y \; .
\ee
Hence the spin operators are
$$
S_{jf}^x = \frac{1}{2} \; \left( b_{jf}^\dgr + b_{jf} \right) \; ,
\qquad
S_{jf}^y = \frac{i}{2} \; \left( b_{jf}^\dgr - b_{jf} \right) \; ,
$$
\be
\label{5.26}
S_{jf}^z = \frac{1}{2} \; - \; \hat n_{jf} \; ,
\ee
where
\be
\label{5.27}
\hat n_{jf} \equiv b_{jf}^\dgr b_{jf}
\ee
is the operator of magnon density. The magnon operators satisfy the commutation 
relations
\be
\label{5.28}
[\; b_{if} , \; b_{jf} \; ] = 0 \; , \qquad   
[\; b_{if} , \; b_{jf}^\dgr \; ] = \dlt_{ij} ( 1 - 2 \hat n_{jf} ) \; , 
\ee
and the property
\be
\label{5.29}
b_{jf}^2 = 0 \; .
\ee

Then Hamiltonian (\ref{5.2}) takes the form
\be
\label{5.30}
H_f = \frac{N}{2} \; w_f^2 \;\left( U \; - \; \frac{J}{2}\right) +
w_f^2 J \sum_j \hat n_{jf} - 
w_f^2 \sum_{i\neq j} J_{ij} \left( \hat n_{if}\hat n_{jf} + 
2 b_{if}^\dgr b_{jf}\right) \;  .
\ee
 
In what follows, we use the causal Green functions 
\cite{Kadanoff_267,Bonch_268,Yukalov_269} also called propagators. The magnon 
propagator is
\be
\label{5.31}
 G_{ijf}(t) = - i \; 
\lgl \; \hat T\; b_{if}(t)\; b_{jf}^\dgr(0) \; \rgl \; ,
\ee
where $\hat{T}$ is the chronological operator. The Fourier transforms for the 
propagator read as
$$
G_{ijf}(t) = 
\frac{1}{\rho} \int G_f(\bk,\om) \; e^{i(\bk\cdot\br_{ij}-\om t)} \;
\frac{d\bk d\om}{(2\pi)^4} \; ,
$$
\be
\label{5.32}
G_f(\bk,\om) = 
\frac{1}{N} \sum_{ij} \; 
\int G_{ijf}(t) \; e^{-i(\bk\cdot\br_{ij}-\om t)}\;dt \; ,
\ee
where $\rho=N/V$ is the spin density and $\br_{ij}\equiv \br_i-\br_ j$.  

In the evolution equation for the magnon propagator we resort to the random-phase 
approximation
\be
\label{5.33}
\lgl \; b_{if}^\dgr\; b_{if}\; b_{jf}^\dgr\; b_{lf} \; \rgl =
\lgl \; b_{if}^\dgr\; b_{if}\; \rgl \lgl \; b_{jf}^\dgr\; b_{lf} \; \rgl \; .
\ee
In this approximation, we get the equation
\be
\label{5.34}
 [\; \om - \om_f(\bk) \; ] \; G_f(\bk,\om) = s_f \; ,
\ee
with the magnon spectrum
\be
\label{5.35}
\om_f(\bk) = w_f^2 \; s_f \; [\; J - J(\bk) \; ] \; .
\ee
Here the Fourier transformation for the interaction is employed,
$$
 J_{ij} = \frac{1}{\rho} 
\int J(\bk) \; e^{i\bk\cdot\br_{ij}}\; \frac{d\bk}{(2\pi)^3} \; ,
\qquad 
J(\bk) = \frac{1}{N} \sum_{i\neq j} J_{ij} e^{-i\bk\cdot\br_{ij} } \; .
$$

If the magnon density is small, magnons can approximately be treated as 
bosons. This can be assumed for the ferromagnetic phase, for which the solution 
to equation (\ref{5.34}) becomes
\be
\label{5.36}
 G_1(\br,\om) = s_1 \left[ \; \frac{1+n_1(\bk)}{\om-\om_1(\bk) + i0} \; - \;
\frac{n_1(\bk)}{\om-\om_1(\bk) - i0} \; \right] \; ,
\ee
with the magnon momentum distribution
\be
\label{5.37}
n_1(\bk) = \frac{1}{2} \; 
\left[ \; \coth\; \frac{\om_1(\bk)}{2T} \; - \; 1 \; \right] \; .
\ee

For the order parameter (\ref{5.5}), according to (\ref{5.26}), we have
\be
\label{5.38}
 s_f =  1 \; - \; \frac{2}{N} \sum_j \lgl \; \hat n_{jf} \; \rgl \;  .
\ee
For the ferromagnetic phase, the relation 
\be
\label{5.39}
\lgl \; \hat n_{j1} \; \rgl = i\; G_{jj1}(-0) = 
\frac{s_1}{\rho} \int n_1(\bk) \; \frac{d\bk}{(2\pi)^3}
\ee
gives the equation for the order parameter
\be
\label{5.40}
 \frac{s_1}{\rho} \int \coth\; 
\left[ \; \frac{\om_1(\bk)}{2T}\;\right] \; \frac{d\bk}{(2\pi)^3} = 1 \; .
\ee

For the paramagnetic phase, where $s_2 = 0$, Eq. (\ref{5.38}) yields
\be
\label{5.41}
\lgl \; \hat n_{j2} \; \rgl  = \frac{1}{2} \; ,
\ee
while the evolution equation (\ref{5.34}) reduces to
\be
\label{5.42}
\om \; G_2(\bk,\om) = 0 \; .
\ee
An approximate solution for the latter is
\be
\label{5.43}
G_2(\bk,\om) = - i \pi \dlt(\om) \;  ,
\ee
which gives
\be
\label{5.44}
G_{ij2}(t) = - \; \frac{i}{2} \; \dlt_{ij} \;  .
\ee
 
The probability of the ferromagnetic phase $w \equiv w_1$ is defined as the 
minimizer of the free energy
\be
\label{5.45}
 F = - \; \frac{T}{N} \; \ln\; {\rm Tr} \; e^{-\bt\widetilde H} \; .
\ee
This yields the condition
\be
\label{5.46}
 \frac{\prt F}{\prt w} = \frac{1}{N} \; \left\lgl \;
\frac{\prt\widetilde H}{\prt w} \; \right\rgl = 0 \; ,
\ee
which gives 
\be
\label{5.47}
w = \frac{U - 2\Phi_2}{2(U-\Phi_1-\Phi_2)} \;  ,
\ee
where the notation for the energy of spin interactions is introduced,
\be
\label{5.48}
\Phi_f \equiv \frac{1}{N} 
\sum_{i\neq j} J_{ij} \; \lgl \; \bS_{if}\cdot\bS_{jf}\; \rgl \;  .
\ee

In the random phase approximation, the interaction energy of the ferromagnetic 
phase is
\be
\label{5.49}
 \Phi_1 = \frac{J}{4} \; - \; \frac{s_1(1+s_1)}{2\rho} 
\int [\; J - J(\bk) \; ] \; n_1(\bk) \; \frac{d\bk}{(2\pi)^3} \; ,
\ee
while the interaction energy of the paramagnetic phase is $\Phi_2\approx 0$. 

Using the notation
\be
\label{5.50}
 u \equiv \frac{U}{J} \;  ,
\ee
and measuring temperature in units of $J$, we find \cite{Yukalov_266} the 
probability of the ferromagnetic phase at low temperature 
\be
\label{5.51}
 w \simeq w_0 \; - \; \frac{6\pi\zeta(5/2)}{u\; w_0^3\;\rho\; a_0^3} \;
\left( \frac{T}{2\pi}\right)^{5/2} \; - \;
\frac{45\pi^2\zeta(7/2)}{2u\; w_0^5\;\rho\; a_0^3} \;
\left( \frac{T}{2\pi}\right)^{7/2} \;  .
\ee
Here $w_0$ is the probability at zero temperature, 
\be
\label{5.52}
w_0 \equiv \frac{2u}{4u-1}
\ee
and $a_0$ is the effective interaction radius defined by the relation 
\be
\label{5.53}
 a_0^2 \equiv \frac{1}{3N} 
\sum_{i\neq j} \frac{J_{ij}}{J} \; r_{ij}^2 \; .
\ee
At zero temperature, the mixed state can exist when $u>1/4$.  

The consideration of the system behavior at the point of the phase transition
from the mixed state to the paramagnetic state shows that in the system with spin 
waves the transition is of first order, while without spin waves it would be of 
second order.

\subsection{Heterophase Ising Model}

Taking account of heterophase fluctuations in the frame of the ferromagnetic Ising
model results in the effective Hamiltonian \cite{Kislinsky_270}
\be
\label{5.54}
 \widetilde H = H_1 \; \bigoplus \; H_2 \; , 
\qquad
H_f = \frac{N}{2} \; w_f^2 \; U \; - \; 
\frac{1}{2} \; w_f^2 \; J \sum_{\lgl ij\rgl} \sgm_{if}\; \sgm_{jf} \; ,
\ee
where the summation is over the nearest neighbors and $\sigma_{jf}=\pm 1$. 

We consider the two-dimensional model that enjoys an exact solution. Employing 
the transfer-matrix approach, we notice that the maximal eigenvalue corresponds 
to the completely ordered phase, while the minimal eigenvalue corresponds to the 
disordered phase \cite{Onsager_271}. Then the order parameters discriminating 
the ordered and disordered phases are
\be
\label{5.55}
 s_f = \frac{1}{N} \sum_j \lgl \; \sgm_{jf}\; \rgl 
\qquad 
( f = 1,2 ) \;  ,
\ee
under condition (\ref{5.5}). 

The free energy reads as
\be
\label{5.56}
F = \frac{1}{2} \; \left[\; w^2 + (1 - w)^2 \; \right]\; U - 
T (\Lbd_1- \Lbd_2) - T \ln \; (2\sinh\al_2) \; ,
\ee
where
$$
\Lbd_f = \frac{1}{2\pi^2} \int_0^\pi \int_0^\pi \ln \left[\;
\cosh^2 \al_f - (\cos\vartheta  + \cos\vartheta') \sinh\al_f\; \right]\;
d\vartheta d\vartheta'
$$
and
$$
\al_f \equiv w_f^2 \;\frac{J}{T}  \;  .
$$

The equation for the ferromagnetic probability has the form (\ref{5.47}) but with
the interaction energies
\be
\label{5.57}
 \Phi_f = \frac{1}{2} + (-1)^{f+1} \;
\frac{\sinh\al_f -1}{\sinh\al_f+1} \; K(\vp_f)\; \cosh\al_f \; ,
\ee
in which
$$
 K(\vp_f) = \int_0^\pi \frac{d\vartheta}{\sqrt{1-\vp_f\sin^2\vartheta} } \; ,
\qquad
\vp_f = \frac{8\sinh\al_f \cdot \cosh^2\al_f}{(1+\sinh\al_f)^4} \;  .
$$

The phase transition of second order from the mixed state to the paramagnetic 
phase occurs at the critical temperature
\be
\label{5.58}
 T_c^* = \frac{1}{4\ln(1+\sqrt{2})} \; .
\ee
The specific heat at the transition temperature behaves as
\be
\label{5.59}
 C_V \simeq \frac{32[\ln(1+\sqrt{2})]^3}{\pi(u+0.348)} \; \ln^2(-\tau) 
\qquad
(\tau \ra -0 ) \; .
\ee

However the mixed state is metastable, since its free energy is larger than the free
energy of the pure ferromagnetic phase.

\subsection{Heterophase Nagle Model}

Nagle \cite{Nagle_272} considered a one-dimensional spin model with the Hamiltonian
containing competing short-range and long-range interactions. The total interaction
can be written \cite{Kislinsky_273} in the form
\be
\label{5.60}
 J_{ij} = \al J_0 \dlt_{|i-j|\; 1} + ( 1 - \al) \overline J_{ij} \;  .
\ee
Here $J_0$ is the intensity of the nearest-neighbor interactions, $\overline J_{ij}$
is a long-range interaction with the properties
\be
\label{5.61}
\lim_{N\ra\infty} \;\overline J_{ij} = 0 \; , 
\qquad
\lim_{N\ra\infty} \; \frac{1}{N} \sum_{i\neq j} \overline J_{ij} \neq 0 \; ,
\ee
and $\alpha$ is a crossover parameter. In what follows, the mean long-range 
interaction is assumed to be positive,
\be
\label{5.62}
 J \equiv \frac{1}{N} \sum_{i\neq j} \overline J_{ij} > 0 \;  .
\ee
 
The heterophase generalization \cite{Yukalov_274} of the Nagle model is characterized 
by the Hamiltonian
\be
\label{5.63}
 \widetilde H = H_1 \; \bigoplus \; H_2 \; , 
\qquad
H_f = \frac{N}{2} \; w_f^2 \; U \; - \; 
\frac{1}{4} \; w_f^2 \sum_{i\neq j} J_{ij} \; \sgm_{if}\; \sgm_{jf} \; ,
\ee
where $\sgm_{jf}=\pm 1$. We consider a mixture of ferromagnetic and paramagnetic 
phases, so that the order parameter is the same as in (\ref{5.55}), with 
$s_1\neq 0$ and $s_2 = 0$. 

Using the transfer-matrix method \cite{Yukalov_249}, we find the free energy of 
the system
$$
F = \left( w^2 - w + \frac{1}{2}\right) \; U - \; 
\frac{1}{4}\; w^2 \; \left[\; \al T - ( 1 - \al) \; J \; s_1^2 \; \right] \; -
$$
\be
\label{5.64}
- \;
T\ln \; \left[\; \cosh\vp + \sqrt{\sinh^2\vp+\exp(-4\vp_1)} \; \right] -
T\ln(2\cosh\vp_2) \;  ,
\ee
where $w \equiv w_1$ and
$$
\vp \equiv w^2 \; \frac{(1-\al)J\; s_1}{2T} \; ,
\qquad
\vp_1 \equiv w^2 \; \frac{\al J_0}{4T} \; , 
\qquad
\vp_2 \equiv ( 1 - w)^2 \;  \frac{\al J_0}{4T} \;   .
$$

In what follows, we introduce the dimensionless parameters
\be
\label{5.65}
u \equiv \frac{U}{J} \; , \qquad g \equiv \frac{J_0}{J}
\ee
and measure temperature in units of $J$. For the ferromagnetic order parameter, 
we have
\be
\label{5.66}
 s_1 = \frac{\sinh\vp}{\sqrt{\sinh^2\vp+\exp(-4\vp_1)} } \;  .
\ee
The ferromagnetic probability is defined as the minimizer of the free energy.
For instance, at zero temperature, this gives
\be
\label{5.67}
w_0 = \frac{2u-|\;\al\;| g}{4u-1+\al-(\al+|\;\al\;|) g } \;   .
\ee
 
Depending on the parameters, the system can be either purely ferromagnetic or 
representing a ferromagnet with paramagnetic fluctuations \cite{Yukalov_274}. 
At low temperature, the ferromagnetic order parameter behaves as
\be
\label{5.68}
s_1 \simeq 1 - 2\exp\left\{ -\; 
\frac{w_0^2}{T} \; ( 1 - \al + 4\al g) \right\}  \;  .
\ee
The critical temperature, defined by the condition $s_1(T_c)=0$, is given by 
the equation
\be
\label{5.69}
 T_c = \frac{1-\al}{8} \; \exp\left( \frac{\al g}{8T_c} \right) \; .
\ee

There exists a tricritical surface, given by a relation $u=u_3(g,\al)$, where 
the order of the phase transition changes between second and first. The 
critical exponents, describing the behavior of the specific heat $C_V$, order 
parameter $s_1$, susceptibility $\chi$, and the ferromagnetic probability $w$, 
defined by the relations
$$
C_V \propto |\;\tau\;|^{-\al} \; , \qquad  
s_1 \propto |\;\tau\;|^\bt \;  , 
\qquad \chi \propto |\;\tau\;|^\gm \;  , \qquad 
w - \; \frac{1}{2} \propto |\;\tau\;|^\ep \;
$$
on the tricritical surface experience a jump, so that outside the surface, one has
\be
\label{5.70}
\al = 0 \; , \qquad \bt = \frac{1}{2} \; , \qquad \gm = 1 \; , \qquad
\ep = 1 \qquad ( u \neq u_3) \;  ,
\ee
while on the tricritical surface, 
\be
\label{5.71}
\al = \frac{1}{2} \; , \qquad \bt = \frac{1}{4} \; , \qquad \gm = 1 \; , 
\qquad \ep = \frac{1}{2} \qquad ( u = u_3) \;   .
\ee
The condition $\al + 2\bt + \gm =2$ is always valid.

\subsection{Model of Heterophase Antiferromagnet}

Antiferromagnets are characterized by two sublattices having opposite average 
magnetizations. In each of sublattices there can occur heterophase fluctuations
representing disordered paramagnetic phase. The model of such a heterophase
antiferromagnet is defined as follows \cite{Kudryavtsev_275,Boky_276}. 

Let us consider two sublattices, $A$ and $B$, the lattice $A$, enumerated 
by the indices $\{i,j\}$, and the lattice $B$, by the indices $\{l,m\}$. The 
particles forming the sublattices interact directly with the strengths $U_A$ 
inside the sublattice $A$ and $U_B$, in the sublattice $B$, respectively. 
The direct interaction between the nodes of different sublattices is denoted by 
$U_{AB}$. Also, there are exchange interactions $J_{ij}^A$ and $J_{lm}^B$ 
between the spins inside each sublattice, as well exchange interactions 
$J_{jl}^{AB}$ between the spins of different sublattices. The corresponding 
spins are denoted as ${\bf S}_{j1}$ and ${\bf S}_{j2}$ for the sublattice $A$, 
and as ${\bf S}_{l1}$ and ${\bf S}_{l2}$ for the sublattice $B$. The probabilities 
of the magnetic and paramagnetic phases are denoted as $w_{A1}$ and $w_{A2}$ 
in the sublattice $A$ and as $w_{B1}$, and $w_{B2}$ in the sublattice $B$. Thus 
the Hamiltonian reads as
$$
\widetilde H = H_1 \; \bigoplus \; H_2 \; , 
$$
$$
H_f = \frac{N_A}{2} \; w_{Af}^2 \; U_A \; - \;  
w_{Af}^2 \sum_{i\neq j} J_{ij}^A \; \bS_{if}\cdot \bS_{jf} +
\frac{N_B}{2}\; w_{Bf}^2 \; U_B  - 
w_{Bf}^2 \sum_{l\neq m} J_{lm}^B \; \bS_{lf}\cdot \bS_{mf} \; +
$$
\be
\label{5.72}
+ \;
N_{AB} \; w_{Af} \; w_{Bf} \; U_{AB} + 
2 w_{Af} \; w_{Bf} \sum_{jl} J_{jl}^{AB} \; \bS_{jf}\cdot \bS_{lf} \; ,
\ee
where
$$
N_{AB} \equiv\frac{1}{2} \; ( N_A + N_B ) \; .
$$

The order parameters for the sublattices are defined as nonzero magnetizations 
for the magnetic phase,
\be
\label{5.73}
{\bf C}_A \equiv \left\lgl \; 
\frac{1}{N_A} \sum_{j=1}^{N_A} \bS_{j1}\; \right\rgl \neq 0 \; ,
\qquad
{\bf C}_B \equiv \left\lgl \; 
\frac{1}{N_B} \sum_{l=1}^{N_B} \bS_{l1}\; \right\rgl \neq 0 \; ,
\ee
and zero magnetizations for the paramagnetic phase,
\be
\label{5.74}
 \left\lgl \; \frac{1}{N_A} \sum_{j=1}^{N_A} \bS_{j2}\; \right\rgl = 0 \; , 
\qquad
\left\lgl \; \frac{1}{N_B} \sum_{l=1}^{N_B} \bS_{l2}\; \right\rgl = 0 \;  .
\ee
Taking into account the normalization conditions
\be
\label{5.75}
 w_{A1} + w_{A2} = 1 \; , \qquad  w_{B1} + w_{B2} = 1 \; ,
\ee
it is possible to simplify the notation by defining
\be
\label{5.76}
w_{A} \equiv w_{A1} \; , \qquad  w_{A2} = 1 -  w_A  \; , 
\qquad
w_{B} \equiv w_{B1} \; , \qquad  w_{B2} = 1 -  w_B \;  .
\ee

Considering the case of the sublattices with the equal number of sites,
\be
\label{5.77}
 N_A = N_B \equiv N \; ,
\ee
we look for the free energy
\be
\label{5.78}
 F = - \; \frac{1}{2N} \; T \ln \; {\rm Tr}\; e^{-\bt\widetilde H} \; .
\ee
We employ the mean-field approximation and define the average exchange 
interactions
\be
\label{5.79}
 J_A \equiv \frac{1}{N} \sum_{i\neq j} J_{ij}^A \; , 
\qquad  
J_B \equiv \frac{1}{N} \sum_{l\neq m} J_{lm}^B \; , 
\qquad
J_{AB} \equiv \frac{1}{N} \sum_{jl} J_{jl}^{AB} \; ,
\ee
and effective fields 
\be
\label{5.80}
h_A \equiv w_A^2 \; J_A\; C_A - w_A \; w_B \; J_{AB} \; C_B \; ,
\qquad
h_B \equiv w_B^2 \; J_B\; C_B - w_A \; w_B \; J_{AB} \; C_A \; .
\ee
We assume that the average spins ${\bf C}_A$ and ${\bf C}_B$ are directed along 
the same axis, say the $z$-axis, and are opposite to each other, so that the 
order parameters are $C_A = C_A^z$ and $C_B = C_B^z$. For these order parameters, 
we get
\be 
\label{5.81}
C_A = S_A \; B_{S_A}(x_A) \; , \qquad C_B = - S_B \; B_{S_B}(x_B) \; ,
\ee 
where $S_A$ and $S_B$ are the corresponding spin values and $B_S(x)$ is the 
Brillouin function
$$
B_S(x) = \frac{2S+1}{2S}\; \coth\left(\frac{2S+1}{2S}\; x\right) - \;
\frac{1}{2S}\; \coth\left( \frac{x}{2S} \right) \;  ,   
$$
with the notation
$$
 x_A \equiv \frac{2S_A}{T} \; h_A \; , 
\qquad 
x_B \equiv \frac{2S_B}{T} \; h_B \;  .   
$$ 

Then for the free energy, we obtain
$$
F = \frac{1}{2} \left( w_A^2 - w_A + \frac{1}{2} \right) \; U_A +
\frac{1}{2} \; w_A^2 \; J_A\;  C_A^2 +
    \frac{1}{2} \left( w_B^2 - w_B + \frac{1}{2} \right) \; U_B +
\frac{1}{2} \; w_B^2 \; J_B\;  C_B^2 \; +
$$
$$
+ \;
\frac{1}{2}\; ( 2w_A w_B - w_A - w_B + 1) \; U_{AB} - 
w_A\; w_B\; J_{AB}\; C_A \;C_B  \; -
$$
$$
- \;
\frac{T}{2} \; \ln \; \frac{\sinh\left(\frac{2S_A+1}{2S_A} \; x_A\right)}
{\sinh\left(\frac{x_A}{2S_A}\right)} \; - \;
\frac{T}{2} \; \ln \; \frac{\sinh\left(\frac{2S_B+1}{2S_B} \; x_B\right)}
{\sinh\left(\frac{x_B}{2S_B}\right)} \; -
$$
\be
\label{5.82}
 - \;
\frac{T}{2} \; \ln (2S_A+1) - \; \frac{T}{2}\; \ln(2S_B+1) \; .
\ee

The analysis of the derived free energy \cite{Boky_276} shows that, depending 
on the system parameters, the mixed antiferromagnetic-paramagnetic state can 
become stable and thermodynamically preferable, as compared to the pure 
antiferromagnetic phase. The phase transition between the mixed state and 
paramagnetic state can be of second as well as of first order.

\subsection{Heterophase Hubbard Model}   
 
A heterophase Hubbard model describing the mixture of ferromagnetic and 
paramagnetic phases is studied in Ref. \cite{Boky_277}. The total Hamiltonian 
of each phase $f$ contains the terms corresponding to delocalized band electrons 
and electrons localized on ions,
$$
\widetilde H = H_1 \; \bigoplus \; H_2 \; ,
$$
\be
\label{5.83}
H_f = H_f^{band} + H_f^{ion} +  H_{fCoul}^{ion-ion} + H_{fexch}^{ion-ion} +
H_{fCoul}^{band-ion} + H_{fexch}^{band-ion} \; .
\ee
Here the first term describes quasi-free band electrons, the second term, the 
electrons localized on ions, the third term, the direct Coulomb interactions of 
ion electrons, the fourth term, the exchange interactions of ion electrons, the 
fifth and sixth terms describe the direct Coulomb and exchange interactions 
between the band and ion electrons. The band electron wave functions are the 
Bloch waves $|nk\rangle$, where $n$ is the band number and $k$ is quasi-momentum. 
The band electron field operators pertaining to the $f$-phase are $c_{f n k\sgm}$, 
with $\sigma$ being the spin index. The ion electrons of an $f$-th phase are 
characterized by the localized wave functions $|f j \gamma \rangle$, where $j$ 
is the site index and $\gamma$ is the set of quantum numbers including the number 
of electrons $\nu$ localized on the ion shell, the total ion spin $S$ (or the 
orbital momentum $J$), and the $z$-projection $m$ of the total spin $S^z$ (or the 
orbital momentum $J^z$). Thus $|f j \gamma \rangle = |f j \nu S m \rangle$. The 
ion electrons are represented by the Hubbard operators
\be
\label{5.84}
\chi_{fj}^{\gm\dlt} \equiv 
|\; f j \gm \; \rgl \lgl \; \dlt j f \; |
\ee
describing the change of the ion state in the $f$-th phase and the site $j$ from
$|f j \gamma \rangle$ to $|f j \delta \rangle$. The exchange interactions of the 
ion electrons localized on an ion $j$ in the $f$-th phase are written through the 
spin operators ${\bf S}_{fj}$ localized at a site $j$.    

The phases are distinguished by their magnetization, being either nonzero for 
ferromagnetic phase or zero for paramagnetic phase. The properties of this model 
are similar to other heterophase models describing the mixture of ferromagnetic 
and paramagnetic phases, although the consideration is essentially more complicated 
\cite{Boky_277}.

\subsection{Heterophase Vonsovsky-Zener Model}    

Another model comprising the mixture of ferromagnetic and paramagnetic phases 
is the heterophase generalization \cite{Boky_278} of the Vonsovsky-Zener model 
\cite{Vonsovsky_279,Irkhin_280}. The system Hamiltonian can be represented as 
the sum
$$
\widetilde H = H_1 \; \bigoplus \; H_2 \; 
$$
\be 
\label{5.85}
H_f = H_f^s + H_f^d + H_f^{dd} + H_{fCoul}^{dd} + H_{fexch}^{dd} +
H_{fCoul}^{sd} + H_{fexch}^{sd} \;   .
\ee
Here: the first term corresponds to quasi-free electrons,
\be
\label{5.86}  
H_f^s = w_f \sum_{nk\sgm} (\ep_{nk} - \mu ) \; c_{fnk\sgm}^\dgr \; c_{fnk\sgm} \; ,
\ee
where $n$ is a band index, $k$ is quasi-momentum, and $\sigma$ is a spin index;
the second term corresponds to single-particle localized electrons,
\be
\label{5.87}
H_f^d = w_f \sum_{mj\sgm} (E_m - \mu ) \; d_{fmj\sgm}^\dgr \; d_{fmj\sgm} \; ,
\ee
where $m$ is a quantum number of a localized electron and $j$ is a lattice index;
the third term describes the on-site interactions of localized electrons,
$$
H_f^{dd} = \frac{w_f^2}{2} \sum_j \left[\; 
\sum_{m\sgm} U_m \; 
\hat n_{fmj\sgm} \; \hat n_{fmj\sgm} +
\sum_{m\neq m'} \sum_{\sgm\sgm'}  V_{mm'} \; 
\hat n_{fmj\sgm} \; \hat n_{fm'j\sgm'} \right. \; +
$$
\be
\label{5.88}
 + \; \left.
\sum_{m\neq m'} \sum_{\sgm\sgm'} I_{mm'} \; 
d_{fmj\sgm}^\dgr \; d_{fm'j\sgm'}^\dgr \; d_{fmj\sgm'}\; d_{fm'j\sgm}\; 
\right]\; ;
\ee
the fourth term, to the intersite direct Coulomb interactions of localized 
electrons,
\be
\label{5.89}
H_{fCoul}^{dd} = \frac{w_f^2}{2} 
\sum_{i\neq j} \sum_{mm'} \sum_{\sgm\sgm'} V_{ij} \;
\hat n_{fmi\sgm}^\dgr \; \hat n_{fm'j\sgm'}^\dgr \;  ;
\ee
the fifth term, to the intersite exchange interactions of localized electrons,
\be
\label{5.90}
H_{fexch}^{dd} = \frac{w_f^2}{2} 
\sum_{i\neq j} \sum_{mm'} \sum_{\sgm\sgm'} I_{ij} \;
d_{fmi\sgm}^\dgr\; d_{fm'j\sgm'}^\dgr\; d_{fmi\sgm'}\; d_{fm'j\sgm} \; ;
\ee
the sixth term, to the direct Coulomb interaction between conducting and localized 
electrons,
\be
\label{5.91}
H_{fCoul}^{sd} = w_f^2 \sum_{nn'} \sum_{kk'} \sum_{mj} \sum_{\sgm\sgm'} 
G_{kk'}^{nn'} \; e^{i(\bk-\bk')\cdot \ba_j} \;
c_{fn'k'\sgm}^\dgr\; d_{fmj\sgm'}^\dgr\; d_{fmj\sgm'}\; c_{fnk\sgm} \; ;
\ee
and the sevenths term describes the exchange interaction of conducting and 
localized electrons,
\be
\label{5.92}
H_{fexch}^{sd} = w_f^2 \sum_{nn'} \sum_{kk'} \sum_{mj} \sum_{\sgm\sgm'} 
J_{kk'}^{nn'} \; e^{i(\bk-\bk')\cdot \ba_j} \;
c_{fn'k'\sgm}^\dgr\; d_{fmj\sgm'}^\dgr\; c_{fnk\sgm'}\; d_{fmj\sgm} \; .
\ee
 
Depending on the system parameters, the model demonstrates the existence of 
pure ferromagnetic phase and mixed ferromagnetic-paramagnetic state. The phase 
transition between the mixed state and paramagnetic phase can be of second as 
well as of first order \cite{Boky_278}.

\subsection{Heterophase Spin Glass}

Spin glass can be described by the Sherrington-Kirkpatrick model 
\cite{Sherrington_281,Kirkpatrick_282}. This model can be generalized by 
taking into account the possible appearance of paramagnetic fluctuations 
inside the spin-glass phase \cite{Yukalov_274,Kislinsky_283}. The Hamiltonian 
of the heterophase model reads as
$$
\widetilde H = H_1 \; \bigoplus \; H_2 \; , \qquad 
H_f = H_f(\{J_{ij}\}) \; ,
$$
\be
\label{5.93}
H_f(\{J_{ij}\}) = \frac{N}{2} \; w_f^2 \; U - 
w_f^2 \sum_{i\neq j} J_{ij} \; \sgm_{if}\; \sgm_{jf} \; ,
\ee
where $\sigma_{jf} \pm 1$ and the exchange interaction is distributed by the 
Gaussian law
\be
\label{5.94}
 p(J_{ij}) = \sqrt{\frac{N}{2\pi} } \;\exp\left\{ -\; \frac{N}{2J^2}
\left( J_{ij} - \; \frac{J_0}{N}\right)^2 \right\} \;  .
\ee
Setting the mean $J_0=0$ excludes the possibility of ferromagnetic phase. 
The randomness of the exchange interactions corresponds to the frozen disorder, 
which implies that the frozen free energy
\be
\label{5.95}
F(\{J_{ij}\} ) = - \; \frac{T}{N} \; 
\ln \;{\rm Tr} \; \exp\{ - \bt \widetilde H( \{ J_{ij} \} ) \}
\ee
has to be averaged over the interactions,
\be
\label{5.96}
F = \lgl\lgl \; F(\{J_{ij}\} ) \; \rgl\rgl \equiv
\int F(\{J_{ij}\} ) \; \prod_{i\neq j} p(J_{ij} )\; \frac{dJ_{ij}}{J} \; .
\ee
       
The phases are distinguished by the Edwards-Anderson \cite{Edwards_284} order 
parameter 
\be
\label{5.97}
q_f \equiv \lgl\lgl \; \left (\lgl \; \sgm_{jf}\; \rgl^2 
\right) \; \rgl\rgl \; .
\ee
The spin glass phase is indexed by $f=1$ and the paramagnetic phase, by $f=2$. 
Then
\be
\label{5.98}
  q_1 \not\equiv 0 \;, \qquad q_2 = 0 \; .
\ee

The calculation of the free energy involves the use of the replica trick
$$
\ln Z = \lim_{n\ra 0} \; \frac{1}{n} \left( Z^n - 1 \right) =
\lim_{n\ra 0} \; \frac{\prt}{\prt n} \; Z^n \;  .
$$
This makes it possible to write
\be
\label{5.99}
F = - \; \frac{T}{N} \; 
\lim_{n\ra 0} \; \frac{1}{n} \left( Z(\{ J_{ij}\} ) - 1 \right) \;  ,
\ee
where
\be
\label{5.100}
Z(\{ J_{ij}\} ) \equiv 
{\rm Tr}\;\exp\{ -\bt \widetilde H(\{ J_{ij}\} ) \} \;  .
\ee
  
Employing the replica trick gives the free energy
$$
F = \frac{1}{2} \; \left[ \; w^2 + (1 - w)^2 \; \right] \; U - \;
\frac{1}{4}\; w^4 (1 - q)^2 \; \frac{J^2}{T} - \; 
\frac{1}{4} \; ( 1 - w)^2 \; \frac{J^2}{T} \; -
$$
\be
\label{5.101}
  - \; 
T \ln 2 - T \int_{-\infty}^\infty p(x) \; \ln \left[ \; 
2\cosh\left( w^2 q^{1/2}\; \frac{J}{T}\; x \right) \; \right]\; dx \; ,   
\ee
in which 
$$
w \equiv w_1 \; , \qquad q \equiv q_1 \; , \qquad 
p(x) = \frac{1}{\sqrt{2\pi}} \;  e^{-x^2/2} \;  .
$$
For the spin glass order parameter, we get
\be
\label{5.102}
q = \int_{-\infty}^\infty p(x) \; 
\tanh^2 \left( w^2 q^{1/2}\; \frac{J}{T}\; x \right)\; dx \;  .
\ee

Minimizing the free energy with respect to $w$ results in the equation
\be
\label{5.103}
w^3 ( 1 - q)^2 - (1 -w)^3 - u (2w -1 )\; T = 0 \; ,
\ee
where $u \equiv U/J$ and temperature is measured in units of $J$. 

At low temperature $T \ra 0$, the spin-glass order parameter is
$$
q \simeq 
1 - \;\frac{u_0}{2} \; T - u_0\; ( u_0 - u) \; T^{4/3} \qquad ( u < u_0 ) \; ,
$$
\be
\label{5.104}
q \simeq 
1 - \;\frac{u_0}{2} \; T - \; \frac{T^2}{\pi} \qquad ( u \geq u_0 ) \; ,
\ee
where
\be
\label{5.105}
 u \equiv \frac{U}{J} \; , \qquad 
u_0 \equiv 2\; \sqrt{\frac{2}{\pi} } = 1.595769 \; .
\ee
At this low temperature, the spin-glass probability reads as  
$$
w \simeq 1 - ( u_0 - u)\; T^{1/3} \qquad ( u < u_0 ) \; ,
$$
\be
\label{5.106}
w \simeq 1  \qquad ( u \geq u_0 ) \;   .
\ee
Specific heat is positive, although divergent, 
$$
C_V \simeq \frac{1}{6}\; ( u - u_0)^{4/3} \; T^{-2/3} 
\qquad 
( u < u_0 ) \; ,
$$
\be
\label{5.107}
C_V \simeq \frac{(\pi^3-6)}{24\pi}\;  u_0 \; T 
\qquad 
( u \geq u_0 ) \;   .
\ee
However, the entropy can become negative and divergent,
$$
S \simeq - \; \frac{1}{4} \; ( u_0 - u)^{4/3}\; T^{-2/3} 
\qquad 
( u < u_0 ) \; ,
$$
\be
\label{5.108}
S \simeq \ln 2 - \; \frac{1}{2\pi} = 0.53399  
\qquad 
( u \geq u_0 ) \;  ,
\ee
similarly to the Sherrington-Kirkpatrick case 
\cite{Sherrington_281,Kirkpatrick_282}, which is connected with the 
instability of the Sherrington-Kirkpatrick solution for the low-temperature 
spin glass \cite{Almeida_285}.  

In the vicinity of the critical point
\be
\label{5.109}
 T_c = \frac{1}{4} \qquad ( q = 0 ) \; ,
\ee
the spin-glass order parameter behaves as
\be
\label{5.110}
q \simeq ( -\tau) \qquad \left( \tau \equiv \frac{T-T_c}{T_c} \right)
\ee
and the phase probability is
\be
\label{5.111}
 w \simeq \frac{1}{2} - \frac{\tau^2}{4(u-3)} \; .
\ee
The stability condition 
\be
\label{5.112}
\frac{\prt^2 f}{\prt w^2} \simeq 2 J ( u - 3) > 0
\qquad
( T \ra T_c )
\ee
shows that the phase transition between the spin-glass and paramagnetic phases 
is of second order, provided that $u>3$. The value $u=3$ corresponds to a 
tricritical point, where the transition changes to first order one. For $u<3$, 
the transition is of first order. 

Thus, the system ground state could be the pure spin glass phase, since $w=1$ 
at $T=0$. However, when $u<u_0$, the system is not stable at low temperature, 
similarly to the Sherrington-Kirkpatrick case 
\cite{Sherrington_281,Kirkpatrick_282}. For $u>u_0$, the mixed state becomes 
partially stable, as far as the specific heat and entropy demonstrate the normal 
behavior. Unfortunately, the corresponding free energy is not minimal. The system 
regains stability for the use of the solution for the order parameter with broken 
replica symmetry \cite{Parisi_286}. By numerical analysis, it is possible to show 
that the free energy with broken replica symmetry, representing the mixed state 
of spin glass with paramagnetic fluctuations, is lower than the free energy of 
the pure spin glass phase \cite{Kislinsky_283}.

\subsection{Systems with Magnetic Reorientations}

Materials, exhibiting the phase transitions of magnetization reorientation, 
are often characterized by the coexistence of phases with different directions 
of magnetization \cite{Belov_36,Belov_37}. A material with coexisting magnetic 
phases, with different orientations of magnetization, is described as follows 
\cite{Yukalov_274,Bakasov_287,Bakasov_288,Yukalov_289}.  

Let us consider a system in zero magnetic field, where four phases can coexist, 
so that $f = 1,2,3,4$. Three of the phases correspond to magnetic states with 
their magnetizations along one of the mutually orthogonal axes and the fourth is 
the paramagnetic phase. The role of the order parameter is played by the set of 
four quantities
\be
\label{5.113}
 s_f = \left\lgl \; \frac{1}{NS} \sum_{j=1}^N S_{jf}^f \; \right\rgl \; ,
\qquad
s_4 = 0 \qquad ( f = 1,2,3,4 ) \; ,
\ee
where $S_{jf}^f$ is an $f$ component of the spin operator localized at a site 
$j$ of the $f$-th phase. The first three quantities are the reduced magnetizations 
$s_f \neq 0$ along the axes $f=1,2,3$, while the fourth phase is paramagnetic with 
$s_4 \equiv 0$.    

In view of the existence of four phases, the space of microscopic states is the 
fiber space
\be
\label{5.114}
\widetilde\cH = 
\cH_1 \; \bigotimes \; \cH_2 \; \bigotimes \; \cH_3 \; \bigotimes \; \cH_4
\ee
of four weighted Hilbert spaces. Respectively, the Hamiltonian is the direct sum 
\be
\label{5.115}
\widetilde H = \bigoplus_{f=1}^4 \; H_f
\ee
of four terms
\be
\label{5.116}
H_f = \frac{w_f^2}{2}\; N\; U - 
w_f^2 \sum_{i\neq j}^N \left( J_{ij}^f \; S_{if}^f \; S_{jf}^f \right) \; .
\ee
The order parameter can be written as the vector
\be
\label{5.117}
 {\bf s} = s_1\bfe_1 + s_2\bfe_2 + s_3\bfe_3 \; , \qquad s_4 = 0 \; ,
\ee 
where ${\bf e}_\alpha$ is a unit vector along the axis $\alpha$. 

If all four phases are present, then the minimization condition
\be
\label{5.118}
\frac{\prt f}{\prt w_f} = 0 \qquad \left( \sum_{f=1}^4 w_f = 1 \right)
\ee
yields the equations for the phase probabilities
\be
\label{5.119}
 w_f = \left( \sum_{\al=1}^4 \frac{U-2\Phi_f}{U-2\Phi_\al}\right)^{-1}
\qquad
( f = 1,2,3,4 ) \; ,
\ee
in which
\be
\label{5.120}
 \Phi_\al = \frac{1}{N} \sum_{i\neq j} J_{ij}^\al \; 
\lgl \; S_{i\al}^\al \; S_{j\al}^\al \; \rgl \;  .
\ee

Similarly, when some of the probabilities are identically zero, say $w_4\equiv 0$,
then the minimization condition
\be
\label{5.121}
\frac{\prt F}{\prt w_f} = 0 
\qquad
\left( \sum_{f=1}^3 w_f = 1 , \: w_4 = 0 \right)
\ee
gives the equations for the phase probabilities
\be
\label{5.122}
 w_f = \left( \sum_{\al=1}^3 \frac{U-2\Phi_f}{U-2\Phi_\al}\right)^{-1} 
\qquad
 ( f = 1,2,3 ) \; .
\ee
Altogether, it is necessary to analyze $15$ possible cases:

\vskip 3mm
\begin{center}
\begin{tabular}{lllll}
1.~& $w_1\neq 0$\; ,~~&~~$w_2\neq 0$\; ,~~&~~$w_3\neq 0$\; ,~~&~~$w_4\neq 0$\; , \\
2.~& $w_1\equiv 0$\; ,~~&~~$w_2\neq 0$\; ,~~&~~$w_3\neq 0$\; ,~~&~~$w_4\neq 0$\; , \\
3.~& $w_1\neq 0$\; ,~~&~~$w_2\equiv 0$\; ,~~&~~$w_3\neq 0$\; ,~~&~~$w_4\neq 0$\; , \\
4.~& $w_1\neq 0$\; ,~~&~~$w_2\neq 0$\; ,~~&~~$w_3\equiv 0$\; ,~~&~~$w_4\neq 0$\; , \\
5.~& $w_1\equiv 0$\; ,~~&~~$w_2\equiv 0$\; ,~~&~~$w_3\neq 0$\; ,~~&~~$w_4\neq 0$\; , \\
6.~& $w_1\equiv 0$\; ,~~&~~$w_2\neq 0$\; ,~~&~~$w_3\equiv 0$\; ,~~&~~$w_4\neq 0$\; , \\
7.~& $w_1\neq 0$\; ,~~&~~$w_2\equiv 0$\; ,~~&~~$w_3\equiv 0$\; ,~~&~~$w_4\neq 0$\; , \\
8.~& $w_1\equiv 0$\; ,~~&~~$w_2\equiv 0$\; ,~~&~~$w_3\equiv 0$\; ,~~&~~$w_4\equiv1$\; , \\
9.~& $w_1\neq 0$\; ,~~&~~$w_2\neq 0$\; ,~~&~~$w_3\neq 0$\; ,~~&~~$w_4\equiv 0$\; , \\
10.~& $w_1\equiv 0$\; ,~~&~~$w_2\neq 0$\; ,~~&~~$w_3\neq 0$\; ,~~&~~$w_4\equiv 0$\; , \\
11.~& $w_1\neq 0$\; ,~~&~~$w_2\equiv 0$\; ,~~&~~$w_3\neq 0$\; ,~~&~~$w_4\equiv 0$\; , \\
12.~& $w_1\neq 0$\; ,~~&~~$w_2\neq 0$\; ,~~&~~$w_3\equiv 0$\; ,~~&~~$w_4\equiv 0$\; , \\
13.~& $w_1\equiv 1$\; ,~~&~~$w_2\equiv 0$\; ,~~&~~$w_3\equiv 0$\; ,~~&~~$w_4\equiv 0$\; , \\
14.~& $w_1\equiv 0$\; ,~~&~~$w_2\equiv 1$\; ,~~&~~$w_3\equiv 0$\; ,~~&~~$w_4\equiv 0$\; , \\
15.~& $w_1\equiv 0$\; ,~~&~~$w_2\equiv 0$\; ,~~&~~$w_3\equiv 1$\; , ~~&~~$w_4\equiv 0$\; .
\end{tabular}
\end{center}

The system state is described by the minimal of the free energies corresponding 
to these cases.

The analysis has been done in the mean-field approximation 
\cite{Yukalov_274,Bakasov_287,Bakasov_288,Yukalov_289}, when the Hamiltonian 
(\ref{5.116}) takes the form
\be
\label{5.123}
 H_f = w_f^2 \; \frac{N}{2} \; U - w_f^2 \; J_f \; N \;
\left( \frac{2}{N} \sum_{j=1}^N S_{jf}^f - C_f\right) \; C_f \;  ,
\ee
in which 
\be
\label{5.124}
C_f \equiv \left\lgl \; \frac{1}{N} \sum_{j=1}^N S_{jf}^f\right\rgl =
S \; s_f \; .
\ee
Then the total free energy, for spin one-half, reads as
\be
\label{5.125}
F = \sum_{f=1}^3 \left\{
w_f^2 \left( \frac{U}{2} + J_f C_f^2\right) -
T \ln \; \left[\; 2\cosh\left( \frac{w_f^2 J_f C_f}{T}\right) \; \right] \right\}
+ w_4^2\; \frac{U}{2} \; - T\ln 2 \;  ,
\ee
where 
\be
\label{5.126}
 J_f \equiv \frac{1}{N} \sum_{i\neq j} J_{ij}^f \;  .
\ee
For concreteness, we set that the average exchange interactions are arranged in 
the order
\be
\label{5.127}
 0 < J_1 < J_2 < J_3 \;  .
\ee
 
The conditions
\be
\label{5.128}
 s_f(T_f) = 0 \qquad ( f = 1,2,3 )  
\ee
define the reorientation temperatures, the largest of which is the critical 
temperature
\be
\label{5.129}
  T_c \equiv \sup_f \; T_f 
\ee
of the ferromagnet-paramagnet phase transition. 

The temperature, where there appears the $f$-th phase, is called the $f$-th phase 
nucleation temperature, where for example
$$
w_f(T) = 0 \qquad ( T < T_{fn} ) \; ,
$$
\be
\label{5.130}
w_f(T) > 0 \qquad ( T > T_{fn} ) \; .
\ee
The average spin along the $f$-th axis is given by the equation
\be
\label{5.131}
C_f = \frac{1}{2}\; \tanh\left( \frac{w_f^2 J_f C_f}{T} \right) \;  .
\ee
 
The overall picture of the transitions between the stable solutions can be 
represented in the following scheme describing the phase transitions an their 
order depending on the value of the parameter 
\be
\label{5.132}
u \equiv \frac{U}{J_3} \;  .
\ee
In the square brackets, zero implies the absence of magnetization along the 
related axis $f = 1,2,3$, and plus means the existence of magnetization along 
the corresponding axis. The arrow shows the increase of temperature.   

For $u < 0$, there exists a single phase transition of second order at the 
temperature $T_c = J_3/2$, which is represented as
$$
[0\;0\;+] \quad 
\overset{T_c}{\underset{2}{\longmapsto}} \quad 
[0\;0\;0]  \; .
$$

For the range $0<u<0.5$, there may happen either the reorientation transitions
$$
[0\;0\;+] \quad
\overset{T_{n_1}}{\underset{2}{\longmapsto}}  \quad
[+\;0\;+] \quad
\overset{T_{n_2}}{\underset{2}{\longmapsto}} \quad
[+\;+\;+] \quad
\overset{T_0}{\underset{1}{\longmapsto}} \quad
[0\;0\;0] \; ,
$$
or the sequence of the transitions
$$
[0\;0\;+] \quad
\overset{T_{n_1}}{\underset{1}{\longmapsto}}  \quad
[+\;0\;+] \quad
\overset{T_{n_2}}{\underset{2}{\longmapsto}} \quad
[+\;+\;+] \quad
\overset{T_0}{\underset{1}{\longmapsto}} \quad
[0\;0\;0] \; ,
$$
or the transitions
$$
[0\;0\;+] \quad
\overset{T_{n_1}=T_{n_2}}{\underset{1}{\longmapsto}}  \quad
[+\;+\;+] \quad
\overset{T_0}{\underset{1}{\longmapsto}} ~[0\;0\;0] \; .
$$

In the range $0.5<u\leq 9/4$, we have the first-order phase transition
$$
[0\;0\;+] \quad 
\overset{T_0}{\underset{1}{\longmapsto}} \quad 
[0\;0\;0]  \; .
$$
And for $u>9/4$, the following sequence of phase transitions occurs,
$$
[+\;+\;+] \quad
\overset{T_{n_1}}{\underset{2}{\longmapsto}}  \quad
[0\;+\;+] \quad
\overset{T_{n_2}}{\underset{2}{\longmapsto}} \quad
[0\;0\;+] \quad
\overset{T_c}{\underset{2}{\longmapsto}} \quad
[0\;0\;0] \; .
$$
More details can be found in Refs. 
\cite{Yukalov_274,Bakasov_287,Bakasov_288,Yukalov_289}.

\subsection{Model of Heterophase Superconductor}

High-temperature superconductors are known to often be heterophase, being 
composed of the mixture of superconducting and normal phases. Actually, it 
is widely accepted that the majority of high-temperature superconductors, 
such as cuprates, possess the principal property distinguishing them from 
the conventional low-temperature superconductors. This property is mesoscopic 
phase separation, implying that not the whole volume of a sample is 
superconducting but it is separated into nanosize regions of superconducting 
and normal phases. There exist numerous experiments confirming the occurrence 
of the phase separation in high-temperature superconductors, as is summarized 
in Refs. \cite{Phillips_172,Benedek_174,Sigmund_175,Kivelson_176}. For instance, 
in high-temperature superconductors there appear the so-called stripes that 
are self-organized networks of charges inside small bubbles of $100-300$\AA \; 
fluctuating at the time scale $10^{-12}$ s \cite{Bianconi_290}. 

A model of a heterophase superconductor was advanced 
\cite{Shumovsky_291,Yukalov_292}, 
yet before the high-temperature superconductors had been discovered 
\cite{Bednorz_293}. The models with isotropic gap 
\cite{Yukalov_292,Shumovsky_294,Shumovsky_295,Yukalov_296,Coleman_297}, 
as well as with anisotropic gap \cite{Yukalov_298,Yukalov_299} have been 
suggested. 

The Hamiltonian of a heterophase superconductor has the general form
\be
\label{5.133}
\widetilde H = H_1 \; \bigoplus \; H_2 \; , \qquad 
H_f = H_f^{kin} + H_f^{int} \;   ,
\ee
consisting of the sum of a kinetic term and interaction term. The kinetic term 
reads as
\be
\label{5.134}
 H_f^{kin} = w_f \sum_s 
\int \psi_{sf}^\dgr(\br) \; [\; \hat K_f(\br) - \mu \;] \; 
\psi_{sf}(\br)\; d\br \; ,
\ee
where $\psi_{sf}({\bf r})$ is the field operator of a charged Fermi particle 
with spin $s$ in the phase $f$. The interaction term has the form
\be
\label{5.135}
H_f^{int} = \frac{w_f^2}{2} \sum_{ss'} 
\int \psi_{sf}^\dgr(\br) \; \psi_{s'f}^\dgr(\br') \; \hat V_f(\br,\br') \;
\psi_{s'f}(\br') \; \psi_{sf}(\br) \; d\br d\br' \; ,
\ee
in which $\hat{V}_f$ is an interaction operator for two particles in the phase 
$f$. This operator is symmetric,
\be
\label{5.136}
  \hat V_f(\br,\br') =  \hat V_f(\br',\br) \; .
\ee

The superconducting and normal thermodynamic phases can be distinguished by 
their order indices \cite{Yukalov_212,Yukalov_213,Coleman_250,Coleman_300} 
or the order parameters. For the superconducting phase, the role of the order 
parameter is played by the non-zero anomalous average
\be
\label{5.137}
\lgl \; \psi_{s1}(\br) \;\psi_{s'1}(\br) \; \rgl \not\equiv 0 \;  ,
\ee
with the anomalous average for the normal phase being identically zero,
\be
\label{5.138}
\lgl \; \psi_{s2}(\br) \;\psi_{s'2}(\br) \; \rgl \equiv 0 \;  .
\ee

The field operator can be expanded over a complete set of functions,
\be
\label{5.139}
 \psi_{sf}(\br) = \sum_k c_{sf}(\bk) \; \vp_k(\br) \;  .
\ee
For a uniform system, one takes the plane waves and for a crystalline structure, 
one uses Bloch functions.

In that way, we meet the following matrix elements: the kinetic transport matrix
\be
\label{5.140}
 K_f(\bk,\bp) \equiv (\vp_k,\;\hat K\vp_p) \;  ,
\ee
the vertex
\be
\label{5.141}
 V_f(\bk,\bk',\bp',\bp) \equiv 
(\vp_k\vp_{k'},\; \hat V_f\vp_{p'}\vp_p) \;  ,
\ee
and the effective interaction
\be
\label{5.142}
 J_f(\bk,\bp) \equiv - V_f(\bk,-\bk,-\bp,\bp) \; .
\ee
Due to the symmetry property (\ref{5.136}), we have
$$
 V_f(\bk,\bk',\bp',\bp) =  V_f(\bk',\bk,\bp,\bp') \; , 
\qquad
 J_f(\bk,\bp) =  J_f(-\bk,-\bp) \; .
$$

The kinetic Hamiltonian acquires the form
\be
\label{5.143}
H_f^{kin} = w_f \sum_s \sum_{kp} [\; K_f(\bk,\bp) - \mu\;\dlt_{kp} \;]
c_{sf}^\dgr(\bk) \; c_{sf}(\bp)
\ee
and the interaction part becomes
\be
\label{5.144}
H_f^{int} = \frac{w_f^2}{2} \sum_{ss'} \sum_{kk'} \sum_{pp'}
 V_f(\bk,\bk',\bp',\bp) \;
c_{sf}^\dgr(\bk)\; c_{s'f}^\dgr(\bk')\; c_{s'f}(\bp')\; c_{sf}(\bp) \; .
\ee

Then we use the Hartree-Fock-Bogolubov approximation \cite{Yukalov_301} and 
consider the restricted space comprising the processes with conserved spin 
and momentum, so that
$$
c_{sf}^\dgr(\bk)\; c_{s'f}(\bk') = \dlt_{ss'}\; \dlt_{kk'} \; 
c_{sf}^\dgr(\bk)\; c_{sf}(\bk) \; ,
$$
\be
\label{5.145}
c_{sf}^\dgr(\bk)\; c_{s'f}^\dgr(\bk') = \dlt_{-ss'}\; \dlt_{-kk'} \; 
c_{sf}^\dgr(\bk)\; c_{-sf}^\dgr(-\bk) \;  .
\ee
Introduce the momentum distribution 
\be
\label{5.146}
n_f(\bk) \equiv \sum_s \lgl \; c_{sf}^\dgr(\bk)\; c_{sf}(\bk) \; \rgl
\ee
and the anomalous average
\be
\label{5.147}
 \sgm_f(\bk) \equiv \lgl \; c_{-sf}(-\bk)\; c_{sf}(\bk) \; \rgl  .
\ee

The resulting Hamiltonian can be diagonalized by means of the Bogolubov 
canonical transformation
\be
\label{5.148}
  c_{sf}(\bk) = u_f(\bk)\; a_{sf}(\bk) + 
v_f(\bk)\; a_{-sf}^\dgr(\bk) \; ,
\ee
in which
\be
\label{5.149}
|\; u_f(\bk)\; |^2 = \frac{1}{2}\; 
\left[ \; 1 + \frac{\om_f(\bk)}{\ep_f(\bk)} \;\right] \; ,
\qquad
|\; v_f(\bk)\; |^2 = \frac{1}{2}\; 
\left[ \; 1 -\; \frac{\om_f(\bk)}{\ep_f(\bk)} \;\right] \; .
\ee
Here we have defined the effective single-particle spectrum
\be
\label{5.150}
\om_f(\bk) = K_f(\bk,\bk) + w_f \; M_f(\bk) - \mu
\ee
and the excitation spectrum
\be
\label{5.151}
\ep_f(\bk) = \sqrt{\Dlt_f^2(\bk) + \om_f^2(\bk)} \; ,
\ee
with the mass operator
\be
\label{5.152}
M_f(\bk) \equiv \sum_p \left[\; V_f(\bk,\bp,\bp,\bk) - \; \frac{1}{2}\;
V_f(\bk,\bp,\bk,\bp) \; \right] \; n_f(\bp)
\ee
and the gap
\be
\label{5.153}
\Dlt_f(\bk) = w_f \sum_p J_p(\bk,\bp) \; \sgm_f(\bp) \;  .
\ee

As a result, the Hamiltonian reduces to 
\be
\label{5.154}
H_f = 
w_f \sum_s \sum_k \ep_f(\bk) \; a_{sf}^\dgr(\bk) \; a_{sf}(\bk) + w_f C_f  
\ee
with the nonoperator term
\be
\label{5.155}
C_f = \sum_k \left[\; \om_f(\bk) - \ep_f(\bk) + 
\Dlt_f(\bk) \; \sgm_f(\bk) - \; 
\frac{1}{2} \; w_f \; M_f(\bk) \; n_f(\bk) \; \right] \; .
\ee
The momentum distribution becomes
\be
\label{5.156}
 n_f(\bk) = 1 - \; \frac{\om_f(\bk)}{\ep_f(\bk)} \; 
\tanh\; \frac{w_f\ep_f(\bk)}{2T}   
\ee
and the anomalous average is
\be
\label{5.157}
\sgm_f(\bk) =  \frac{\Dlt_f(\bk)}{2\ep_f(\bk)} \; 
\tanh\; \frac{w_f\ep_f(\bk)}{2T} \;    .
\ee

The anomalous average is the order parameter distinguishing the superconducting 
and normal phases, so that
\be
\label{5.158}
 \sgm_1(\bk) \not\equiv 0 \; , \qquad  
\sgm_2(\bk) \equiv 0 \; .
\ee
This yields the distinction for the gap in the superconducting and normal phases,
\be
\label{5.159}
\Dlt_1(\bk) \not\equiv 0 \; , \qquad  
\Dlt_2(\bk) \equiv 0 \; .
\ee
  
\subsection{Stability of Heterophase States}

The existence of a heterophase state implies the stability of the state with 
respect to the variation of phase probabilities. For grand thermodynamic 
ensemble, we need to consider the grand thermodynamic potential
\be
\label{5.160}
\Om = - T\ln \; {\rm Tr}\; e^{-\bt\widetilde H} \; ,
\ee
where $\beta \equiv 1/T$. The extremum of the grand potential with respect 
to $w \equiv w_1$, with $w_2 = 1-w$, yields the equation defining the phase 
probability $w$,
\be
\label{5.161}
 \frac{\prt\Om}{\prt w} = 
\left\lgl \; \frac{\prt\widetilde H}{\prt w}\; \right\rgl = 0 \; .
\ee
This extremum is a minimum provided that 
\be
\label{5.162}
\frac{\prt^2\Om}{\prt w^2} = \left[\; \left\lgl \; 
\frac{\prt^2\widetilde H}{\prt w^2}\; \right\rgl - \bt\; 
\left\lgl \; \left(\frac{\prt\widetilde H}{\prt w}\right)^2 \;\right\rgl \; 
\right] ~ > ~ 0 \; .
\ee
Since the second term in the last inequality is positive, the necessary 
condition for the minimum is
\be
\label{5.163}
\left\lgl \; \frac{\prt^2\widetilde H}{\prt w^2}\; \right\rgl ~ > ~ 0 \; .
\ee

Introducing the notations 
\be
\label{5.164}
 \lgl \; H_f^{kin}\; \rgl \equiv w_f Q_f \; , \qquad
\lgl \; H_f^{int}\; \rgl \equiv \frac{w_f^2}{2} \; \Phi_f \; ,
\ee
from (\ref{5.161}), we get the equation for the phase probability
\be
\label{5.165}
 w = \frac{\Phi_2 + Q_2 - Q_1}{\Phi_1+\Phi_2}  
\ee
and from (\ref{5.163}), we find the stability condition
\be
\label{5.166}
 \Phi_1+\Phi_2 > 0 \;  .
\ee
  
For the heterophase superconductor, we have 
\be
\label{5.167}
   \Phi_f = \sum_k M_f(\bk) \; n_f(\bk) -
 2 \sum_{kp} J_p(\bk,\bp) \; \sgm_f(\bk) \; \sgm_f(\bp) \; .
\ee
Then the stability condition (\ref{5.166}) yields
\be
\label{5.168}
\sum_f \sum_k M_f(\bk) \; n_f(\bk) > 
2 \sum_{kp} J_1(\bk,\bp) \; \sgm_1(\bk) \; \sgm_1(\bp) \; .
\ee

The gap equation (\ref{5.154})reads as
\be
\label{5.169}
 \Dlt_f(\bk)= w_f \sum_p J_f(\bk,\bp) \; 
\frac{\Dlt_f(\bp)}{2\ep_f(\bp)}\; \tanh\; \frac{w_f\ep_f(\bp)}{2T} \; .
\ee
A positive solution for the gap requires that 
\be
\label{5.170}
  J_f(\bk,\bp)  > 0
\ee 
in the region of momenta making the main contribution to the summation. Therefore
a necessary condition for (\ref{5.168}) is
\be
\label{5.171}
 \sum_f \sum_k M_f(\bk) \; n_f(\bk) > 0 \; .
\ee
This implies that there should exist in the system sufficiently strong repulsive 
interactions. The first candidate for this is the Coulomb interaction of charge 
carriers. Thus we come to the requirement:

{\bf Conclusion 1}. {\it Heterophase state in a superconductor can exist only in 
the presence of repulsive interactions, such as Coulomb interactions}.

\subsection{Uniform Heterophase Superconductor}

For the case of a uniform system, it is reasonable to deal with the plane waves
\be
\label{5.172}
\vp_k(\br) = \frac{1}{\sqrt{V}} \; e^{i\bk\cdot\br} \; ,
\ee
where $V$ is the system volume. Since the gap for the normal phase is trivial, 
$\Delta_2({\bf k}) \equiv 0$, it is necessary to consider only the gap of the 
superconducting phase. It is convenient to define
\be
\label{5.173}
 J_1(\bk,\bp) \equiv \frac{1}{\sqrt{V}} \; J(\bk,\bp) \;  .
\ee
To simplify notations, we can omit the index $1$ in the expressions related to 
the superconducting phase, such as 
$$
 \Dlt_1(\bk) \equiv \Dlt(\bk) \; , \qquad  
 \ep_1(\bk) \equiv \ep(\bk) \; , \qquad 
 \om_1(\bk) \equiv \om(\bk) \; .
$$
Taking into account that in the thermodynamic limit the replacement 
$$
 \frac{1}{V} \sum_p \longmapsto \int \frac{d\bp}{(2\pi)^3}  
$$
is valid, where the integration is over all values of the momentum, from Eq. 
(\ref{5.169}) we come to the equation for the gap of a uniform heterophase 
superconductor
\be
\label{5.174}
\Dlt(\bk) = w \int J(\bk,\bp) \; \frac{\Dlt(\bp)}{2\ep(\bp)} \;
\tanh \; \frac{w\ep(\bp)}{2T} \; \frac{d\bp}{(2\pi)^3} \; .
\ee
 
The charge carriers interact with each other in two ways, through attractive 
interactions induced by phonon exchange and repulsive Coulomb interactions 
\cite{Pines_302,Ziman_303}, which can be denoted as
\be
\label{5.175}
J(\bk,\bp) = J_{ph}(\bk,\bp) + J_C(\bk,\bp) \; .
\ee
The phonon-exchange part is
\be
\label{5.176}
 J_{ph}(\bk,\bp) = 
\frac{|\;\al\;|^2}{\om_{ph}^2 -[\om(\bk)-\om(\bp)]^2} \;  ,
\ee
with the electron-phonon coupling
$$
\al = - \; \frac{4\pi\; i\;Z_{ion}e_0^2}{k_F(1+\varkappa^2/k_F^2)} \;
\left( \frac{\rho_{ion}}{m_{ion}}\right)^{1/2}
$$
and the characteristic phonon frequency $\omega_{ph}$. Here $e_0$ is the 
charge of the carriers, $k_F$ is the Fermi momentum defined by the equation 
$\om(k_F)=0$, $\varkappa^{-1}$ is the screening radius of charge interactions, 
and $\rho_{ion}$, $m_{ion}$, and $Z_{ion}$ are the density, mass, and charge 
of an ion. The Fermi momentum approximately is $k_F\approx(3\pi^2\rho_e)^{1/3})$, 
where $\rho_e$ is electron density. 

The term due to the screened Coulomb interactions is 
\be
\label{5.177}
J_C(\bk,\bp) = - \;
\int \Phi_C(\br) \; e^{-i(\bk-\bp)\cdot\br}\; d\br \;  ,
\qquad
\Phi_C(\br) = \frac{e_0^2}{r} \; e^{-\varkappa r} \; ,
\ee
which gives
\be
\label{5.178}
J_C(\bk,\bp) = - \; \frac{4\pi e_0^2}{|\bk-\bp|^2+\varkappa^2} \; ,
\ee
where the notation (\ref{5.142}) with the sign minus is taken into account. 
The screening radius $\varkappa^{-1}$ is defined by the expression
$$
 \varkappa^2 = \frac{4}{a_B} \; 
\left( \frac{3}{\pi}\; \rho_e\right)^{1/3} \; ,
$$
where $a_B=1/(m_e e_0^2)$ is the Bohr radius. Thus we have 
\be
\label{5.179}
J(\bk,\bp) = 
\frac{|\;\al\;|^2}{\om_{ph}^2 -[\om(\bk)-\om(\bp)]^2} \; - \; 
\frac{4\pi e_0^2}{|\bk-\bp|^2+\varkappa^2} \;  .
\ee

The main contribution to integral (\ref{5.174}) comes form the momentum 
$p$ that is close to the Fermi momentum $k_F$. To simplify the gap equation 
(\ref{5.174}), let us average the effective interactions over spherical 
angles,
\be
\label{5.180}
 J(\bk) \equiv \lim_{p\ra k_F} \; \frac{1}{4\pi} \int
J(\bk,\bp)\; d\Om(\bp) \;  .
\ee
This gives
\be
\label{5.181}
J(\bk)= \frac{|\;\al\;|^2}{\om_{ph}^2-\om^2(\bk)} \; - \; 
J_C(\bk) \;  ,
\ee
where
\be
\label{5.182}
J_C(\bk) = \frac{\pi e_0^2}{k k_F} \; \ln\; \left| \;
\frac{ (k+k_F)^2+\varkappa^2}{(k-k_F)^2+\varkappa^2} \; \right| \;  .
\ee

The gap equation (\ref{5.174}) can be approximated by the equation
\be
\label{5.183}
  \Dlt(\bk) = w J(\bk) \int \frac{\Dlt(\bp)}{2\ep(\bp)} \;
\tanh\; \frac{w\ep(p)}{2T} \; \frac{d\bp}{(2\pi)^3} \; .
\ee
The condition for the existence of a nontrivial gap is
\be
\label{5.184}
\frac{|\;\al\;|^2}{\om_{ph}^2-\om^2(\bk)} \; - \; J_C(\bk) > 0 \; .
\ee
At the Fermi surface, this reduces to the condition
\be
\label{5.185}
\frac{|\;\al\;|^2}{\om_{ph}^2} \; - \; J_C(k_F) > 0 \; ,
\ee
where
\be
\label{5.186}
 J_C(k_F) = \frac{\pi e_0^2}{k_F^2}\; 
\ln\left( 1 + 4 \; \frac{k_F^2}{\varkappa^2}\right) \;  .
\ee
This is similar to the Bardeen-Cooper-Schriffer criterion of 
superconductivity,\cite{Bardeen_304,Bardeen_305}, however in our case 
\cite{Yukalov_292,Yukalov_296,Coleman_297} the characteristic phonon 
frequency depends on the probability of superconducting phase,
\be
\label{5.187}
 \om_{ph} = \sqrt{w}\; \om_0 \;  ,
\ee 
where $\om_0$ is the characteristic phonon frequency for a pure superconducting 
phase.

Defining the ion plasma frequency
\be
\label{5.188}
\om_{ion} \equiv 2 Z_{ion} \; e_0 \;
\left( \frac{\pi\rho_{ion}}{m_{ion}}\right)^{1/2} 
\ee    
translates condition (\ref{5.185}) to the form
\be
\label{5.189}
 \frac{\om_{ion}^2}{\om_{ph}^2} > \frac{1}{4}\; 
\left( 1 + \frac{\varkappa^2}{k_F^2}\right) \; 
\ln \; \left( 1 + 4\;\frac{k_F^2}{\varkappa^2}\right) \; .
\ee
For $\varkappa\sim k_F$, the right-hand side of the above inequality is 
close to one, hence giving $\omega_{ion} > \omega_{ph}$. Taking into account 
the phonon frequency softening, defined in (\ref{5.187}), we come to the 
criterion
\be
\label{5.190}
\frac{\om_{ion}}{\sqrt{w}\;\om_0} > 1 \; .
\ee

In that way, we see that the mesoscopic phase separation facilitates the 
appearance of superfluidity. Despite that in a pure bad conductor, without 
heterophase states, there may be no superfluidity because of the inequality 
$\om_{ion}<\om_0$, in a heterophase system the criterion (\ref{5.190}) can
become valid, since $w<1$. Thus we can conclude:

\vskip 2mm
{\bf Conclusion 2}. {\it Phase separation enables the appearance of 
superconductivity in a heterophase sample even if it were impossible in 
pure-phase matter}.

\vskip 2mm 
Considering the gap equation (\ref{5.183}) at the Fermi surface, we define 
the gap
\be
\label{5.191}
 \Dlt \equiv \Dlt(k_F) \; ,
\ee  
the density of states
\be
\label{5.192}
N(\om) \equiv \frac{1}{(2\pi)^3} \int
\frac{dS(\om)}{|\;\prt\om(\bk)/\prt\bk\;|_{k=k(\om)} } \;  ,
\ee
where the integration is over the Fermi surface given by the equation $\om(\bk)= 0$, 
and the effective coupling parameter
\be
\label{5.193}
\Lbd \equiv w \; N(0) \; \left[\; \frac{|\;\al\;|^2}{w\om_0^2} \; - \;
J_C(k_F) \; \right] \; .
\ee
Assuming that the main contribution to the integral (\ref{5.183}) comes from 
a narrow region around the Fermi surface, where $\omega$ varies between zero 
and $\omega_{ph}$, we get the gap equation 
\be
\label{5.194}
 \int_0^{\om_{ph}} \frac{\Lbd}{\sqrt{\Dlt^2+\om^2} } \;
\tanh \left( \frac{w\sqrt{\Dlt^2+\om^2}}{2T}\right)\; d\om = 1 \; .
\ee

The equation for the critical temperature, where $\Dlt=0$, is
\be
\label{5.195}
\Lbd \int_0^{\om_{ph}} \frac{1}{\om}\;  
\tanh \left( \frac{w\om}{2T_c}\right)\; d\om =1 \; .
\ee
In the case of weak coupling, the critical temperature reads as
\be
\label{5.196}
 T_c \simeq 1.14 \; w^{3/2}\; \om_0 \; 
\exp\left( -\; \frac{1}{\Lbd} \right)  
\qquad 
( \Lbd \ll 1 )
\ee
and for strong effective coupling, it becomes
\be
\label{5.197}
T_c \simeq \frac{1}{2} \; w^{3/2} \; \Lbd \; \om_0 
\qquad 
( \Lbd \gg 1 ) \;  .
\ee

In order to clearly show the dependence of the critical temperature on the 
phase probability, let us define the coupling parameter $\lbd$ and the Coulomb 
parameter $\mu^*$ corresponding to the pure superconducting phase,
\be
\label{5.198}
\lbd \equiv N(0) \; \frac{|\;\al\;|^2}{\om_0^2} \; , 
\qquad
\mu^* \equiv N(0) \; J_C(k_F) \;  .
\ee
Then the effective coupling (\ref{5.193}) reads as
\be
\label{5.199}
\Lbd = \lbd - w \mu^* \; .
\ee
Define the dimensionless critical temperature
\be
\label{5.200}
 t_c \equiv \frac{T_c}{\om_0} \; .
\ee
Then equation (\ref{5.195}) for the critical temperature takes the form
\be
\label{5.201}
( \lbd - w \mu^* ) \int_0^1 
\tanh\left( \frac{w^{3/2}}{2t_c}\; x\right) \; \frac{dx}{x} = 1 \; .
\ee

The probability of the superconducting phase depends on thermodynamic 
parameters, such as temperature, and on the system parameters, such as the 
electron-phonon coupling, characteristic phonon frequency, density of states, 
effective Coulomb interaction, and doping of the material with admixtures. If 
the crystalline structure does not change at the phase transition, then the 
phase probabilities, defined as the minimizers of the thermodynamic potential, 
are equal at the transition point, where the gap becomes zero. However, if the 
superconducting transition is accompanied by a structural transition, the 
situation can be more complicated, so that the phase probabilities might be 
not equal at the transition point due to different effective interactions 
$V_f(\bk,\bk',\bp',\bp)$. Keeping in mind this possibility, it is possible to 
consider the phase probability $w$ as a variable between $0$ and $1$. The 
variation of $w$, for instance due to doping, changes the critical temperature 
in a nonmonotonic way. As is seen from (\ref{5.201}), the critical temperature 
tends to zero when $w$ tends to zero and also when $w$ tends to $\lbd/\mu^*$. 
For good conductors, one has $\mu^*\ll\lbd$, hence the limit $\lbd/\mu^*$ 
for $w$ cannot be achieved. However, for bad conductors, corresponding to 
high-temperature superconductors, the limit $\lbd/\mu^*$ can be reached, since 
for them $\mu^*>\lbd$. The typical for high-temperature superconductors bell 
shape of the critical temperature, as a function of the superconducting fraction 
$w$, is demonstrated in Fig. 9, Fig. 10, and Fig. 11.   

\begin{figure}[ht]
\centerline{
\includegraphics[width=10cm]{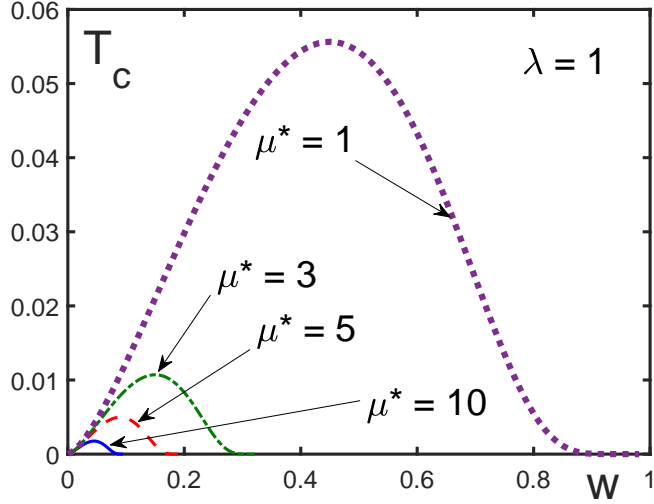} }
\caption{Critical temperature $T_c$ of superconducting phase transition 
as a function of the superconducting fraction $w$ for the electron-phonon 
coupling $\lbd=1$ and different effective Coulomb couplings: $\mu^*=1$
(dotted line), $\mu^*=3$ (dashed-dotted line), $\mu^*=5$ (dashed line), 
and $\mu^*=10$ (solid line). 
}
\label{fig:Fig.9}
\end{figure}

\begin{figure}[ht]
\centerline{
\includegraphics[width=10cm]{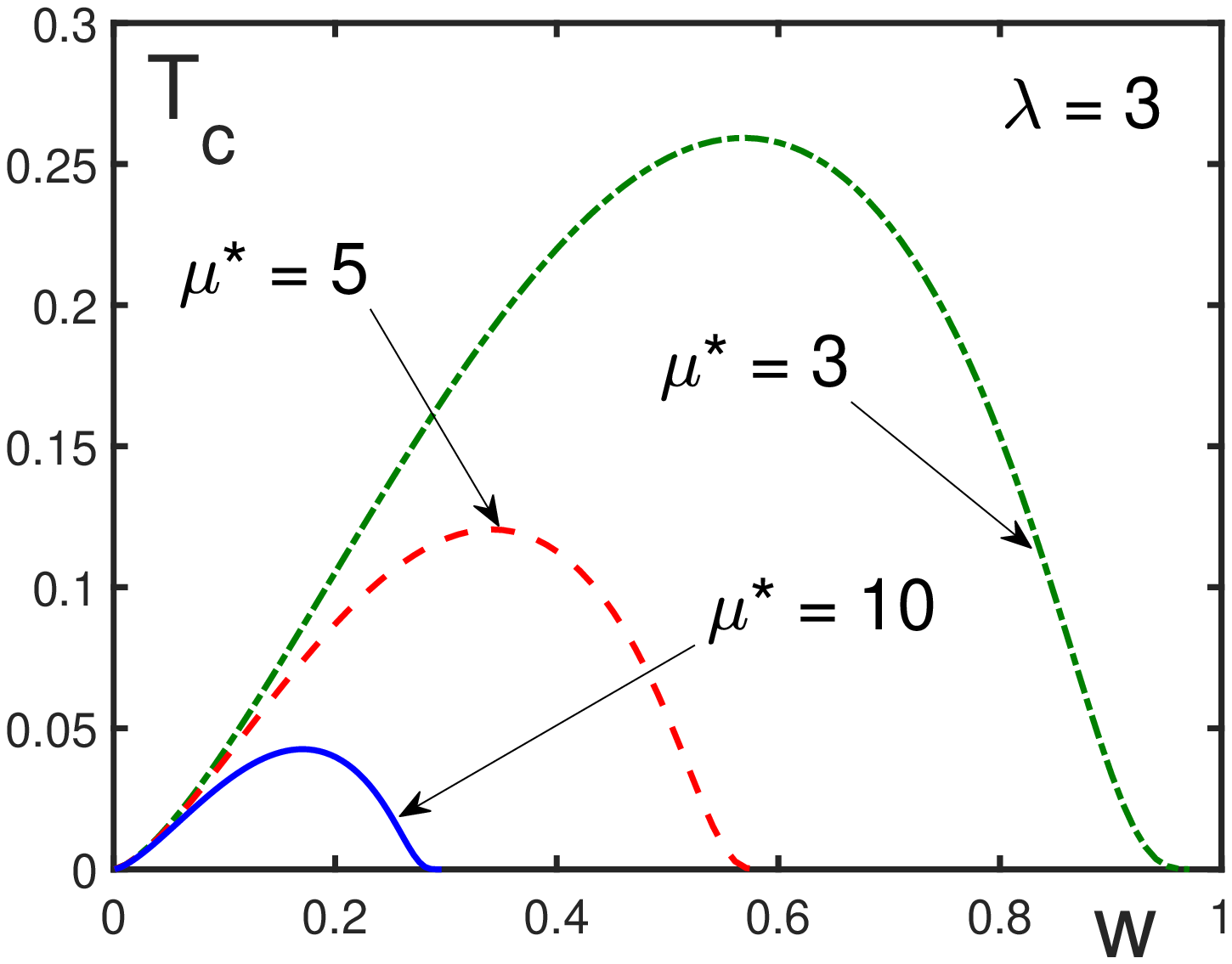} }
\caption{Critical temperature $T_c$ of superconducting transition 
as a function of the superconducting fraction $w$ for the electron-phonon 
coupling $\lbd=3$ and different effective Coulomb couplings: $\mu^*=3$ 
(dashed-dotted line), $\mu^*=5$ (dashed line), and $\mu^*=10$ (solid line).
}
\label{fig:Fig.10}
\end{figure}

\begin{figure}[ht]
\centerline{
\includegraphics[width=10cm]{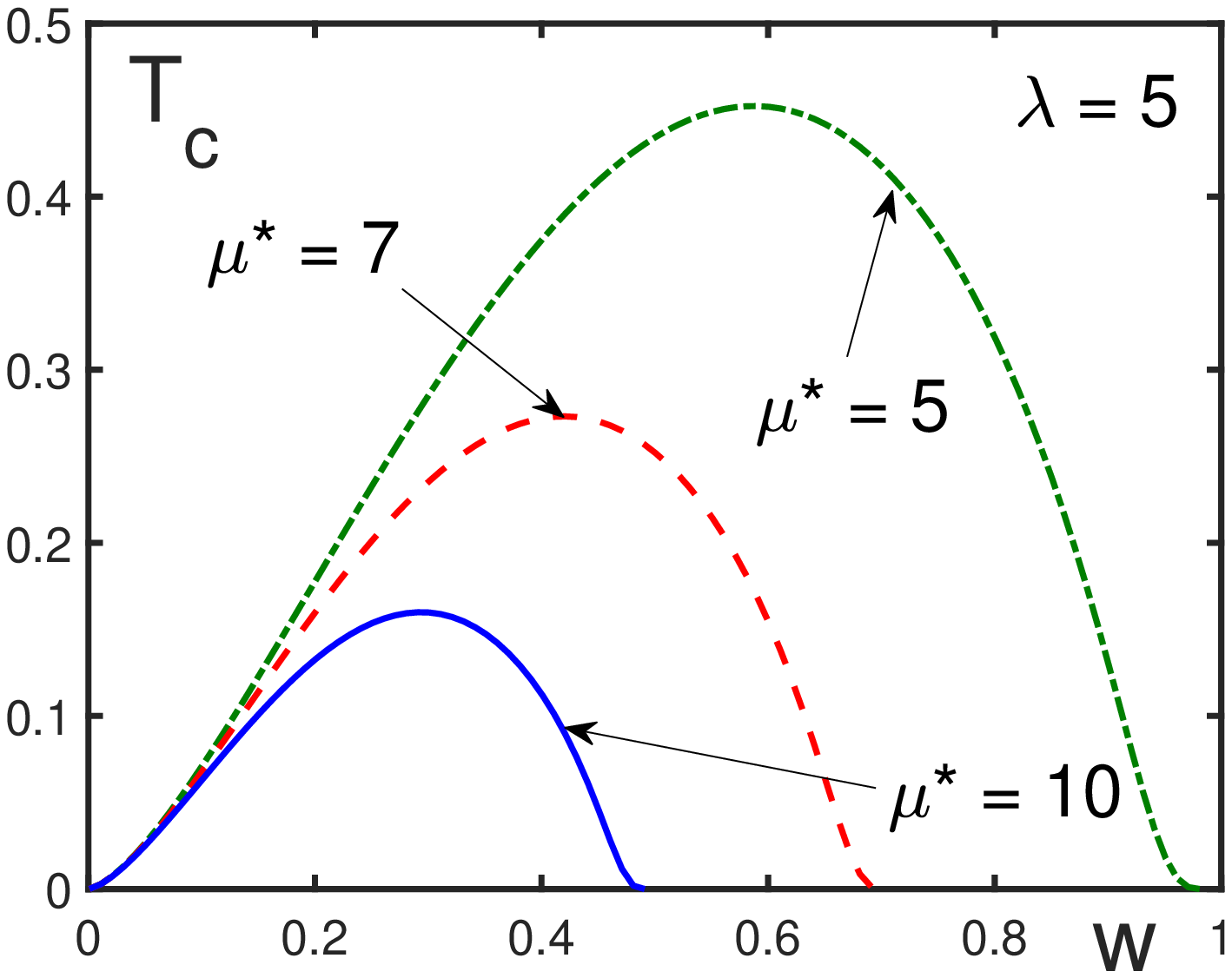} }
\caption{Critical temperature $T_c$ of superconducting transition as 
a function of the superconducting fraction $w$ for the electron-phonon 
coupling $\lbd=5$ and different effective Coulomb couplings: $\mu^*=5$ 
(dashed-dotted line), $\mu^*=7$ (dashed line), and $\mu^*=10$ (solid 
line).
}
\label{fig:Fig.11}
\end{figure}

\subsection{Anisotropic Heterophase Superconductor}

The gap of a high-temperature superconductor is known to often display 
anisotropic dependence on momentum. Numerous experiments point at the 
predominantly $d-$wave symmetry of the superconducting order parameter 
\cite{Harlingen_306,Tsuei_307}, though in some cases one claims that the 
isotropic $s-$wave symmetry can be dominant \cite{Kabanov_308}. The majority 
of experiments evidence the existence of the mixed $s+d$ pairing in cuprates, 
which describes well experimental data \cite{Tsuei_307}.

To describe the anisotropy \cite{Yukalov_298,Yukalov_299}, one may introduce 
a basis $\{ \chi(\bf k) \}$ of functions characterizing the lattice symmetry, 
with the index $i=1,2,\ldots$ enumerating irreducible representations of the 
symmetry group. The basis is assumed to be orthonormal and complete,
$$
\sum_k \chi_i^*(\bk) \; \chi_j(\bk) = \dlt_{ij}\; ,
\qquad
\sum_i \chi_i^*(\bk) \; \chi_i(\bp) = \dlt_{kp}\; .
$$
The effective interaction $J_1({\bf k}, {\bf p})$ can be expanded over this 
basis,
\be
\label{5.202}
 J_1(\bk,\bp) = \sum_{ij} J_{ij} \; \chi_i(\bk) \; \chi_j^*(\bp)  ,
\ee
together with the gap,
\be
\label{5.203}
 \Dlt(\bk) = \sum_i \Dlt_i \; \chi_i(\bk) \;  .
\ee
Then the gap equation (\ref{5.169}) reduces to 
\be
\label{5.204}
 \Dlt_i = \sum_j A_{ij} \;\Dlt_j \;  ,   
\ee
where
\be
\label{5.205}
A_{ij} \equiv \sum_p \frac{w J_{ij}}{2\ep(\bp)} \;
\tanh\left( 
\frac{w\ep(\bp)}{2T}\right) \; \chi_i^*(\bp) \; \chi_j(\bp) \;  ,
\ee
with the summation over the Brillouin zone.

The system of uniform algebraic equations (\ref{5.204}) enjoys nontrivial 
solutions, provided that
\be
\label{5.206}
 {\rm det} \; ( \hat 1 - \hat A ) = 0 \;  ,
\ee
with $\hat{1}$ being the unity matrix and the elements of the matrix $\hat{A}$
being given in (\ref{5.205}).

The effective interaction (\ref{5.202}) includes the attractive part caused by 
the electron-phonon interaction and a repulsive part due to the direct Coulomb 
interaction. This means that $J_{ij}$ has the structure
\be
\label{5.207}
 J_{ij} = \left( \frac{|\;\al\;|^2}{w\om_0^2} \; - \; 
J_C(k_F) \right) \; b_{ij} \; .
\ee
 
At the critical temperature, we have 
\be
\label{5.208}
A_{ij}(T_c) = \sum_p \frac{w J_{ij}}{2\om(\bp)} \; \tanh\left( 
\frac{w\om(\bp)}{2T_c}\right) \; \chi_i^*(\bp) \; \chi_j(\bp) \;  .
\ee
The density of states can be written as
\be
\label{5.209}
N_{ij}(\om) \equiv \sum_p \dlt(\om - \om(\bp) )\; 
\chi_i^*(\bp) \; \chi_j(\bp) \; ,
\ee
with the normalization condition
$$
 \int_{-\infty}^\infty N_{ij}(\om) \; d\om = \dlt_{ij} \; .
$$
Replacing the summation over momenta by the integration over the Brillouin 
zone,
$$
 \sum_{p\in\mathbb{B} } \longmapsto 
\frac{1}{\rho_e} \int_\mathbb{B} \frac{d\bp}{(2\pi)^3} \; ,
$$
we have
\be
\label{5.210}
N_{ij}(\om) = \frac{1}{\rho}  \int_\mathbb{B} \dlt(\om - \om(\bp) ) \;
 \chi_i^*(\bp) \; \chi_j(\bp) \; \frac{d\bp}{(2\pi)^3} \;  .
\ee
Then Eq. (\ref{5.208}) can be represented as
\be
\label{5.211}
 A_{ij}(T_c) = w\;  J_{ij}  
\int_{-\infty}^\infty \frac{N_{ij}(\om)}{2\om} \;
\tanh\left( \frac{w\om}{2T_c}\right) \; d\om \; .
\ee

Assuming again that the density of states is the largest on the Fermi 
surface and quickly diminishes outside it after the effective phonon 
frequency $\om_{ph}$, we come to the equation
\be
\label{5.212}
A_{ij}(T_c) = w\;  J_{ij}  N_{ij}(0) \int_0^{\om_{ph}} 
\tanh\left( \frac{w\om}{2T_c}\right) \; \frac{d\om}{\om} \; .
\ee
Introducing the electron-phonon coupling and effective Coulomb interaction 
for the pure anisotropic superconductor,
\be
\label{5.213}
\lbd_{ij} \equiv 
N_{ij}(0) \; \frac{|\;\al\;|^2}{\om_0^2} \; b_{ij} \; , 
\qquad 
\mu_{ij} \equiv N_{ij}(0) \; J_C(k_F) \; b_{ij} \;  ,  
\ee
reduces the effective coupling matrix to the form
\be
\label{5.214}
\Lbd_{ij} = \lbd_{ij} - w \mu_{ij} \;   .
\ee
Taking account of relation (\ref{5.187}) gives
\be
\label{5.215}
 A_{ij}(T_c) = \Lbd_{ij} \int_0^1 
\tanh\left( \frac{w^{3/2}\om_0}{2T_c} \; x\right) \; \frac{dx}{x} \; .
\ee
The critical temperature is defined by equation (\ref{5.206}), with the matrix
$\hat{A}$ having the elements (\ref{5.215}).
 
Defining the quantity $A_{eff}(T_c)$ by the equation
\be
\label{5.216}
1 - A_{eff}(T_c) \equiv {\rm det}\; ( \hat 1 - \hat A)
\ee
and the effective coupling $\Lbd_{eff}$ by the relation
\be
\label{5.217}
 A_{eff}(T_c) = \Lbd_{eff} \int_0^1
\tanh\left( \frac{w^{3/2}\om_0}{2T_c} \; x\right) \; \frac{dx}{x} \; ,
\ee
we come to the equation for the critical temperature
\be
\label{5.218}
\Lbd_{eff} \int_0^1 
\tanh\left( \frac{w^{3/2}}{2t_c} \; x\right) \; \frac{dx}{x} = 1 
\ee
having the same form as equation (\ref{5.201}) for an isotropic 
superconductor, except that now, instead of the coupling $\Lambda$, we 
have $\Lbd_{eff}$. 

The analysis of the properties of anisotropic heterophase superconductors 
\cite{Yukalov_298,Yukalov_299} allows us to make the following conclusions. 

(i) Mesoscopic phase separation in superconductors can be thermodynamically 
stable only in the presence of repulsive Coulomb interactions.

(ii) Phase separation enables the appearance of superconductivity in a 
heterophase sample even if it were impossible in pure-phase matter.

(iii) Phase separation is crucial for the occurrence of superconductivity
in bad conductors.

(iv) The critical temperature for a mixture of gap waves is higher than the 
critical temperature related to any pure gap wave from this mixture.

(v) In bad conductors, the critical temperature as a function of the 
superconducting fraction has the bell shape.

(vi) Phase separation softens the single-particle energy dispersion.

(vii) Mesoscopic phase separation suppresses the contribution of $d-$wave 
superconductivity and enhances that of $s-$wave superconductivity.

\subsection{Model of Heterophase Ferroelectric}

Dipolar ferroelectrics, exhibiting order-disorder phase transitions, 
mathematically are similar to magnetic systems, except that instead of 
magnetic moments they possess electric dipolar moments 
\cite{Bruce_72,Blinc_309}. Taking account of paraelectric fluctuations 
in ferroelectrics is accomplished in the same way as taking into account 
paramagnetic fluctuations in ferromagnets 
\cite{Bashkirov_310,Bashkirov_311,Yukalov_312,Yukalov_313}.    

It is illustrative to show how the ferroelectric model can be derived. 
Let us start with the standard Hamiltonian of a heterophase system
$$
\widetilde H = H_1 \; \bigoplus \; H_2 \;
$$
$$
H_f = w_f \int \psi_f^\dgr(\br) \; H_0(\br) \; \psi_f(\br)\; d\br \; +
$$
\be
\label{5.219}
+\; \frac{w_f^2}{2} \int \psi_f^\dgr(\br) \; \psi_f^\dgr(\br') \;
\Phi(\br-\br') \; \psi_f(\br') \; \psi_f(\br) \; d\br d\br' \; ,
\ee
where the field operators pertain to either boson or fermion statistics, 
which is not of great importance for well localized particles. Here 
\be
\label{5.220}
H_0(\br) = -\; \frac{\nabla^2}{2m} + U(\br)
\ee
is a single-particle term, with an external potential $U({\bf r})$. Keeping 
in mind a periodic solid, we expand the field operators over Wannier functions,
\be
\label{5.221}
\psi_f(\br) = \sum_{nj} c_{njf}\; w_n(\br-\br_j) \; ,
\ee
with $n$ being the band index and $\br_j$, a vector of a particle location.
Intending to consider an insulating state, we assume that Wannier functions 
are well localized \cite{Marzari_314}. 

Substituting expansion (\ref{5.221}) into Hamiltonian (\ref{5.219}) gives
$$
H_f = w_f \sum_{ij} \sum_{mn} E_{ij}^{mn} c_{mif}^\dgr \; c_{njf} \; +
$$
\be
\label{5.222}
\frac{w_f^2}{2} \sum_{ \{j\} } \sum_{ \{n\} }
\Phi_{j_1j_2j_3j_4}^{n_1n_2n_3n_4}\; 
c_{n_1j_1}^\dgr \; c_{n_2j_2}^\dgr \; c_{n_3j_3} \; c_{n_4j_4} \; ,
\ee
where
\be
\label{5.223}
E_{ij}^{mn} \equiv \int w_m(\br-\br_i) \; H_0(\br) \; w_n(\br-\br_j)\; d\br
\ee
and $\Phi^{n_1 n_2 n_3 n_4}_{j_1 j_2 j_3 j_4}$ are the matrix elements of 
the interaction potential over Wannier functions. 

Considering a lattice with one particle in each lattice site, it is necessary
to impose the no-double-occupancy constraint
\be
\label{5.224}
 \sum_n c_{nj}^\dgr \; c_{nj} = 1 \; 
\qquad
c_{mj} \; c_{nj} = 0 \;  .
\ee
In the case of an insulating lattice, with well localized particles, the 
no-hopping condition is valid,
\be
\label{5.225}
 c_{mi}^\dgr \; c_{nj} = \dlt_{ij} \; c_{mj}^\dgr \; c_{nj} \; .
\ee
Then Hamiltonian (\ref{5.222}) reads as
$$
H_f = w_f \sum_j \sum_{mn} E_{jj}^{mn} \;  c_{mj}^\dgr \; c_{nj} \; +
$$
\be
\label{5.226}
+ \; \frac{w_f^2}{2} \sum_{ij} \sum_{ \{n\} } V_{ij}^{n_1n_2n_3n_4} \;
c_{n_1i}^\dgr \; c_{n_2j}^\dgr \; c_{n_3j} \; c_{n_4i} \; ,
\ee
with
$$
V_{ij}^{n_1n_2n_3n_4} \equiv \Phi_{ijji}^{n_1n_2n_3n_4} \pm
\Phi_{ijij}^{n_1n_2n_3n_4} \; ,
$$ 
where the upper sign is for the Bose and the lower, for Fermi statistics. 

In ferroelectrics with the order-disorder phase transition, a lattice site is 
formed by a double well. To describe the motion of a particle between the wells 
of a double well, it is necessary to consider at least two lowest energy levels, 
corresponding to the ground state and the first excited state, so that in what 
follows we take $n=1,2$. The typical situation is when the ground-state wave 
function is symmetric with respect to spatial inversion, while the wave function 
of the first excited state is antisymmetric,
$$
 w_1(-\br) = w_1(\br) \; , \qquad  w_2(-\br) = - w_2(\br) \; .
$$
Taking into account that the interaction potential is symmetric, such that
$$
\Phi(-\br) = \Phi(\br) \; ,
$$ 
we see that the matrix elements where the energy indices enter in odd numbers, 
are zero, for instance
$$
 V_{ij}^{mmmn}  = V_{ij}^{mmnm}  =  V_{ij}^{mnmm}  = V_{ij}^{nmmm}  = 0
$$ 
for $m\neq n$. The nonzero elements enter the following expressions in the 
combinations
$$
A_{ij} \equiv \frac{1}{4} \; \left(  
V_{ij}^{1111} + V_{ij}^{2222} + 2 V_{ij}^{1221} \right) \; ,
\qquad
B_{ij} \equiv \frac{1}{4} \; \left(  
V_{ij}^{1111} + V_{ij}^{2222} - 2 V_{ij}^{1221} \right) \; ,
$$
\be
\label{5.227}
C_{ij} \equiv \frac{1}{2} \; \left(  
 V_{ij}^{2222} - V_{ij}^{1111} \right) \; ,
\qquad
I_{ij} =- 2 V_{ij}^{1122} \; .
\ee
 
Introducing the notations for the matrix elements of kinetic energy,
\be
\label{5.228}
K_{ij}^{mn} \equiv \int w_m^*(\br-\br_i) \; 
\left( -\;\frac{\nabla^2}{2m}\right) \; w_n(\br-\br_j) \; d\br
\ee
and of the external potential,
\be
\label{5.229}
U_{ij}^{mn} \equiv \int w_m^*(\br-\br_i) \; U(\br) \;  w_n(\br-\br_j) \; d\br   ,
\ee
and using the symmetry properties, we find
\be
\label{5.230}
E_{ij}^{mn} = \dlt_{mn}\; K_{jj}^{nn} + U_{jj}^{mn} \; , 
\qquad
K_{jj}^{mn}  = \dlt_{mn} \; K_{jj}^{nn} \;  .
\ee
Also, let us introduce the notations
$$
\frac{\bp_j^2}{2m} \equiv \frac{1}{2} \; 
\left( K_{jj}^{11} + K_{jj}^{22} \right) \; ,
\qquad
U_0 \equiv \frac{1}{2N} 
\sum_j \left( U_{jj}^{11} + U_{jj}^{22} \right) \; ,
$$
\be
\label{5.231}
E_0 \equiv \frac{1}{2} 
\sum_j \left( E_{jj}^{11} + E_{jj}^{22} \right) = 
\sum_j \frac{\bp_j^2}{2m}  + U_0 \; N \; .
\ee
The quantity
\be
\label{5.232}
\Om_j \equiv E_{jj}^{22} - E_{jj}^{11} + 
w_f \sum_i C_{ij} \; \cong \; E_{jj}^{22} - E_{jj}^{11}
\ee
is the frequency of tunneling between the wells of a double well. The value 
of $\Om_j$ depends on the shape of the double well and can be varied in a 
wide range \cite{Yukalov_315}. The expression
\be
\label{5.233}
B_j \equiv - E_{jj}^{12} - E_{jj}^{21} = - U_{jj}^{12} - U_{jj}^{21}
\ee
plays the role of an external field acting on a particle in the $j$-th lattice 
site.

Employing the transformation
$$
c_{1jf}^\dgr \;c_{1jf} = \frac{1}{2} + S_{jf}^x \; ,
\qquad
c_{2jf}^\dgr \;c_{2jf} = \frac{1}{2}\; - S_{jf}^x \; ,
$$
\be
\label{5.234}
 c_{1jf}^\dgr \; c_{2jf} = S_{jf}^z - i \; S_{jf}^y \; ,
\qquad
 c_{2jf}^\dgr \; c_{1jf} = S_{jf}^z + i \; S_{jf}^y \; ,
\ee
we obtain the operators
$$
S_{jf}^x = \frac{1}{2} \; \left( c_{1jf}^\dgr \; c_{1jf} -
c_{2jf}^\dgr \; c_{2jf} \right) \; ,
\qquad
S_{jf}^y = \frac{i}{2} \; \left( c_{1jf}^\dgr \; c_{2jf} -
c_{2jf}^\dgr \; c_{1jf} \right) \; ,
$$
\be
\label{5.235}
S_{jf}^z = \frac{1}{2} \; \left( c_{1jf}^\dgr \; c_{2jf} +
c_{2jf}^\dgr \; c_{1jf} \right) \; ,
\ee
satisfying the algebra of spin one-half operators, because of which they are 
called pseudospin operators. The above transformations are valid for Bose as 
well as for Fermi statistics. In that way, Hamiltonian (\ref{5.226}) reduces 
to the form
$$
H_f = w_f E_0 - 
w_f \sum_j \left( \Om_j \; S_{jf}^x + B_j \; S_{jf}^z \right) \; +
$$
\be
\label{5.236}
+ \; w_f^2 \sum_{i\neq j} \left( \frac{1}{2}\; A_{ij} +
 B_{ij}\; S_{if}^x\; S_{jf}^x - 
I_{ij} \; S_{if}^z \; S_{jf}^z \right) \; .
\ee
 
If the tunneling frequency $\Om_j$ and the transverse interaction $B_{ij}$ are 
small and can be neglected, then the Hamiltonian $H_f$ reduces to the Heisenberg 
form considered in Sec. 5.1. The similar form of the Hamiltonian $H_f$ occurs 
for double-well optical lattices \cite{Yukalov_316,Yukalov_317}, macromolecular 
systems \cite{Yukalov_318}, and granular Bose condensate \cite{Yukalov_319}. 
The role of phonon degrees of freedom on the order-disorder phase transition 
has also been studied \cite{Akhmeteli_320,Yukalov_321}.

\subsection{Heterophase Crystalline Structure}

In addition to phonon vibrations, crystalline structures can exhibit 
structural phase transitions and heterophase structural fluctuations 
\cite{Yukalov_322,Yukalov_323,Yukalov_324,Yukalov_325}. The Hamiltonian of a 
heterophase system, where two crystalline structures coexist, can be written 
as in (\ref{5.219}). The following idea is to consider the ground state of 
the system characterized by this Hamiltonian and to describe the collective 
excitations by means of phonon variables \cite{Yukalov_326,Yukalov_327}. For 
this purpose, we expand the field operators over Wannier functions of the 
lowest-level band 
\be
\label{5.237} 
 \psi_f(\br) = \sum_j c_{jf}\; w(\br-\br_{jf}) \;  ,
\ee
omitting the band index. This reduces Hamiltonian (\ref{5.219}) to the form 
similar to (\ref{5.226}),
\be
\label{5.238}
 H_f = w_f \sum_j E_{jjf} \; c_{jf}^\dgr\; c_{jf} + \frac{w_f^2}{2}
\sum_{ij} V_{ijf} \; c_{if}^\dgr\; c_{jf}^\dgr \; c_{jf}\; c_{if} \; .
\ee
Then we employ the no-double-occupancy constraint (\ref{5.224}) and no-hopping 
condition (\ref{5.225}) that now read as
\be
\label{5.239}
 c_{jf}^\dgr \; c_{jf} =1 \; , \qquad 
c_{jf}\; c_{jf} = 0 \; , \qquad
c_{if}^\dgr \; c_{jf} = \dlt_{ij} \; .
\ee
As a result, we come to the Hamiltonian
\be
\label{5.240}
 H_f = w_f \sum_j \left( \frac{\bp_{jf}^2}{2m} + U_{jf} \right) +
\frac{w_f^2}{2} \sum_{ij} V_{ijf} \; ,
\ee
in which
$$
\bp_{jf}^2 \equiv \int w^*(\br-\br_{jf}) \; (- \nabla^2 ) w(\br-\br_{jf}) \; d\br \; , 
$$
\be
\label{5.241}
U_{jf} \equiv \int w^*(\br-\br_{jf}) \; U(\br) \; w(\br-\br_{jf}) \; d\br \; .
\ee

The phase probabilities are defined as the minimizers of thermodynamic 
potential. We again can simplify the formulas by using the notation $w_1\equiv w$ 
and $w_2=1-w$. The Hamiltonian (\ref{5.240}) depends on the phase probabilities 
explicitly as well as implicitly through the values under summation. If we 
neglect the implicit dependence on the phase probabilities and consider the 
case without external fields, then the minimization of the free energy gives 
the equation
\be
\label{5.242}
 w = \frac{\Phi_2+ K_2 - K_1}{\Phi_1 + \Phi_2} \;  ,
\ee
where
\be
\label{5.243}
K_f \equiv \left\lgl \; \frac{1}{N} \sum_{j=1}^N 
\frac{\bp_{jf}^2}{2m} \; \right\rgl \; ,
\qquad 
\Phi_f \equiv \left\lgl \; \frac{1}{N} \sum_{j=1}^N V_{ijf} \; 
\right\rgl \; .
\ee

Excitations above the ground state are described by phonon degrees of freedom. 
For this purpose, the vectors $\br_{jf}$, showing the location of a particle 
in the vicinity of the vector of a lattice site $\ba_{jf}$ in the $f$-th phase, 
is represented as
\be
\label{5.244}
 \br_{jf} = \ba_{jf} + \bu_{jf} \;  ,
\ee
so that 
\be
\label{5.245}
 \ba_{jf} \equiv \lgl \; \br_{jf} \; \rgl \; ,
\qquad  
\lgl \; \bu_{jf} \; \rgl = 0 \; .
\ee
Expanding the interaction $V_{ijf}=V(\br_{if}-\br_{jf})$ in powers of the 
deviation $\bu_{jf}$ up to the second order transforms Hamiltonian (\ref{5.240}) 
into the expression
\be
\label{5.246}
 H_f = w_f \sum_{j=1}^N \frac{\bp_{jf}^2}{2m} + 
\frac{w_f^2}{2} \sum_{ij}^N \sum_{\al\bt}^3 \Phi_{ijf}^{\al\bt} \; u_{if}^\al \; u_{jf}^\bt +
\frac{w_f^2}{2} \sum_{i\neq j}^N V(\ba_{ijf}) \; ,
\ee
in which
\be
\label{5.247}
\Phi_{ijf}^{\al\bt} \equiv 
\frac{\prt^2 V(\ba_{ijf})}{\prt a_{if}^\al \prt a_{jf}^\bt} \; , 
\qquad
\ba_{ijf} \equiv \ba_{if} - \ba_{jf} \; .
\ee
   
Introducing phonon operators $b_{ksf}$ by the relations
$$
\bu_{jf} =\frac{1}{\sqrt{2N}} 
\sum_{ks} \frac{\bfe_{ksf}}{\sqrt{m\om_{ksf}} } \;
\left( b_{ksf} + b_{-ksf}^\dgr \right) \; e^{i\bk\cdot\ba_{jf} } \; ,
$$
\be
\label{5.248}
\bp_{jf} = - \;\frac{i}{\sqrt{2N}} 
\sum_{ks} \sqrt{m\om_{ksf} } \; \bfe_{ksf} \; 
\left( b_{ksf} - b_{-ksf}^\dgr \right)\; e^{i\bk\cdot\ba_{jf} } \;   ,
\ee
where ${\bf e}_{ksf}$ are the polarization vectors of the crystalline lattice 
in the $f$-th phase, reduces Hamiltonian (\ref{5.246}) to the form
\be
\label{5.249}
H_f = w_f \sum_{ks} \om_{ksf} \left( b_{ksf}^\dgr \; b_{ksf} +
\frac{1}{2} \right) + \frac{w_f^2}{2} \; N A_f \; ,
\ee
where
\be
\label{5.250}
 A_f \equiv \frac{1}{N} \sum_{i\neq j} V(\ba_{ijf} )  
\ee
and the phonon frequency is defined by the equation
\be
\label{5.251}
\frac{w_f}{m} \sum_{j=1}^N \; \sum_{\bt=1}^3 \Phi_{ijf}^{\al\bt} \;
e^{i\bk\cdot\ba_{ijf} } \; e_{ksf}^\bt = \om_{ksf}^2 e_{ksf}^\al \; .
\ee

To separate the explicit dependence on the phase probabilities, let us introduce
the frequency 
\be
\label{5.252}
\ep_{ksf}^2 \equiv \frac{1}{m} 
\sum_{j=1}^N \; \sum_{\al\bt}^3 
\Phi_{ijf}^{\al\bt} \; e_{ksf}^\al \; e_{ksf}^\bt \;
e^{i\bk\cdot\ba_{ijf} } \; .
\ee
Then the phonon frequency reads as
\be
\label{5.253}
 \om_{ksf} = \sqrt{w_f} \; \ep_{ksf} \; .
\ee
Hamiltonian (\ref{5.249}) becomes
\be
\label{5.254}
H_f = w_f^{3/2} \sum_{ks} \ep_{ksf} \; \left( b_{ksf}^\dgr \; b_{ksf} +
\frac{1}{2} \right) + \frac{w_f^2}{2} \; N A_f \; .
\ee

With Hamiltonian (\ref{5.254}), it is straightforward to find the kinetic 
energy $K_f$,
\be
\label{5.255}
K_f = \frac{1}{4N} \sum_{ks} \om_{ksf} \; 
\cosh\left( \frac{w_f \om_{ksf}}{2T}\right)
\ee
and the potential energy $\Phi_f$ ,
\be
\label{5.256}
\Phi_f = A_f + \frac{1}{2N} \sum_{ij} \; 
\sum_{\al\bt} \Phi_{ijf}^{\al\bt} \;
\lgl \; u_{if}^\al \; u_{jf}^\bt \; \rgl
\ee
defined in (\ref{5.243}). Here the deviation-deviation correlation function is
\be
\label{5.257}
\lgl \; u_{if}^\al \; u_{jf}^\bt \; \rgl = \frac{\dlt_{ij}}{2N} \sum_{ks}
\frac{e_{ksf}^\al e_{ksf}^\bt}{m\om_{ksf} } \;
\coth\left( \frac{w_f\om_{ksf}}{2T} \right) \;  .
\ee

The average of Hamiltonian (\ref{5.246}) reads as
\be
\label{5.258}
\lgl \; H_f \; \rgl = w_f \; K_f \; N + 
\frac{w_f^2}{2} \sum_{ij} \sum_{\al\bt}^3 \Phi_{ijf}^{\al\bt} \;
 \lgl \; u_{if}^\al \; u_{jf}^\bt \; \rgl +
\frac{w_f^2}{2} \; A_f \; N \; ,
\ee
while (\ref{5.254}) gives
\be
\label{5.259}
 \lgl \; H_f \; \rgl = 2 w_f \; K_f \; N + 
\frac{w_f^2}{2} \; A_f \; N \;  .
\ee
Comparing these two forms results in the relations
\be
\label{5.260}
w_f \sum_{ij} \sum_{\al\bt}^3 \Phi_{ijf}^{\al\bt} \; 
\lgl \; u_{if}^\al \; u_{jf}^\bt \; \rgl = 2 K_f
\ee
and
\be
\label{5.261}
 w_f \; \Phi_f = w_f \; A_f + 2 K_f \;  .
\ee

The free energy is $F = F_1 + F_2$, where
\be
\label{5.262}
 F_f = \frac{w_f^2}{2} \; A_f + \frac{T}{N} \sum_{ks} \ln\; 
\left[ \; 2\sinh\left( \frac{w_f\om_{ksf}}{2T}\right) \; \right] \; .
\ee
Minimizing the free energy with respect to the probability $w \equiv w_1$ yields
\be
\label{5.263}
 w = \frac{A_2 + 3(K_2 - K_1)}{A_1 + A_2} \;  .
\ee
As is easy to check, expression (\ref{5.263}) coincides with (\ref{5.242}), if
relation (\ref{5.261}) is taken into account.

\subsection{Structural Phase Transition}

The existence of heterophase fluctuations of competing structures is especially 
noticeable in the vicinity of structural phase transitions. Several quantities 
that can be measured are connected with the mean-square deviation that for the 
$f$-th phase reads as 
\be
\label{5.264}
r_f^2 \equiv \sum_{\al=1}^3 
\lgl \; u_{jf}^\al \; u_{jf}^\bt \; \rgl \;  .
\ee
With Eq. (5.257), the mean-square deviation becomes
\be
\label{5.265}
 r_f^2 = \frac{1}{2N} \sum_{ks} \frac{1}{m\om_{ksf}} \;
\coth\left( \frac{w_f\om_{ksf}}{2T} \right) \; .
\ee

Defining the frequency averaged over polarizations gives
\be
\label{5.266}
 \om_{kf}^2 = \frac{1}{3} \sum_{s=1}^3 \om_{ksf}^2 = 
w_f \; \ep_{kf}^2 \; ,
\ee
where
\be
\label{5.267}
 \ep_{kf}^2 = \frac{1}{3m} \sum_{j=1}^N \sum_\al
 \Phi_{ijf}^{\al\al} \; e^{i\bk\cdot\ba_{ijf} } \; .
\ee

In the Debye approximation \cite{Guyer_328}, for the mean-square deviation 
we obtain \cite{Yukalov_323,Yukalov_324,Yukalov_326,Yukalov_327} the form
\be
\label{5.268}
 r_f^2 = \frac{9w_f}{2m\Theta_f} \int_0^1 
x\; \coth\left( \frac{\Theta_f}{2T} \; x \right) \; dx \; ,
\ee
in which the effective Debye temperature 
\be
\label{5.269}
 \Theta_f \equiv w_f^{3/2} \; T_{Df}  
\ee
is expressed through the Debye temperature $T_{Df}$ of a pure phase $f$. Due 
to the existence of heterophase fluctuations, the mean-square deviation in each 
phase depends on the phase probability $w_f$. At the point of a structural phase 
transition, the probability of structures is close to $w_f\approx 1/2$, which 
results in the increase of the mean-square deviations in each structure 
\cite{Yukalov_323,Yukalov_324}.

The mean-square deviation is related to the M\"{o}ssbauer factor or Debye-Waller 
factor for an $f$-th phase,
\be
\label{5.270}
 f_{Mf} = \exp\left( - q^2 \; r_f^2\right) \; ,
\ee
in which $q$ is the momentum of a gamma-quantum in the case of the M\"ossbauer 
effect and the momentum of a R\"ontgen quantum or a neutron momentum in the case 
of the Debye-Waller factor. The averaged factor for the whole sample is
\be
\label{5.271}
f_M = w_1 \; f_{M_1} + w_2 \; f_{M_2} \;  .
\ee

The M\"ossbauer and Debye-Waller factors exhibit anomalous behaviour at the 
points of phase transitions, such as cusps and saggings 
\cite{Bhide_329,Bhide_330,Bhide_331,Bhide_332,Owens_333,Thosar_334,Bishop_335,
Egami_336,Muller_337,Sirdeshmukh_338}. It was shown \cite{Meissner_339,Binder_340} 
that these anomalies cannot be explained by the appearance of soft modes. But 
such anomalies can be explained by the existence of heterophase fluctuations 
increasing in the vicinity of phase transitions 
\cite{Yukalov_341,Yukalov_342,Yukalov_343,Yukalov_344,Yukalov_345,Yukalov_346,
Yukalov_347}.        

Depending on temperature, the mean-square deviation can be written as
\be
\label{5.272}
r_f^2 = \frac{9}{4m w_f^{1/2} T_{Df}} \qquad
( T = 0 )
\ee
at zero temperature and as
\be
\label{5.273}
r_f^2 \simeq \frac{9T}{m w_f^2 T_{Df}^2} \qquad
( T \gg \Theta_f )
\ee
at finite temperature. These expressions show that the appearance of heterophase 
fluctuations, when the phase probability $w_f$ becomes less than one, increases 
the mean-square deviation, hence diminishes the M\"ossbauer factor. The sharp 
decrease of the M\"ossbauer factor at phase transitions explains the cusp-shape 
anomaly observed at different phase-transition points.  

Sound velocity also exhibits anomalous behavior in the vicinity of phase 
transitions \cite{Yukalov_324}. The sound velocity of a pure phase $f$ in the 
Debye approximation can be defined by the relation
\be
\label{5.274}
 T_{Df} = c_f k_D \qquad \left( k_D^3 = 6\pi^2 \rho\right) \;  ,
\ee
where $k_D$ is the Debye momentum and $\rho$ is the average density of 
the sample. In this approximation, the phonon spectrum $\omega_{ksf}$ of a 
heterophase system takes the form
\be
\label{5.275}
 \om_{ksf}= s_f k \qquad ( 0 \leq k \leq k_D ) \;  ,
\ee
with the sound velocity
\be
\label{5.276}
 s_f = \sqrt{w_f}\; c_f \;  .
\ee
The average sound velocity through a heterophase sample is
\be
\label{5.277}
 s = w_1 \; s_1 + w_2 \; s_2 = 
w_1^{3/2} c_1 + w_2^{3/2} c_2 \; .
\ee

To estimate the variation of the sound velocity caused by heterophase 
fluctuations, let us consider the case where the sound velocities in competing 
pure heterophase structures are close to each other, so that $c_f\approx c$. 
Then the average sound velocity in a heterophase system is
\be
\label{5.278}
 s \approx \left( w_1^{3/2} + w_2^{3/2} \right) \; c \;  .
\ee
If at the point of the phase transition, $w_f$ approximately equals $1/2$, then 
the average sound velocity diminishes to $s\approx 0.7 c$.

\subsection{Stability of Heterophase Solids}

Solids are characterized by localized particles. In the case of crystals, 
particles form crystalline lattices, while in amorphous solids the particle 
location in a sample is random. Particle localization is described by mean-square 
deviation. When this deviation becomes close to half-distance between the nearest 
neighbors, the solid gets unstable. The condition 
\be
\label{5.279} 
\frac{r_f}{a} < \frac{1}{2} 
\ee
is called the Lindemann criterion of stability \cite{Lindemann_348}. This 
criterion defines the stability boundary of solid state 
\cite{Yukalov_326,Yukalov_327,Yukalov_349,Yukalov_350}. For heterophase 
materials, this criterion includes the phase probability $w_f$, so that the 
stability condition can essentially change, as compared to that of a pure 
phase. 

Studying the mean-square deviation in the Debye approximation, it is instructive 
to consider a sample in arbitrary space dimensionality $d=1,2,3,\ldots$. Then 
the sum over momenta can be reduced to the integral,
\be
\label{5.280}
 \frac{1}{N} \sum_k \; \longmapsto \;
\frac{2}{(4\pi)^{d/2}\Gm(d/2)\rho} \;
\int_0^{k_D} k^{d-1} \; dk \; ,
\ee
where $k_D$ is the Debye momentum, or Debye radius,
\be
\label{5.281}
 k_D = \sqrt{4\pi} \; \left[ \;
\frac{d}{2}\; \Gm\left( \frac{d}{2}\right) \; \rho \; \right]^{1/d} \; ,
\ee
with $\rho a^d = 1$, $\rho$ being the average density and $a$, the average distance 
between the nearest neighbors. 

In a pure $f$-th phase, the mean-square deviation is expressed through the 
Debye temperature
\be
\label{5.282}
 T_{Df} = c_f \; k_D =\sqrt{4\pi\; \frac{D_f}{m} } \; 
\left[\; 
\frac{d}{2} \; \Gm\left( \frac{d}{2}\right) \; \right]^{1/d} \; ,
\ee
in which 
\be
\label{5.283}
c_f = \sqrt{ \frac{D_f}{m} } \; a
\ee
is the sound velocity in a pure phase and 
\be
\label{5.284}
D_f = \frac{1}{d} \sum_{\al=1}^d 
\frac{\prt^2 V(\ba_{ij})}{\prt a_{jf}^\al \prt a_{jf}^\al}
\ee
is the dynamic parameter equal to the dynamical matrix in the nearest-neighbor 
approximation. 

In a heterophase system, the role of Debye temperature is played by expression 
(\ref{5.269}). Then Eq. (\ref{5.265}) takes the form
\be
\label{5.285}
r_f^2 = \frac{w_f d^2}{2m\Theta_f} 
\int_0^1 x^{d-2}\; \coth\left( \frac{\Theta_f}{2T}\; x\right) \; dx \; .
\ee

At low temperature, this gives
\be
\label{5.286}
r_f^2 \simeq \frac{d^2}{2(d-1) m T_{Df}\;\sqrt{w_f}} \qquad 
( T \ll \Theta_f) \; .
\ee
Introducing the characteristic kinetic energy
\be
\label{5.287}
 E_K \equiv \frac{1}{2ma^2}  \; ,
\ee
we get the stability criterion 
\be
\label{5.288}
 \frac{E_K}{T_{Df} } \; < \; \frac{d-1}{4d^2} \; \sqrt{w_f} 
\qquad 
( T \ll \Theta_f) \;  .
\ee
This shows that an infinite one-dimensional crystal cannot exist. For a 
two-dimensional crystal, we have
\be
\label{5.289}
\frac{E_K}{T_{Df} } \; < \; \frac{\sqrt{w_f}}{16}  
\qquad 
( T \ll \Theta_f\; , ~ d = 2 ) 
\ee
and for a three-dimensional crystal,
\be
\label{5.290}
\frac{E_K}{T_{Df} } \; < \; \frac{\sqrt{w_f}}{18}  
\qquad 
( T \ll \Theta_f\; , ~ d = 3 ) \; .
\ee

At high temperature, the mean-square deviation reads as 
\be
\label{5.291}
 r_f^2 = \frac{T d^2}{(d-2) m T_{Df}^2 \; w_f^2} \qquad 
( T \gg \Theta_f ) \;  ,
\ee
which yields the stability criterion
\be
\label{5.292}
\frac{E_K}{T_{Df} } \; < \; \frac{(d-2)T_{Df}}{8T d^2}  \; w_f^2
\qquad 
( T \gg \Theta_f ) \;   .
\ee
Hence, there are no one- and two-dimensional infinite crystals, while for 
a three-dimensional crystal, we have
\be
\label{5.293}
\frac{E_K}{T_{Df} } \; < \;  \frac{T_{Df}}{72T}  \; w_f^2
\qquad 
( T \gg \Theta_f\; , ~ d = 3 ) \;  .
\ee
Since in a heterophase system $w_f < 1$, the stability boundary diminishes as compared 
to a pure system.
    
The above criteria are derived for spatially infinite solids. For finite 
solids, it is necessary to take into account finite-size effects that limit 
the smallest wave vector by
\be
\label{5.294}
 k_{min} = \frac{\pi}{L} \qquad ( L =N^{1/d}\; a ) \; .
\ee
Then in the mean-square deviation (\ref{5.285}), the integral has to be limited 
from below by the value
\be
\label{5.295}
 x_{min} = \frac{k_{min}}{k_D} =
\frac{\sqrt{\pi}}{2[\;(d/2)\Gm(d/2)\;]^{1/d} N^{1/d} } \;  .
\ee
For one-dimension, the latter is
$$
 x_{min} = \frac{1}{N} \qquad ( d = 1) \; ,
$$
while for two dimensions,
$$ 
  x_{min} = \frac{1}{2} \; \sqrt{\frac{\pi}{N} }
\qquad 
( d = 2) \;  .
$$  

At low temperature, for a one-dimensional crystalline chain, it follows
\be
\label{5.296}
r_f^2 \simeq \frac{\ln N}{2m T_{Df}\; \sqrt{w_f} } 
\qquad
 ( T \ll \Theta_f\; , ~ d = 1 ) \;  .
\ee
Then the stability criterion gives the largest number of particles that are 
able to form a finite chain,
\be
\label{5.297}
N \; < \; \exp\left( \frac{T_{Df}\; \sqrt{w_f}}{4E_K} \right) 
\qquad
 ( T \ll \Theta_f\; , ~ d = 1 ) \;  .
\ee

At high temperature, for a one-dimensional chain, we have
\be
\label{5.298}
r_f^2 \simeq \frac{NT}{m T_{Df}^2\;w_f^2} 
\qquad
 ( T \gg \Theta_f\; , ~ d = 1 ) \;  .
\ee
The stability criterion results in the inequality
\be
\label{5.299}
N \; < \; \frac{T_{Df}^2 w_f^2}{8T E_k} 
\qquad
 ( T \gg \Theta_f\; , ~ d = 1 ) \;  .
\ee

For a two-dimensional solid at high temperature, we find the mean-square 
deviation
\be
\label{5.300} 
 r_f^2 \simeq \frac{T\ln N}{m T_{Df}^2\;w_f^2} 
\qquad
 ( T \gg \Theta_f\; , ~ d = 2 ) \;   ,
\ee
which limits the largest number of particles by the value
\be
\label{5.301}
N < \exp\left( \frac{T_{Df}^2\; w_f^2}{8T E_K} \right) 
\qquad
 ( T \gg \Theta_f\; , ~ d = 2 ) \;   .
\ee

The obtained stability conditions demonstrate that heterophase fluctuations 
diminish the possible size of finite solids.

\subsection{Solids with Nanoscale Defects}

There exists a large class of solids containing nanoscale defects, such as 
pores, cracks, dislocations, heterophase embryos, and polymorphic inclusions
\cite{Bakai_101,Cottrell_351,Friedel_352,Ziman_353,Hirth_354,Hull_355}. Such 
solids can be considered as heterophase systems composed of two phases with 
different density
\cite{Yukalov_356,Kad_357,Yukalov_358,Yukalov_359,Yukalov_360,Yukalov_361}. 
The system Hamiltonian has the standard form
$$
\widetilde H = H_1 + H_2 \; ,
$$
$$
H_f = w_f \int \psi_f^\dgr(\br)\; 
\left( -\; \frac{\nabla^2}{2m} - \mu \right) \; \psi_f(\br) \; d\br \; +
$$
\be
\label{5.302}
+ \; 
\frac{1}{2}\; w_f^2 
\int \psi_f^\dgr(\br) \; \psi_f^\dgr(\br')\; 
\Phi(\br-\br') \; \psi_f(\br') \; \psi_f(\br) \; d\br d\br'\; .
\ee

The number of particles in an $f$-th phase of the heterophase mixture is
\be
\label{5.303}
N_f = 
w_f \int \lgl \; \psi_f^\dgr(\br) \; \psi_f(\br) \;\rgl \; d\br
\ee
occupying the volume $V_f$, so that the density of an $f$-th phase is
\be
\label{5.304}
\rho_f \equiv \frac{N_f}{V_f} = \frac{1}{V} \int
\lgl \; \psi_f^\dgr(\br) \; \psi_f(\br) \;\rgl \; d\br \;  .
\ee
The average density of the total system is
\be
\label{5.305}
 \rho \equiv \frac{N}{V} = w_1 \; \rho_1 + w_2 \; \rho_2 \; .
\ee
The phases differ from each other by their densities, such that
\be
\label{5.306}
 \rho_1 > \rho_2 \;  .
\ee
Thus the first phase can be called dense, while the second, rarefied. 

It is also convenient to introduce the lattice filling factor
\be
\label{5.307}
 \nu \equiv \frac{N}{N_L} = \frac{\rho}{\rho_L} \;  ,
\ee
the density of lattice sites
\be
\label{5.308}
 \rho_L \equiv \frac{N_L}{V} \; ,
\ee
and the dimensionless phase density
\be
\label{5.309}
 x_L \equiv \frac{\rho_f}{\rho_L} \; .
\ee
Then the equality (\ref{5.305}) reduces to
\be
\label{5.310}
 \nu = w_1 \; x_1 + w_2 \; x_2   
\ee
and the condition (\ref{5.306}) becomes
\be
\label{5.311}
 x_1 > x_2 \;  .
\ee

The field operators can be represented as expansions over the localized 
orbitals $\vp_{nj}(\br)$ \cite{Coleman_251}. Here $n$ is a band index, 
$j=1,2,\ldots,N_L$ is the label enumerating lattice sites. Let $c_{njf}$ 
be the annihilation operator of a particle in a band $n$ at a lattice site $j$, in 
a phase $f$, and the variable $e_{jf}$ take the values $0$ or $1$, depending 
on whether the site $j$ is free or occupied by a particle. 

Supposing that the considered temperatures are much lower than the energy gap 
between the lowest and excited levels allows us to resort to the single-band 
approximation writing down the field-operator expansion as
\be
\label{5.312}
 \psi_f(\br) = \sum_{j=1}^{N_L} e_{jf} \; c_{jf}\; \vp_j(\br) \;  .
\ee
Assuming that each lattice site can host not more than one particle imposes 
the unipolarity condition
\be
\label{5.313}
 c_{jf}^\dgr \; c_{jf} = 1 \; , \qquad  c_{jf}\; c_{jf} = 0 \; .
\ee
We consider a good isolator, where the intersite transitions are suppressed, 
so that the only surviving matrix elements are the diagonal elements
$$
\ep_0 \equiv \int \vp_j^*(\br) \; \left( - \; 
\frac{\nabla^2}{2m} \right) \; \vp_j(\br) \; d\br \; ,
$$
\be
\label{5.314}
 \Phi_{ij}\equiv \int \vp_i^*(\br) \; \vp_j^*(\br') \; 
\Phi(\br-\br') \; \vp_j(\br') \; \vp_i(\br) \; d\br d\br' \; .
\ee
Then we get the Hamiltonian
\be
\label{5.315}
H_f = \frac{1}{2} \; w_f^2 \sum_{i\neq j}^{N_L} \Phi_{ij} \;
e_{if}  e_{jf} \;  - \; w_f\; ( \mu-\ep_0)  \sum_{j=1}^{N_L} e_{jf} \; .
\ee
      
The phase density (\ref{5.304}) reads as
\be
\label{5.316}
\rho_f \equiv \frac{N_f}{V_f} = \rho_L \; \frac{1}{N_L} \;
\left\lgl \; \sum_{j=1}^{N_L} e_{jf} \; \right\rgl
\ee
and the dimensionless density (\ref{5.309}) becomes
\be
\label{5.317}
 x_f \equiv \frac{\rho_f}{\rho} = 
\left\lgl \; \frac{1}{N_L} \sum_{j=1}^{N_L} e_{jf} \; \right\rgl \; .
\ee
 
Minimizing the grand potential
\be
\label{5.318}
\Om = - T\; \ln \; {\rm Tr}\; e^{-\bt\widetilde H}
\ee
with respect to the phase probabilities $w_1\equiv w$ and $w_2=1-w$ gives 
the probability of a dense phase
\be
\label{5.319}
w = \frac{\Phi_2+(\ep_0-\mu)(x_2-x_1)}{\Phi_1 + \Phi_2} \; ,
\ee
where
$$
\Phi_f \equiv \frac{1}{N_L} \int 
\lgl \; \psi_f^\dgr(\br) \; \psi_f^\dgr(\br') \; \Phi(\br-\br') \;
\psi_f(\br') \; \psi_f(\br) \; \rgl \; d\br d\br' \; =
$$
\be
\label{5.320}
= \; \frac{1}{N_L} \sum_{i \neq j}^{N_L} \Phi_{ij} \; 
\lgl \; e_{if} \; e_{jf} \; \rgl \; .
\ee
 
By a canonical transformation, it is possible to introduce pseudospin operators,
\be
\label{5.321}
e_{jf} = \frac{1}{2} + S_{jf}^z \; , 
\qquad
S_{jf}^z = e_{jf} - \; \frac{1}{2} \; .
\ee
Then Hamiltonian (\ref{5.315}) acquires the pseudospin form
$$
H_f = \frac{N_L}{8} \; \left[ \; w_f^2 \; \Phi - 
4 w_f \; ( \mu - \ep_0 ) \; \right] \ + 
\frac{1}{2} \; \left[ \; w_f^2 \; \Phi - 
2 w_f \; (\mu - \ep_0 ) \; \right] \; \sum_{j=1}^{N_L} S_{jf}^z \; + 
$$
\be
\label{5.322}
+ \; \frac{1}{2} \; w_f^2 
\sum_{i\neq j}^{N_L} \Phi_{ij} \; S_{if}^z \; S_{jf}^z \;  ,
\ee
in which 
\be
\label{5.323}
 \Phi \equiv \frac{1}{N_L} \sum_{i\neq j}^{N_L} \Phi_{ij} \;  .
\ee
This shows that the model is similar to a model of a magnetic system with the 
order parameters
\be
\label{5.324}
 s_f \equiv 2 \lgl \; S_{jf}^z \; \rgl = 
\frac{2}{N_L} \sum_{j=1}^{N_L} \lgl \; S_{jf}^z \; \rgl  
\ee 
satisfying the inequality
\be
\label{5.325}
 s_1 > s_2  
\ee
because of the relation
\be
\label{5.326}
x_f = \frac{1}{2} \; ( 1 + s_f )
\ee
and condition (\ref{5.311}). 

The probability (\ref{5.319}) is derived from the equation
\be
\label{5.327}
 \frac{\prt\Om}{\prt w} = \left\lgl \;
\frac{\prt\widetilde H}{\prt w} \; \right\rgl = 0 \; .
\ee
This equation defines a minimum, provided that
\be
\label{5.328}
 \frac{\prt^2\Om}{\prt w^2} = \left\lgl \;
\frac{\prt^2\widetilde H}{\prt w^2} \; \right\rgl - \bt \;
\left\lgl \; 
\left( \frac{\prt\widetilde H}{\prt w}\right)^2 \; 
\right\rgl ~ > ~ 0 \; .
\ee
The latter inequality yields the necessary condition
\be
\label{5.329}
\left\lgl \; \frac{\prt^2\widetilde H}{\prt w^2} \; \right\rgl ~ > ~ 0 \; .
\ee

Note that on the mean-field level, one has
\be
\label{5.330}
 \left\lgl \; 
\left( \frac{\prt\widetilde H}{\prt w}\right)^2 \; \right\rgl =
\left\lgl \;
\frac{\prt\widetilde H}{\prt w} \; \right\rgl^2 = 0 \; .
\ee
Thus the condition of $w$ to be a minimizer for $\Omega$ agrees with 
(\ref{5.329}).

In the mean-field approximation,
\be
\label{5.331}
 \Phi_f = \Phi \; x_f^2 \; ,
\ee
hence condition (\ref{5.329}) reduces to the inequality
\be
\label{5.332}
 \Phi > 0 \;  .
\ee
 
Hamiltonian (\ref{5.315}) in the mean-field approximation reads as
\be
\label{5.333}
 \frac{H_f}{N_L\Phi} = - \; \frac{1}{2} \; w_f^2 \; x_f^2 +
\left( w_f^2 \; x_f - \mu^* \; w_f \right) \; 
\frac{1}{N_L} \sum_{j=1}^{N_L} e_{jf} \; ,
\ee
with the effective chemical potential
\be
\label{5.334}
 \mu^* = \frac{\mu-\ep_0}{\Phi} \; .
\ee
Equation (\ref{5.319}) yields
\be
\label{5.335}
w = \frac{x_2^2+\mu^*(x_1-x_2)}{x_1^2+x_2^2} \;  ,
\ee
from where we find
\be
\label{5.336}
\mu^* = \frac{w_1 x_1^2-w_2 x_2^2}{x_1-x_2} \;  .
\ee
Equality (\ref{5.310}) can be used for getting the probability of the dense 
phase
\be
\label{5.337}
w = \frac{\nu-x_2}{x_1-x_2} \;   .
\ee
Measuring temperature in units of $\Phi$, for the grand potential (\ref{5.318}), 
we have
$$
\frac{\Om}{N_L\Phi} = \frac{1}{2} \; w_1^2 \; x_1 \; ( 1 - x_1) +
\frac{1}{2}\; w_2^2 \; x_2 \; ( 1 - x_2) - \; \frac{1}{2} \; \mu^* \; -
$$
\be
\label{5.338}
- \; 
T \ln \; \left\{ 4\cosh\left( \frac{w_1^2 x_1-\mu^*\; w_1}{2T}\right) \;
\cosh\left( \frac{w_2^2 x_2-\mu^*\; w_2}{2T}\right) \right\} \; .
\ee
Minimizing the grand potential with respect to the order parameters $x_f$, we 
obtain
$$
 2x_1 = 1 + 
\tanh\left\{ \frac{w_1 x_2(w_1 x_1 - w_2 x_2)}{2(x_1-x_2)T} \right\} \; ,
$$
\be
\label{5.339}
2x_2 = 1 + 
\tanh\left\{ \frac{w_2 x_1(w_1 x_1 - w_2 x_2)}{2(x_1-x_2)T} \right\} \; .
\ee  

Numerical investigations show that the two-density heterophase system can be 
stable only for small filling factor
\be
\label{5.340}
  0 < \nu < \frac{1}{2} 
\ee
in the region between the lower nucleation temperature $T_n$, depending on 
the filling factor $\nu$, and the upper nucleation temperature
\be
\label{5.341}
 T_n^* = \frac{\nu}{(1-2\nu)\;\ln[(1-\nu)/\nu]} \;  .
\ee

\subsection{Theory of Melting and Crystallization}

In crystals below the melting point there can exist regions of disorder, such 
as pores, cracks, dislocations, and heterophase fluctuations that play the role 
of liquid-like embryos \cite{Frenkel_2,Frenkel_3,Hirshfelder_362,Ubbelohde_363}. 
In their turn, above the melting point, there exist fluctuating crystal-like 
clusters representing the germs of crystalline state. The fluctuating coexistence 
of the corresponding crystalline and liquid states can be treated in the frame 
of the theory of heterophase systems \cite{Yukalov_238,Yukalov_364}.    

Spatial densities play the role of order parameters, so that the crystalline 
density is 
periodic over the lattice vectors,
\be
\label{5.342}
\rho_1(\br+\ba) = \rho_1(\br) \;  ,
\ee
while the liquid density is constant,
\be
\label{5.343}
 \rho_2(\br) = \rho \; .
\ee
It is also possible to define the order parameter
\be
\label{5.344}
\Dlt \;\rho_f \equiv \max_\br \; \frac{\rho_f(\br)}{\rho} \; - \; 1
\ee
that is nonzero for the crystalline phase and zero for the liquid phase. 
 
Starting with the standard Hamiltonian of a two-phase mixture
$$
\widetilde H = H_1 \; \bigoplus \; H_2 \; ,
$$
$$ \;
H_f = w_f \int \psi_f^\dgr(\br) \; \left( - \; 
\frac{\nabla^2}{2m} - \mu \right) \; \psi_f(\br) \; d\br \; +
$$
\be
\label{5.345} 
+ \; 
\frac{1}{2} \; w_f^2 \int \psi_f^\dgr(\br) \; \psi_f^\dgr(\br') \;
V(\br-\br') \; \psi_f(\br') \; \psi_f(\br) \; d\br d\br' \; ,
\ee
where $V({\bf r} - {\bf r}')$ is the bare interaction potential, we meet the 
problem of dealing with divergences caused by the fact that bare interaction 
potentials are usually represented by strongly singular functions, such that 
the integral 
$$
\int V(\br) \; d\br \longrightarrow \infty
$$    
diverges. 

This problem can be avoided by employing correlated approximations 
\cite{Yukalov_301}. Consider, for instance, the system potential energy
\be
\label{5.346}
 E_{pot} = \left\lgl \; \frac{w^2}{2} \int 
\psi_f^\dgr(\br) \; \psi_f^\dgr(\br') \; V(\br-\br') \; 
\psi_f(\br') \; \psi_f(\br) \; d\br d\br' \; \right\rgl \;  .
\ee
If we resort to the Hartree-Fock approximation, while the interaction potential 
is not integrable, the potential energy diverges. The way out of this pitfall 
is the use, e.g., of the Kirkwood approximation \cite{Kirkwood_365}
\be
\label{5.347}
 \lgl \; \psi_f^\dgr(\br) \; \psi_f^\dgr(\br') \;
\psi_f(\br') \; \psi_f(\br) \; \rgl = g(\br-\br') \;
 \lgl \; \psi_f^\dgr(\br) \; \psi_f(\br) \; \rgl \;
 \lgl \; \psi_f^\dgr(\br') \; \psi_f(\br') \; \rgl \; ,
\ee
where $g({\bf r})$ is the pair correlation function that can be found from 
the measured structural factor
\be
\label{5.348}
 g(\br) = 1 + \frac{1}{\rho} \int [ \; S(\bk) - 1 \; ] \; 
e^{i\bk\cdot\br} \; \frac{d\bk}{(2\pi)^3} \; .
\ee
The other way was suggested by Bogolubov \cite{Bogolubov_366} by defining 
the pair correlation function through the modulus squared of the wave function 
describing the relative motion of the pair of particles. In the Kirkwood 
approximation the potential energy
\be
\label{5.349}
 E_{pot} = \left\lgl \; \frac{w_f^2}{2} 
\int \rho_f(\br) \; \rho_f(\br') \; \Phi(\br-\br') \; d\br d\br' \right\rgl
\ee
is finite, since the correlated potential
\be
\label{5.350}
\Phi(\br-\br') \equiv g(\br - \br') \; V(\br -\br')
\ee
is integrable,
\be
\label{5.351}
 \int \Phi(\br) \; d\br \; < \; \infty \; .
\ee

Beginning from the Kirkwood approximation, it is possible to develop an 
iterational procedure for Green functions, where all approximations contain 
only the correlated potential (\ref{5.350}) and no divergences arise 
\cite{Yukalov_367,Yukalov_368,Yukalov_369,Yukalov_370}. 

The chemical potential is defined by the total number of particles
\be
\label{5.352}
 N = N_1 + N_2 \;  ,
\ee
with the number of particles in each phase 
\be
\label{5.353}
N_f = w_f \int \rho_f(\br) \; d\br \; .
\ee
Minimizing the grand potential and introducing the notations 
$$
K_f \equiv \frac{1}{N} \int \lgl \; \psi_f^\dgr(\br) \;
\left( - \; \frac{\nabla^2}{2m} \right) \; \psi_f(\br) \; \rgl \; d\br \; ,
$$
$$
\Phi_f \equiv \frac{1}{N} \int 
\lgl \; \psi_f^\dgr(\br) \; \psi_f^\dgr(\br') \;
V(\br-\br') \; \psi_f(\br') \; \psi_f(\br) \; \rgl \; d\br  d\br' \; ,
$$
\be
\label{5.354}
x_f \equiv \frac{\rho_f}{\rho} = \frac{1}{N} \int \rho_f(\br) \; d\br
\ee
gives us the probability of the crystalline phase
\be
\label{5.355}
 w = \frac{\Phi_2 + K_2 - K_1 + \mu(x_1 - x_2)}{\Phi_1+\Phi_2} \; .
\ee

Then the quantities related to the crystalline phase are to be calculated with 
the methods appropriate for a periodic structure, while those for the liquid 
phase, by the methods describing a uniform system. The behavior of the 
crystalline probability (\ref{5.355}) characterizes the solid state and the 
probability $1-w$ describes the liquid state \cite{Yukalov_238,Yukalov_364}. 
The typical behavior of the solid-state probability, as a function of temperature 
in energy units is shown in Fig. 12.

\begin{figure}[ht]
\centerline{
\includegraphics[width=10cm]{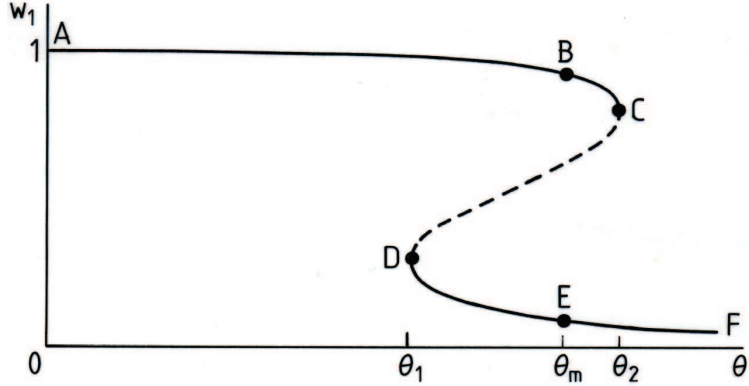} }
\caption{Solid-state probability as a function of temperature in energy 
units. The lines correspond to the following states: $AB-$stable crystal 
with small admixture of liquid droplets; $BC-$metastable overheated 
crystal; $CD-$unstable state; $DE-$overcooled liquid state; $EF-$stable 
liquid state with crystal clusters.
}
\label{fig:Fig.12}
\end{figure}

\subsection{Model of Superfluid Solid}

If there exist liquid-like embryos inside a solid composed of Bose particles, 
then it may happen that these liquid-like regions could exhibit the property 
of superfluidity. Andreev and Lifshits \cite{Andreev_371} assumed that 
superfluidity in quantum crystals could be due to the motion of delocalized 
vacancies. Chester \cite{Chester_372} argued that superfluidity should arise 
in any quantum Bose crystal at sufficiently low temperatures. 

Formally, a superfluid crystal can be connected with the simultaneous breaking 
of translational and gauge symmetries in a system. If this double breaking 
could be allowed, then, in the frame of the same system, crystalline diagonal 
order, typical of solids, could coexist with off-diagonal order related to 
superfluid flow. That is, an unusual object "superfluid solid" could arise. 
An ideal crystalline state is known to prohibit such a spontaneous double 
symmetry breaking. But superfluidity can appear if inside a crystal there 
occur mesoscopic regions of disorder, such as grain boundaries and screw 
dislocations.   

In order to describe a superfluid object it has been natural to turn to a 
coherent system characterized by the coherent-field equation first published 
by Bogolubov \cite{Bogolubov_373} in 1949 and since then republished numerous 
times, e.g., \cite{Bogolubov_214,Bogolubov_215,Bogolubov_216}. Gross 
\cite{Gross_374,Gross_375,Gross_376,Gross_377,Gross_378,Gross_379} investigated 
this equation and showed that it possesses periodic solutions imitating solid 
state. At the same time, due to its coherent nature, this equation corresponds 
to a Bose-condensed system, hence enjoying superfluidity. Periodic solutions 
describing a coherent quantum crystal have been studied by Kirzhnits and 
Nepomnyashchii \cite{Kirzhnits_380}.     

There have been numerous attempts to observe superfluidity in quantum 
crystals, especially in solid $^4$He, where it could be due to the presence 
of dislocations (see reviews \cite{Prokofev_381,Kuklov_382,Boninsegni_383,
Yukalov_384,Fil_385}). Trapped quantum gases with dipolar interactions can 
exhibit the simultaneous breaking of translational and gauge symmetries 
\cite{Boninsegni_383,Yukalov_384}, although periodically modulated gases are
anyway gases but not solids.

Strictly speaking, the simultaneous breaking of translational and gauge 
symmetries is neither a necessary nor sufficient condition for the existence 
of a superfluid solid. Thus this double breaking happens in metastable quantum 
Bose gases, while they, nevertheless, are gases but not solids. From the other 
side, a solid is not compulsorily ideally periodic, but localized molecules 
can be randomly distributed in space, as in amorphous solids. In addition, 
superfluidity can occur without gauge symmetry breaking \cite{Yukalov_384}.    

The assumption that Bose-condensed, hence superfluid, local states can arise 
in crystals with regions of disorder was advanced in Refs. 
\cite{Yukalov_238,Yukalov_364,Yukalov_384}. The coexistence of solid-like 
and liquid-like regions in the frame of the theory of heterophase systems is 
described as follows.

Let us consider a system of $N$ particles, among which $N_{sol}$ particles 
compose a crystalline structure, while $N_{liq}$ particles form liquid-like 
nanosize embryos, so that
\be
\label{5.356}
 N_{sol} + N_{liq} = N \; .
\ee
Respectively, the volumes occupied by the phases are $V_{sol}$ and $V_{liq}$, 
such that
\be
\label{5.357}
  V_{sol} + V_{liq} = V \;  .
\ee
The related geometric probabilities are
\be
\label{5.358}
 w_{sol} \equiv \frac{V_{sol}}{V} \; , \qquad
 w_{liq} \equiv \frac{V_{liq}}{V} \; ,
\ee
thence they are normalized,
\be
\label{5.359}
  w_{sol} + w_{liq} = 1 \;  .
\ee
The particle fractions are 
\be
\label{5.360}
 n_{sol} \equiv \frac{N_{sol}}{N} \; , \qquad
 n_{liq} \equiv \frac{N_{liq}}{N} \;   ,
\ee
also being normalized,
\be
\label{5.361}
  n_{sol} + n_{liq} = 1 \;  .
\ee
The densities of the corresponding phases are
\be
\label{5.362}
 \rho_{sol} \equiv \frac{N_{sol}}{V_{sol}} \; , \qquad
 \rho_{liq} \equiv \frac{N_{liq}}{V_{liq}} \;  ,
\ee
which gives the relations
\be
\label{5.363}
 w_{sol}\; \rho_{sol} = n_{sol} \; \rho \; ,
\qquad
 w_{liq}\; \rho_{liq} = n_{liq} \; \rho \; .
\ee
The average system density is
\be
\label{5.364}
\rho \equiv \frac{N}{V} = 
w_{sol}\; \rho_{sol} +   w_{liq}\; \rho_{liq} \; .
\ee
When the phases differ by their structure but not by their density, the 
probabilities and fractions coincide,
\be
\label{5.365}
 n_{sol} = w_{sol} \; , \qquad n_{liq} = w_{liq} 
\qquad
( \rho_{sol} = \rho_{liq} = \rho ) \; .
\ee
The numbers of particles forming the solid and liquid phases can be written as
$$
N_{sol} = \lgl \; \hat N_{sol} \;\rgl \; , 
\qquad 
\hat N_{sol} = 
w_{sol} \int \psi_{sol}^\dgr(\br)\; \psi_{sol}(\br) \; d\br \; ,
$$
\be
\label{5.366}
N_{liq} = \lgl \; \hat N_{liq} \;\rgl \; , 
\qquad 
\hat N_{liq} = 
w_{liq} \int \psi_{liq}^\dgr(\br)\; \psi_{liq}(\br) \; d\br \; .
\ee
Here and in what follows, we consider boson field operators. Keeping in mind 
$^4$He, we study spinless particles.  
 
If in the liquid phase there occurs Bose-Einstein condensation, then the 
global gauge symmetry becomes broken 
\cite{Yukalov_169,Yukalov_369,Yukalov_386,Yukalov_387}. The symmetry breaking 
is conveniently described by the Bogolubov shift
\be
\label{5.367}
 \psi_{liq}(\br) = \eta(\br) + \psi_1(\br) \;  ,
\ee
in which 
\be
\label{5.368}
\eta(\br) \equiv \lgl \; \psi_{liq}(\br) \; \rgl
\ee
is the condensate wave function, while $\psi_1({\bf r})$ is the field operator 
of the normal phase, such that
\be
\label{5.369}
 \lgl \; \psi_1(\br) \; \rgl = 0 \; .
\ee
The condensate function and the field operator of the normal phase are 
independent variables orthogonal to each other,
\be
\label{5.370}
 \int \eta^*(\br) \; \psi_1(\br) \; d\br = 0 \;  .
\ee

The gauge symmetry breaking defines the condensate density
\be
\label{5.371}
\rho_0(\br) = |\; \eta(\br)\; |^2 =
|\; \lgl\; \psi_{liq}(\br) \; \rgl\; |^2
\ee
and, respectively, the number of condensed particles
\be
\label{5.372}
 N_0 = w_{liq} \int \rho_0(\br) \; d\br =
w_{liq} \int |\; \eta(\br)\; |^2 \; d\br \;  .
\ee
The number of uncondensed particles is
\be
\label{5.373}
 N_1 =  \lgl\; \hat N_1 \; \rgl \; , \qquad
\hat N_1 = w_{liq} \int \psi_1^\dgr(\br) \; \psi_1(\br) \; d\br \; .
\ee
 
Thus the number of particles in the liquid-like phase reads as
\be
\label{5.374}
 N_{liq} = N_0 +N_1 \;  ,
\ee
so that the total number of particles in the system is
\be
\label{5.375}
 N = N_{sol} + N_{liq} = N_{sol} +  N_0 +N_1 \; .
\ee
The density of condensed and uncondensed particles in the liquid phase is 
defined as
\be
\label{5.376}
 \rho_0 \equiv \frac{N_0}{V_{liq}} \; , \qquad 
\rho_1 \equiv \frac{N_1}{V_{liq}} \;  ,
\ee
respectively. Hence
\be
\label{5.377}
 \rho_0 + \rho_1 = \rho_{liq} \;  .
\ee

The corresponding particle fractions inside the liquid phase are
\be
\label{5.378}
n_0 \equiv \frac{N_0}{N_{liq}} = \frac{\rho_0}{\rho_{liq}} \; , 
\qquad 
n_1 \equiv \frac{N_1}{N_{liq}} = \frac{\rho_1}{\rho_{liq}} \; ,
\ee
so that
\be
\label{5.379}
  n_0 + n_1 = 1 \; .
\ee
It is possible to define the particle fractions with respect to the total 
number of particles
\be
\label{5.380}
 \overline n_0 \equiv \frac{N_0}{N} = n_{liq} \; n_0 \; ,
\qquad
 \overline n_1 \equiv \frac{N_1}{N} = n_{liq} \; n_1 \; ,
\ee
for which
\be
\label{5.381}
 \overline n_0 + \overline n_1 = n_{liq} \;  .
\ee

\subsection{Relations between Chemical Potentials}

For a system with several components there exist the corresponding chemical 
potentials whose relations with each other are prescribed by the condition 
of equilibrium and the imposed constraints. For example, let us consider the 
equilibrium between the component consisting of $N_{sol}$ particles forming 
the solid phase and the component of $N_{liq}$ particles constituting the 
liquid phase, with the total number of particles $N=N_{sol}+N_{liq}$ being 
fixed. The latter implies that
\be
\label{5.382}
 \dlt \; N_{sol} + \dlt \; N_{liq} = 0 \;  .
\ee
The condition of equilibrium, under fixed temperature and volume, tells us 
that
\be
\label{5.383}
 \dlt \; F = \frac{\prt F}{\prt N_{sol} } \; \dlt N_{sol} + 
\frac{\prt F}{\prt N_{liq} } \; \dlt N_{liq} = 0  \; .
\ee
Introducing the corresponding chemical potentials
\be
\label{5.384}
 \mu_{sol} = \frac{\prt F}{\prt N_{sol} } \; , \qquad 
\mu_{liq} = \frac{\prt F}{\prt N_{liq} } \; ,
\ee
and using (\ref{5.382}) yields
\be
\label{5.385}
( \mu_{sol} - \mu_{liq} ) \; \dlt N_{sol} = 0   
\ee
for arbitrary $N_{sol}$. This requires the equality of the chemical potentials,
\be
\label{5.386}
 \mu_{sol} = \mu_{liq} \equiv \mu \; .
\ee

Now, let us $N_{sol}$ and $N_{liq}$ be fixed and consider the equilibrium 
between $N_0$ condensed particles and $N_1$ uncondensed particles. The latter 
means that
\be
\label{5.387}
\dlt N_{sol} = 0 \; , \qquad \dlt N_0 + \dlt N_1 = 0 \; .
\ee
Therefore the condition of equilibrium reads as
\be
\label{5.388}
 \dlt \; F = \frac{\prt F}{\prt N_0}\; \dlt N_0 + 
          \frac{\prt F}{\prt N_1}\; \dlt N_1 = 0 \;  .
\ee
With the notation
\be
\label{5.389}
 \mu_0 = \frac{\prt F}{\prt N_0} \; , \qquad
  \mu_1 = \frac{\prt F}{\prt N_1} \; ,
\ee
we have
\be
\label{5.390}
 (\mu_0 - \mu_1) \; \dlt N_0 = 0 \;  .
\ee

If $N_0$ would be arbitrary, the chemical potentials would coincide. However 
$N_0$ is not arbitrary, but it is fixed by the condition of gauge symmetry 
breaking (\ref{5.368}), according to which the number of condensed particles 
is given by normalization (\ref{5.372}), hence $\dlt\;N_0=0$. Actually, the 
normalization condition (\ref{5.372}) is another form of the gauge-symmetry 
breaking condition (\ref{5.368}). Therefore the chemical potentials are 
not obliged to coincide and, generally, can be different 
\cite{Yukalov_388,Yukalov_389,Yukalov_390,Yukalov_391}.

Thus the inequality of the chemical potentials $\mu_0$ and $\mu_1$ is 
the direct consequence of the gauge-symmetry breaking. If, accepting the 
gauge-symmetry breaking, one sets by force the equality of these potentials, 
then there arises the well-known Hohenberg-Martin \cite{Hohenberg_392} 
dilemma, when either there appears a gap in the spectrum of excitations, which 
contradicts the condition of condensate existence and the Hugenholtz-Pines 
\cite{Hugenholtz_393} relation, or thermodynamic equalities become invalid. 
In both these cases, the system becomes unstable 
\cite{Yukalov_388,Yukalov_389,Yukalov_390,Yukalov_391}. 
On the other hand, equating by force these chemical potentials results in the 
zero anomalous self-energy \cite{Nepomnyashchii_394}, which is equivalent to the 
absence of gauge-symmetry breaking 
\cite{Yukalov_388,Yukalov_389,Yukalov_390,Yukalov_391}.      

The relation between the chemical potentials $\mu$, $\mu_0$, and $\mu_1$ can 
be found from the definition
\be
\label{5.391}
\mu = \frac{\prt F}{\prt N_{liq}} \;   .
\ee
Expanding the derivative
$$
\frac{\prt F}{\prt N_{liq}} = \frac{\prt F}{\prt N_0}\;
\frac{\prt N_0}{\prt N_{liq}} + \frac{\prt F}{\prt N_1}\;
\frac{\prt N_1}{\prt N_{liq}}
$$
and using the notation for the chemical potentials (\ref{5.389}) gives
$$
\mu = \mu_0 \; \frac{\prt N_0}{\prt N_{liq}} + 
\mu_1 \; \frac{\prt N_1}{\prt N_{liq}} \;  .
$$
From the equalities $N_0=n_0 N_{liq}$ and $N_1=n_1 N_{liq}$, under given $n_0$ 
and $n_1$, we get
\be
\label{5.392}
 \frac{\prt N_0}{\prt N_{liq}} = n_0 \; , \qquad
 \frac{\prt N_1}{\prt N_{liq}} = n_1 \; .
\ee
Thus we come to the relation
\be
\label{5.393}
\mu = \mu_0 \; n_0 + \mu_1 \; n_1 \; .
\ee

\subsection{Hamiltonian of Superfluid Solid}

The grand Hamiltonian of a superfluid solid has the form
\be
\label{5.394}
\widetilde H = H_{sol} \; \bigoplus \; H_{liq} \; .
\ee
Here the Hamiltonian part describing the solid phase is
$$
H_{sol} = \hat H_{sol} - \mu \; \hat N_{sol} \; ,
$$
$$
\hat H_{sol} = w_{sol} \int \psi_{sol}^\dgr(\br) \; \left( - \; 
\frac{\nabla^2}{2m} \right) \; \psi_{sol}(\br) \; d\br \; +
$$
\be
\label{5.395}
+ \;
\frac{w_{sol}^2}{2} \int 
\psi_{sol}^\dgr(\br) \; \psi_{sol}^\dgr(\br') \; \Phi(\br-\br') \;
\psi_{sol}(\br') \; \psi_{sol}(\br) \; d\br d\br'
\ee
and the part corresponding to the liquid phase reads as
$$
H_{liq} = \hat H_{liq} - \mu_0\; N_0 - \mu_1\; \hat N_1 - \hat \Lbd \; ,
$$
$$
\hat H_{liq} = w_{liq} \int \psi_{liq}^\dgr(\br) \; 
\left( -\; \frac{\nabla^2}{2m} \right) \; \psi_{liq}(\br)\; d\br \; +
$$
\be
\label{5.396}
+ \;
\frac{w_{liq}^2}{2} 
\int \psi_{liq}^\dgr(\br)\; \psi_{liq}^\dgr(\br') \; \Phi(\br-\br') \;
\psi_{liq}(\br') \; \psi_{liq}(\br) \; d\br d\br' \; .
\ee
The interaction potential $\Phi(\br)$ is assumed to represent an effective 
potential taking account of pair correlations, so that it is integrable.

The expression 
\be
\label{5.397}
\hat\Lbd = \int \left[\; \lbd(\br) \; \psi_1^\dgr(\br) +
\lbd^*(\br) \; \psi_1(\br) \; \right] \; d\br
\ee
is introduced for canceling in the Hamiltonian the terms linear in the field 
operators $\psi_1$, which is required for satisfying condition (\ref{5.369}).

The energy per particle of the solid phase consists of the kinetic energy
\be
\label{5.398}
K_{sol} = \frac{w_{sol}}{N} \int \lgl \; \psi_{sol}^\dgr(\br) \;
\left( -\; \frac{\nabla^2}{2m} \right)\; \psi_{sol}^\dgr(\br)\; \rgl \; d\br 
\ee
and the potential interaction energy  
\be
\label{5.399}
\Pi_{sol} = \frac{w_{sol}^2}{2N} \int \Phi(\br - \br') \;  
\lgl \; \psi_{sol}^\dgr(\br) \; \psi_{sol}^\dgr(\br') \;
\psi_{sol}(\br') \; \psi_{sol}(\br) \; \rgl \; d\br d\br' \; .
\ee

Similarly, for the liquid phase, the kinetic energy is
\be
\label{5.400}
K_{liq} = \frac{w_{liq}}{N} \int \lgl \; \psi_{liq}^\dgr(\br) \;
\left( -\; \frac{\nabla^2}{2m} \right)\; \psi_{liq}(\br)\; \rgl \; d\br
\ee
and the potential energy is
\be
\label{5.401}
\Pi_{liq} = \frac{w_{liq}}{2N} \int \Phi(\br - \br') \;  
\lgl \; \psi_{liq}^\dgr(\br) \; \psi_{liq}^\dgr(\br') \;
\psi_{liq}(\br') \; \psi_{liq}(\br) \; \rgl \; d\br d\br' \;  .
\ee
The kinetic energy of the liquid phase is the sum 
\be
\label{5.402}
K_{liq}= K_0 + K_1
\ee
of the condensate energy
\be
\label{5.403}
 K_0 = \frac{w_{liq}}{N} \int \eta^*(\br) \; \left( -\; 
\frac{\nabla^2}{2m}\right) \; \eta(\br) \; d\br 
\ee
and the kinetic energy of uncondensed particles
\be
\label{5.404}
 K_1 = \frac{w_{liq}}{N} \int \lgl \; \psi_1^\dgr(\br) \; 
\left( -\; \frac{\nabla^2}{2m} \right) \; 
\psi_1(\br) \; \rgl \; d\br \; .
\ee
The internal energy per particle of the whole system is
\be
\label{5.405}
E_0 = \frac{\lgl \;\widetilde H\;\rgl}{N} + \mu =
\frac{\lgl \;\hat H_{sol}\;\rgl}{N} + \frac{\lgl \;\hat H_{liq}\;\rgl}{N}\;   .
\ee

The averages of the grand Hamiltonians are
$$
 \frac{\lgl \; H_{sol}\;\rgl}{N} = K_{sol} + \Pi_{sol} - \mu\; n_{sol} \;
$$
\be
\label{5.406}
 \frac{\lgl \; H_{liq}\;\rgl}{N} = K_{liq} + \Pi_{liq} - \mu\; n_{liq} \; ,
\ee
where
$$
\mu\; n_{liq} = \mu_0 \; \overline n_0 +  \mu_1 \; \overline n_1 \; .
$$
The latter expression is equivalent to (\ref{5.393}). In that way, the 
internal energy becomes
\be
\label{5.407}
E_0 = K_{sol} + \Pi_{sol} + K_{liq} + \Pi_{liq} \;  .
\ee
 
The grand potential 
\be
\label{5.408}
\Om = - T \; \ln \; {\rm Tr}\; \exp(-\bt\; \widetilde H ) =
\Om_{sol} + \Om_{liq}
\ee
is the sum of the terms
\be
\label{5.409}
 \Om_{sol} = - T \; \ln \; {\rm Tr}\; \exp(-\bt\; H_{sol} ) \; ,
\qquad
 \Om_{liq} = - T \; \ln \; {\rm Tr}\; \exp(-\bt\; H_{liq} ) \; .
\ee
The phase probabilities $w_{sol}$ and $w_{liq}$ are defined as the minimizers 
of the thermodynamic potential.

\subsection{Possibility of Superfluid Crystals}

Low temperatures favor the appearance of superfluidity. Therefore, trying to 
understand whether this property could arise in quantum crystals, it is reasonable 
to study first of all the case of zero temperature.

The crystalline state can be characterized using the self-consistent harmonic 
approximation \cite{Yukalov_324,Yukalov_326,Guyer_328,Yukalov_349}. Then for 
the kinetic energy (\ref{5.398}) in the Debye approximation, we find
\be
\label{5.410}
 K_{sol} = \frac{3\rho_{sol}}{4\rho} \; T_D \; w_{sol}^{3/2} \;  ,
\ee  
with the Debye temperature
\be
\label{5.411}
 T_D = \left[ \; \frac{2\nu\rho_{sol}}{3m\rho} 
\sum_j \; \sum_\al
\frac{\prt^2 \Phi(\ba_j)}{\prt a_j^\al\prt a_j^\al} \; 
\right]^{1/2}\; ,
\ee
with $\nu$ being the filling factor, that is the number of particles in a 
lattice site. The potential energy of the crystal reads as
\be
\label{5.412}
\Pi_{sol} = \frac{\rho_{sol}}{2\rho} \; \left( u_0 \; w_{sol}^2 +
\frac{3}{4} \; T_D \; w_{sol}^{3/2} \right) \;  ,
\ee
where 
\be
\label{5.413}
u_0 = \nu \; \frac{\rho_{sol}}{\rho} \sum_j \Phi(\ba_j) \; .
\ee

The average density of the crystalline state practically coincides with the 
density of the liquid-like state, differing from it only by the structure. 
The crystalline state is periodic, while the liquid state is uniform. In what 
follows, we keep in mind the closeness of these densities setting
$$
\rho_{sol} = \rho_{liq}  = \rho \;    .
$$

The liquid Bose-condensed state is well described by the self-consistent 
Hartree-Fock-Bogolubov theory developed in Refs. \cite{Yukalov_169,Yukalov_170,
Yukalov_388,Yukalov_389,Yukalov_391,Yukalov_395,Yukalov_396,Yukalov_397,
Yukalov_398,Yukalov_399,Yukalov_400,Yukalov_401}. The interaction strength 
is characterized by the gas parameter
\be
\label{5.414}
 \gm \equiv \rho^{1/3} \; a_s = 
\frac{m}{4\pi} \; \rho^{1/3} \; \Phi_0 \; ,
\ee
where $a_s$ is a scattering length and
\be
\label{5.415}
 \Phi_0 \equiv \int \Phi(\br) \; d\br = 4\pi \; \frac{a_s}{m} \; .
\ee
The gas parameter is proportional to the ratio of a typical potential energy 
to the characteristic kinetic energy,
\be
\label{5.416}
 \frac{\rho\Phi_0}{E_K} = 8\pi\;\gm \qquad
\left( E_K \equiv \frac{\rho^{2/3}}{2m} \right) \; .
\ee
The latter is also called the zero-point energy. The dimensionless sound 
velocity is denoted 
by 
\be
\label{5.417}
 s \equiv \frac{mc}{\rho^{1/3}} \;  ,
\ee
where $c$ is a dimensional sound velocity. 

For the fraction of uncondensed particles, we have
\be
\label{5.418}
  n_1 = \frac{s^3}{3\pi^2}\; w_{liq}^{3/2} \; ,
\ee
so that the condensate fraction is
\be
\label{5.419}
 n_0 = 1 - \; \frac{s^3}{3\pi^2}\; w_{liq}^{3/2} \;  .
\ee

The sound velocity satisfies the equation
\be
\label{5.420}
 s^2 = 4\pi \gm \; (n_0 + \sgm ) \;  ,
\ee
where $\sigma$ is the anomalous average
\be
\label{5.421}
 \sgm = \frac{8}{\sqrt{\pi}} \; (\gm w_{liq} )^{3/2} \;
\left[\; n_0 + 
\frac{8}{\sqrt{\pi}} \; (\gm w_{liq} )^{3/2} \; \sqrt{n_0} \; 
\right]^{1/2} \;  .
\ee   

For the liquid phase, we have the kinetic energy
\be
\label{5.422}
K_{liq} = \frac{16s^5}{15\pi^2} \; w_{liq}^{7/2} \; E_K
\ee
and the potential energy
\be
\label{5.423}
 \Pi_{liq} = 4\pi \gm \; w_{liq}^2 \left( 1 + n_1^2 - 2n_1 \sgm -
\sgm^2 \right) \; E_K \;  .
\ee

In what follows, it is convenient to pass to dimensionless quantities, 
measuring the energy in units of $E_K$. Thus the dimensionless internal 
energy is
\be
\label{5.424}
 E \equiv \frac{E_0}{E_K} = E(w) \;  ,
\ee
where we use the notation 
\be
\label{5.425}
w_{sol} \equiv w \; , \qquad
w_{liq} = 1 - w \;   .
\ee
The dimensionless Debye temperature reads as
\be
\label{5.426}
 t_D \equiv \frac{T_D}{E_K} \;  .
\ee
The depth of the potential well (\ref{5.413}) in dimensionless form becomes
\be
\label{5.427}
 u \equiv \frac{|\;u_0\;|}{E_K} \;  .
\ee
Summarizing the energy parts gives the dimensionless internal energy 
\be
\label{5.428}
 E = \frac{9}{8} \; t_D \; w^{3/2} - \; \frac{u}{2} \; w^2 
+ \frac{16s^5}{15\pi^2} \; (1 - w)^{7/2} + 4\pi \gm \; (1-w)^2 \; 
 \left( 1 + n_1^2 - 2n_1 \sgm - \sgm^2 \right) \; .
\ee

The phase probabilities are the minimizers of the free energy under the 
normalization condition $w_{sol}+w_{liq}=1$. Our aim is to find out whether 
the regions of disorder arising inside a crystal can support superfluidity. 
Since the latter is a low-temperature phenomenon, its appearance is most 
probable at low temperatures. Therefore, we shall concentrate at zero 
temperature. In that case, the free energy reduces to the internal energy. 
Thus the probability of the solid phase is defined by the conditions
\be
\label{5.429}
 \frac{\prt E}{\prt w} = 0 \; , \qquad 
\frac{\prt^2 E}{\prt w^2} > 0 \; .
\ee
The energy $E = E(w)$ of the heterophase crystal with superfluid regions 
of disorder has to be compared with the energy $E(w \equiv 1)$ of the pure 
crystalline phase, when $w\equiv 1$, and with the energy $E(w\equiv 0)$ of 
the pure liquid phase, when $w\equiv 0$.  
 
To accomplish numerical investigation, we need to fix the following 
quantities: the number of particles at a lattice site $\nu$, the interaction 
potential, the Debye temperature $t_D$, the depth of the potential well at 
the lattice site $u$, and the effective interaction strength $\gamma$. Let 
us consider solid $^4$He forming the hexagonal closest packed (hcp) lattice 
with $12$ nearest neighbors for each atom and one atom at a lattice site. 
The interaction between atoms can be described \cite{Guyer_328} by the 
Lennard-Jones potential with the parameters $\ep=10.2$ K and $\sgm_0=2.556$\AA. 
More often, one uses the Aziz \cite{Aziz_402,Aziz_403,Aziz_404} potential. 
The properties of hcp solid $^4$He have been studied in Monte Carlo numerical 
simulations and in experiments \cite{Hodgdon_405,Ceperley_406,Diallo_407,
Maris_408,Casorla_409,Vitiello_410,Chan_411,Casorla_412}. For the pressure 
$25.3$ bar at zero temperature, the density of solid $^4$He along the melting 
line is $\rho_{sol}=0.0288$\AA$^{-3}$. The density of liquid $^4$He near the 
freezing line is $\rho_{liq}=0.0262$\AA$^{-3}$. This shows that at the 
solid-liquid transition the density does not change much, since 
$\rho_{sol}/\rho_{liq}=1.1$. Hence it is possible to accept that 
$\rho_{sol}\approx\rho_{liq}\approx\rho$. The Debye temperature is $T_D=25$ K. 
The zero-point energy is $E_K=0.572$ K. The Debye temperature in units of $E_K$ 
is $t_D=T_D/E_K=43.7$. The scattering length is $a_s=2.203$ \AA, which gives 
$\gm=0.677$. The potential well (\ref{5.413}) is connected with the static 
potential energy
$$
 E_{sol}^{pot}= \frac{1}{2N} \sum_{i\neq j} \Phi(\ba_i - \ba_j) =
\frac{1}{2} \; u_0 \; ,
$$
which equals $-31.3$ K \cite{Casorla_412}, hence $u_0 = -62.6$ K. In the units 
of $E_K$, this gives $u = |u_0|/E_K = 109$. 

We solve numerically the system of equations (\ref{5.417}) to (\ref{5.421}), 
with the solid phase probability $w$ being the minimizer of the internal energy 
(\ref{5.428}). We fix the Debye temperature $t_D=43.7$, the interaction strength 
$\gm=0.677$, and vary the potential depth $u$. Figure 13 presents the energy 
$E=E(u)$ of the heterophase crystal as a function of $u$, compared to the energy 
of the pure crystal $E_{w=1}(u)$ and the energy $E_{w=0}(u)$ of the pure 
superfluid liquid. The heterophase crystal is stable for $u<75.58$. In the 
region $75.78 < u < 76.39$ it is metastable. And it cannot exist for more deep 
wells with $u > 76.39$, where the most stable is the pure crystalline phase. 
Figure 14 shows the Bose-condensed fraction $n_0 = N_0/N_{liq}$ with respect 
to the number of particles in the liquid state, while $\overline{n}_0=N_0/N$ is the 
Bose-condensed fraction with respect to the total number of particles $N$ in 
the system. The anomalous averages normalized to the number of particles in the 
liquid phase, $\sigma(u)$, and normalized to the total number of particles, 
$\overline\sgm(u)$, are shown in Fig. 15.   

\begin{figure}[ht]
\centerline{
\hbox{ \includegraphics[width=7.5cm]{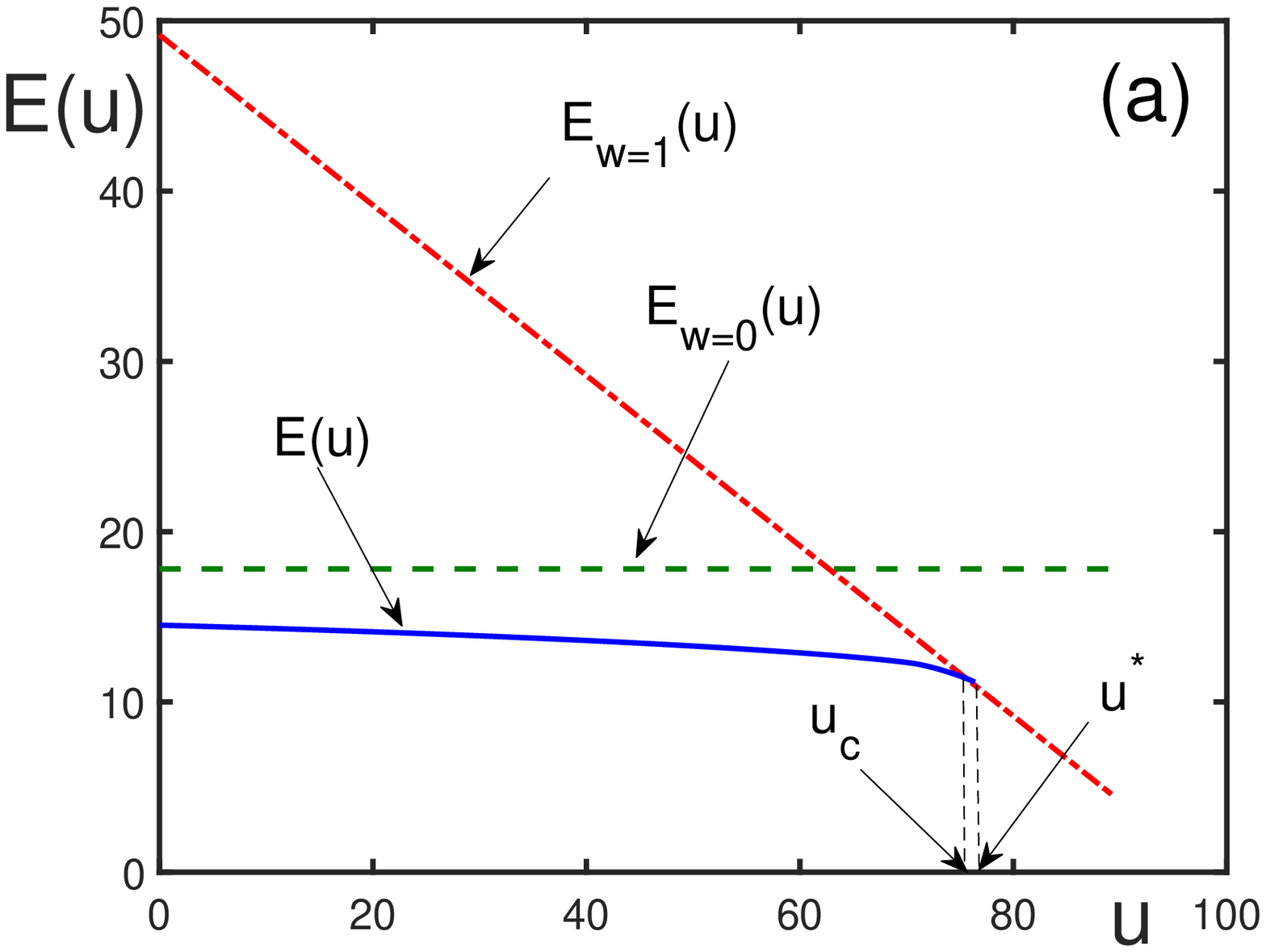} \hspace{1cm}
\includegraphics[width=7.5cm]{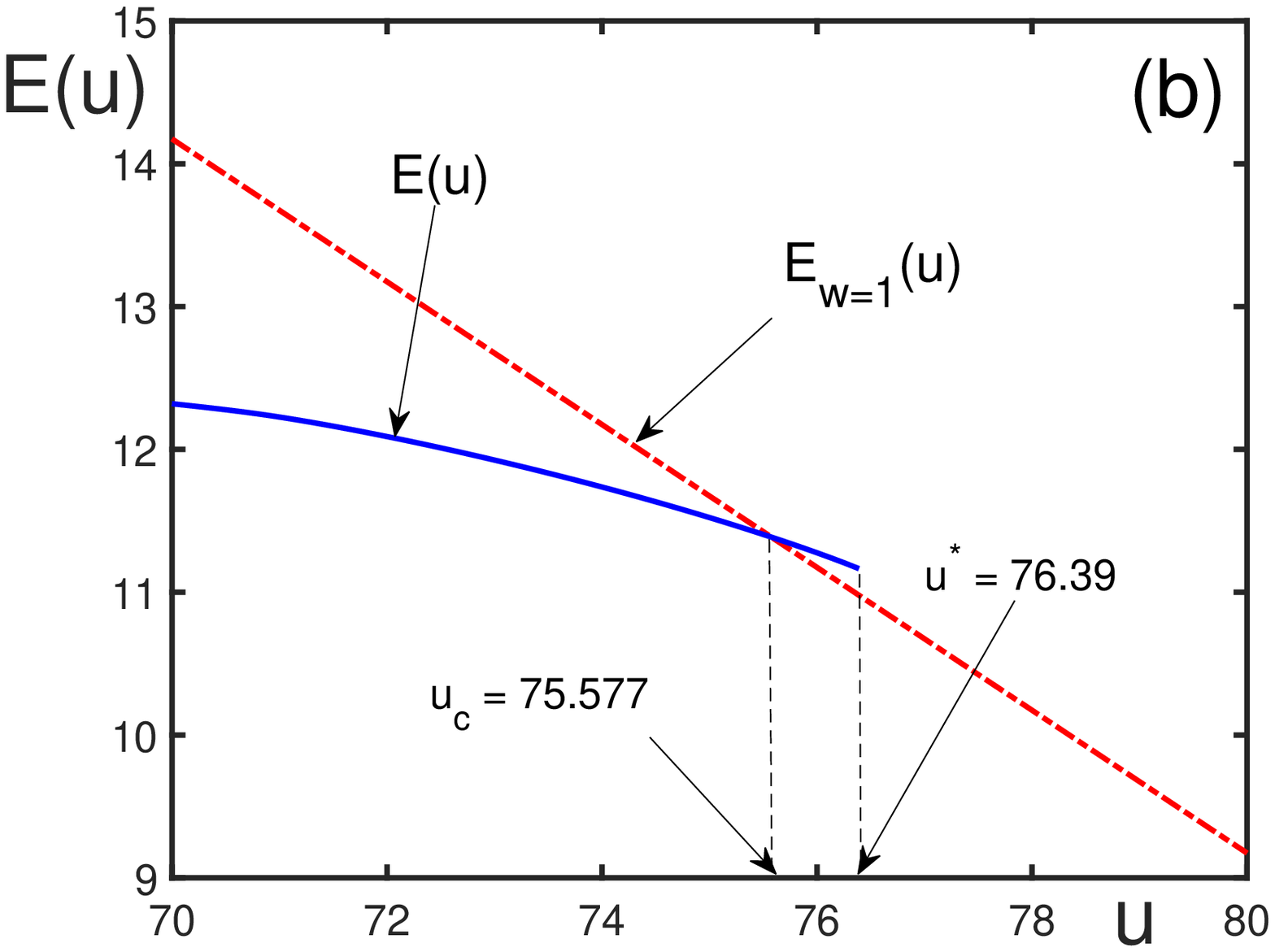} } }
\caption{Energy $E(u)$ of a heterophase crystal with superfluid 
fluctuations (solid line); the energy $E_{w=1}(u)$ of the pure crystalline 
phase (dash-dotted line); the energy $E_{w=0}(u)$ of the pure liquid phase
(dashed line). The critical region of the transition between heterophase 
superfluid crystal and the pure crystalline state is detailed in Fig. 13b.
}
\label{fig:Fig.13}
\end{figure}

\begin{figure}[ht]
\centerline{
\includegraphics[width=10cm]{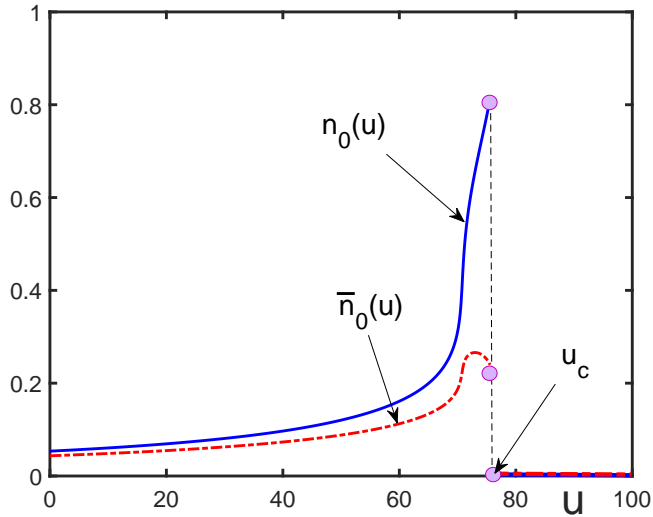} }
\caption{Bose-condensed fractions with respect to the number of particles 
in the liquid phase, $n_0(u)$ (solid line), and with respect to the total
number of particles, ${\overline n}_0(u)$ (dash-dotted line).
}
\label{fig:Fig.14}
\end{figure}

\begin{figure}[ht]
\centerline{
\includegraphics[width=10cm]{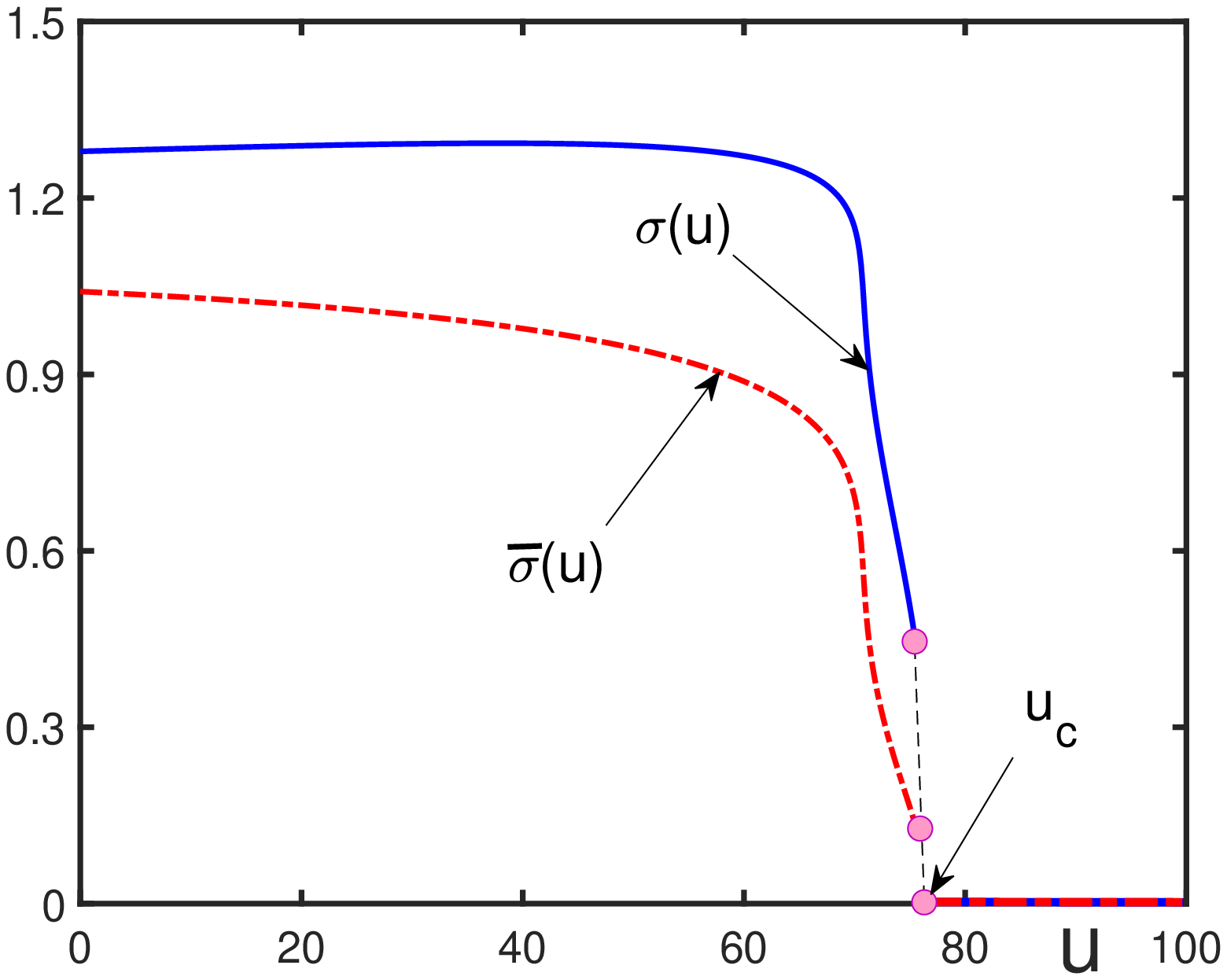} }
\caption{Anomalous averages, normalized to the number of particles in the
liquid phase, $\sgm(u)$ (solid line), and normalized to the total number
of particles ${\overline\sgm}(u)$ (dash-dotted line).
}
\label{fig:Fig.15}
\end{figure}

Since for the solid $^4$He, with the melting density, the depth of the 
potential well $u=109$ is far above the critical depth $u_c=75.58$ above 
which superfluid properties in a crystal are not able to arise, we have to 
conclude that, in the frame of the present model, solid $^4$He cannot contain 
any Bose-condensed fraction, hence it does not support a superfluid fraction. 
However, in principle, heterophase superfluid solids could exist, provided the 
potential well $u$ is sufficiently shallow.

\section{Mixture of Microscopic Components}

As is discussed in Sec. 1, there are three types of mixed matter, macroscopic 
Gibbs mixture, mesoscopic heterophase mixture, and the mixture of several 
microscopic components, such as atoms, molecules, or other particles that 
could be treated as elementary. In this type of mixture, all constituents are 
uniformly distributed in space, as is shown in Fig. 3, so that neither of the 
components forms a separate phase. In some cases, the components can separate 
in space, as a result of which the macroscopic phases could be formed, as in 
Fig. 1. But here we will pay the main attention to the situation when the 
constituents are uniformly intermixed in space, exactly as in Fig 3. For 
concreteness, we consider the mixture of quarks and hadrons that can be created 
in heavy-ion or nuclear collisions.

\subsection{Mixed Quark-Hadron Matter}

At high temperatures and/or densities, hadronic matter is expected to undergo 
a transition to quark-gluon plasma, where quarks and gluons are no longer 
confined inside hadrons but can propagate much further in extent than the 
typical sizes of hadrons. This is called deconfinement transition. It is 
assumed to be possible under heavy-ion or nuclear collisions. It is supposed 
to exist in the early universe at a time on the order of  microseconds after 
Big Bang, when temperature was high enough for the elementary degrees of 
freedom of quantum chromodynamics to be in a deconfined state. The quark-gluon 
plasma can also exist in the interior of neutron stars. More details can be 
found in the review articles \cite{Shuryak_413,Rafelsky_414,Satz_415,
Hagedorn_416,Cleymans_417,Stocker_418,Clare_419,Mclerran_420,Reeves_421,
Adami_422,Yukalov_423,Yukalov_424,Glendenning_425,Satz_426,Soltz_427,Ayala_428,
Busza_429,Guenther_430}.  

An important question is: How does the deconfinement transition occur? If 
it is a phase transition between pure phases of hadron matter and quark-gluon 
plasma, then, it seems, quark degrees of freedom should be unobservable in 
hadron matter. But if it is a gradual crossover, then quark degrees of freedom 
could somehow show up in nuclear matter and, probably, even in nuclei. 
Blokhintsev \cite{Blokhintsev_431} suggested that even in cold nuclear matter 
there can fluctuationally arise dense objects, called fluctons, composed of 
more than three quarks. Dense multiquark formations in nuclei were assumed to 
be the cause of the cumulative effect advanced by Baldin 
\cite{Baldin_432,Strikman_433,Efremov_434,Baldin_435,Shimanskiy_436}. 
Multiquark clusters can appear in nuclei even at normal nuclear density 
\cite{Kobushkin_437,Makarov_438,Burov_439}. These works suggest that hadron 
and quark degrees of freedom could coexist forming a kind of quark-hadron 
mixture. 

As has been explained above, there can exist three types of mixtures depending 
on the relation between the spatial sizes of the three characteristic length 
scales, mean interparticle distance $a$, the size $l_{het}$ of regions occupied 
by a competing phase inside a host phase, and the size of the system $l_{exp}$. 
The Gibbs phase separation, as shown in Fig. 1, corresponds to a macroscopic 
mixture, where $a\ll l_{het}\sim l_{exp}$. Figure 2, where 
$a\ll l_{het}\ll l_{exp}$, illustrates a mesoscopic mixture, and Fig. 3 
represents a uniformly mixed system of microscopic particles.      

The principal question is: What kind of mixture do we keep in mind with 
respect to mixed quark-hadron matter? Generally, one talks about the possible 
existence of a mixed quark-hadron phase that can arise in colliding heavy ions 
or nuclei, in the interior of neutron stars, or in the early Universe, a few 
microseconds after the Big Bang. Below we shall follow the ideas of clustering 
nuclear matter we started in the collaboration with Baldin \cite{Baldin_440,
Baldin_441,Baldin_442,Baldin_443,Chizhov_444}. This approach has also been 
developed and summarized in the articles \cite{Yukalov_423,Yukalov_424,
Yukalov_445,Yukalov_446,Yukalov_447,Yukalov_448,Yukalov_449}.  

When ions of nuclei are relativistically accelerated towards each other, 
they resemble not spheres but rather disks due to the Lorentz contraction. 
After they collide, a fireball is formed, as is shown in Fig. 16. The first 
question that arises is: Can the methods of statistical equilibrium (or 
quasi-equilibrium) physics be used for the description of a fireball? For 
this, the fireball lifetime has to be much longer than the time of local 
equilibration. The fireball lifetime is of order $10^{-22}$ s. The local 
equilibration time can be estimated as $t_{loc}\sim\lbd/c$, where $\lambda$ 
is mean free path and $c$ is light velocity \cite{Yukalov_423}. The mean 
free path is of the order of $1$ fm. Therefore the local equilibration time 
is $t_{loc}\sim 10^{-24}-10^{-23}$ s. This is much shorter then the fireball 
lifetime $10^{-22}$ s, hence equilibrium (quasi-equilibrium) statistical 
mechanics is applicable. 

\begin{figure}[ht]
\centerline{
\includegraphics[width=10cm]{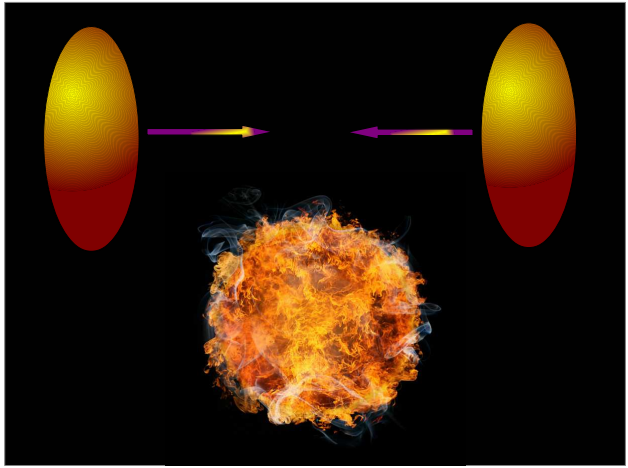} }
\caption{Formation of a fireball under heavy-ion or nuclear collisions.
}
\label{fig:Fig.16}
\end{figure}    

When talking about mixed quark-hadron phase, one usually assumes the 
situation typical of first-order phase transitions resulting in a macroscopic 
phase separation, as in Fig. 1. However, this phase separation requires the 
use of the Maxwell construction (see \cite{Yukalov_449}), when in the equation 
of state, defining the dependence of pressure $P(V)$ on volume, there appears 
a horizontal line where the pressure derivative is zero, or close to zero, if 
the separating surface energy is taken into account, $\prt P/\prt V\ra 0$. 
Because of this, the isothermal compressibility diverges.

This implies that the state with the macroscopic phase separation is unstable, 
so that very weak fluctuations quickly destroy it and intermix the phases 
\cite{Yukalov_449}.

One often mentions the case of water-vapor separation in a test-tube as 
an example of the situation under a first-order phase transition, where the 
macroscopic mixture looks stable. However, one should not forget that, in 
addition to the surface energy, the macroscopic water-vapor mixture is 
stabilized by gravity. Also, being inside a test-tube, the mixture is isolated 
from external perturbations, while the real systems, such as fireballs, are 
not isolated from surrounding.  

The other type of spatial separation could be due to the stratification 
of different kinds of hadrons. However the stratification is a slow process 
requiring the time of the order of $(R/a)t_{loc}\sim 10^{-22}$ s. Here 
$R\sim 10$ fm is the fireball radius and $a$ means interparticle distance. 
This stratification time equals the fireball lifetime. Hence, it seems, the 
stratification of different hadron components will not occur during the 
fireball lifetime. In this way, it is possible to assume that in a fireball 
there is neither first-order phase separation nor component stratification. 
Thus under mixed quark-hadron matter we keep in mind a uniformly mixed system, 
as shown in Fig. 3.

\subsection{Stability of Multicomponent Mixture}

Even when the fireball lifetime is much longer than the time of the components 
separation, anyway these components can form a stable equilibrium mixture. 
The condition of mixture stability can be derived in the following way. 

Consider $N_{com}$ different components, which are enumerated by the indices 
$i,j=1,2,\ldots,N_{com}$. The physical details of the components will be 
specified a bit later. For a while, we can keep in mind that an $i$-th 
component is composed of a particular type of quark clusters. The number of 
clusters in an $i$-th component is $N_i$.

The interaction energy of a mixture of the components, written in the Hartree 
approximation, is
\be
\label{6.1}
E_{mix} = \frac{1}{2} \sum_{ij} \int_V \;
\overline\rho_i(\br)\; \Phi_{ij}(\br-\br')\; \overline\rho_j(\br')\; 
d\br d\br' \; ,
\ee
where the density of the $i$-th component clusters for a uniform system is
\be
\label{6.2}
 \overline\rho_i(\br) = \frac{N_i}{V} \;  .
\ee
With the notation
\be
\label{6.3}
  \Phi_{ij} \equiv \int_V \Phi_{ij}(\br) \; d\br \; ,
\ee
we have
\be
\label{6.4}
E_{mix} = \frac{1}{2} \sum_{ij} \Phi_{ij} \; \frac{N_i N_j}{V} \; .
\ee
Adding the single-particle (single-cluster) energy
\be
\label{6.5}
E_{mix}' = \sum_i \int_V U_i(\br) \; \overline\rho_i(\br)\; d\br
\ee
yields the total internal energy
\be
\label{6.6}
 E_{mix}^{tot} = E_{mix} + E_{mix}' \; .
\ee
Assuming that the external fields are approximately uniform, $U_i(\br)=U_i$ 
gives the single-cluster energy
\be
\label{6.7}
 E_{mix}' = \sum_i U_i \; N_i \;  .
\ee
 
The mixture free energy writes as
\be
\label{6.8}
 F_{mix} = E_{mix}^{tot} - T S_{mix} \;  ,
\ee
with the entropy 
\be
\label{6.9}
 S_{mix} = S_0 + \Dlt \; S_{mix} \;  ,
\ee
where $S_0$ is the entropy of a non-mixed system, while the entropy of mixing 
is
\be
\label{6.10}
 \Dlt \; S_{mix} = N \; \Dlt \; s_{mix} \qquad 
\left( N = \sum_i N_i\right) \; ,
\ee
and the reduced entropy of mixing per particle being
\be
\label{6.11}
 \Dlt \; s_{mix} = - \sum_i n_i \; \ln n_i \qquad
\left( n_i \equiv \frac{N_i}{N} \right) \; .
\ee
    
When the components are separated, the interaction energy has the form
\be
\label{6.12}
E_{sep} = \frac{1}{2} \sum_i \int_{V_i} \rho_i(\br) \; \Phi_{ii}(\br-\br') \;
\rho_i(\br') \; d\br d\br' \; .
\ee
In a uniform system, the density is
\be
\label{6.13}
\rho_i(\br) = \frac{N_i}{V_i} \;   .
\ee
Keeping in mind the separation into macroscopic Gibbs phases allows us to 
accept that
$$
 \int_V \Phi_{ii}(\br) \; d\br = 
\int_{V_i} \Phi_{ii}(\br) \; d\br \;  .
$$
Therefore the interaction energy of separated components reads as
\be
\label{6.14}
 E_{sep} = \frac{1}{2} \sum_i \Phi_{ii} \; \frac{N_i^2}{V_i} \; .
\ee
The total internal energy of separated components is
\be
\label{6.15}
 E_{sep}^{tot} = E_{sep} + E_{sep}' \;  ,
\ee
with the single-cluster energy
\be
\label{6.16}
E_{sep}' = \sum_i \int_{V_i} U_i(\br)\; \rho_i(\br)\; d\br \; .
\ee
For uniform fields, we get
\be
\label{6.17}
E_{sep}' = \sum_i U_i \; N_i = E_{mix}' \; .
\ee
Thus the free energy of a separated system is
\be
\label{6.18}
F_{sep} = E_{sep}^{tot} - T\; S_{sep} \;  ,
\ee
with $S_{sep}\approx S_0$.   

The mixed state is stable, as compared to the separated state, provided that
\be
\label{6.19}
 F_{mix} < F_{sep} \;  .
\ee
From here, it follows that
\be
\label{6.20}
E_{mix} - T \; \Dlt \; S_{mix}  < E_{sep} \; .
\ee
Using the identity
$$
\frac{V}{V_i} = \sum_j \frac{V_j}{V_i}
$$
transforms (\ref{6.14}) into
\be
\label{6.21}
 E_{sep} = \frac{1}{2} 
\sum_{ij} \Phi_{ii} \; \frac{N_i^2 V_j}{V V_i} \; .
\ee

The separated state is mechanically stable, if the partial pressures coincide,
\be
\label{6.22}
P_i = P_j \; , \qquad P_i = -\; \frac{\prt F_{sep}}{\prt V_i} \;   .
\ee
Since
\be
\label{6.23}
P_i = \frac{1}{2} \; 
\Phi_{ii} \; \left( \frac{N_i}{V_i} \right)^2 \; ,
\ee
we have
\be
\label{6.24}
 \frac{P_i}{P_j} = \frac{\Phi_{ii}}{\Phi_{jj}}\;
\left( \frac{N_iV_j}{N_jV_i} \right)^2 = 1 \; ,
\ee
from where
\be
\label{6.25}
\frac{V_j}{V_i} = 
\frac{N_j}{N_i}\; \sqrt{ \frac{\Phi_{jj}}{\Phi_{ii}} } \; .
\ee

In that way, we obtain
\be
\label{6.26}
 E_{sep} = \frac{1}{2} \sum_{ij} \frac{N_i N_j}{V} \;
\sqrt{\Phi_{ii} \; \Phi_{jj} } \;  .
\ee
So that the condition of mixture stability (\ref{6.20}) becomes
\be
\label{6.27}
\sum_{ij} \frac{N_i N_j}{2V} \; \left( \Phi_{ij} - \;
\sqrt{\Phi_{ii}\; \Phi_{jj} } \right) < N\; T\; \Dlt\; s_{mix} \;  .
\ee
Applying the identity
$$
 N = \frac{1}{N} \sum_{ij} N_i \; N_j  
$$
results in the inequality
\be
\label{6.28}
\sum_{ij} \frac{N_i N_j}{2V} \left( \Phi_{ij} -\;
\sqrt{\Phi_{ii}\; \Phi_{jj} } \; - \; 
\frac{2T}{\rho} \; \Dlt \; s_{mix} \right) < 0 \; .
\ee
Assuming that the latter inequality is to be valid for any $N_i$ and $N_j$, 
we obtain the condition of mixture stability
\be
\label{6.29}
 \Phi_{ij} < \sqrt{\Phi_{ii}\;\Phi_{jj} } + 
\frac{2T}{\rho}\; \Dlt\; s_{mix} \; ,
\ee
where $\rho\equiv N/V$.

Generally, this is a sufficient condition for the mixture stability. As is 
evident, this condition requires that both $\Phi_{ii}$ and $\Phi_{jj}$ be 
of the same sign. Temperature makes mixing easier. The components may be 
immiscible at zero temperature, but can become miscible at finite temperature.

\subsection{Theory of Clustering Matter}

Clusters composed of particles forming bound states can be treated as separate 
types of particles \cite{Ter_450}. Detailed mathematical representation for 
clustering matter has been suggested in Refs. \cite{Tani_451,Weinberg_452,
Weinberg_453,Weinberg_454,Girardeau_455,Girardeau_456,Girardeau_457,Girardeau_458}. 
The approach describing quark-hadron matter as a type of clustering matter has 
been developed \cite{Yukalov_423,Yukalov_424}. 

Let us consider a system composed of different types of particles, say quarks, 
antiquarks, gluons, and various multiquark clusters (or bags) formed of two 
quarks (mesons), three quarks (nucleons), and multiquark clusters (including 
anti-multiquark clusters). Let the elementary particles, that is, quarks, 
antiquarks, and gluons, be enumerated by an index $\al$, while the composite 
particles be enumerated by the index $i$. Thus the Fock space of the considered 
system is the {\it cluster space}
\be
\label{6.30}
 {\cal F} = \bigotimes_\al \; {\cal F}_\al \; 
\bigotimes_i \; {\cal F}_i \; .
\ee
In many cases, it is convenient to introduce the general enumeration of all 
particles by means of a multi-index $n=\al,i$. 

The system constituents are characterized by several specific numbers. The 
compositeness number $z_n$ shows the number of elementary particles composing 
the bound cluster of the $n$-th type. The other numbers characterizing a cluster 
of the $n$-th type are the baryon number $B_n$, strangeness $S_n$, and charge 
$Q_n$.  
 
The balance between the particles composing the system is regulated by the 
chemical potentials. If the average density
\be
\label{6.31}
\rho = \sum_n z_n \; \rho_n 
\qquad 
\left( \rho_n \equiv \frac{N_n}{V} \right)
\ee
is given, the chemical potentials $\mu_i$ of the $i$-th type quark clusters 
are connected with the free quarks and antiquark chemical potentials through 
the relation 
\be
\label{6.32}
\mu_q = - \mu_q = \frac{\mu_i}{z_i} \; , \qquad \mu_g = 0 \; ,
\ee
while the gluon chemical potential being zero. When the baryon density of 
matter
\be
\label{6.33}
\rho_B \equiv \frac{N_B}{V} = \sum_i B_i \; \rho_i
\ee
is known, this defines the baryon potential $\mu_B$. If the strangeness 
density
\be
\label{6.34}
\rho_S \equiv \frac{N_S}{V} = \sum_i S_i \; \rho_i
\ee
is conserved, the strangeness potential $\mu_S$ is defined. Similarly, the 
fixed density of charge
\be
\label{6.35}
\rho_Q \equiv \frac{N_Q}{V} = \sum_i Q_i \; \rho_i
\ee
sets the charge potential $\mu_Q$. The general relation between the chemical 
potentials reads as
$$
\sum_i \mu_i \; \rho_i = \mu_B \sum_i B_i \; \rho_i + 
\mu_S \sum_i S_i \; \rho_i + \mu_Q \sum_i Q_i \; \rho_i \; =
$$
\be
\label{6.36}
 = \;
\mu_B \; \rho_B + \mu_S \; \rho_S +  \mu_Q \; \rho_Q \; ,
\ee  
which implies the relation between the chemical potentials
\be
\label{6.37}
 \mu_i = \mu_B \; B_i + \mu_S \; S_i + \mu_Q \; Q_i \; .
\ee

Defining the corresponding baryon, strangeness, and charge numbers
\be
\label{6.38}
 N_B = \sum_i B_i\; N_i \; , \qquad N_S = \sum_i S_i\; N_i \; , 
\qquad N_Q = \sum_i Q_i\; N_i \; , 
\ee
it is straightforward to introduce the related number operators
\be
\label{6.39}
 \hat N_B = \sum_i B_i\; \hat N_i \; , \qquad 
\hat N_S = \sum_i S_i\; \hat N_i \; , 
\qquad \hat N_Q = \sum_i Q_i\; \hat N_i \;  ,
\ee
so that 
\be
\label{6.40}
\sum_i \mu_i \hat N_i = \mu_B \hat N_B +  \mu_S \hat N_S + 
\mu_Q \hat N_Q \; .
\ee

Finally, it is important to keep in mind that when constructing the effective 
Hamiltonian of the clustering matter, one often employs the terms containing 
thermodynamic parameters, such as temperature and density. Then it is necessary 
to supplement the Hamiltonian with a correcting term providing the validity of 
the thermodynamic relations for the pressure, internal energy, and entropy,
$$
P = - \; \frac{\Om}{V} = - \; \frac{\prt \Om}{\prt V} \; ,
$$
$$
E = \frac{1}{V} \; \lgl \; \hat H \; \rgl = T \; \frac{\prt P}{\prt T} -
P + \sum_i \mu_i \; \rho_i \; ,
$$
\be
\label{6.41}
 S = \frac{\prt P}{\prt T} = \frac{1}{T} \;
\left( E + P - \sum_i \mu_i \; \rho_i\right) \; ,
\ee
as well as for the densities
$$
\rho_i = \frac{1}{V} \; \lgl \; \hat N_i \; \rgl =
\frac{\prt P}{\prt\mu_i} \; , 
$$
\be
\label{6.42}
\rho_B = \frac{\prt P}{\prt\mu_B} \; ,  \qquad
\rho_S = \frac{\prt P}{\prt\mu_S} \; ,  \qquad
\rho_Q = \frac{\prt P}{\prt\mu_Q} \; .
\ee
In this way, if the energy Hamiltonian is $\hat{H}$, then the grand Hamiltonian 
is
\be
\label{6.43}
H = \hat H - \sum_\al \mu_\al\; \hat N_\al - \sum_i \mu_i\; \hat N_i +
CV \;  ,
\ee
where $CV$ is a correcting term. 

The probability of an $n$-th component is
\be
\label{6.44}
w_n \equiv \frac{z_n\rho_n}{\rho} \; , 
\ee
which implies
$$
w_\al \equiv \frac{\rho_\al}{\rho} \; , \qquad 
w_i \equiv \frac{z_i\rho_i}{\rho} \;  .
$$
As is clear, the normalization condition is valid:
\be
\label{6.45}
 \sum_n w_n = 1 \; , \qquad 0 \leq w_n < 1 \;  .
\ee
In addition, the system thermodynamic stability has to be checked, such as the 
positiveness and finiteness of specific heat and isothermal compressibility:
$$
0 \leq C_V < \infty \qquad 
\left( C_V \equiv \frac{\prt E}{\prt T}\right) \; ,
$$
\be
\label{6.46}
0 \leq \varkappa_T < \infty \qquad 
\left( \varkappa_T \equiv -\; 
 \frac{1}{V} \; { \displaystyle /  } \; \frac{\prt P}{\prt V}  \right)  \; .
\ee

Often, when modeling Hamiltonian (\ref{6.43}), one includes the terms containing 
thermodynamic parameters, e.g. the density of clusters $\rho_n$ or temperature 
$T$. Then, in order that thermodynamic relations (\ref{6.41}) and (\ref{6.42}) be 
valid, it is necessary to impose the {\it conditions of statistical correctness} 
\cite{Yukalov_423,Yukalov_424} having the form
\be
\label{6.47}
\left\lgl \; \frac{\prt H}{\prt\rho_n}\; \right\rgl = 0 \; ,
\qquad
\left\lgl \; \frac{\prt H}{\prt T}\; \right\rgl = 0 \;   .
\ee
This defines the correcting term $CV$ prescribing the equations for the 
function $C=C(\rho_n,T)$. Without imposing these conditions, the model 
thermodynamics is not reliable. For instance, one constantly employs the 
excluded-volume approximation without using the correcting term, which makes 
this approximation incorrect.

\subsection{Clustering Quark-Hadron Mixture}

The energy Hamiltonian of quark-hadron matter can be written as a sum of 
two terms,
\be
\label{6.48}
 \hat H = \hat H_1 + \hat H_z \;  ,
\ee
where the first term describes the unbound particles with the compositeness 
number $z_\al=1$, i.e. quarks, antiquarks, and gluons, while the second term 
describes hadron clusters with the compositeness number $z_i\geq 2$. Particles 
are characterized by the field operators 
$$
\psi_\al(\br) = [\;\psi_\al(\br,\sgm_\al)\; ] \; , 
\qquad
\psi_i(\br) = [\; \psi_i(\br,\sgm_i)\; ] \;  ,
$$
which are the columns with respect to the internal degrees of freedom 
$\sgm_\al$ and $\sgm_i$.     

The Hamiltonian of the unbound particles is
\be
\label{6.49}
 \hat H_1 = \sum_\al \int \psi_\al^\dgr(\br) \; 
\left( \sqrt{-\nabla^2+m_\al^2}\; + U_\al \right) \;
\psi_\al(\br) \; d\br \; ,
\ee
with the potential $U_\alpha$ describing the effective interactions between 
the unbound particles and between the unbound particles and clusters. This 
potential should satisfy the {\it color-confinement condition}
\be
\label{6.50}
U_\al \ra \infty \qquad (\rho \ra 0 )
\ee
and the {\it asymptotic-freedom condition}
\be
\label{6.51}
 U_\al \ra 0 \qquad (\rho \ra \infty ) \; .
\ee
Thus it can be represented by the expression
\be
\label{6.52}
 U_\al = \frac{A^{1+\nu}}{\rho^{\nu/3}}  
\ee
parameterized with $A$ and $\nu$. 

The cluster Hamiltonian reads as
$$
\hat H_z = \sum_i \int \psi_i^\dgr(\br) \; 
\left( \sqrt{-\nabla^2+m_i^2}\; + U_i \right) \;
\psi_i(\br) \; d\br \; +
$$
\be
\label{6.53}
+\; \frac{1}{2} \sum_{ij} \int \psi_i^\dgr(\br)\; \psi_j^\dgr(\br') \;
\Phi_{ij}(\br - \br') \;
\psi_j(\br') \; \psi_i(\br)\; d\br d\br' \; .
\ee
The effective field $U_i$ takes into account the interactions between the 
clusters and unbound particles, hence it has to be zero when the latter are 
absent, that is
\be
\label{6.54}
 U_i \ra 0 \qquad (\rho_1 \ra 0 ) \; .
\ee
For example, this effective field can be accepted in the form
\be
\label{6.55}
 U_i = z_i A^{1+\nu} \; \left( \frac{1}{\rho^{\nu/3}} \; - \;
\frac{1}{\rho_z^{\nu/3}} \right) \; .
\ee
Here and in what follows, we take into account that the average density $\rho$ 
is the sum of the density of unbound particles and of the cluster density, so 
that
\be
\label{6.56}
 \rho_1 \equiv \sum_\al \rho_\al \; , \qquad 
\rho_z \equiv \sum_i z_i\; \rho_i \qquad ( \rho = \rho_1 + \rho_z) \; .
\ee

The cluster interactions, resorting to the potential scaling 
\cite{Yukalov_423,Yukalov_424}
\be
\label{6.57}
 \frac{\Phi_{ij}(\br)}{z_i z_j} = \frac{\Phi_{ab}(\br)}{z_a z_b} \; ,
\ee    
can be expressed through the nucleon-nucleon interactions, 
\be
\label{6.58}
 \Phi_{ij}(\br) = \frac{z_i z_j}{9} \; \Phi_{NN}(\br) \;  ,
\ee
taken, e.g., in the form of the Bonn potential \cite{Machleidt_459}. 

Another type of the potential scaling can be represented with the relation
$$
\frac{\Phi_{ij}(\br)}{m_i m_j} = \frac{\Phi_{ab}(\br)}{m_a m_b} \; .
$$

Note that under the validity of the potential scaling the condition of mixture 
stability (\ref{6.29}) reduces to the positiveness of the entropy of mixing, which 
is certainly true.  

The correcting term is defined by the conditions of statistical correctness 
(\ref{6.47}), which for the considered case read as
$$
\left\lgl \;\frac{\prt H}{\prt\rho} \; \right\rgl = 0 \; ,
\qquad
 \left\lgl \;\frac{\prt H}{\prt\rho_z} \; \right\rgl = 0 \; .
$$
This yields
\be
\label{6.59}
 C = \frac{\nu}{3-\nu} \; A^{1+\nu} \;\left( \rho^{1-\nu/3} - 
\rho^{1-\nu/3}_z \right) \;  .
\ee
In the particular case of $\nu = 2$, we have
$$
 C = 2A^3 \; \left( \rho^{1/3} - \rho^{1/3}_z \right) \; ,
\qquad U_\al = \frac{A^3}{\rho^{2/3}} \; .
$$

\subsection{Thermodynamics of Quark-Hadron Matter}

Keeping in mind that the system made of quark-hadron matter is sufficiently large 
to be treated as uniform, it is possible to resort to the Fourier transformation 
for the field operators,
\be
\label{6.60}
\psi_n(\br)= \frac{1}{\sqrt{V}} \sum_k a_n(\bk) \; e^{i\bk\cdot\br} \; ,
\ee
where the operators
$$
a_n(\bk) = [\; a_n(\bk,\sgm_n)\;]
$$
are the columns with respect to the internal degrees of freedom. The potentials 
of cluster interactions also are assumed to allow for the Fourier transformation,
\be
\label{6.61}
 \overline\Phi_{ij}(\bk) = 
\int \Phi_{ij}(\br)\; e^{-i\bk\cdot\br} \; d\br \; , 
\qquad
\Phi_{ij}(\br) = 
\frac{1}{V} \sum_k \overline\Phi_{ij}(\bk)\; e^{i\bk\cdot\br} \;  .
\ee
In particular, the notation
\be
\label{6.62}
\Phi_{ij} \equiv \int \Phi_{ij}(\br)\; d\br = \overline\Phi_{ij}(0)
\ee
will be used. 

The Hamiltonian (\ref{6.49}) of unbound particles takes the form
\be
\label{6.63}
 \hat H_1 - \sum_\al \mu_\al \; N_\al = 
\sum_\al \sum_k \om_\al(\bk)\; a_\al^\dgr(\bk)\; a_\al(\bk) \; ,
\ee
with the spectrum
\be
\label{6.64}
 \om_\al(\bk) = \sqrt{k^2 + m_\al^2} \; + U_\al - \mu_\al \; .
\ee
Recall that the chemical potentials for gluons and quarks are $\mu=0$ and 
$\mu_q=-\mu_{\overline q}$. 

The momentum distribution of unbound particles is 
\be
\label{6.65}
 n_\al(\bk) \equiv \lgl \; a_\al^\dgr(\bk)\; a_\al(\bk) \; \rgl =
\frac{\zeta_\al}{\exp\{\bt\om_\al(\bk)\}\mp 1} \;  ,
\ee
with $\zeta_\alpha$ being the degeneracy factor and, in the right-hand side, 
the upper sign is for Bose-Einstein statistics, while the lower, for Fermi-Dirac 
statistics. For gluons and quarks,
\be
\label{6.66}
 \zeta_g = 2 \times (N_c^2 -1 ) \; , \qquad 
\zeta_q = 2\times N_f \times N_c \;  ,
\ee
where $N_c$ is the number of colors and $N_f$, number of flavors. For three 
colors, we have
$$
 \zeta_g = 16 \; , \qquad \zeta_q = 6N_f   \qquad (N_c = 3 ) \;  .
$$

The cluster Hamiltonian (\ref{6.53}) can be simplified in the Hartree-Fock 
approximation yielding
\be
\label{6.67}
\hat H_z - \sum_i \mu_i \; N_i = 
\sum_i \sum_k \om_i(\bk)\; a_i^\dgr(\bk)\; a_i(\bk) + R_{HF} \;  ,
\ee
with the spectrum
\be
\label{6.68}
 \om_i(\bk) = \sqrt{k^2 + m_i^2} \; + U_i + \sum_j \Phi_{ij} \;\rho_j
\pm \frac{1}{V} \sum_p n_i(\bp) \; \overline\Phi_{ii}(\bk+\bp) - \mu_i
\ee
and a nonoperator term
\be
\label{6.69}
R_{HF} = -\; \frac{V}{2} \sum_i \left[\; 
\rho_i \sum_j \Phi_{ij}\; \rho_j  \pm  \frac{1}{V^2} 
\sum_{kp} \overline\Phi_{ii}(\bk+\bp) \; n_i(\bk) \; n_i(\bp) \;
\right] \; .
\ee
The momentum distribution of clusters reads as
\be
\label{6.70}
n_i(\bk) \equiv \lgl \; a_i^\dgr(\bk) \; a_i(\bk) \; \rgl =
\frac{\zeta_i}{\exp\{\bt\om_i(\bk)\}\mp 1} \; ,
\ee
where again the upper sign is for Bose-Einstein statistics, while the lower, for 
Fermi-Dirac statistics.

The average densities for unbound particles and clusters are
$$
\rho_\al = \frac{1}{V} \int 
\lgl \; \psi_\al^\dgr(\br) \; \psi_\al(\br)\;\rgl \; d\br =
\frac{1}{V} \sum_k n_\al(\bk) \;
$$
\be
\label{6.71}
\rho_i = \frac{1}{V} \int 
\lgl \; \psi_i^\dgr(\br) \; \psi_i(\br)\;\rgl \; d\br =
\frac{1}{V} \sum_k n_i(\bk) \;  .
\ee
  
Employing the central-peak approximation \cite{Yukalov_170,Yukalov_460}
$$
\sum_p n_i(\bp) \; \overline\Phi_{ii}(\bk+\bp) \cong 
\overline\Phi_{ii}(\bk) \sum_p n_i(\bp) \; ,
$$
we get
\be
\label{6.72} 
 \frac{1}{V} \sum_p n_i(\bp) \; \overline\Phi_{ii}(\bk+\bp) \cong 
\rho_i \;  \overline\Phi_{ii}(\bk) \; .
\ee
Then the cluster spectrum becomes
\be
\label{6.73}
 \om_i(\bk) = \sqrt{k^2 + m_i^2} \; + U_i + \sum_j \Phi_{ij} \; \rho_j
\pm \rho_i \overline\Phi_{ii}(\bk) -\mu_i  
\ee
and the nonoperator remainder (\ref{6.69}) takes the form
\be
\label{6.74}
 R_{HF} = -\; \frac{V}{2} \sum_i \rho_i \; 
\left( \sum_j \Phi_{ij} \; \rho_j \pm \rho_i \Phi_{ii} \right) \;  .
\ee

For the grand thermodynamic potential 
$$
 \Om = - T\; \ln \; {\rm Tr}\; e^{-\bt H} \;  ,
$$
with the Hamiltonian
\be
\label{6.75}
 H = \hat H_1 - \sum_\al \mu_\al \; \hat N_\al + \hat H_z - 
\sum_i \mu_i \;\hat N_i + CV \;  ,
\ee
we have
$$
\Om = \mp T \sum_\al \zeta_\al \; 
\sum_k \ln \; \left[\; 1 \pm \frac{n_\al(\bk)}{\zeta_\al}\; \right] \; \mp
$$
\be
\label{6.76}
\mp \; T \sum_i \zeta_i \; 
\sum_k \ln \; \left[\; 1 \pm \frac{n_i(\bk)}{\zeta_i}\; \right] +
R_{HF} + CV \; .
\ee
Hence the pressure is
$$
P = -\; \frac{\Om}{V} = \pm T \sum_\al \zeta_\al \;
\int \ln \; \left[\; 1 \pm \frac{n_\al(\bk)}{\zeta_\al}\; \right] \; 
\frac{d\bk}{(2\pi)^3} \;  \pm  
$$
\be
\label{6.77}
\pm \; T \sum_i \zeta_i  
\int \ln \; \left[\; 1 \pm \frac{n_i(\bk)}{\zeta_i}\; \right] \; 
\frac{d\bk}{(2\pi)^3} - R_{HF} - CV \; .
\ee

From here, it is straightforward to find all system thermodynamic 
characteristics and the weights of all particles (\ref{6.44}). The quark-hadron 
mixture is stable. Deconfinement is found to be a sharp crossover, in good 
agreement with the QCD lattice simulations \cite{Soltz_427,Guenther_430,Aoki_461,
Karsch_462,Borsanyi_463,Bazavov_464,Philipsen_465}. In the case of a crossover, 
there is no a precise point where the phase transition occurs. The deconfinement 
temperature can be conditionally defined as the point where the derivatives of 
observables have a maximum, which defines the deconfinement temperature around 
170 MeV. Of course, considering different observables can result in slightly 
different deconfinement temperatures, which is the common situation for 
crossovers. Numerical simulations \cite{Yukalov_423,Yukalov_424} show that 
pion clusters survive till around $2T_c$. 

The deconfinement transition being a continuous crossover, becomes smoother 
with increasing baryon density. The deconfinement at a fixed low temperature 
and rising baryon density is due to the disintegration of hadrons into unbound 
quarks. When both the temperature and baryon density increase, the deconfinement 
is a result of the hadron disintegration combined with the generation from the 
vacuum of quarks and gluons.

Multiquark clusters, especially six-quark clusters, can exist at normal nuclear 
density. Since the quark clusters with even number of quarks behave as bosons, 
they can experience Bose-Einstein condensation. Thus at low temperature, six-quark 
clusters are shown \cite{Yukalov_423,Yukalov_424} to Bose-condense. Strictly 
speaking, the actual Bose condensation phase transition, requiring global gauge 
symmetry breaking, cannot occur in fireballs that are finite objects. However, 
even for finite quantum systems one can define {\it asymptotic symmetry breaking} 
\cite{Birman_466}, when the properties of a finite system asymptotically approach 
those of a system in the thermodynamic limit. 

The crossover nature of the deconfinement transition rules out those predictions 
that have been based on a sharp first-order phase transition. This concerns the 
interpretation of signals of the quark-gluon plasma in heavy-ion collisions and 
the hadronization scenario related to the evolution of the early universe after 
the Big Bang. More details can be found in Refs. \cite{Yukalov_423,Yukalov_424}.

\section{Conclusion}

There exist three types of mixed statistical systems that, for brevity, can be 
called macroscopic, mesoscopic, and microscopic mixtures. The first type is the 
well known Gibbs macroscopic mixture consisting of several spatially separated 
phases of macroscopic size. The description of this type of mixture has been 
developed quite a time ago, yet since Gibbs, and does not meet principal 
theoretical problems.

Mesoscopic mixture consists of a matrix of one phase inside which there appear 
the regions of mesoscopic size of another phase. The term "mesoscopic" implies 
that the characteristic linear size of the appearing germs of a competing phase 
is much larger than the mean interparticle distance, but much smaller than the 
linear system size. Usually, the characteristic size of a competing phase, at 
least in one direction, is of nanoscale. Generally, the embryos of a competing 
phase can vary their size in time, even can disappear and then appear again, 
thus fluctuating in space and time, because of which they are called heterophase 
fluctuations. The description of this type of mixture is of great difficulty,
since the system is nonuniform and, strictly speaking, nonequilibrium. The 
review presents the theory allowing for the statistical description of 
heterophase systems and illustrates this theory for several typical statistical 
models as well as for such exotic systems as superfluid crystals.    

In the third case, a microscopic mixture is considered consisting of several 
components of microscopic objects that can be termed particles. In general, 
this microscopic mixture can disintegrate into spatially separated macroscopic 
phases. The condition of stability of a microscopic mixture at finite temperature 
is derived. The theory of clustering matter composed of microscopic objects is 
developed. The approach is applied to quark-hadron mixture that can arise in 
heavy-ion and nuclear collisions. It is assumed to occur in the early universe 
at a time on the order of microseconds after Big Bang. And it is also supposed 
to exist in the interior of neutron stars.   

It is useful to mention that there can arise systems combining two types of 
mixtures. For example, a microscopic multicomponent mixture can exhibit inside 
itself the mesoscopic droplets of phases composed of a single kind of particles, 
thus presenting a multicomponent heterophase matter. As another kind of illustration, 
we can imagine that inside a mixed quark-hadron matter there can appear fluctuating 
mesoscopic regions of quark-gluon plasma. The investigation of such complex 
multicomponent heterophase systems is the goal for future studies.

\section*{Acknowledgements}

One of the authors (V.I.Y.) remembers with gratitude numerous highly useful 
discussions on the problems touched in this review he has had in former years 
with A.M. Baldin, J.L. Birman,  A.J. Coleman, and D. Ter Haar. We are grateful 
to our co-authors for collaboration.

\newpage

\newpage

\newpage

\newpage

\newpage


\newpage

\begin{thebibliography}{99}

\bibitem{Kubo_1}
R. Kubo,
{\it Thermodynamics} (North-Holland, Amsterdam, 1968).

\bibitem{Frenkel_2}
J. Frenkel,
{\it Kinetic Theory of Liquids} (Clarendon, Oxford, 1946).

\bibitem{Frenkel_3}
J.I. Frenkel,
{\it Statistical Physics} (Academy of Sciences, Moscow, 1948). 

\bibitem{Bakai_101}
A.S. Bakai,
{\it Polycluster Amorphous Solids} (Sinteks, Kharkov, 2013). 

\bibitem{Khait_4}
Y.L. Khait, 
Kinetic many-body theory of short-lived large energy fluctuations of small numbers 
of particles in solids and its applications,
Phys. Rep. {\bf 99}, 237--340 (1983).

\bibitem{Yukalov_5}
V.I. Yukalov,
Phase transitions and heterophase fluctuations,
Phys. Rep. {\bf 208}, 395--489 (1991).

\bibitem{Yukalov_6}
V.I. Yukalov,
Mesoscopic phase fluctuations: General phenomenon in condensed matter,
Int. J. Mod. Phys. B {\bf 17}, 2333--2358 (2003). 

\bibitem{Yukalov_7}
V.I. Yukalov,
Systems with symmetry breaking and restoration,
Symmetry, {\bf 2}, 40--68 (2010). 

\bibitem{Bakai_178}
A.S. Bakai,
Heterophase liquid states: Thermodynamics, structure, dynamics,
Condens. Matter Phys. {\bf 17}, 43701 (2014).

\bibitem{Bakai_179}
O. Bakai,
Viscous flow of glass-forming liquids and glasses,
Springer Proc. Phys. {\bf 171}, 103--137 (2015).

\bibitem{Kagan_261}
M.Y. Kagan, K.I. Kugel, and A.L. Rakhmanov,
Electronic phase separation: Recent progress in the old problem,
Phys. Rep. {\bf 916}, 1--105 (2021). 

\bibitem{Wollan_8}
E.O. Wollan and W.C. Koehler,
Neutron diffraction study of the magnetic properties of the series of 
Perovskite-type compounds,
Phys. Rev. {\bf 100}, 545--563 (1955).

\bibitem{Grazhdankina_9}
N.P. Grazhdankina, Y.S. Bersenev, I.V. Medvedeva, M.A. Novikov, and A.R. Koemets,
Investigation of mictomagnetic MnBi alloys,
J. Exp. Theor. Phys. {\bf 52}, 297--303 (1980). 

\bibitem{Nagaev_10}
E.L. Nagaev,
{\it Physics of Magnetic Semiconductors} (Mir, Moscow, 1983).

\bibitem{Suzuki_11}
H. Suzuki and J. Harada,
Local order and magnetic properties of disordered Au$_4$Mn and Cu$_3$Mn,
J. Magn. Magn. Mater. {\bf 31}, 69--70 (1983).

\bibitem{Moriya_12}
T. Moriya,
{\it Spin Fluctuations in Itinerant Electron Magnetism}
(Springer, Berlin, 1985).

\bibitem{Marder_13}
M. Marder, N. Papanicolaou, and G.C. Psaltakis,
Phase separation in a $t-J$ model,
Phys. Rev. B {\bf 41}, 6920--6932 (1990).

\bibitem{Allodi_14}
G. Allodi, R. De Renzi, G. Guidi, F. Licci, and M.W. Pieper,
Electronic phase separation in lantanum manganites: Evidence from $^{55}$Mn NMR,
Phys. Rev. B {\bf 56}, 6036--6046 (1997). 

\bibitem{Allodi_15}
G. Allodi, R. De Renzi, and G. Guidi,
$^{139}$La NMR in lantanum manganites: Indication of the presence of magnetic 
polarons from spectra and nuclear relaxation,
Phys. Rev. B {\bf 57}, 1024--1034 (1998).

\bibitem{Yi_16}
H. Yi and J. Yu,
Double-exchange model with background superexchange interaction: Phase diagram of
La$_{1-x}$A$_x$MnO$_3$ manganites,
Phys. Rev. B {\bf 58}, 11123--11126 (1998).

\bibitem{Yunoki_17}
S. Yunoki, J. Hu, A.L. Malvezzi, A. Moreo, N. Furukawa, and E. Dagotto,
Phase separation in electronic models for manganites,
Phys. Rev. Lett. {\bf 80}, 845--848 (1998). 

\bibitem{Yunoki_18}
S. Yunoki and A. Moreo,
Static and dynamic properties of the ferromagnetic Kondo model with direct 
antiferromagnetic coupling between the localized t$_{2g}$ electrons,
Phys. Rev. B {\bf 58}, 6403--6413 (1998).

\bibitem{Dagotto_19}
E. Dagotto, S. Yunoki, A.L. Malvezzi, A. Moreo, J. Hu, S. Capponi, D. Poilblanc,
and N. Furukawa,
Ferromagnetic Kondo model for manganites: Phase diagram, charge segregation, and
influence of quantum localized spins,
Phys. Rev. B {\bf 58}, 6414--6427 (1998).

\bibitem{Balagurov_20}
A.M. Balagurov, V.Y. Pomjakushin, D.V. Sheptyakov, V.L. Aksenov, N.A. Babushkina, 
L.M. Belova, A.N. Taldenkov, A.V. Inyushkin, P. Fischer, M. Gutmann, L. Keller, 
O.Y. Gorbenko, and A.R. Kaul,
Effect of oxigen isotope substitution on the magnetic structure of 
(La$_x$Pr$_{1-x}$)$_{0.7}$Ca$_{0.3}$MnO$_3$,
Phys. Rev. B {\bf 60}, 383--387 (1999).

\bibitem{Yi_21}
H. Yi, J. Yu, and S.I. Lee,
Coexistence of antiferromagnetic and ferromagnetic phase for ferromagnetic Kondo 
lattice model,
Eur. Phys. J. B {\bf 7}, 509--512 (1999).

\bibitem{Kapusta_22}
C. Kapusta, P.C. Riedi, M. Sikora, and M.R. Ibarra,
NMR probe of phase segregation in electron doped mixe valence manganites,
Phys. Rev. Lett. {\bf 84}, 4216 (2000).

\bibitem{Podzorov_23}
V. Podzorov, M. Uehara, M.E. Gershenson, T.Y. Koo, and S.W. Cheong,
Giant $1/f$ noise in perovskite manganites: Evidence of the percolation threshold,
Phys. Rev. B {\bf 61}, 3784--3787 (2000).

\bibitem{Krupicka_24}
S. Krupicka, Z. Jirak, J. Hejtmanek, M. Marysko, P. Novak, M.M. Savosta, and R. Sonntag,
Phase separation in structural and magnetic transitions in 
Pr$_{0.5}$Ca$_{0.5-x}$Sr$_x$MnO$_3$,
J. Phys. Condens. Matter {\bf 13}, 6813--6834 (2001). 

\bibitem{Sha_25}
H. Sha, F. Ye, P. Dai, J.A. Fernandez-Baca, and D. Mesa,
Signature of magnetic phase separation in the ground state of 
Pr$_{1-x}$Ca$_{x}$MnO$_3$,
Phys. Rev. B {\bf 78}, 052410 (2008). 

\bibitem{Pramanik_26}
A.K. Pramanik and A. Banerjee, 
Phase separation and the effect of quenched disorder in Pr$_{0.5}$Sr$_{0.5}$MnO$_3$,
J. Phys. Condens. Matter {\bf 20}, 275207 (2008).

\bibitem{Nagaev_27}
E.L. Nagaev,
Colossal-magnetoresistance materials: Manganites and conventional ferromagnetic 
semiconductors,
Phys. Rep. {\bf 346}, 387--531 (2001).

\bibitem{Sudheendra_28}
I. Sudheendra and C.N.R. Rao,
Electronic phase separation in the rare-earth manganites 
(La$_{1-x}$Ln$_x$)$_{0.7}$Ca$_{0.3}$Mno$_3$ (Ln = Nd, Gd, and Y), 
J. Phys. Condens. Matter {\bf 15}, 3029--3040 (2003). 

\bibitem{Shenoy_29}
V.B. Shenoy and C.N.R. Rao,
Electronic phase separation and other novel phenomena and properties exhibited 
by mixed-valent rare-earth manganites and related materials,
Phil. Trans. Roy. Soc. A {\bf 366}, 63--82 (2008).

\bibitem{Bertelsen_30}
U. Bertelsen, J. Knudsen, and H. Krogh,
M\"{o}ssbauer effect in FeF$_3$,
Phys. Status Solidi {\bf 22}, 59--64 (1967).

\bibitem{Yamamoto_31}
H. Yamamoto, T. Osaka, H. Watanabe, and M. Fukase,
M\"{o}ssbauer effect study of spin relaxation in CaFe$_2$O$_4$,
J. Phys. Soc. Jap. {\bf 24}, 275--279 (1968).

\bibitem{Eibschutz_32}
M. Eibsch\"{u}tz, S. Shtrikman, and D. Treves,
M\"{o}ssbauer studies of Fe$^{57}$ in orthoferrites,
Phys. Rev. {\bf 156}, 562--577 (1967).

\bibitem{Levinson_33}
L. Levinson, M. Luban, and S. Shtrikman,
M\"{o}ssbauer studies of Fe$^{57}$ near the Curie temperature,
Phys. Rev. {\bf 177}, 864--870 (1969).

\bibitem{Krivoglaz_34}
M.A. Krivoglaz,
Fluctuon states of electrons,
Phys. Usp. {\bf 16}, 856--877 (1974).

\bibitem{Nagaev_35}
E.L. Nagaev,
Ferromagnetic and antiferromagnetic semiconductors,
Phys. Usp. {\bf 18}, 863--892 (1975). 

\bibitem{Belov_36}
K.P. Belov,
{\it Rare-Earth Magnets and Their Applications} (Nauka, Moscow, 1980).

\bibitem{Belov_37}
K.P. Belov, Y.D. Tretyakov, I.V. Gordeev, L.I. Koroleva, and Y.A. Kesler,
{\it Magnetic Semiconductors -- Chalcogenide Spinels} 
(Moscow State University, Moscow, 1981).

\bibitem{Reissner_36}
M. Reissner, W. Steiner, J. Kappler, P. Bauer, and M. Besnus,
Magnetic behaviour of Y(Fe$_x$Al$_{1-x}$)$_2$ alloys,
J. Phys. F {\bf 14}, 1249--1260 (1984). 

\bibitem{Kumeishin_37}
V.F. Kumeishin and O.A. Ivanov,
Investigation of relaxation processes in nickel near the Curie temperature by means of
nuclear gamma-resonance, 
Phys. Met. Metallogr. {\bf 40}, 1295--1299 (1975).

\bibitem{Baryakhtar_38}
V.G. Baryakhtar, I.M. Vitebsky, and D.A. Yablonsky,
Theory of creation of nuclei at magnetic first order phase tarnsition between paramagnetic 
and magneto-ordered phases,
Phys. Solid State. {\bf 19}, 347--352 (1977).

\bibitem{Goldman_39}
M. Goldman,
Nuclear dipolar magnetic ordering,
Phys. Rep. {\bf 32}, 1--47 (1977).

\bibitem{Reimann_40}
H. Reimann, H. Hagen, F. Waldner, and H. Arend,
Observation of excitation of the antiferromagnetic mode in the paramagnetic state
of (C$_2$H$_5$NH$_3$)$_2$CuCl$_4$, 
Phys. Rev. Lett. {\bf 40}, 1344--1346 (1978).

\bibitem{Bhargava_41}
S. Bhargava and N. Zeeman,
M\"{o}ssbauer study of Ni$_{0.25}$Zn$_{0.75}$Fe$_2$O$_4$: spin fluctuations,
Phys. Rev. B {\bf 21}, 1717--1724 (1980).

\bibitem{Uen_42}
T.M. Uen and P.K. Tseng,
M\"{o}ssbauer-effect studies on the magnetic properties of the Ni-Zn-ferrite system,
Phys. Rev. B {\bf 25}, 1848--1859 (1982).

\bibitem{Srivastava_43}
J. Srivastava, K. Muraleedharan, and R. Vijayaraghavan,
On anomalous M\"{o}ssbauer spectra in spinel ferrites,
Phys. Lett. A {\bf 104}, 482--486 (1984).

\bibitem{Halg_44}
B. H\"{a}lg, A. Furrer, and O. Vogt,
Coexistence of different short-range-ordered spin fluctuations in Ce$_{1-x}$(LaY)$_x$Sb,
Phys. Rev. Lett. {\bf 54}, 1388--1391 (1985).

\bibitem{Lynn_45}
J.W. Lynn, 
Temperature dependence of the magnetic excitations in iron,
Phys. Rev. B {\bf 11}, 2624--2637 (1975).

\bibitem{Liu_46}
S.H. Liu,
Magnetic excitations above the critical temperature,
Phys. Rev. B {\bf 13}, 2979--2985 (1976). 

\bibitem{Lynn_47}
J.W. Lynn and H.A. Mook,
Temperature dependence of the dynamic susceptibility of nickel,
Phys. Rev. B {\bf 23}, 198--206 (1981). 
 
\bibitem{Cable_48}
J.W. Cable, R.M. Nicklow, and N. Wakabayashi,
Temperature dependence of the magnetic excitations in gadolinium,
Phys. Rev. B {\bf 32}, 1710--1719 (1986).

\bibitem{Lynn_49}
J.W. Lynn and H.A. Mook,
Nature of the magnetic excitations above $T_c$ in Ni anf Fe,
J. Magn. Magn. Mater. {\bf 54}, 1169--1170 (1986). 

\bibitem{Cable_50}
J.W. Cable and R.M. Nicklow,
Spin dynamics of Gd at high temperatures,
Phys. Rev. B {\bf 39}, 11732--11741 (1989).

\bibitem{Goto_51}
T. Goto, T. Sakakabara, and M. Yamaguchi,
Coexistence of nonmagnetic and ferromagnetic Co in Y$_2$Co$_7$ and YCo$_3$ hydrides,
J. Magn. Magn. Mater. {\bf 54}, 1085--1086 (1986).

\bibitem{Shinogi_52}
A. Shinogi, T. Saito, and K. Endo,
Coexistence of nonmagnetic and magnetic Co in cubic Laves phase compounds 
Lu(Co$_{1-x}$Al$_x$)$_2$,
J. Phys. Soc. Jap. {\bf 56}, 2633--2636 (1987).

\bibitem{Jaime_53}
M. Jaime, P. Lin, S.H. Chun, M.B. Salamon, P. Dorsey, and M. Rubinshtein,
Coexistence of localized and itinerant carriers near $T_C$ in calcium-doped manganites,
Phys. Rev. B {\bf 60}, 1028--1032 (1999).

\bibitem{Merithew_54}
R.D. Merithew, M.B. Weissman, F.M. Hess, P. Spradling, E.R. Nowak, J. O'Donnell,
J.M. Ekstein, Y. Tokura, and Y. Tomioka,
Mesoscopic thermodynamics of an inhomogeneous colossal-magnetoresistive phase,
Phys. Rev. Lett. {\bf 84}, 3442--3445 (2000).

\bibitem{Baio_55}
J. Baio, G. Barucca, R. Caciuffo, D. Rinaldi, J. Mira, J. Rivas, M.A. Senaris-Rodriguez,
and D. Fiorani,
Phase separation, thermal history and magnetic behavior of Sr dopped LaCoO$_3$,
J. Phys. Condens. Matter {\bf 12}, 9761--9770 (2000).  

\bibitem{Sun_56}
J.R. Sun, B.G. Shen, J. Gao, Y. Fei, and Y.P. Nie,
Presence of a paramagnetic phase well below the ferromagnetic onset in 
La$_{0.67-x}$Bi$_x$Ca$_{0.33}$MnO$_3$,
Eur. Phys. Lett. {\bf 62}, 732--738 (2003).

\bibitem{Batko_68}
I. Batko, M. Batkova, V.H. Tran, U. Keiderling, and V.B. Filipov,
Evidence for magnetic phase separation in colossal magnetoresistence compound 
EuB$_{5.99}$C$_{0.01}$,
Solid State Commun. {\bf 190}, 23--27 (2014).

\bibitem{Gabay_57}
M. Gabay and G. Toulouse,
Coexistence of spin-glass and ferromagnetic orderings,
Phys. Rev. Lett. {\bf 47}, 201--203 (1981).

\bibitem{Cheung_58}
T.D. Cheung and J.S. Kouvel,
Spin-glass-like distribution of interaction fields in Pd-Ni alloys,
Phys. Rev. B {\bf 28}, 3831--3837 (1983).

\bibitem{Abdul_59}
W. Abdul-Razzaq and J.S. Kouvel,
Spin glassiness and ferromagnetism in disordered Ni-Mn, 
J. Appl. Phys. {\bf 55}, 1623--1627 (1984).

\bibitem{Koroleva_60}
L.I. Koroleva and A.I. Kuzminich,
Spin-glass state and its suppression by indirect exchange through current carriers 
in the system of solid solutions,
J. Exp. Theor. Phys. {\bf 84}, 1882--1886 (1983).

\bibitem{Matsuda_61}
M. Matsuda, M. Fujita, K. Yamado, R. Birgeneau, Y. Endoh, and G. Shirane,
Electronic phase separation in lightly doped La$_{2-x}$Sr$_x$CuO$_4$,
Phys. Rev. B {\bf 65}, 134515 (2002). 

\bibitem{Li_62}
R.W. Li, J.R. Sun, Q.A. Li, S.Y. Zhang, and B.G. Shen,
Structure, magnetic properties and phase separation of 
Nd$_{0.5}$Ca$_{0.5}$Mn$_{1-x}$Ga$_x$O$_3$ $(0 \leq x \leq 0.1)$,
J. Magn. Magn. Mater. {\bf 265}, 248--256 (2003). 

\bibitem{Emmerich_68}
K. Emmerich,  E. Lippelt, R. Neuhaus, H. Pinkvos, C. Schwink, F.N. Gygax, 
A. Hintermann, A. Schenck, W. Studer, and A.J. van der Wal,
Experimental evidence for spatial inhomogeneous spin freezing in CuMn,
Phys. Rev. B {\bf 31}, 7226--7232 (1988). 

\bibitem{Belov_63}
K.P. Belov, A.K. Zvezdin, A.M. Kadomtseva, and R.Z. Levitin,
Spin-reorientation transitions in rare-earth magnets,
Phys. Usp. {\bf 19}, 574--596 (1976). 

\bibitem{Miwa_64}
H. Miwa,
Phase diagram of random mixtures of antiferromagnets with competing anisotropy,
J. Magn. Magn, Mater. {\bf 34}, 1455--1456 (1983).

\bibitem{Balestrino_65}
G. Balestrino, F. Scarinci, and A. Tucciarone,
Magnetic phase transitions in YIG(Ru$^{4+}$),
J. Magn. Magn. Mater. {\bf 44}, 249--253 (1984). 

\bibitem{Long_66}
M.W. Long,
Spontaneous local symmetry breaking in frustrated antiferromagnets: Non-collinear 
impurities in MnCu alloys,
J. Phys. Condens. Matter {\bf 2}, 5383--5402 (1990).

\bibitem{Belyaeva_67}
A.I. Belyaeva, Y.N. Stelmakov, and V.A. Potakova,
Visual study of spin reorientation phenomenon in DyFeO$_3$ near the Morin temperature,
Phys. Solid State {\bf 19}, 3124--3129 (1977).

\bibitem{Cook_69}
H.E. Cook, 
Droplet model for central phonon peaks,
Phys. Rev. B {\bf 15}, 1477--1488 (1977).

\bibitem{Rigamonti_70}
A. Rigamonti and G. Brookeman, 
$^{35}$Cl nuclear quadrupole resonance study of the ferroelectric transition in HCl and in 
the mixed crystals HCl-DCl,
Phys. Rev. B {\bf 21}, 2681--2694 (1980).

\bibitem{Brookeman_71}
J. Brookeman and A. Rigamonti, 
Pretransitional clusters and heterophase fluctuations at first-order phase transitions 
in crystals,
Phys. Rev. B {\bf 24}, 4925--4930 (1981).

\bibitem{Bruce_72}
A.D. Bruce and R.A. Cowley, 
{\it Structural Phase Transitions} 
(Taylor and Francis, London, 1981).

\bibitem{Gordon_73}
A. Gordon and J. Genossar,
Precursor order clusters at ferroelectric phase transitions,
Physica B, {\bf 125}, 53--62 (1984).

\bibitem{Gordon_74}
A. Gordon,
Heterophase fluctuations in ferroelectrics,
J. Phys. C {\bf 20}, L111--L114 (1987).

\bibitem{Muller_75}
K.A. M\"{u}ller and T. Thomas, Eds., 
{\it Structural Phase Transitions} (Springer, Berlin, 1991).

\bibitem{Bhide_76}
V.G. Bhide and M.S. Multani,
M\"{o}ssbauer effect in ferroelectric BaTiO$_3$,
Phys. Rev. A {\bf 139}, 1983--1990 (1965).

\bibitem{Bokov_77}
V.A. Bokov, V.P. Romanov, and V.V. Chekin, 
The M\"{o}ssbauer effect on Sn$^{119}$ in case of ferroelectric phase transition in
Ba(Ti$_{0.8}$Sn$_{0.2}$)O$_3$ solid solution,
Phys. Solid State {\bf 7}, 1886--1891 (1965).

\bibitem{Gleason_78}
T.G. Gleason and J.C. Walker,
M\"{o}ssbauer effect study of the ferroelectric transitions in Potassium Ferrocyanide
Trihydrate and Ferric Ammonium Sulfate Dodecahydrate,
Phys. Rev. B {\bf 188}, 893--897 (1969).

\bibitem{Jain_79}
A.P. Jain, S.N. Shringi, and M.L. Sharma,
Temperature-dependent optical mode in antiferrpelectric PbZrO$_3$ by the M\"{o}ssbauer 
effect,
Phys. Rev. B {\bf 2}, 2756--2759 (1970).

\bibitem{Canner_80}
J. Canner, C. Yagnik, R. Gerson, and W. James,
M\"{o}ssbauer-effect studies of the phase transition in antiferroelectric PbZrO$_3$ and
ferroelectric PbTi$_{0.02}$Zr$_{0.2}$O$_3$,
J. Appl. Phys. {\bf 42}, 4708--4712 (1971).  

\bibitem{Bhide_81}
V.G. Bhide and M.S. Hegde,
M\"{o}ssbauer effect for Fe$^{57}$ in ferroelectric Lead Titanate,
Phys. Rev. B {\bf 5} 3488--3498 (1972).

\bibitem{Suzdalev_82}
I.P. Suzdalev,
{\it Dynamical Effects in Gamma-Resonance Spectroscopy}
(Atomizdat, Moscow, 1979). 

\bibitem{Nikolov_83}
O. Nikolov, T. Ruskov, T. Tomov, A. Kadomtseva, I. Krinetskii, and M. Lukina,
Gallium substitution for Fe$^{3+}$ ions and its influence on the phase transitions in
dysprosium orthoferrite,
J. Magn. Magn. Mater. {\bf 44}, 181--186 (1984). 

\bibitem{Dezsi_84}
I. Dezsi, L. Keszthelyi, B. Molnar, and L. Pocs,
Mössbauer effect study of phase transition in ice,
Phys. Lett. {\bf 18}, 28--29 (1965).

\bibitem{Tsurin_85}
V.A. Tsurin, N.P. Filippova, A.M. Sorkin, L.Y. Kobelev, L.L. Nugaeva, and A.P. Stepanov, 
Mossbauer-spectrum temperature-dependence anomalies of YBa$_2$Cu$_3$O$_{9-y}$,
JETP Lett. {\bf 46}, 459--461 (1987).

\bibitem{Cherepanov_86}
V.M. Cherepanov, M.A. Chuev, S.S. Yakimov, V.Y. Goncharov, and S.A. Smirnov, 
Anomalous behavior of the temperature dependence of the parameters of the Mossbauer 
spectra of the superconducting ceramics YBa$_2$Cu$_{2.95}$Fe$_{0.05}$O$_{7-y}$,
JETP Lett. {\bf 47}, 424--427 (1988).

\bibitem{Cherepanov_87}
V.M. Cherepanov, M.A. Chuev, S.S. Yakimov, V.Y. Goncharov, S.A. Smirnov, and A.A. Bush, 
Anomalies in the temperature dependence of the Mossbauer-effect probability for impurity 
tin nuclei in a $1-2-3$ superconducting ceramic,
JETP Lett. {\bf 49}, 435--438 (1989).

\bibitem{Cherepanov_88}
V.M. Cherepanov, M.A. Chuev, S.S. Yakimov, and V.Y. Goncharov,
Phonon spectrum softening near temperatures of superconducting and structural transitions 
in HTSC $1-2-3$ Ceramics,
Hyperf. Interact. {\bf 65}, 1257--1260 (1990).

\bibitem{Egorushkin_89}
V.E. Egorushkin, A.I. Lotkov, and S.V. Anokhin, 
Physical nature of the anomalies in the temperature dependence of the probability of the 
Mössbauer effect near phase transitions,
Russ. Phys. J. {\bf 34}, 997--999 (1991).

\bibitem{Nowik_90}
I. Nowik, E. Bauminger, S. Cohen, and S. Ofer,
Spectral shapes of M\"{o}ssbauer absorption and incoherent neutron scattering from 
harmonically bound nuclei in Brownian motion: Applications to macromolecular systems,
Phys. Rev. A {\bf 31}, 2291--2299 (1985).

\bibitem{Eselson_91}
B.N. Eselson, V.N. Grigoriev, V.G. Ivantsov, E.Y. Rudavsky, D.G. Sanikidze, 
and I.A. Serbin,
{\it Solutions of Quantum Liquids He$^3$-He$^4$} (Nauka, Moscow, 1973).
   
\bibitem{Eselson_92}
B.N. Eselson, V.G. Ivantsov, V.A. Koval, E.Y. Rudavsky, and I.A. Serbin,
{\it Properties of Liquid and Solid Helium} (Naukova Dumka, Kiev, 1982). 

\bibitem{Kosilov_93}
A.T. Kosilov,
Dissipative properties of materials with thermoelastic martensite conversion,
Russ. Phys. J. {\bf 28}, 380--389 (1985). 

\bibitem{Pushin_94}
V.G. Pushin, V.V. Kondrat'ev, and V.N. Khachin,
Pretransformation phenomena and martensitic transformations in titanium nickelide alloys,
Russ. Phys. J. {\bf 28}, 341--355 (1985).

\bibitem{Gibaud_95}
A. Gibaud, A. Bulou, A. Le Bail, J. Nouet, and C. Zeyen,
A premartensitic phase in KAlF$_4$: Neutron and x-ray scattering evidence,
J. Phys. (Paris) {\bf 48}, 1521--1532 (1987).

\bibitem{Van_96}
G. Van Tendeloo, D. Broddin, C. Leroux, D. Schryvers, L. Tanner, J. Van Landuyt, and 
S. Amelindex,
Electron microscopy observations of pretransitional effects in alloys,
Phase Trans. {\bf 27}, 61--71 (1990).  

\bibitem{Emery_97}
V.J. Emery, S.A. Kivelson, and H.Q. Lin,
Phase separation in the t-J model,
Phys. Rev. Lett. {\bf 64}, 475--478 (1990).

\bibitem{Morris_98}
J.R. Morris and R.J. Gooding,
Exactly solvable heterophase fluctuations at a vibrational-entropy-driven first-order
transition,
Phys. Rev. Rett. {\bf 65}, 1769--1772 (1990). 

\bibitem{Zhu_99}
X. Zhu, H. Zabel, I. Robinson, E. Vieg, J. Dura, and C. Flynn,
Surface-induced heterophase fluctuations,
Phys. Rev. Lett. {\bf 65}, 2692--2695 (1990).

\bibitem{Gibaud_100}
A. Gibaud, S. Shapiro, J. Nouet, and H. Yoo,
Phase diagram of KMn$_{1-x}$Ca$_x$F$_3$ ($x < 0.05$) determined by high-resolution 
x-ray scattering, 
Phys. Rev. B {\bf 44}, 2437--2443 (1991).

\bibitem{Kizel_102}
V.A. Kizel and S.I. Panin,
Fluctuation phenomena under melting of holesteric mesophase of liquid crystals,
J. Exp. Theor. Phys. {\bf 66}, 105--112 (1987). 

\bibitem{Volmer_103}
M. Volmer, 
{\it Kinetik der Phasenbildung} 
(Steinkopff, Dresden, 1939).

\bibitem{Akulichev_104}
V.A. Akulichev,
{\it Cavitation in Cryogenic and Boiling Liquids} 
(Nauka, Moscow, 1978).

\bibitem{Kertez_105}
J. Kertez,
Existence of weak singularities when going around the liquid-gas critical point,
Physica A {\bf 161}, 58--62 (1989).

\bibitem{Bakai_180}
O. Bakai,
Mesoscopic equation of state of the heterophase fluid and its application
to description of the liquid-gas asymmetry,
J. Mol. Liq. {\bf 235}, 135--148 (2017).

\bibitem{Bakai_106}
O. Bakai, M. Bratchenko, and S. Dyuldya,
Three-state mesoscopic model of a heterophase fluid in application to the gas-liquid 
and dielectric-semiconductor-metal transformations in expanded mercury,
J. Mol. Liq. {\bf 260}, 245--260 (2018).

\bibitem{Bakai_181}
O. Bakai, M. Bratchenko, and S. Dyuldya,
On the singularity of the liquid-gas coexistence curve diameter,
Ukr. J. Phys. {\bf 65}, 802--809 (2020). 

\bibitem{Bakai_182}
O. Bakai,
The van der Waals idea of pseudo associations and the critical compressibility factor,
Condens. Matter Phys. {\bf 23}, 13603 (2020). 

\bibitem{Rice_107}
T.M. Rice, J.C. Hensel, T.G. Phillips, and G.A. Thomas,
{\it Electron-Hole Liquid in Semiconductors} (Academic, New York, 1977). 

\bibitem{Exerowa_108}
D. Exerowa and D. Kashchiev,
Hole-mediated stability and permeabiity of bilayers,
Contemp. Phys. {\bf 27}, 429--461 (1986).

\bibitem{Kelton_109}
K. Kelton and A.L. Greer, 
{\it Nucleation in Condensed Matter: Applications in Materials and Biology} 
(Elsevier, Amsterdam, 2010).

\bibitem{Dubrovskii_110}
V.G. Dubrovskii, 
{\it Nucleation Theory and Growth of Nanostructures} (Springer, Berlin, 2014).

\bibitem{Rontgen_111}
W.C. R\"{o}ntgen,
Ueber die Constitution des fl\"{u}ssigen Wassers,
Ann. Phys. (Berlin) {\bf 45}, 91--97 (1892).

\bibitem{Brody_112}
V.E. Brody,
On the theory of specific heat in the vicinity of transition points,
Phys. Zeit. {\bf 23}, 197--202 (1922).

\bibitem{Bernal_113}
J.D. Bernal and R.H. Fowler,
A theory of water and ionic solution with particular reference to Hydrogen and 
Hydroxyl ions,
J. Chem. Phys. {\bf 1}, 515--548 (1933).

\bibitem{Bartenev_114}
G.M. Bartenev,
On the theory of anomalous phenomena in the vicinity of melting points,
Russ. J. Phys. Chem. {\bf 22}, 587--592 (1948).
 
\bibitem{Bartenev_115}
G.M. Bartenev,
Thermal expansion near melting points,
Russ. J. Phys. Chem. {\bf 23}, 1075--1079 (1948).
 
\bibitem{Bartenev_116}
G.M. Bartenev,
Specific heat near melting points,
Russ. J. Phys. Chem. {\bf 24}, 1016--1020 (1950).

\bibitem{Bartenev_117}
G.M. Bartenev,
Elasticity of crystals near melting points,
Russ. J. Phys. Chem. {\bf 24}, 1437--1441 (1950).

\bibitem{Fisher_118}
I.Z. Fisher,
{\it Statistical Theory of Liquids} (Chicago University, Chicago, 1964).

\bibitem{Samoilov_119}
O.Y. Samoilov,
{\it Structure of Aqueous Electrolyte Solutions and the Hydration of Ions}
(Consultants Bureau, New York, 1965). 

\bibitem{Dass_120}
N. Dass and N. Gilra,
Refractive index of liquids on supercooling water,
J. Phys. Soc. Jap. {\bf 21}, 2039--2042 (1966).

\bibitem{Gilra_121}
N.K. Gilra,
Precrystallization theory applied to ultrasonic velocity in supercooled water,
J. Phys. Soc. Jap. {\bf 23}, 1431--1431 (1967).

\bibitem{Gilra_122}
N. Gilra and N. Dass,
Precrystallization theory applied to surface-free energy of the water-ice interface,
J. Phys. Soc. Jap. {\bf 24}, 910--912 (1968).

\bibitem{Gilra_123}
N.K. Gilra,
Homogeneous nucleation temperature of supercooled water,
Phys. Lett. A {\bf 28}, 51--52 (1968). 

\bibitem{Fletcher_124}
N.H. Fletcher,
{\it The Chemical Physics of Ice} (Cambridge University, Cambridge, 1970).

\bibitem{Skripov_125}
V.P. Skripov and V.P. Koverda, 
{\it Spontaneous Crystallization of Supercooled Liquids} (Nauka, Moscow, 1984).

\bibitem{Hayes_126}
W. Hayes,
Premelting,
Contemp. Phys. {\bf 27}, 519--532 (1986).

\bibitem{Gabuda_127}
S.P. Gabuda and A.G. Lundin,
{\it Internal Mobility of Solids} (Nauka, Novosibirsk, 1986).

\bibitem{Rasmussen_128}
D.H. Rasmussen and A.P. MacKenzie,
Cluctering in supercooled water,
J. Chem. Phys. {\bf 59}, 5003--5013 (1973).

\bibitem{Speedy_129}
R.J. Speedy and C.A. Angell,
Isothermal compressibility of supercooled water and evidence for a thermodynamic 
singularity at $-45$ C,
J. Chem. Phys. {\bf 65}, 851--858 (1976).

\bibitem{Trinh_130}
E. Trinh and R. Apfel,
The sound velocity in metastable liquid water under atmospheric pressure,
J. Chem. Phys. {\bf 69}, 4245--4251 (1978).

\bibitem{Arrigo_131}
G. D'Arrigo,
Sound propagation in moderately supercooled liquids: A comparison with water,
J. Chem. Phys. {\bf 75}, 921--928 (1981).

\bibitem{Arrigo_132}
G. D'Arrigo,
Heterophase fluctuations and thermodynamic properties in supercooled water,
Nuovo Cimento B {\bf 61}, 123--140 (1981).

\bibitem{Ubbelohde_133}
A.R. Ubbelohde,
{\it The Molten State of Matter} (Wiley, New York, 1978).

\bibitem{Ziman_134}
J.M. Ziman,
{\it Models of Disorder} (Camridge University, Cambridge, 1979).

\bibitem{Wadati_135}
M. Wadati,
Quantum field theory of crystals and extended objects,
Phys. Rep. {\bf 50}, 87--155 (1979).

\bibitem{Borsa_136}
F. Borsa, M. Corti, A. Rigamonti, and S. Torre,
Diffusion and phononlike excitations in intercalated graphite from $^{133}$Cs NMR
and relaxation,
Phys. Rev. Lett. {\bf 53}, 2102--2105 (1984). 

\bibitem{Steffen_137}
B. Steffen and R. Hosemann,
Paracrystalline microdomains in monoatomic liquids: Three-dimensional structure of 
microdomains in liquid lead,
Phys. Rev. B {\bf 13}, 3232--3238 (1976).

\bibitem{Anisimov_138}
M.A. Anisimov,
{\it Critical Phenomena in Liquids and Liquid Crystals} 
(Gordon and Breach, Philadelphia, 1991).

\bibitem{Bilgram_139}
J.H. Bilgram,
Dynamics at the liquid transition: Experiments at the freezing point,
Phys. Rep. {\bf 153}, 1--89 (1987).

\bibitem{Fontana_140}
M. Fontana, F. Brehat, and B. Wyncke,
A central peak as a premelting feature in NaNO$_3$ spectra,
J. Phys. Condens. Matter {\bf 2}, 9125--9132 (1990).

\bibitem{Rousset_141}
J. Rousset, E. Duval, and A Boukenter,
Dynamical structure of water: Low-frequency Raman scattering from a disordered network 
and aggregates,
J. Chem. Phys. {\bf 92}, 2150--2154 (1990).

\bibitem{Barker_142}
J. Barker and D. Henderson,
What is liquid? Understanding the states of matter,
Rev. Mod. Phys. {\bf 48}, 587--672 (1976).

\bibitem{Lagarkov_143}
A.N. Lagarkov and V.M. Sergeev,
Molecular dynamics method in statistical physics,
Phys. Usp. {\bf 21}, 566--588 (1978). 

\bibitem{Choquard_144}
P. Choquard and J. Clerouin,
Cooperative phenomena below melting of the one-component two-dimensional plasma,
Phys. Rev. Lett. {\bf 50}, 2086--2089 (1983). 

\bibitem{Hiwatari_145}
Y. Hiwatary,
Free volumes and liquidlike clusters in soft-core liquids and glasses,
J. Chem. Phys. {\bf 76}, 5502--5507 (1982).

\bibitem{Stishov_146}
S.M. Stishov,
Entropy, disorder, melting,
Phys. Usp. {\bf 31}, 52--67 (1988). 

\bibitem{Zollweg_147}
J.A. Zollweg and G.V. Chester,
Melting in two dimensions,
Phys. Rev. B {\bf 46}, 11186--11189 (1992). 

\bibitem{Kosterlitz_148}
J.M. Kosterlitz and D.J. Thouless,
Two-dimensional physics,
Prog. Low Temp. Phys. B {\bf 7}, 373--433 (1978).

\bibitem{Barber_149}
M.N. Barber,
Phase transitions in two dimensions,
Phys. Rep. {\bf 59}, 375--409 (1980). 

\bibitem{Stern_150}
E.A. Stern and K. Zhang,
Local premelting about impurities,
Phys. Rev. Lett. {\bf 60}, 1872--1875 (1988).

\bibitem{Schechter_151}
H. Schechter, E.A. Stern, Y. Yacoby, R. Brener, and Z. Zhang,
Anomalous local hopping of Sn impurities in lead,
Phys. Rev. Lett. {\bf 63}, 1400--1403 (1989).

\bibitem{Stern_152}
E.A. Stern, P. Livins, and Z. Zhang,
Thermal vibration and melting from a local perspective,
Phys. Rev. B {\bf 43}, 8850--8860 (1991). 

\bibitem{Cohen_153}
M. Cohen and J. Jortner,
Inhomogeneous transport regime in disordered materials,
Phys. Rev. Lett. {\bf 30}, 699--702 (1973). 

\bibitem{Cohen_154}
M. Cohen and J. Jortner,
Conduction regimes in expanded liquid mercury,
Phys. Rev. A {\bf 10}, 978--996 (1974).

\bibitem{Jortner_155}
J. Jortner and M. Cohen,
Metal-nonmetal transition in metal-ammonia solutions,
Phys. Rev. B {\bf 13}, 1548--1568 (1976).

\bibitem{Cohen_156}
M. Cohen and J. Jortner,
Electronic structure and transport in liquid Te,
Phys. Rev. B {\bf 13}, 5255--5260 (1976).

\bibitem{Tsuchiya_157}
Y. Tsuchya, S. Takeda, S. Tamaki, Y. Waseda, and E. Seymour,
Evidence for structural inhomgeneity in liquid In$_2$Te$_3$,
J. Phys. C {\bf 15}, 2561--2575 (1982). 
  
\bibitem{Tsuchiya_158}
Y. Tsuchiya and E. Seymour, 
Thermodynamic properties of the selenium-tellurium system,
J. Phys. C {\bf 15}, L687--L695 (1982).

\bibitem{Tsuchiya_159}
Y. Tsuchya, S. Takeda, S. Tamaki, and E. Seymour,
Structural inhomogeneity and valence fluctuations in IIIb-Te liquid semiconductors,
J. Phys. C {\bf 15}, 6497--6512 (1982).

\bibitem{Chandra_160}
D. Chandra,
Anomalous volume expansion in Hg$_{1-x}$Cd$_x$Te melts: An analysis employing the 
inhomogeneous structure model,
Phys. Rev. B {\bf 31}, 7206--7212 (1985). 

\bibitem{Takeda_161}
S. Takeda, H. Okazaki, and S. Tamaki,
Shottky-type specific heat in liquid Se-Te alloys,
Phys. Rev. B {\bf 31}, 7452--7454 (1985).

\bibitem{Regel_162}
A.R. Regel, V.M. Galzov, and S.G. Kim,
Acoustic study of structural variations under the heating of semiconductor and 
semimetal melts,
Phys. Techn. Semicond. {\bf 20}, 1353--1357 (1986).

\bibitem{Loseva_163}
G.V. Loseva, S.G. Ovchinnikov, and G.A. Petrakovsky,
{\it Metal-Dielectric Transition in Sulfides of 3d-Metals}
(Nauka, Novosibirsk, 1983).

\bibitem{Lopez_164}
G.E. Lopez and D.L. Freeman,
A study of low temperature heat capacity anomalies in bimetallic alloy clusters 
using J-walking Monte Carlo methods,
J. Chem. Phys. {\bf 98}, 1428--1435 (1993).

\bibitem{Gao_165}
P. Gao, T.A. Tyson, Z. Liu, M.A. DeLeon, and C. Dubourdieu,
Optical evidence for mixed phase behavior in manganites films,
Phys. Rev. B {\bf 78}, 220404 (2008).

\bibitem{Gorter_166}
C.J. Gorter,
The two fluid model for superconductors and Helium II,
Prog. Low Temp. Phys. {\bf 1}, 1--16 (1955). 

\bibitem{Schrieffer_167}
J.R. Schrieffer,
{\it Theory of Superconductivity} (Benjamin, New York, 1964). 

\bibitem{Lynton_168}
E.A. Lynton,
{\it Superconductivity} (Methuen, London, 1969).

\bibitem{Yukalov_169}
V.I. Yukalov,
Basics of Bose-Einstein condensation,
Phys. Part. Nucl. {\bf 42}, 460--513 (2011).

\bibitem{Yukalov_170}
V.I. Yukalov,
Theory of cold atoms: Bose-Einstein statistics,
Laser Phys. {\bf 26}, 062001 (2016).

\bibitem{Hizhnyakov_171}
V. Hizhnyakov and E. Sigmund,
High-T$_c$ superconductivity induced by ferromagnetic clustering,
Physica C {\bf 156}, 655--666 (1988). 

\bibitem{Phillips_172}
J.C. Phillips, 
{\it Physics of High-T$_c$ Superconductors} 
(Academic, Boston, 1989).

\bibitem{Hizhnyakov_173}
V. Hizhnyakov, E. Sigmund, and M. Schneider,
Magnetic interactions and dynamics of holes in CuO$_2$ planes of high-T$_c$ 
superconducting materials,
Phys. Rev. B {\bf 44}, 795--800 (1991).  

\bibitem{Benedek_174}
G. Benedek and K.A. M\"{u}ller, Eds.,
{\it Phase Separation in Cuprate Superconductors} 
(World Scientific, Singapore, 1992).

\bibitem{Sigmund_175}
E. Sigmund and K.A. M\"{u}ller, Eds.,
{\it Phase Separation in Cuprate Superconductors} 
(Springer, Berlin, 1994).

\bibitem{Kivelson_176}
S.A. Kivelson, I.P. Bindloss, E. Fradkin, V. Oganesyan, J.M. Tranquada, A. Kapitulnik, 
and C. Howald,
How to detect fluctuating stripes in the high-temperature superconductors,
Rev. Mod. Phys. {\bf 75}, 1201--1242 (2003).

\bibitem{Gorkov_177}
L.P. Gorkov and A.V. Sokol,
Phase stratification of an electron liquid in the new superconductors,
JETP Letters {\bf 46}, 420--423 (1987).

\bibitem{Spielberg_183}
J.I. Spielberg and E. Gelerinter,
Further studies on the molecular dynamics of the glass transition and the glass state 
using EPR probes,
Phys. Rev. B {\bf 30}, 2319--2323 (1984). 

\bibitem{Klinger_184}
M.I. Klinger,
Soft atomic configurations, medium-range order and universal properties of glasses,
Comments Condens. Matter Phys. {\bf 16}, 137--152 (1992). 

\bibitem{Bakai_185}
A.S. Bakai and E.W. Fischer,
Nature of long-range correlations of density fluctuations in glass-forming liquids,
J. Chem. Phys. {\bf 120}, 5235--5252 (2004).

\bibitem{Bakai_186}
A.S. Bakai,
On low-temperature polyamorphous transformations,
Low Temp. Phys. {\bf 32}, 868--876 (2006).

\bibitem{Lazarev_187}
N. Lazarev and A. Bakai,
Theoretical strength and homogeneous sliding in metallic glass: Exactly solvable model,
J. Mech. Behav. Mater. {\bf 22}, 119--128 (2013).

\bibitem{Ivlev_188}
B.I. Ivlev and N.B. Kopnin,
Theory of current states in narrow superconductor channels,
Phys. Usp. {\bf 27}, 206--227 (1984).

\bibitem{Rabinovich_189}
M.I. Rabinovich and M.M. Sushchik,
The regular and chaotic dynamics of structures in fluid flows,
Phys. Usp. {\bf 33}, 1--35 (1990).

\bibitem{Yukalov_190}
V.I. Yukalov,
Turbulent superfluid as continuous vortex mixture,
Laser Phys. Lett. {\bf 7}, 467--476 (2010). 

\bibitem{Shiozaki_191}
R.F. Shiozaki, G.D. Telles, V.I. Yukalov, and V.S. Bagnato,
Transition to quantum turbulence in finite-size superfluids,
Laser Phys. Lett. {\bf 8}, 393--397 (2011).

\bibitem{Seman_192}
J.A. Seman, E.A.L. Henn, R.F. Shiozaki, G. Roati, F.J. Poveda-Cuevas, K.M.F. Magalh\"{a}es, 
V.I. Yukalov, M. Tsubota, M. Kobayashi, K. Kasamatsu, and V.S. Bagnato,
Route to turbulence in a trapped Bose-Einstein condensate,
Laser Phys. Lett. {\bf 8}, 691--696 (2011). 

\bibitem{Bagnato_193}
V.S. Bagnato and V.I. Yukalov,
From coherent modes to turbulence and granulation of trapped gases,
Porg. Opt. Sci. Photon. {\bf 1}, 377--401 (2013).

\bibitem{Yukalov_194}
V.I Yukalov, A.N. Novikov, and V.S. Bagnato,
Formation of granular structures in trapped Bose-Einstein condensates under oscillatory
excitations,
Laser Phys. Lett. {\bf 11}, 095501 (2014).  

\bibitem{Yukalov_195}
V.I Yukalov, A.N. Novikov, and V.S. Bagnato,
Strongly nonequilibrium Bose-condensed atomic systems,
J. Low Temp. Phys. {\bf 180}, 53--67 (2015).

\bibitem{Yukalov_196}
V.I Yukalov, A.N. Novikov, and V.S. Bagnato,
Realization of inverse Kibble-Zurek scenario with trapped Bose gases,
Phys. Lett. A {\bf 379}, 1366--1371 (2015). 

\bibitem{Yukalov_197}
V.I Yukalov, A.N. Novikov, and V.S. Bagnato,
Characterization of nonequilibrium states of trapped Bose-Einstein condensates,
Laser Phys. Lett. {\bf 15}, 065501 (2018).

\bibitem{Yukalov_198}
V.I. Yukalov, A.N. Novikov, E.P. Yukalova, and V.S. Bagnato,
Characteristic quantities for nonequilibrium Bose systems,
J. Phys. Conf. Ser. {\bf 1508}, 012006 (2020). 

\bibitem{Boltzmann_199}
L. Boltzmann,
{\it Lectures on Gas Theory} (University of California, Berkeley, 1964).

\bibitem{Neimark_200}
Y.I. Neimark and P.S. Landa,
{\it Stochastic and Chaotic Oscillations} (Springer, Dordrecht, 1992).

\bibitem{Neimark_244}
J.I. Neimark,
{\it Mathematical Models in Natural Science and Engineering}
(Springer, Berlin, 2003).

\bibitem{Yukalov_201}
V.I. Yukalov,
Stochastic instability of quasi-isolated systems,
Phys. Rev. E {\bf 65}, 056118 (2002).

\bibitem{Yukalov_202}
V.I. Yukalov,
Irreversibility of time for quasi-isolated systems,
Phys. Lett. A {\bf 308}, 313--318 (2003).

\bibitem{Yukalov_203}
V.I. Yukalov,
Expansion exponents for nonequilibrium systems,
Physica A {\bf 320}, 149--168 (2003).

\bibitem{Yukalov_204}
V.I. Yukalov,
Decoherence and equilibration under nondestructive measurements,
Ann. Phys. (N.Y.) {\bf 327}, 253--263 (2012).

\bibitem{Yukalov_205}
V.I. Yukalov,
Existence of a wave function for a subsystem,
Moscow Univ. Phys. Bull. {\bf 25}, 49--53 (1970).

\bibitem{Yukalov_206}
V.I. Yukalov,
Concept of distinctness for quantum subsystems,
Moscow Univ. Phys. Bull. {\bf 26}, 22--26 (1971). 

\bibitem{Yukalov_207}
V.I. Yukalov,
Spontaneous restoration of broken symmetry,
Phys. Lett. A {\bf 85}, 68--71 (1981).

\bibitem{Ruelle_208}
D. Ruelle, 
{\it Statistical Mechanics} 
(Benjamin, New York, 1969).

\bibitem{Dixmier_209}
J. Dixmier, 
{\it Les $C^*$-Algebres et Leurs Representations} 
(Gauthier-Villars, Paris, 1969).

\bibitem{Emch_210}
G.G. Emch, 
{\it Algebraic Methods in Statistical Mechanics and Quantum Field Theory} 
(Wiley, New York, 1972).

\bibitem{Bratteli_211}
O. Bratteli and D. Robinson, 
{\it Operator Algebras and Quantum Statistical Mechanics} 
(Springer, New York, 1979).

\bibitem{Yukalov_212}
V.I. Yukalov,
Matrix order indices in statistical mechanics,
Physica A {\bf 310}, 413--434 (2002).

\bibitem{Yukalov_213}
V.I. Yukalov,
Order indices and entanglement production in quantum systems,
Entropy {\bf 22}, 565 (2020).  

\bibitem{Bogolubov_214}
N.N. Bogolubov, 
{\it Lectures on Quantum Statistics} 
(Gordon and Breach, New York, 1967), Vol. {\bf 1}.

\bibitem{Bogolubov_215}
N.N. Bogolubov, 
{\it Lectures on Quantum Statistics} 
(Gordon and Breach, New York, 1970), Vol. {\bf 2}.

\bibitem{Bogolubov_216}
N.N. Bogolubov, 
{\it Quantum Statistical Mechanics} 
(World Scientific, Singapore, 2015).

\bibitem{Yukalov_217}
V.I. Yukalov,
Statistical theory of heterophase fluctuations,
Physica A {\bf 108}, 402--416 (1981). 

\bibitem{Yukalov_218}
V.I. Yukalov,
Method of thermodynamic quasiaverages,
Int. J. Mod. Phys. B {\bf 5}, 3235--3253 (1991).

\bibitem{Brout_219}
R. Brout,
{\it Phase Transitions} 
(Benjamin, New York, 1965).

\bibitem{Sewell_220}
G.L. Sewell,
Stability, equilibrium and metastability in statistical mechanics,
Phys. Rep. {\bf 57}, 307--342 (1980).

\bibitem{Sinai_221}
Y.G. Sinai,
{\it Theory of Phase Transitions} 
(Pergamon, Oxford, 1982).

\bibitem{Shumovsky_222}
A.S. Shumovsky and V.I. Yukalov,
Spontaneous symmetry breaking and critical phenomena,
in {\it International School on High Energy Physics}, Ed. by N.N. Bogolubov
(JINR, Dubna, 1983), pp. 223--313.

\bibitem{Bogolubov_223}
N.N. Bogolubov, A.S. Shumovsky, and V.I. Yukalov,
The concept of quasiaverages and spaces of states,
Theor. Math. Phys. {\bf 60}, 921--931 (1984).

\bibitem{Shumovsky_224}
A.S. Shumovsky and V.I. Yukalov,
{\it Phase States and Transitions} (JINR, Dubna, 1985).  

\bibitem{Yukalov_225}
V.I. Yukalov, 
Methods for breaking the symmetry of statistical systems,
in {\it Current Group Analysis}, Ed. by F.G. Maksudov and K.A. Rustamov
(Elm, Baku, 1989), pp. 250--258.

\bibitem{Gibbs_226}
J.W. Gibbs, 
{\it Collected Works} (Longmans, New York, 1928).

\bibitem{Ono_227}
S. Ono and S. Kondo, 
{\it Molecular Theory of Surface Tension in Liquids} 
(Springer, Berlin, 1960).

\bibitem{Rusanov_228}
A.I. Rusanov,
Thermodynamics of solid surfaces,
Surf. Sci. Rep. {\bf 23}, 173--247 (1996).

\bibitem{Rusanov_229}
A.I. Rusanov,
Surface thermodynamics revisited,
Surf. Sci. Rep. {\bf 37}, 111--239 (2005).

\bibitem{Bourbaki_230}
N. Bourbaki, 
{\it Th\'{e}orie des Ensembles} (Hermann, Paris, 1958).

\bibitem{Kullback_231}
S. Kullback and R.A. Leibler,
On information and sufficiency, 
Ann. Math. Stat., {\bf 22}, 79--86 (1951).

\bibitem{Kullback_232}
S. Kullback, 
{\it Information Theory and Statistics} (Wiley, New York, 1959).

\bibitem{Yukalov_233}
V.I. Yukalov,
Remarks on quasiaverages,
Theor. Math. Phys. {\bf 26}, 274--281 (1976).

\bibitem{Yukalov_234}
V.I. Yukalov,
Model of a hybrid crystal,
Theor. Math. Phys. {\bf 28}, 652--660 (1976).

\bibitem{Yukalov_235}
V.I. Yukalov,
Quantum crystal with jumps of particles,
Physica A {\bf 89}, 363--372 (1977).

\bibitem{Yukalov_236}
V.I. Yukalov,
Phase transitions and spontaneous symmetry breaking,
in {\it Selected Topics in Statistical Mechanics}, Ed. by N.N. Bogolubov
(JINR, Dubna, 1978), pp. 437--444. 

\bibitem{Yukalov_237}
V.I. Yukalov,
A new method in the theory of phase transitions,
Phys. Lett. A {\bf 81}, 249--251 (1981).

\bibitem{Yukalov_238}
V.I. Yukalov,
A method to consider metastable states,
Phys. Lett. A {\bf 81}, 433--435 (1981).

\bibitem{Yukalov_239}
V.I. Yukalov,
Spaces of states for heterophase systems,
Physica A {\bf 110}, 247--256 (1982).

\bibitem{Yukalov_240}
V.I. Yukalov,
Effective Hamiltonians for systems with mixed symmetry,
Physica A {\bf 136}, 575--587 (1986).

\bibitem{Yukalov_241}
V.I. Yukalov,
Renormalization of quasi-Hamiltonians under heterophase averaging,
Phys. Lett. A {\bf 125}, 95--100 (1987).

\bibitem{Yukalov_242}
V.I. Yukalov,
Procedure of quasiaveraging for heterophase mixtures,
Physica A {\bf 141}, 352--374 (1987).

\bibitem{Yukalov_243}
V.I. Yukalov,
Lattice mixtures of fluctuating phases,
Physica A {\bf 144}, 369--389 (1987).

\bibitem{Rusanov_244}
A.I. Rusanov,
Problems of surface thermodynamics,
Pure Appl. Chem. {\bf 64}, 111--124 (1992).

\bibitem{Kjelstrup_245}
S. Kjelstrup and D. Bedeaux, 
{\it Non-Equilibrium Thermodynamics of Heterogeneous Systems}
(World Scientific, Singapore, 2008).

\bibitem{Bedeaux_246}
D. Bedeaux and S. Kjelstrup,
Fluid-fluid interfaces of multi-component mixtures in local equilibrium,
Entropy {\bf 20}, 250 (2018). 

\bibitem{Yukalov_247}
V.I. Yukalov,
Phase probabilities as order parameters,
in {\it Current Topics in Statistical Physics}, Ed. by I.R. Yukhnovsky
(Naukova Dumka, Kiev, 1989), Vol. 2, pp. 114--120.

\bibitem{Yukalov_248}
V.I. Yukalov,
Additional order parameters for heterogeneous systems,
in {\it Selected Topics in Statistical Mechanics}, Ed. by A.A. Logunov
(World Scientific, Sinapore, 1990), pp. 298--312. 

\bibitem{Yukalov_249}
V.I. Yukalov and A.S. Shumovsky,
{\it Lectures on Phase Transitions} (World Scientific, Singapore, 1990).

\bibitem{Coleman_250}
A.J. Coleman and V.I. Yukalov, 
Order indices and mid-range order, 
Mod. Phys. Lett. B {\bf 5}, 1679--1686 (1991).

\bibitem{Coleman_251}
A.J. Coleman and V.I. Yukalov,
{\it Reduced Density Matrices} (Springer, Berlin, 2000).

\bibitem{Heisenberg_252}
W. Heisenberg,  
On the theory of ferromagnetism, 
Zeitschr. Phys. {\bf 49}, 619--636 (1928).

\bibitem{Shumovsky_253}
A.S. Shumovsky and V.I. Yukalov,
An exactly solvable model of a ferromagnet with paramagnetic nuclei,
Dokl. Phys. {\bf 25}, 361--363 (1980).

\bibitem{Shumovsky_254}
A.S. Shumovsky and V.I. Yukalov, 
On specific heat anomalies in magnets,
Chem. Phys. Lett. {\bf 83}, 582--584 (1981).

\bibitem{Shumovsky_255}
A.S. Shumovsky and V.I. Yukalov,
Problem of description of heterophase fluctuations at phase transitions,
in {\it Selected Topics in Statistical Mechanics}, Ed. by N.N. Bogolubov
(JINR, Dubna, 1981), pp. 238--260. 

\bibitem{Shumovsky_256}
A.S. Shumovsky and V.I. Yukalov,
Exact solutions for heterophase ferromagnets,
Physica A {\bf 110}, 518--534 (1982).

\bibitem{Shumovsky_257}
A.S. Shumovsky and V.I. Yukalov,
Equilibrium nucleation: A new type of phase transition,
Chem. Phys. Lett. {\bf 117}, 617--621 (1985). 

\bibitem{Shumovsky_258}
A.S. Shumovsky and V.I. Yukalov,
Heterophase states in physical systems,
Phys. Part. Nucl. {\bf 16}, 569--592 (1985).

\bibitem{Lawrie_259}
I. Lawrie and S. Sarbach,
Theory of tricritical points,
Phase Trans. Crit. Phenom. {\bf 9}, 2--163 (1984).

\bibitem{Yukalov_260}
V.I. Yukalov,
Tricritical phenomena in strongly fluctuating systems,
in {\it Statistical Mechanics and Theory of Phase Transitions}, Ed. by N.N. Bogolubov
(Samara University, Samara, 1989), pp. 12--20.

\bibitem{Yukalov_262}
V.I. Yukalov and E.P. Yukalova,
Zeroth-order nucleation transition under nanoscale phase separation,
Symmetry {\bf 13}, 2379 (2021). 

\bibitem{Yukalov_263}
V.I. Yukalov,
On the difference between homophase and heterophase fluctuations,
in {\it Problems of Statistical Physics and Field Theory}, Ed. by Y.I. Zaparovanny
(Moscow University, Moscow, 1987).

\bibitem{Tyablikov_264}
S.V. Tyablikov,
{\it Methods in the Quantum Theory of Magnetism}
(Springer, New York, 1967).

\bibitem{Akhiezer_265}
A.I. Akhiezer, V.G. Baryakhtar, and S.V. Peletminskii,
{\it Spin Waves}
(North-Holland, Amsterdam, 1968).  

\bibitem{Yukalov_266}
V.I. Yukalov,
Interplay between mesoscopic and microscopic fluctuations in ferromagnets,
Physica A {\bf 262}, 467--482 (1999).

\bibitem{Kadanoff_267}
L.P. Kadanoff and G. Baym,
{\it Quantum Statistical Mechanics}
(Benjamin, New York, 1962).

\bibitem{Bonch_268}
V.L. Bonch-Bruevich and S.V. Tyablikov,
{\it The Green Function Method in Statistical Mechanics}
(North-Holland, Amsterdam, 1962).

\bibitem{Yukalov_269}
V.I. Yukalov,
{\it Statistical Green's Functions}
(Queen's University, Kingston, 1998).

\bibitem{Kislinsky_270}
V.B. Kislinsky, A.S. Shumovsky, and V.I. Yukalov,
Metastable heterophase system of the Ising type.
Phys. Lett. A {\bf 109}, 254--256 (1985).

\bibitem{Onsager_271}
L. Onsager, 
Crystal statistics: A two-dimensional model with an order-disorder transition,
Phys. Rev. {\bf 65}, 117--149, (1944).

\bibitem{Nagle_272}
J.F. Nagle,
Ising chain with competing interactions,
Phys. Rev. A {\bf 2}, 2124--2128 (1970). 

\bibitem{Kislinsky_273}
V.B. Kislinsky and V.I. Yukalov,
Crossover between short- and long-range interactions in the one-dimensional Ising 
model,
J. Phys. A {\bf 21}, 227--232 (1988).

\bibitem{Yukalov_274}
V.I. Yukalov,
Stabilizing role of mesoscopic fluctuations in spin systems,
Physica A {\bf 261}, 482--498 (1998).  

\bibitem{Kudryavtsev_275}
I.K. Kudryavtsev, A.S. Shumovsky, and V.I. Yukalov,
On a model of hybrid antiferromagnet,
in {\it Selected Topics in Statistical Mechanics},
Ed. by N.N. Bogolubov (JINR, Dubna, 1981), pp. 318--325.

\bibitem{Boky_276}
M.A. Boky, I.K. Kudryavtsev, A.S. Shumovsky, and V.I. Yukalov,
Model of antiferromagnet with heterophase fluctuations,
Fizika {\bf 19}, 263--284 (1987). 

\bibitem{Boky_277}
M.A. Boky, I.K. Kudryavtsev, and V.I. Yukalov,
Critical temperature in heterophase Hubbard model,
Solid State Commun. {\bf 63}, 731--735 (1987).

\bibitem{Boky_278}
M.A. Boky and V.I. Yukalov,
Generalization of the Vonsovsky-Ziener model for heterogeneous systems,
in {\it Selected Topics in Statistical Mechanics}, 
Ed. by N.N. Bogolubov (JINR, Dubna, 1984), Vol. {\bf 1}, pp. 170--176.

\bibitem{Vonsovsky_279}
S.V. Vonsovsky,
{\it Magnetism} 
(Wiley, New York, 1974).

\bibitem{Irkhin_280}
V.Y. Irchin,
S.V. Vonsovsky and a contemporary model description of magnetism,
Phys. Met. Metallogr. {\bf 110}, 602--641 (2010). 

\bibitem{Sherrington_281}
D. Sherrington and S. Kirkpatrick,
Solvable model of a spin-glass,
Phys. Rev. Lett. {\bf 35}, 1792--1795 (1975).

\bibitem{Kirkpatrick_282}
S. Kirkpatrick and D. Sherrington,
Infinite-ranged models of spin-glasses,
Phys. Rev. B {\bf 17}, 4384--4403 (1978).  

\bibitem{Kislinsky_283}
V.B. Kislinsky and V.I. Yukalov,
Modified Sherrington-Kirkpatrick model for heterophase spin glass,
in {\it Selected Topics in Statistical Mechanics}, 
Ed. by N.N. Bogolubov (JINR, Dubna, 1984), Vol. {\bf 1}, pp. 344--349.

\bibitem{Edwards_284}
S.F. Edwards and P.W. Anderson, 
Theory of spin glasses,
J. Phys. F {\bf 5}, 965--974 (1975). 

\bibitem{Almeida_285}
J.R.L. De Almeida and D.J. Thouless,
Stability of the Sherrington-Kirkpatrick solution of a spin glass model,
J. Phys. A {\bf 11}, 983--990 (1978).

\bibitem{Parisi_286}
G. Parisi, 
The order parameter for spin glasses: A function on the interval $0-1$,
J. Phys. A {\bf 13}, 1101--1112 (1980).

\bibitem{Bakasov_287}
A.A. Bakasov and V.I. Yukalov,
Microscopic theory of spin reorientations: Heterophase approach and basic model,
Physica A {\bf 157}, 1203--1226 (1989).

\bibitem{Bakasov_288}
A.A. Bakasov and V.I. Yukalov,
Microscopic theory of spin reorientations: Thermodynamics and nucleation phenomenon,
Physica A {\bf 162}, 31--66 (1989). 

\bibitem{Yukalov_289}
V.I. Yukalov,
Microscopic theory of spin reorientations: General analysis,
Physica A {\bf 167}, 833--860 (1990). 

\bibitem{Bianconi_290}
A. Bianconi,
Superstripes,
Int. J. Mod. Phys. B {\bf 14}, 3289--3297 (2000).

\bibitem{Shumovsky_291}
A.S. Shumovsky and V.I. Yukalov,
Microscopic model of a superconductor with normal-state nuclei,
Dokl. Phys. {\bf 27}, 709--711 (1982).

\bibitem{Yukalov_292}
V.I. Yukalov,
{\it On the model of heterophase superconductor}
(JINR, Dubna, 1985). 

\bibitem{Bednorz_293}
J.G. Bednorz and K.A. M\"{u}ller,  
Possible high $T_c$ super-conductivity in the Ba-La-Cu-O system, 
Z. Phys. B {\bf 64}, 189--193 (1986).

\bibitem{Shumovsky_294}
A.S. Shumovsky and V.I. Yukalov,
Two-fluid superconductor as an example of heterophase systems,
Phys. Many-Part. Syst. {\bf 14}, 24--28 (1988).

\bibitem{Shumovsky_295}
A.S. Shumovsky and V.I. Yukalov,
On a new formula for temperature of superconducting transition,
in {\it Selected Topics in Statistical Mechanics}, Ed. by N.N. Bogolubov
(JINR, Dubna, 1988), pp. 434--443. 

\bibitem{Yukalov_296}
V.I. Yukalov,
Heterostructural fluctuations in superconductors,
Int. J. Mod. Phys. B {\bf 6}, 91--107 (1992).

\bibitem{Coleman_297}
A.J. Coleman, E.P. Yukalova, and V.I. Yukalov,
Superconductors with mesoscopic phase separation,
Physica C {\bf 243}, 76--92 (1995). 

\bibitem{Yukalov_298}
V.I. Yukalov and E.P. Yukalova,
Mesoscopic phase separation in anisotropic superconductors,
Phys. Rev. B {\bf 70}, 224516 (2004).

\bibitem{Yukalov_299}
V.I. Yukalov and E.P. Yukalova,
Statistical theory of materials with nanoscale phase separation,
J. Supercond. Nov. Magn. {\bf 27}, 919--924 (2014). 

\bibitem{Coleman_300}
A.J. Coleman and V.I. Yukalov,
Relation between microscopic and macroscopic characteristics of statistical systems,
Int. J. Mod. Phys. B {\bf 10}, 3505--3515 (1996).
 
\bibitem{Yukalov_301}
V.I. Yukalov,
Theory of cold atoms: Basics of quantum statistics,
Laser Phys. {\bf 23}, 062001 (2013).

\bibitem{Pines_302}
D. Pines, 
{\it Elementary Excitations in Solids}
(Benjamin, New York, 1963).

\bibitem{Ziman_303}
J.M. Ziman, 
{\it Principles of the Theory of Solids} 
(Cambridge University, Cambridge, 1972).

\bibitem{Bardeen_304}
J. Bardeen, L.N. Cooper, and J.R. Schrieffer,
Microscopic theory of superconductivity,
Phys. Rev. {\bf 106}, 162--163 (1957).

\bibitem{Bardeen_305}
J. Bardeen, L.N. Cooper, and J.R. Schrieffer, 
Theory of superconductivity,
Phys. Rev. {\bf 108}, 1175--1203 (1957).

\bibitem{Harlingen_306}
J. Van Harlingen, 
Phase-sensitive tests of the symmetry of the pairing state in the high-temperature 
superconductors: Evidence for $d_{x^2-y^2}$ symmetry,
Rev. Mod. Phys. {\bf 67}, 515--535 (1995).

\bibitem{Tsuei_307}
C. Tsuei and J.R. Kirtley, 
Pairing symmetry in cuprate superconductors,
Rev. Mod. Phys. {\bf 72}, 969--1016 (2000).

\bibitem{Kabanov_308}
V. Kabanov, J. Demsar, B. Podobnik, and D. Mihailovic, 
Quasiparticle relaxation dynamics in superconductors with different gap structures: 
Theory and experiments on YBa$_2$Cu$_3$O$_{7-\delta}$,
Phys. Rev. B {\bf 59}, 1497--1506 (1999).

\bibitem{Blinc_309}
R. Blinc and B. Zeks, 
{\it Soft Modes in Ferroelectrics and Antiferroelectrics}
(North-Holland, Amsterdam, 1974).

\bibitem{Bashkirov_310}
E.K. Bashkirov and V.I. Yukalov,
On microscopic theory of heterophase states in ferroelectrics,
in {\it Selected Topics in Classical and Quantum Physics},
Ed. by Y.I. Granovsky (Samara University, Samara, 1983), pp. 99--107.

\bibitem{Bashkirov_311}
E.K. Bashkirov and V.I. Yukalov,
Heterophase phenomena in ferroelectrics,
in {\it Selected Topics in Statistical Mechanics},
Ed. by N.N. Bogolubov (JINR, Dubna, 1984), Vol. 1, pp. 76--82.

\bibitem{Yukalov_312}
V.I. Yukalov,
Heterophase fluctuations in ferroelectrics,
Ferroelectrics {\bf 82}, 11--24 (1988). 

\bibitem{Yukalov_313}
V.I. Yukalov and E.P. Yukalova,
Nanoscale phase separation in ferroelectric materials,
J. Supercond. Nov. Magn. {\bf 29}, 3119--3126 (2016).

\bibitem{Marzari_314}
N. Marzari, A.A. Mostofi, J.R. Yates, I. Souza, and D. Vanderbilt,
Maximally localized Wannier functions: Theory and applications,
Rev. Mod. Phys. {\bf 84}, 1419--1475 (2012). 

\bibitem{Yukalov_315}
V.I. Yukalov and E.P. Yukalova, 
Asymptotic properties of eigenvalues in variational calculations for double-well
oscillators,
J. Phys. A {\bf 29}, 6429--6442 (1996).

\bibitem{Yukalov_316}
V.I. Yukalov and E.P. Yukalova, 
Mesoscopic disorder in double-well optical lattices,
Laser Phys. {\bf 21}, 1448--1458 (2011).

\bibitem{Yukalov_317}
V.I. Yukalov and E.P. Yukalova, 
Double-well optical lattices with atomic vibrations and mesoscopic disorder,
Laser Phys. {\bf 22}, 1070--1080 (2012).

\bibitem{Yukalov_318}
V.I. Yukalov and E.P. Yukalova,
Statistics of multiscale fluctuations in macromolecular systems,
J. Phys. Chem. B {\bf 116}, 8435--8448 (2012).

\bibitem{Yukalov_319}
V.I. Yukalov and E.P. Yukalova,
Statistical models of nonequilibrium Bose gases,
Rom. Rep. Phys. {\bf 67}, 159--185 (2015).

\bibitem{Akhmeteli_320}
A.M. Akhmeteli, A.S. Shumovsky, and V.I. Yukalov,
Spin-phonon interaction in a model of hybrid ferromagnet,
in {\it Selected Topics in Statistical Mechanics}, 
Ed. by N.N. Bogolubov (JINR, Dubna, 1981), pp. 300--306. 

\bibitem{Yukalov_321}
V.I. Yukalov,
Spin-phonon interactions in heterophase ferromagnets,
Physica A {\bf 155}, 519--544 (1989).

\bibitem{Yukalov_322}
V.I. Yukalov,
Properties of crystals with local symmetry breaking,
in {\it Symmetry and Structural Properties of Condensed Matter},
Ed. by W. Florek, T. Lulek, and M. Mucha 
(World Scientific, Singapore, 1991), pp. 141--152.

\bibitem{Yukalov_323}
V.I. Yukalov,
Influence of structural fluctuations on the dynamical characteristics of solids,
Chem. Phys. Lett. {\bf 229}, 239--243 (1994).
 
\bibitem{Yukalov_324}
V.I. Yukalov,
Statistical mechanics of structural fluctuations,
Physica A {\bf 213}, 500--524 (1995). 

\bibitem{Yukalov_325}
V.I. Yukalov and E.P. Yukalova,
Crystal symmetry and time scales,
in {\it Symmetry and Structural Properties of Condensed Matter},
Ed. by T. Lulek, B. Lulek, and A. Wal 
(World Scientific, Singapore, 2001), pp. 383--393.

\bibitem{Yukalov_326}
V.I. Yukalov,
Destiny of optical lattices with strong intersite interactions,
Laser Phys. {\bf 30}, 015501 (2020).

\bibitem{Yukalov_327}
V.I. Yukalov,
From optical lattices to quantum crystals,
J. Phys. Conf. Ser. {\bf 1508}, 012008 (2020).  

\bibitem{Guyer_328}
R.A. Guyer,
The physics of quantum crystals,
Solid State Phys. {\bf 23}, 413--499 (1969).

\bibitem{Bhide_329}
V.G. Bhide and M.S. Multani,
M\"{o}ssbauer effect in ferroelectric BaTiO$_3$,
Phys. Rev. A {\bf 139}, 1983--1990 (1965).

\bibitem{Bhide_330}
V.G. Bhide and G.K. Shenoy, 
Temperature dependent lifetimes of nonequilibrium Fe$^{57}$ ions in CoO from the 
M\"{o}ssbauer effect,
Phys. Rev. {\bf 147}, 306--310 (1966).

\bibitem{Bhide_331}
V.G. Bhide and M.S. Hegde,
M\"{o}ssbauer effect for Fe$_{57}$ in ferroelectric lead titanate,
Phys. Rev. B {\bf 5}, 3488--3499 (1972).

\bibitem{Bhide_332}
V.G. Bhide, 
{\it M\"{o}ssbauer Effect and its Applications}
(McGraw-Hill, New Delhi, 1973).

\bibitem{Owens_333}
F.G. Owens, C.P. Poole, and H.A. Farrach, 
Eds. {\it Magnetic Resonance of Phase Transitions}
(Academic, New York, 1979).

\bibitem{Thosar_334}
B.V. Thosar and P.K. Iyengar, 
Eds. {\it Advances in M\"{o}ssbauer Spectroscopy: Application to Physics, Chemistry and Biology}
(Elsevier, Amsterdam, 1983).

\bibitem{Bishop_335}
A.R. Bishop, S.R. Shenoy, and S. Sridhar, 
Eds. {\it Intrinsic Multiscale Structure and Dynamics in Complex Electronic Oxides}
(World Scientific, Singapore, 2002). 

\bibitem{Egami_336}
T. Egami, and S.J. Billinge, 
{\it Underneath the Bragg Peacks} (Pergamon, Amsterdam, 2003).

\bibitem{Muller_337}
K.A. M\"{u}ller and A. Bussmann-Holder, 
Eds. {\it Superconductivity in Complex Systems} (Springer, Berlin, 2004).

\bibitem{Sirdeshmukh_338}
D. Sirdeshmukh, L. Sirdeshmukh, and K.G. Subhadra, 
{\it Micro- and Macro-Properties of Solids} (Springer, Berlin, 2006).

\bibitem{Meissner_339}
G. Meissner and K. Binder, 
Debye-Waller factor, compressibility sum rule, and central peak at structural phase transitions,
Phys. Rev. B {\bf 12} 3948--3955 (1975).

\bibitem{Binder_340}
K. Binder, G. Meissner, and H. Mais, 
Equation of state, Debye-Waller factor, and electrical resistivity of ferroelectrics near 
their critical point,
Phys. Rev. B {\bf 13}, 4890--4898 (1976).

\bibitem{Yukalov_341}
V.I. Yukalov,
M\"{o}ssbauer effect in magnetic materials,
in {\it Selected Topics in Statistical Mechanics},
Ed. by N.N. Bogolubov (JINR, Dubna, 1988), pp. 444--467.

\bibitem{Yukalov_342}
V.I. Yukalov,
M\"{o}ssbauer-effect probability for heterogeneous matters,
Solid State Commun. {\bf 69}, 393--395 (1989). 

\bibitem{Yukalov_343}
V.I. Yukalov,
Magnetic anomalies under M\"{o}ssbauer effect,
in {\it Physics of Transition Metals},
Ed. by V.G. Baryakhtar (Naukova Dumka, Kiev, 1989), Vol. 2, pp. 165--168.

\bibitem{Yukalov_344}
V.I. Yukalov,
Collective effects during nuclear gamma-resonance,
in {\it Proceedings of Workshop on Gravitational Wave Emitter and Detector},
Ed. by A.F. Pisarev (JINR, Dubna, 1989), pp. 66--73.

\bibitem{Yukalov_345}
V.I. Yukalov,
Anomalous saggings of M\"{o}ssbauer effect probability at phase transitions,
Hyperf. Interact. {\bf 55}, 1165--1168 (1990).

\bibitem{Yukalov_346}
V.I. Yukalov,
Influence of radiation damage on M\"{o}ssbauer effect probability,
Hyperf. Interact. {\bf 56}, 1657--1660 (1990). 

\bibitem{Yukalov_347}
V.I. Yukalov,
M\"{o}ssbauer spectroscopy of heterophase systems,
in {\it Topics in Application of Nuclear Raiation},
Ed. by Y.F. Babikova (Energoatomizdat, Moscow, 1991), pp. 33--38.

\bibitem{Lindemann_348}
F.A. Lindemann,
The calculation of molecular vibration frequencies,
Z. Phys. {\bf 11}, 609--612 (1910).

\bibitem{Yukalov_349}
V.I. Yukalov and K. Ziegler,
Instability of insulating states in optical lattices due to collective phonon excitations,
Phys. Rev. A {\bf 91}, 023628 (2015).

\bibitem{Yukalov_350}
V.I. Yukalov and K. Ziegler,
Phonon instability of insulating states in optical lattices,
J. Phys. Conf. Ser. {\bf 691}, 012014 (2016).

\bibitem{Cottrell_351}
A.H. Cottrell, 
{\it Dislocations and Plastic Flow in Crystals} (Oxford University, London, 1953).

\bibitem{Friedel_352}
J. Friedel, 
{\it Dislocations} (Pergamon, Oxford, 1967).

\bibitem{Ziman_353}
J.M. Ziman, 
{\it Models of Disorder} (Cambridge University, Cambridge, 1979).

\bibitem{Hirth_354}
J.P. Hirth and J. Lothe, 
{\it Theory of Dislocations} (Wiley, New York, 1982).

\bibitem{Hull_355}
D. Hull and D. Bacon, 
{\it Introduction to Dislocations} (Elsevier, Oxford, 2001).

\bibitem{Yukalov_356}
V.I. Yukalov,
Properties of solids with pores and cracks,
Int. J. Mod. Phys. B {\bf 3}, 311--326 (1989).  

\bibitem{Kad_357}
E.P. Kadantseva and V.I. Yukalov,
Thermodynamics of solids with regions of disorder,
Int. J. Mod. Phys. B {\bf 3}, 465--472 (1989).

\bibitem{Yukalov_358}
V.I. Yukalov and E.P. Yukalova,
Chaotic lattice-gas model,
Physica A {\bf 213}, 482--499 (1995).

\bibitem{Yukalov_359}
V.I. Yukalov and E.P. Yukalova,
Possibility of turbulent crystals,
Int. J. Mod. Phys. B {\bf 15}, 2433--2453 (2001).

\bibitem{Yukalov_360}
V.I. Yukalov and E.P. Yukalova,
Optical lattice with heterogeneous atomic density,
Laser Phys. {\bf 25}, 035501 (2015).

\bibitem{Yukalov_361}
V.I. Yukalov and E.P. Yukalova,
Statistical theory of structures with extended defects,
in {\it Mechanics and Physics of Structured Media},
Ed. by I. Andrianov, S. Gluzman, and V. Mityushev 
(Elsevier, London, 2022), pp. 417--443.

\bibitem{Hirshfelder_362}
J.O. Hirschfelder, C.F. Curtiss, and R.B. Bird,
Molecular Theory of Gases and Liquids
(Wiley, New York, 1954).

\bibitem{Ubbelohde_363}
A.R. Ubbelohde,
Melting and Crystal Structure, (Oxford University, London, 1965).

\bibitem{Yukalov_364}
V.I. Yukalov,
Theory of melting and crystallization,
Phys. Rev. B {\bf 32}, 436--446 (1985).

\bibitem{Kirkwood_365}
J.G. Kirkwood,
{\it Quantum Statistics and Cooperative Phenomena} 
(Gordon and Breach, New York, 1965).

\bibitem{Bogolubov_366}
N.N. Bogolubov, 
On the theory of superfluidity,
J. Phys. (Moscow) {\bf 11}, 23--32 (1947).

\bibitem{Yukalov_367}
V.I. Yukalov,
On the description of quasiparticles by Green functions,
Theor. Math. Phys. {\bf 17}, 1244--1248 (1973).

\bibitem{Yukalov_368}
V.I. Yukalov,
Strongly interacting particles with strongly singular potentials,
Int. J. Theor. Phys. {\bf 28}, 1237--1254 (1989).

\bibitem{Yukalov_369}
V.I. Yukalov,
Principal problems in Bose-Einstein condensation of dilute gases,
Laser Phys. Lett. {\bf 1}, 435--461 (2004).

\bibitem{Yukalov_370}
V.I. Yukalov,
Statistical systems with nonintegrable interaction potentials,
Phys. Rev. E {\bf 94}, 012106 (2016).  

\bibitem{Andreev_371}
A.F. Andreev and I.M. Lifshits, 
Quantum theory of defects in crystals,
J. Exp. Theor. Phys. {\bf 29}, 1107--1113 (1969).

\bibitem{Chester_372}
G.V. Chester, 
Speculations on Bose-Einstein condensation and quantum crystals,
Phys. Rev. A {\bf 2}, 256--258 (1970).

\bibitem{Bogolubov_373}
N.N. Bogolubov, 
{\it Lectures on Quantum Statistics} (Ryadyanska Shkola, Kiev, 1949).

\bibitem{Gross_374}
E.P. Gross,  
Unified theory of interacting bosons,
Phys. Rev. {\bf 106}, 161--162 (1957). 

\bibitem{Gross_375}
E.P. Gross, 
Classical theory of boson wave fields,
Ann. Phys. (N.Y.) {\bf 4}, 57--74 (1958).

\bibitem{Gross_376}
E.P. Gross, 
Quantum theory of interacting bosons,
Ann. Phys. (N.Y.) {\bf 9}, 292--324 (1960).

\bibitem{Gross_377}
E.P. Gross, 
Periodic ground states in the many-body problem,
Phys. Rev. Lett. {\bf 4}, 599--601 (1960). 

\bibitem{Gross_378}
E.P. Gross, 
Structure of a quantized vortex in boson systems,
Nuovo Cimento {\bf 20}, 454--477 (1961). 

\bibitem{Gross_379}
E.P. Gross, 
Hydrodynamics of a superfluid condensate, 
J. Math. Phys. {\bf 4}, 195--207 (1963).

\bibitem{Kirzhnits_380}
D.A. Kirzhnits and Y.A. Nepomnyashchii, 
Coherent crystallization of quantum liquid, 
J. Exp. Theor. Phys. {\bf 32}, 1191--1197 (1971).

\bibitem{Prokofev_381}
N. Prokof’ev, 
What makes a crystal supersolid? 
Adv. Phys. {\bf 56}, 381--402 (2007).

\bibitem{Kuklov_382} 
A.B. Kuklov, N.V. Prokof’ev, and B.V. Svistunov, 
How solid is supersolid? 
Physics {\bf 4}, 109 (2011).

\bibitem{Boninsegni_383}
M. Boninsegni and N.V. Prokof’ev,  
Supersolids: What and where are they? 
Rev. Mod. Phys. {\bf 84}, 759--776 (2012).

\bibitem{Yukalov_384}
V.I. Yukalov,
Saga of superfluid solids,
Physics {\bf 2}, 49--66 (2020).

\bibitem{Fil_385}
D.V. Fil and S.I. Shevchenko,
Supersolid induced by dislocations with superfluid cores,
Low Temp. Phys. {\bf 48}, 429--451 (2022).

\bibitem{Yukalov_386}
V.I. Yukalov,
Nonequivalent operator representations for Bose-condensed systems,
Laser Phys. {\bf 16}, 511--525 (2006).

\bibitem{Yukalov_387}
V.I. Yukalov,
Bose-Einstein condensation and gauge symmetry breaking,
Laser Phys. Lett. {\bf 4}, 632--647 (2007). 

\bibitem{Yukalov_388}
V.I. Yukalov,
Fluctuations of composite observables and stability of statistical systems,
Phys. Rev. E {\bf 72}, 066119 (2005).

\bibitem{Yukalov_389}
V.I. Yukalov,
Self-consistent theory of Bose-condensed systems,
Phys. Lett. A {\bf 359}, 712--717 (2006).

\bibitem{Yukalov_390}
V.I. Yukalov,
Representative ensembles in statistical mechanics,
Int. J. Mod. Phys. B {\bf 21}, 69--86 (2007).

\bibitem{Yukalov_391}
V.I. Yukalov,
Representative statistical ensembles for Bose systems with broken gauge symmetry,
Ann. Phys. (N.Y.) {\bf 323}, 461--499 (2008).  

\bibitem{Hohenberg_392}
P.C. Hohenberg and P.C. Martin, 
Microscopic theory of superfluid helium,
Ann. Phys. (N.Y.) {\bf 34}, 291--359 (1965).

\bibitem{Hugenholtz_393}
N.M. Hugenholtz and D. Pines,
Ground-state energy and excitation spectrum of a system of interacting bosons,
Phys. Rev. {\bf 116}, 489--506 (1959).

\bibitem{Nepomnyashchii_394}
A.A. Nepomnyashchii and Y.A. Nepomnyashchii,
Contribution to the theory of a spectrum of a Bose system with condensate at small momenta,
JETP Lett. {\bf 21}, 1--2 (1975). 

\bibitem{Yukalov_395}
V.I. Yukalov and E.P. Yukalova,
Bose-Einstein condensed gases with arbitrary strong interactions, 
Phys. Rev. A {\bf 74}, 063623 (2006)

\bibitem{Yukalov_396}
V.I. Yukalov and E.P. Yukalova,
Condensate and superfluid fractions for varying interactions and temperature,
Phys. Rev. A {\bf 76}, 013602 (2007).

\bibitem{Yukalov_397}
V.I. Yukalov and E.P. Yukalova,
Bose-Einstein condensation in self-consistent mean-field theory,
J. Phys. B {\bf 47}, 095302 (2014)

\bibitem{Yukalov_398}
V.I. Yukalov and E.P. Yukalova,
Ground state of a homogeneous Bose gas of hard spheres,
Phys. Rev. A {\bf 90}, 013627 (2014).  

\bibitem{Yukalov_399}
V.I. Yukalov and E.P. Yukalova,
Bose-Einstein condensation temperature of weakly interacting atoms,
Laser Phys. Lett. {\bf 14}, 073001 (2017).

\bibitem{Yukalov_400}
V.I. Yukalov and E.P. Yukalova,
Local condensate depletion at trap center under strong interactions,
J. Phys. B {\bf 51}, 085301 (2018).

\bibitem{Yukalov_401}
V.I. Yukalov and E.P. Yukalova,
Hartree-Fock-Bogolubov method in the theory of Bose-condensed systems,
Phys. Part. Nucl. {\bf 51}, 823--828 (2020).

\bibitem{Aziz_402}
R.A. Aziz, V.P.S. Nain, and J.S. Carley   
An accurate intermolecular potential for helium,
J. Chem. Phys. {\bf 70}, 4330--4342 (1979).

\bibitem{Aziz_403}
R.A. Aziz, W.J. Meath, and A.R. Allnatt,
On the Ne-Ne potential-energy curve and related properties,
Chem. Phys. {\bf 78}, 295--309 (1983).

\bibitem{Aziz_404}
R.A. Aziz, F.R.W. McCourt, and C.C.K. Wong,
A new determination of the ground state interatomic potential for He$_2$,
Mol. Phys. {\bf 61}, 1487--1511 (1987).

\bibitem{Hodgdon_405}
J.A. Hodgdon and F.H. Stillinger,
Inherent structures in the potential energy landscape of solid $^4$He,
J. Chem. Phys. {\bf 102}, 457--464 (1995). 

\bibitem{Ceperley_406}
D.M. Ceperley, R.O. Simmons, and R.C. Blasdell,
Kinetic energy of liquid and solid $^4$He,
Phys. Rev. Lett. {\bf 77}, 115--118 (1996).

\bibitem{Diallo_407}
S.O. Diallo, J.V. Pearce, R.T. Azuah, and H.R. Glyde,
Quantum momentum distribution and kinetic energy in solid $^4$He,
Phys. Rev. Lett. {\bf 93}, 075301 (2004).

\bibitem{Maris_408}
H.J. Maris and S. Balibar,
Supersolidity and the thermodynamics of solid Helium,
J. Low Temp. Phys. {\bf 147}, 539--547 (2007).

\bibitem{Casorla_409}
C. Cazorla, G.E. Astrakharchik, J. Casulleras, and J. Boronat,
Bose-Einstein quantum statistics and the ground state of solid $^4$He,
New J. Phys. {\bf 11}, 013047 (2009).

\bibitem{Vitiello_410}
S.A. Vitiello,
Helium atoms kinetic energy at temperature $T = 0$,
J. Low Temp. Phys. {\bf 162}, 154--159 (2011).

\bibitem{Chan_411}
M.H.W. Chan, R.B. Hallock, and L. Reatto,
Overview on solid $^4$He and the issue of supersolidity,
J. Low Temp. Phys. {\bf 172}, 317--363 (2013).

\bibitem{Casorla_412}
C. Casorla and J. Boronat,
Simulation and understanding of atomic and molecular quantum crystals,
Rev. Mod. Phys. {\bf 89}, 035003 (2017).

\bibitem{Shuryak_413} 
E.V. Shuryak, 
Quantum chromodynamics and the theory of superdense matter,
Phys. Rep. {\bf 61}, 71--158 (1980).

\bibitem{Rafelsky_414}
J. Rafelsky,
Formation and observation of the quark-gluon plasma,
Phys. Rep. {\bf 88}, 331--348 (1982).

\bibitem{Satz_415}
H. Satz, 
Critical behaviour in finite temperature QCD,
Phys. Rep. {\bf 88}, 349--364 (1982).

\bibitem{Hagedorn_416}
R. Hagedorn, 
Multiplicities, $p_T$ distributions and the expected hadron $\ra$ quark-gluon phase transition,
Riv. Nuovo Cimento {\bf 6}, 1--50 (1983).

\bibitem{Cleymans_417}
J. Cleymans, R. Gavai, and E. Suhonen, 
Quarks and gluons at high temperatures and densities,
Phys. Rep. {\bf 130}, 217--292 (1986).

\bibitem{Stocker_418}
H. St\"{o}cker and W. Greiner, 
High-energy heavy ion collisions: Probing the equation of state of 
highly excited hadronic matter,
Phys. Rep. {\bf 137}, 277--392 (1986).

\bibitem{Clare_419}
R. Clare and D. Strottman, 
Relativistic hydrodynamics and heavy ion reactions,
Phys. Rep. {\bf 141}, 177--280 (1986).

\bibitem{Mclerran_420}
L. McLerran, 
The physics of the quark-gluon plasma,
Rev. Mod. Phys. {\bf 58}, 1021--1064 (1986).

\bibitem{Reeves_421}
H. Reeves, 
Big Bang nucleosynthesis and the quark-hadron phase transition,
Phys. Rep. {\bf 201}, 335--354 (1991).

\bibitem{Adami_422}
C. Adami and G.E. Brown,
Matter under extreme conditions,
Phys. Rep. {\bf 234}, 1--71 (1993).

\bibitem{Yukalov_423}
V.I. Yukalov and E.P. Yukalova, 
Thermodynamics of strong interactions,
Phys. Part. Nucl. {\bf 28}, 37--65 (1997).

\bibitem{Yukalov_424}
V.I. Yukalov and E.P. Yukalova, 
Multichannel approach to clustering matter,
Physica A {\bf 243}, 382--414 (1997).

\bibitem{Glendenning_425}
N.K. Glendenning,
Phase transitions and crystalline structures in neutron star cores,
Phys. Rep. {\bf 342}, 393--447 (2001).

\bibitem{Satz_426}
H. Satz,  
States of strongly interacting matter,
Lect. Notes Phys. {\bf 616}, 126--137 (2003).

\bibitem{Soltz_427}
R.A. Soltz, C. DeTar, F. Karsch, S. Mukherjee, and P. Vranas,
Lattice QCD thermodynamics with physical quark masses,
Annu. Rev. Nucl. Part. Sci. {\bf 65}, 379--402 (2015).

\bibitem{Ayala_428}
A. Ayala,
Hadronic matter at the edge: A survey of some theoretical approaches to the physics 
of the QCD phase diagram,
J. Phys. Conf. Ser. {\bf 761}, 012066 (2016).

\bibitem{Busza_429}
W. Busza, K. Rajagopal, and W. van der Schee,
Heavy ion collisions: The big picture, and the big questions,
Ann. Rev. Nucl. Part. Sci. {\bf 68}, 1--49 (2018).

\bibitem{Guenther_430}
J.N. Guenther,
Overview of the QCD phase diagram,
Eur. Phys. J. A {\bf 57}, 136 (2021).

\bibitem{Blokhintsev_431}
D.I. Blokhintsev, 
On the fluctuations of nuclear matter,
J. Exp. Theor. Phys. {\bf 6}, 995--999 (1958).

\bibitem{Baldin_432}
A.M. Baldin, 
Physics of relativistic nuclei,
Phys. Part. Nucl. {\bf 8}, 429--477 (1977).

\bibitem{Strikman_433}
M.I. Strikman and L.L. Frankfurt, 
Probing few nucleon correlations in deuteron and nuclei in high energy scattering,
Phys. Part. Nucl. {\bf 11}, 571--629 (1980).

\bibitem{Efremov_434}
A.V. Efremov, 
Quark-parton picture of the cumulative production,
Phys. Part. Nucl. {\bf 13}, 613--634 (1982).

\bibitem{Baldin_435}
A.M. Baldin and A.A. Baldin, 
Relativistic nuclear physics: Relative $4$-velocity space, symmetries of solutions,
correlation depletion principle, similitude, intermediate asymptotics,
Phys. Part. Nucl. {\bf 29}, 577--630 (1998).

\bibitem{Shimanskiy_436}
S.S. Shimanskiy, 
in {\it Relativistic Nuclear Physics: From Hundreds of MeV to Tev}
(JINR, Dubna, 2019), pp. 179--186.

\bibitem{Kobushkin_437}
A.P. Kobushkin and V.P. Shelest, 
Problems of relativistic quark dynamics and quark structure of deuteron, 
Phys. Part. Nucl. {\bf 14}, 1146--1192 (1983).

\bibitem{Makarov_438}
M.M. Makarov, 
Dibaryon resonanses,
Phys. Part. Nucl. {\bf 15}, 941--981 (1984).

\bibitem{Burov_439}
V.V. Burov, V.K. Lukyanov, and A.I. Titov, 
Multiquark systems in nuclear processes,
Phys. Part. Nucl. {\bf 15}, 1249--1295 (1984).

\bibitem{Baldin_440}
A.M. Baldin, R.G. Nazmitdinov, A.V. Chizhov, A.S. Shumovsky, and V.I. Yukalov, 
Coexistence of a hadron phase and a hexaquark phase in nuclear matter,
Dokl. Phys. {\bf 29}, 952--954 (1984).

\bibitem{Baldin_441}
A.M. Baldin, R.G. Nazmitdinov, A.V. Chizhov, A.S. Shumovsky, and V.I. Yukalov, 
On heterogeneous states in nuclear matter,
in {\it Multiquark Interactions and Quantum Chromodynamics},
Ed. by A.M. Baldin (JINR, Dubna, 1984), pp. 531--543.

\bibitem{Baldin_442}
A.M. Baldin, A.S. Shumovsky, and V.I. Yukalov, 
{\it Statistical Methods of Describing Quark Degrees of Freedom} 
(JINR, Dubna, 1985).

\bibitem{Baldin_443}
A.M. Baldin, A.S. Shumovsky, and V.I. Yukalov, 
Quark matter as a statistical  system,
Phys. Many--Part. Syst. {\bf 10}, 10--18 (1986).

\bibitem{Chizhov_444}
A.V. Chizhov, R.G. Nazmitdinov, A.S. Shumovsky, and V.I. Yukalov,
Statistical model of coexisting multiquark clusters,
Nucl. Phys. A {\bf 449}, 660--672 (1986).

\bibitem{Yukalov_445}
V.I. Yukalov,
Conditions for nuclear matter lasers,
Laser Phys. {\bf 8}, 1249--1256 (1998).

\bibitem{Yukalov_446}
V.I. Yukalov and E.P. Yukalova,
Equation of state in quantum chromodynamics,
in {\it Relativistic Nuclear Physics and Quantum Chromodynamics},
Ed. by A.M. Baldin and V.V. Burov 
(JINR, Dubna, 2000), Vol. 2, pp. 238--245.

\bibitem{Yukalov_447}
V.I. Yukalov and E.P. Yukalova,
Do we understand what is deconfinement?
in {\it Relativistic Nuclear Physics and Quantum Chromodynamics},
Ed. by A.M. Baldin, V.V. Burov, and A.I. Malakhov 
(JINR, Dubna, 2001), Vol. 1, pp. 109--126.

\bibitem{Yukalov_448}
V.I. Yukalov and E.P. Yukalova,
Coherent nuclear radiation,
Phys. Part. Nucl. {\bf 35}, 348--382 (2004).

\bibitem{Yukalov_449}
V.I. Yukalov and E.P. Yukalova,
Models of mixed hadron-quark-gluon matter,
Proc. Sci. (ISHEPP) {\bf 2012}, 046 (2012).

\bibitem{Ter_450}
D. Ter Haar, 
{\it Elements of Statistical Mechanics} 
(Rinehart, New York, 1956).

\bibitem{Tani_451}
S. Tani, 
Scattering involving a bound state,
Phys. Rev. {\bf 117}, 252--260 (1960).

\bibitem{Weinberg_452}
S. Weinberg, 
Elementary particle theory of composite particles,
Phys. Rev. {\bf 130}, 776--782 (1963).

\bibitem{Weinberg_453}
S. Weinberg, 
Systematic solution of multiparticle scattering problems,
Phys. Rev. B {\bf 133}, 232--256 (1964).

\bibitem{Weinberg_454}
S. Weinberg, 
Phenomenological Lagrangians,
Physica A {\bf 96}, 327--340 (1979).

\bibitem{Girardeau_455}
M.D. Girardeau, 
Second-quantization representation for systems of atoms, nuclei, and electrons,
Phys. Rev. Lett. {\bf 27}, 1416--1418 (1971).

\bibitem{Girardeau_456}
M.D. Girardeau, 
Second‐quantization representation for a nonrelativistic system of composite particles:
Generalized Tani transformation and its iterative evaluation,
J. Math. Phys. {\bf 16}, 1901--1919 (1975).

\bibitem{Girardeau_457}
M.D. Girardeau,
Second-quantization representation for a nonrelativistic system of composite particles: 
Kinematical properties of the multispecies Tani transformation,
J. Math. Phys. {\bf 19}, 2605--2613 (1978). 

\bibitem{Girardeau_458}
M.D. Girardeau and J.D. Hilbert, 
Fock-Tani representation for the quantum theory of reactive collisions,
Physica A {\bf 97}, 42--74 (1979).

\bibitem{Machleidt_459}
R. Machleidt, K. Holinde, and C. Elster, 
The Bonn meson-exchange model for the nucleon-nucleon interaction,
Phys. Rep. {\bf 149}, 1--89 (1987).

\bibitem{Yukalov_460}
V.I. Yukalov and E.P. Yukalova,
Bose-condensed atomic systems with nonlocal interaction potentials,
Laser Phys. {\bf 26}, 045501 (2016). 

\bibitem{Aoki_461}
Y. Aoki, G. Endrodi, Z. Fodor, S. Katz, and K. Szabo, 
The order of the quantum chromodynamics transition predicted by the standard model of 
particle physics,
Nature {\bf 443} 675--678 (2006).

\bibitem{Karsch_462}
F. Karsch, 
Lattice results on QCD at high temperature and non-zero baryon number density,
Prog. Part. Nucl. Phys. {\bf 62}, 503--511 (2009).

\bibitem{Borsanyi_463}
S. Borsanyi, Z. Fodor, C. Hoelbling, S.D. Katz, S. Krieg, C. Ratti, and K. Szabo, 
Is there still any $T_c$ mystery in lattice QCD? Results with physical masses in the 
continuum limit,
J. High Energy Phys. {\bf 2010}, 73 (2010).

\bibitem{Bazavov_464}
A. Bazavov et al., 
Chiral and deconfinement aspects of the QCD transition,
Phys. Rev. D 85, 054503 (2012).

\bibitem{Philipsen_465}
O. Philipsen, 
The QCD equation of state from the lattice,
Prog. Part. Nucl. Phys. {\bf 70}, 55--107 (2013).

\bibitem{Birman_466}
J.L. Birman, R.G. Nazmitdinov, and V.I. Yukalov,
Effects of symmetry breaking in finite quantum systems,
Phys. Rep. {\bf 526}, 1--91 (2013).  

\end{thebibliography}
\end{document}